\documentclass[letterpaper,11pt]{article}
\pdfoutput=1

\usepackage{hyperref}
\hypersetup{
    colorlinks=true,       
    linkcolor=blue,          
    citecolor=blue,        
    filecolor=blue,      
    urlcolor=blue           
}

\usepackage{float}
                     
\usepackage{braket,slashed,bm}
\usepackage{array,multirow}
\usepackage[normalem]{ulem}
\usepackage{xcolor,cancel,youngtab}

\usepackage[T1]{fontenc} 

\usepackage{mathrsfs}
\usepackage{booktabs}
\usepackage{adjustbox}
\usepackage{mathtools}

\usepackage{soul}

\usepackage{tikz}
\usetikzlibrary{arrows,decorations.pathmorphing,backgrounds,positioning,fit,petri,automata,shadows,calendar,mindmap,
decorations.markings,calc}

\definecolor{labelkey}{rgb}{0,0.5,0.0}

\usepackage{xspace}
\usepackage{slashed}
\usepackage{array,multirow}
\usepackage{graphicx}
\usepackage{graphics}
\usepackage{epsfig}
\usepackage{amsfonts}
\usepackage{amsmath}
\usepackage{amssymb}
\usepackage{dsfont}
\usepackage{color}
\usepackage{xcolor}
\usepackage{cite}
\usepackage{pdflscape}
\usepackage{pifont}
\usepackage{pgfplots}
\usepackage{tikz} 
\usepackage{pgfplotstable}
\usepackage{bbold}

\newcommand{\hc}{\mathrm{h.c.}}
\newcommand{\ep}{\epsilon}

\newcommand{\al}{\alpha}
\newcommand{\bt}{\beta}
\newcommand{\g}{\gamma}

\newcommand{\dt}{\delta}

\newcommand{\la}{\lambda}
\newcommand{\simu}{\sigma^{\mu\nu}}

\newcommand{\vL}{\ensuremath{\mathcal{L}}}

\newcommand{\ga}{\gamma}

\usepackage{bm}  %

\newcommand{\beq}{\begin{equation}}
\newcommand{\eeq}{\end{equation}}
\newcommand{\be}{\begin{equation}}
\newcommand{\ee}{\end{equation}}
\newcommand{\bea}{\begin{eqnarray}}
\newcommand{\eea}{\end{eqnarray}}
\newcommand{\ben}{\begin{eqnarray*}}
\newcommand{\een}{\end{eqnarray*}}

\newcommand{\boldtau}{\mbox{\boldmath $\tau$}}
\newcommand{\boldsigma}{\mbox{\boldmath $\sigma$}}
\newcommand{\boldpi}{\mbox{\boldmath $\pi$}}

\renewcommand{\vec}[1]{{\mathbf #1}} 

\newcommand{\bma}{\begin{pmatrix}}
\newcommand{\ema}{\end{pmatrix}}

\newcommand{\epc}{\epsilon_\chi}

\def\lixo#1{}

\def\slashchar#1{\setbox0=\hbox{$#1$}           
  \dimen0=\wd0                                    
  \setbox1=\hbox{/} \dimen1=\wd1                  
  \ifdim\dimen0>\dimen1                           
    \rlap{\hbox to \dimen0{\hfil/\hfil}}            
    #1                                             
  \else                                          
    \rlap{\hbox to \dimen1{\hfil$#1$\hfil}}        
    /                                           
 \fi}                                           %

\newcommand{\nnpp}{$nn \rightarrow p p\, e e$ }

\newcommand{\Or}{\mathcal O}

\newcommand{\sq}{^{2}}

\newcommand{\dslash}[1]{#1 \llap{/\kern-0.5pt}}
\newcommand{\Dslash}[1]{#1 \llap{/\kern+1.5pt}}
\newcommand{\DDslash}[1]{#1 \llap{/\kern+2.3pt}}
\newcommand{\dslashh}[1]{#1 \llap{/\kern+1pt}}

\newcommand{\nn}{\nonumber}

\newcommand{\NLDBD}{$0 \nu \beta \beta$}
\newcommand{\textoverline}[1]{$\overline{\mbox{#1}}$}

\definecolor{cadmiumgreen}{rgb}{0.0, 0.42, 0.24}
\definecolor{darkpastelgreen}{rgb}{0.01, 0.75, 0.24}
\definecolor{darkspringgreen}{rgb}{0.09, 0.45, 0.27}
\definecolor{forestgreen(web)}{rgb}{0.13, 0.55, 0.13}
\definecolor{forestgreen(traditional)}{rgb}{0.0, 0.27, 0.13}
\definecolor{cobalt}{rgb}{0.0, 0.28, 0.67}
\definecolor{darkblue}{rgb}{0.0, 0.0, 0.75}
\definecolor{darkred}{rgb}{0.55, 0.0, 0.0}
\definecolor{palatinatepurple}{rgb}{0.41, 0.16, 0.38}
\definecolor{burntorange}{rgb}{0.8, 0.33, 0.0}

\begin{document}

\begin{titlepage}

\begin{flushright}
 LA-UR-20-21376
\\
\end{flushright}

\vspace{2.0cm}

\begin{center}
{\LARGE  \bf 
Sterile neutrinos and neutrinoless double beta decay\\ 
\vspace{3mm}
in effective field theory
\vspace{3mm}
}
\vspace{2cm}

{\large \bf  W. Dekens$^{a}$, J. de Vries$^{b,c}$, K. Fuyuto$^{b,d}$, E. Mereghetti$^d$, G. Zhou$^b$} 
\vspace{0.5cm}

\vspace{0.25cm}

\vspace{0.25cm}
{\large 
$^a$ 
{\it 
Department of Physics, University of California at San Diego, La Jolla, CA 92093, USA}}

\vspace{0.25cm}
{\large 
$^b$ 
{\it 
Amherst Center for Fundamental Interactions, Department of Physics, University of Massachusetts, Amherst, MA 01003}}

\vspace{0.25cm}
{\large 
$^c$ 
{\it 
RIKEN BNL Research Center, Brookhaven National Laboratory,
Upton, New York 11973-5000, USA}}

{\large 
$^d$ 
{\it Theoretical Division, Los Alamos National Laboratory,
Los Alamos, NM 87545, USA}}

\end{center}

\vspace{0.2cm}

\begin{abstract}
\vspace{0.1cm}

We investigate neutrinoless double beta decay ($0\nu\beta\beta$) in the presence of sterile neutrinos with Majorana mass terms. 
These gauge-singlet fields are allowed to interact with Standard-Model (SM) fields via renormalizable Yukawa couplings as well as higher-dimensional gauge-invariant operators up to dimension seven in the Standard Model Effective Field Theory extended with sterile neutrinos. At the GeV scale, we use Chiral effective field theory involving sterile neutrinos to connect the operators at the level of quarks and gluons to hadronic interactions involving pions and nucleons. 
This allows us to derive an expression for \NLDBD\ rates for various isotopes in terms of phase-space factors, hadronic low-energy constants, nuclear matrix elements, the neutrino masses, and the Wilson coefficients of higher-dimensional operators.  The required hadronic low-energy constants and nuclear matrix elements depend on the neutrino masses, for which we obtain interpolation formulae grounded in QCD and chiral perturbation theory that improve existing formulae that are only valid in a small regime of neutrino masses. The resulting framework can be used directly to assess the impact of \NLDBD\ experiments on scenarios with light sterile neutrinos and should prove useful in global analyses of sterile-neutrino searches.
We perform several phenomenological studies of \NLDBD\ in the presence of sterile neutrinos with and without higher-dimensional operators. We find that non-standard interactions involving sterile neutrinos have a dramatic impact on \NLDBD\ phenomenology, and next-generation experiments can probe such interactions up to scales of  $\mathcal O(100)$ TeV.

\end{abstract}

\vfill
\end{titlepage}

\tableofcontents

\section{Introduction}
The observation of neutrino oscillations implies that neutrinos are massive particles. The absence of a right-handed neutrino and $SU(2)_L \times U(1)_Y$ gauge invariance forbid a renormalizable neutrino mass term within the context of the Standard Model (SM) of particle physics. Beyond-the-SM (BSM) physics is thus required to account for neutrino masses. A minimal solution is to extend the SM with a right-handed gauge-singlet neutrino field, often called a sterile neutrino, that can couple to the left-handed neutrino field and the Higgs field via a Yukawa interaction. Electroweak symmetry breaking then generates a neutrino mass term in the same way the SM generates mass terms for the charged fermions. However, no gauge symmetry forbids a Majorana mass term for the sterile neutrino. Adding this term, in combination with the Yukawa interaction, leads to Majorana mass eigenstates and the violation of Lepton number ($L)$ by two units. Such scenarios involving sterile neutrinos can affect neutrinoless double beta decay (\NLDBD) experiments, which are excellent probes of lepton number violation (LNV). Current experimental limits on \NLDBD\ half lives are at the level of $10^{26}$ years \cite{Arnaboldi:2002te,Umehara:2008ru,Barabash:2010bd,Gando:2012zm,Agostini:2013mzu,Albert:2014awa,Andringa:2015tza,Arnold:2015wpy,Arnold:2016ezh,KamLAND-Zen:2016pfg,Elliott:2016ble,Arnold:2016qyg,Arnold:2016bed,Agostini:2017iyd,Aalseth:2017btx, Albert:2017owj,Alduino:2017ehq,Agostini:2018tnm, Azzolini:2018dyb,Arnold:2018tmo,Adams:2019jhp,Alvis:2019sil,CANDLES_TAUP2019,Agostini:2019hzm,Azzolini:2019tta,Alenkov:2019jis,Anton:2019wmi}
and next-generation ton-scale experiments aim for one or two order-of-magnitude improvements \cite{Iida:2016vfi,Abgrall:2017syy,Patrick:2017eso,Salvio:2019agg,Adams:2018nek,Paton:2019kgy,Albert:2017hjq,Gomez-Cadenas:2019sfa,Han:2017fol,CUPIDInterestGroup:2019inu}. 

If sterile neutrinos are heavy  with respect to the electroweak scale $v\simeq 246$ GeV, they can be integrated out and their contributions to LNV processes can be described by local gauge-invariant effective operators that appear in the SM effective field theory (SMEFT). LNV operators have odd dimension \cite{Kobach:2016ami} and start at dimension five. The single dimension-five operator, the Weinberg operator \cite{Weinberg:1979sa}, provides, after electroweak symmetry breaking (EWSB), the first contribution to the neutrino Majorana mass. 
In the well-known  ``type-I seesaw'' scenario \cite{Minkowski:1977sc,GellMann:1980vs,Mohapatra:1980yp}, the  Weinberg operator originates from integrating out a heavy right-handed neutrino. Assuming Yukawa couplings of $\mathcal O(1)$ indicates a scale of BSM physics around $\Lambda \sim 10^{15}$ GeV. Higher-dimensional operators are then greatly suppressed by additional powers of $v/\Lambda \simeq 10^{-13}$ and thus negligible. 

This is not necessarily the end of the story. In several BSM scenarios, the dimension-five operator is forbidden by imposing additional symmetries or is suppressed by small  couplings and higher-dimensional operators can become competitive or even dominant.  These scenarios include radiative neutrino models \cite{Zee:1980ai, Zee:1985id, Babu:1988ki, Babu:1988ig, Babu:1988wk, Babu:2001ex}, 
where the Weinberg operator is only generated at loop level, models with flavor symmetries, reviewed for example in Ref.\ \cite{Meloni:2017cig}, and models such as the left-right symmetric model, 
in which the light neutrino mass is proportional to small Yukawa couplings \cite{Mohapatra:1980yp}. 
In recent work, we studied \NLDBD\ arising from such higher-dimensional operators up to dimension nine \cite{Cirigliano:2017djv,Cirigliano:2018yza}, see also Ref.~\cite{Pas:1999fc,Pas:2000vn,Graf:2018ozy}. Starting from the gauge-invariant SMEFT operators at the scale $\Lambda$, nuclear \NLDBD\ rates are calculated in a systematic way. In a first step, the SMEFT operators are evolved to the electroweak scale where heavy SM particles (top, Higgs, W, Z) are integrated out of the EFT. The resulting operators are evolved to the GeV scale, after which they are matched to LNV hadronic operators in chiral perturbation theory ($\chi$PT). The $\chi$PT Lagrangian is used to calculate the \nnpp\ transition operator, which, once inserted into many-body nuclear wave functions, gives rise to \NLDBD\ of atomic nuclei. The final result is a so-called `Master formula' that relates SMEFT operators to  \NLDBD\ decay rates in a systematic expansion in $(v/\Lambda)^\alpha (\Lambda_\chi/v)^\beta(m_\pi/\Lambda_\chi)^\gamma$, where $m_\pi$ is the pion mass and $\Lambda_\chi \simeq 1$ GeV the chiral-symmetry-breaking scale, and $\alpha,\,\beta,\,\gamma$ exponents that depend on the original LNV source. The formula expresses \NLDBD\ rates in terms of a set of phase-space factors, nuclear matrix elements, hadronic low-energy constants, QCD evolution factors, and the original LNV Wilson coefficients. With this formula, any BSM model for which the SMEFT framework is applicable ($\Lambda \gg v$ and no light BSM degrees of freedom) can be directly connected to  \NLDBD\ rates.

In this work, we extend the above-sketched framework to an important class of BSM scenarios: models with additional light sterile neutrinos. Such models have been considered in light of low-scale leptogenesis \cite{Ghiglieri:2017gjz,Hernandez:2016kel,Akhmedov:1998qx,Asaka:2005an,Asaka:2005pn,Shaposhnikov:2006nn,Canetti:2012vf}, the possibility of sterile neutrinos as a dark matter candidate \cite{Asaka:2005an,Asaka:2005pn,Shaposhnikov:2006nn,Canetti:2012vf,Boyarsky:2018tvu,Adhikari:2016bei}, and to account for anomalies in neutrino-oscillation experiments \cite{Boser:2019rta}. More generally, the presence of neutrino masses hints towards the existence of sterile neutrinos but not towards a specific mass scale. As such, it is important to extend the framework developed in Refs.~\cite{Cirigliano:2017djv,Cirigliano:2018yza} to include the option of light sterile neutrinos and allow for non-standard interactions that could originate at scales not too far from the EW scale, $\Lambda\sim 1$-$100$ TeV. 
\NLDBD\ in presence of light sterile neutrinos is not a new topic and has been investigated extensively in the literature, see e.g. Refs.~\cite{Blennow:2010th,Mitra:2011qr,Li:2011ss,deGouvea:2011zz,Barea:2015zfa,Giunti:2015kza,Asaka:2005pn,Asaka:2011pb,Asaka:2013jfa,Asaka:2016zib}. Here we wish to go beyond these studies in several directions. 
First of all, we perform a systematic study in the framework of the sterile-neutrino-extended SMEFT \cite{delAguila:2008ir,Cirigliano:2012ab,Liao:2016qyd}. We extend the SM not only with sterile neutrinos and the usual renormalizable interactions with SM fields, but we also include higher-dimensional operators 
arising from integrating out non-neutrino states that are assumed to be heavy compared to the electroweak scale. This is relevant to describe a vast class of models,
from left-right symmetric models \cite{Pati:1974yy, Mohapatra:1974hk, Senjanovic:1975rk}, where sterile neutrinos interact with heavy $SU(2)_R$ gauge bosons, to leptoquark models \cite{Perez:2013osa,Dorsner:2016wpm}  and Grand Unified Theories \cite{Grinstein:2006cg}. 

To describe physics at the EW and lower scales, the heavy mediators can be integrated out, leading to the appearance of effective operators such as right-handed Fermi-like interactions. 
We extend the SM with a full set of dimension-six and -seven gauge-invariant operators including the light gauge-singlet sterile neutrinos, where light means a mass of order of the electroweak scale or below. Depending on the mass $m_{\nu_R}$ we proceed in different ways. For $\Lambda_\chi  < m_{\nu_R} \leq v$ we can integrate out sterile neutrinos before matching to hadronic operators. After integrating out $\nu _R$ we obtain effective dim-3, -6, -7, and -9 operators that have already been studied in Refs.~\cite{Cirigliano:2017djv,Cirigliano:2018yza}. The resulting \NLDBD\ rates can then be readily read from the Master formula in those works. 

The situation becomes more complicated for $m_{\nu_R} \lesssim \Lambda_\chi$. In this case, $\nu_R$ remains a propagating degree of freedom at hadronic scales and needs to be considered explicitly in calculations of \nnpp\ transition operators. In this paper, we derive the $0\nu\beta\beta$ transition operators induced by light sterile neutrino in the framework of chiral EFT.
In particular:
\begin{itemize}
\item We systematically construct the chiral Lagrangian in the presence of light sterile neutrinos with non-standard interactions with SM degrees of freedom. 
The Lagrangian includes terms with explicit neutrinos, couplings to nucleons and pions via vector, axial, scalar, pseudoscalar, and tensor currents. In addition, 
the Lagrangian contains LNV operators induced by the exchange of virtual sterile neutrinos, which in several cases contribute to the transition operator at leading order.
We organize these interactions in the chiral EFT power counting, and study the neutrino mass dependence of the associated low-energy constants (LECs)
\item We derive the transition operators in a consistent power counting, which guarantees that LNV scattering amplitudes are properly renormalized. 
We further identify the set of nuclear matrix elements (NMEs) required to calculate the $0\nu\beta\beta$ half-life.
\item The NMEs and the LECs in the chiral Lagrangian depend on the mass of the sterile neutrinos. We derive effective interpolation formulae grounded in QCD and $\chi$PT
which allow us to smoothly interpolate between the $m_{\nu_R} \ll \Lambda_\chi$ and $m_{\nu_R} \gg \Lambda_\chi$ regimes. These formulae can be systematically improved 
by calculating pion, nucleon, and two-nucleon LNV matrix elements with nonperturbative methods for different neutrino masses.
\item We address sources of theoretical uncertainties on the $0\nu\beta\beta$ half-lives. In addition to the large uncertainty
on the NMEs, the often-neglected uncertainties of the LECs, which originate when matching the EFT at the quark/gluon level to Chiral EFT,  is very significant.
We estimate this uncertainty by conservatively varying the unknown LECs.
\end{itemize}
Our main result is an extension of the master formula obtained in Refs.~\cite{Cirigliano:2017djv,Cirigliano:2018yza} that now includes the contributions from light sterile neutrinos.  As we provide a direct matching to the UV scale, the applied framework can be matched to any BSM scenario involving sterile neutrinos and be readily connected to other probes of sterile neutrinos such as LHC searches, oscillation experiments, and meson decays \cite{Alcaide:2019pnf,Butterworth:2019iff}. In particular, our results can be used in models where sterile neutrinos play a role as dark matter or in producing the universal matter/antimatter asymmetry via leptogenesis. To illustrate the use of the framework, we end by studying several simple scenarios involving sterile neutrinos and the associated \NLDBD\ phenomenology.

The organization of the paper is as follows. In Sect.~\ref{nuSMEFT} we introduce the sterile-neutrino-extended SMEFT framework and discuss its evolution to the GeV scale. We discuss the matching to the low-energy EFT where heavy SM fields are integrated out, and the effects of integrating out sterile neutrinos with masses between the GeV and electroweak scale. In Sect.~\ref{ChPTneut} we match the operators at the quark level to the hadronic level using $\chi$PT, the low-energy EFT of QCD. We discuss the hadronic input required to describe LNV processes at low-energies. In Sect.~\ref{potentials} we derive the resulting \nnpp\ transition operators by considering soft- and hard-neutrino exchange between nucleons. In Sect.~\ref{MasterFormula} we present our formulae for \NLDBD\ decay rates as a function of phase-space factors, hadronic low-energy constants, nuclear matrix elements, the neutrino mass eigenvalues, and the Wilson coefficients of higher-dimensional operators. The nuclear matrix elements and their neutrino-mass dependence are discussed in Sect.~\ref{sectionNME}. The neutrino-mass dependence of hadronic low-energy constants and so-called subamplitudes are studied in Sect.~\ref{interLEC}. In Sect.~\ref{pheno} we illustrate some applications of the developed framework by considering several scenarios involving light sterile neutrinos. We summarize and conclude in Sect.~\ref{conclusions}. Several appendices are devoted to technical issues. 

\section{The Lagrangian in the Standard Model Effective Field Theory}\label{nuSMEFT}
We consider a Lagrangian at the scale of BSM physics $\Lambda \gg v$ that consists of the SM Lagrangian supplemented by a right-handed gauge-singlet neutrino and higher-dimensional operators. In this work, we consider operators up to dimension seven. To be precise, when discussing gauge-invariant operators in the SMEFT, we follow Ref.~\cite{Cirigliano:2018yza} and denote their dimensions by \textoverline{dim-n}  with $n=5,6,7$. After EWSB, the EFT operators are only $SU(3)_c\times U(1)_{\rm em}$ invariant and we refer to them, without the overline, as dim-n operators where $n=3,6,7,9$. \textoverline{dim-9} LNV operators play an important role in the phenomenology of \NLDBD\, but the relevant operators do not involve neutrinos. Their contribution has been studied in detail in Ref.~\cite{Cirigliano:2018yza} and is not affected by the inclusion of light sterile neutrinos. The Lagrangian we consider is then
\begin{eqnarray}\label{eq:smeft}
\mathcal L &=&  \mathcal L_{SM} - \left[ \frac{1}{2} \bar \nu^c_{R} \,\bar M_R \nu_{R} +\bar L \tilde H Y_\nu \nu_R + \rm{h.c.}\right]\nn \\
&&+  \mathcal L^{(\bar 5)}_{\nu_L}+  \mathcal L^{(\bar 5)}_{\nu_R}+  \mathcal L^{(\bar 6)}_{\nu_L}+  \mathcal L^{(\bar 6)}_{\nu_R} +   \mathcal L^{(\bar 7)}_{\nu_L} +   \mathcal L^{(\bar 7)}_{\nu_R}\,,
\end{eqnarray}
in terms of the lepton doublet $L=(\nu_L,\, e_L)^T$, while $\tilde H = i \tau_2 H^*$ with $H$ the Higgs doublet 
\begin{equation}
H = \frac{v}{\sqrt{2}} U(x) \left(\begin{array}{c}
0 \\
1 + \frac{h(x)}{v}
\end{array} \right)\,,
\end{equation}
where $v=246$ GeV is the Higgs vacuum expectation value (vev),  $h(x)$ is the Higgs field, and $U(x)$ is a $SU(2)$ matrix encoding the Goldstone modes. $\nu_{R}$ is a column vector of $n$ right-handed sterile neutrinos. $Y_\nu$ is a $3\times n$ matrix of Yukawa couplings and $\bar M_R$ a general symmetric complex $n \times n$ mass matrix. Without loss of generality we will work in the basis where the charged leptons $e^i_{L,R}$ and quarks $u^i_{L,R}$ and $d^i_R$ are mass eigenstates ($i=1,2,3$). This implies $d^i_L = V^{ij} d_L^{j,\,\rm mass}$, where $V$ is the CKM matrix. The relation between the mass and weak eigenstates for the neutrinos will be discussed below. We define $\Psi^c = C \bar \Psi^T$ for a field $\Psi$ in terms of the charge conjugation matrix $C = - C^{-1} = -C^T = - C^\dagger$. We use the definition for chiral fields $\Psi_{L,R}^c = (\Psi_{L,R})^c =  C \overline{\Psi_{L,R}}^T= P_{R,L} \Psi^c$, with $P_{R,L}=(1\pm\gamma_5)/2$.

We now turn to the higher-dimensional operators. In general they contain all generations of quarks, but for \NLDBD\ the most important operators are those involving the first generation of quarks. We therefore focus on operators with just $u$ and $d$ quarks, which implies that the Wilson coefficients will carry indices in lepton flavor only. We make one further truncation of the set of effective operators by focusing on interactions containing just one neutrino field. The only exception are operators that contribute to neutrino masses after EWSB and thus contain two neutrino fields. 

The \textoverline{dim-5} operators obeying the above criteria are written as
\be
 \mathcal L^{(\bar 5)}_{\nu_L} = \ep_{kl}\ep_{mn}(L_k^T\, C^{( 5)}\,CL_m )H_l H_n\,,\qquad  \mathcal L^{(\bar 5)}_{\nu_R}=- \bar \nu^c_{R} \,\bar M_R^{(5)} \nu_{R} H^\dagger H\,,
\ee
which after EWSB contribute to Majorana mass terms for active and sterile neutrinos. For $n\geq 2$ there appears a \textoverline{dim-5} transition dipole operator but it does not play an important role in \NLDBD . See e.g.~Ref.\ \cite{Caputo:2017pit} for the more general phenomenology of these \textoverline{dim-5} operators.
The number of operators grows when going to higher dimensions, but the operators that match at tree level to \NLDBD\ operators is not that large. 
In Tables~\ref{tab:O6L} and \ref{tab:O7L} we list the operators in $ \mathcal L^{(\bar 6)}_{\nu_L}$ and $\mathcal L^{(\bar 7)}_{\nu_L}$, which involve active left-handed neutrinos and were constructed in Ref.\ \cite{Grzadkowski:2010es} and \cite{Lehman:2014jma,Liao:2016hru}, respectively. 
The operators appearing in  $ \mathcal L^{(\bar 6)}_{\nu_R}$ and $\mathcal L^{(\bar 7)}_{\nu_R}$ involve sterile neutrinos and were first constructed in Ref.\ \cite{Liao:2016qyd}, they are given in Tables~\ref{tab:O6R} and \ref{tab:O7R}. We use the convention of Ref.\ \cite{Cirigliano:2018yza} for the covariant derivative. 

Operators with even dimensions are $L$-conserving (LNC) and must be combined with LNV interactions to induce $0\nu\beta\beta$. This might lead one to believe that the \textoverline{dim-6} interactions can only give small corrections to the effects of the LNV $\overline{\text{ dim-3}}$, \textoverline{-5}, and \textoverline{-7} operators. However, the LNV interactions themselves often require the insertions of small couplings. For example, the Majorana mass $\bar M_R$ contributes through the weak interaction and $Y_\nu$ couplings which can be bypassed if \textoverline{dim-6} interactions are active. As we will see explicitly in Sect.\ 8.3, this means that the LNC  $\overline{\text{ dim-6}}$ interactions can significantly affect the $0\nu\beta\beta$ rate.

{\renewcommand{\arraystretch}{1.3}\begin{table}[t!]\small
\center
\begin{tabular}{||c|c||c|c||c|c||}
\hline Class $1$& $\psi^2 H X$ & Class $2$ & $\psi^2 H^2 D$        & Class $3$ &  $\psi^4 $\\
\hline
$\mathcal O_{eW}^{(6)}$  &        $(\bar L  \sigma^{\mu\nu} e) \tau^I H W_{\mu\nu}^I$ & 
 $ \mathcal O_{H L\,3}^{(6)}$ &   $(H^\dag i\overleftrightarrow{D}^I_\mu H)(\bar L \tau^I \gamma^\mu L)$   &
$\mathcal O_{LeQu\,1}^{(6)}$&  $(\bar L ^j e) \epsilon_{jk} (\bar Q^k u)$  \\
$\mathcal O_{uW}^{(6)}$  &        $(\bar Q \sigma^{\mu\nu} u) \tau^I \widetilde H \, W_{\mu\nu}^I$ &
$\mathcal O_{H Q\,3}^{(6)}$      & $(H^\dag i\overleftrightarrow{D}^I_\mu H)(\bar Q \tau^I \gamma^\mu Q)$ &
$\mathcal O_{Lequ\,3}^{(6)}$ &    $(\bar L^j \sigma_{\mu\nu} e) \epsilon_{jk} (\bar Q^k \sigma^{\mu\nu} u)$\\
$\mathcal O_{dW}^{(6)}$ & $(\bar Q \sigma^{\mu\nu} d) \tau^I H\, W_{\mu\nu}^I$ &
$\mathcal O_{H u d}^{(6)}$   & $i(\widetilde H ^\dag D_\mu H)(\bar u \gamma^\mu d)$ &
$\mathcal O_{LQ\,3}^{(6)}$     & $(\bar L \gamma^\mu \tau^I L)(\bar Q \gamma_\mu \tau^I Q)$ \\
 &  & & & $\mathcal O_{LedQ}^{(6)}$ & $(\bar L^j e)(\bar d Q^{j})$ \\\hline
\hline 
\end{tabular}
\caption{LNC \textoverline{dim-6} operators \cite{Grzadkowski:2010es} involving active neutrinos that affect \NLDBD\ at tree level.}   \label{tab:O6L}
\end{table}}

{\renewcommand{\arraystretch}{1.3}
\begin{table}[t!]\small
\center
\begin{tabular}{||c|c||c|c||}
\hline Class $1$& $\psi^2 H^4$  & Class $5$ &  $\psi^4 D$\\
\hline
$\mathcal O^{(7)}_{LH}$  & $\ep_{ij}\ep_{mn}(L_i^TCL_m )H_j H_n (H^\dagger H)$ & $\mathcal O^{(7)}_{LL\bar d u D\,1}$&  $\ep_{ij} (\bar d \g_\mu u)(L_i^T C (D^\mu L)_j)$  \\\hline
Class $2$&  $\psi^2 H^2 D^2$ & Class $6$ & $\psi^4 H$\\\hline
$\mathcal O^{(7)}_{LHD\,1}$  & $\ep_{ij}\ep_{mn}(L_i^TC (D_\mu L)_j )H_m  (D^\mu H)_n$ & $\mathcal O^{(7)}_{L e u \bar d  H}$   & $\ep_{ij}( L_i^T C\g_\mu e)(\bar d\g^\mu  u )H_j$ \\
$\mathcal O^{(7)}_{LHD\,2}$  & $\ep_{im}\ep_{jn}(L_i^TC (D_\mu L)_j )H_m  (D^\mu H)_n$  & $\mathcal O^{(7)}_{LL Q\bar d  H\,1}$  &  $\ep_{ij}\ep_{mn}(\bar d L_i)(Q_j^TC L_m )H_n$ \\
\cline{1-2} Class $3$& $\psi^2 H^3 D$  & $\mathcal O^{(7)}_{LL Q\bar d  H\,2}$  &  $\ep_{im}\ep_{jn}(\bar d L_i)(Q_j^TC L_m )H_n$\\\cline{1-2} 
$\mathcal O^{(7)}_{LHDe}$ & $\ep_{ij}\ep_{mn}(L_i^TC \g_\mu e )H_j H_m  (D^\mu H)_n$ & $\mathcal O^{(7)}_{LL \bar Q u  H}$    &$\ep_{ij}(\bar Q_m u)(L_m^TC L_i )H_j$\\
\cline{1-2} \cline{1-2} Class $4$& $\psi^2 H^2 X $ & &\\\cline{1-2} 
$\mathcal O^{(7)}_{LHW}$  & $\ep_{ij} (\ep\tau^I)_{mn} g(L_i^TC \simu L_m ) H_j H_n W^I_{\mu\nu}$ & & \\
\hline
\end{tabular}
\caption{LNV \textoverline{dim-7} operators \cite{Lehman:2014jma} involving active neutrinos that affect \NLDBD\ at tree level.  
} \label{tab:O7L}
\end{table}}

{\renewcommand{\arraystretch}{1.3}\begin{table}[t!]\small
\center
\begin{tabular}{||c|c||c|c||}
\hline Class $1$& $\psi^2 H^3$  & Class $4$ &  $\psi^4 $\\
\hline
$\mathcal{O}^{(6)}_{L\nu H}$ & $(\bar{L}\nu_R)\tilde{H}(H^\dagger H)$ & $\mathcal{O}^{(6)}_{du\nu e}$ & $ (\bar{d}\gamma^\mu u)(\bar{\nu_R }\gamma_\mu e)$  \\ \cline{1-2}
 Class $2$&  $\psi^2 H^2 D$ &  $\mathcal{O}^{(6)}_{Qu\nu L}$ & $(\bar{Q}u)(\bar{\nu}_RL)$  \\ \cline{1-2}
$\mathcal{O}^{(6)}_{H\nu e}$ & $(\bar{\nu }_R\gamma^\mu e)({\tilde{H}}^\dagger i D_\mu H)$ & $\mathcal{O}^{(6)}_{L\nu Qd}$ & $(\bar{L}\nu_R )\epsilon(\bar{Q}d))$ \\ \cline{1-2}
Class $3$ & $\psi^2 H X$  & $\mathcal{O}^{(6)}_{LdQ\nu }$ & $(\bar{L}d)\epsilon(\bar{Q}\nu_R )$ \\ \cline{1-2}
$\mathcal{O}^{(6)}_{\nu W}$ &$(\bar{L}\sigma_{\mu\nu}\nu_R )\tau^I\tilde{H}W^{I\mu\nu}$  & &\\
\hline
\end{tabular}
\caption{LNC \textoverline{dim-6} operators \cite{Liao:2016qyd} involving a sterile neutrino that affect \NLDBD\ at tree level.
} \label{tab:O6R}
\end{table}}

{\renewcommand{\arraystretch}{1.3}\begin{table}[H]\small
\center
\begin{tabular}{||c|c||c|c||}
\hline Class $1$& $\psi^2 H^4$  & Class $5$ &  $\psi^4 D$\\
\hline
$\mathcal{O}_{\nu H}^{(7)}$ &$ (\nu^T_R C\nu_R)(H^\dagger H)^2$ & $\mathcal{O}_{du\nu eD}^{(7)} $&$(\bar{d}\gamma_\mu u)(\nu^T_R CiD_\mu e)$  \\\cline{1-2}
Class $2$&  $\psi^2 H^2 D^2$ &  $\mathcal{O}_{QL\nu uD}^{(7)}$&$(\bar{Q}\gamma_\mu L)(\nu^T_RCiD_\mu u)$\\\cline{1-2}
$\mathcal{O}_{\nu eD}^{(7)}$&$\epsilon_{ij}(\nu^T_R CD_\mu e)(H^iD^\mu H^j)$ & $\mathcal{O}_{d\nu QLD}^{(7)}$&$\epsilon_{ij}(\bar{d}\gamma_\mu \nu_R)(Q^iCiD_\mu L^j)$ \\\hline
 Class $3$& $\psi^2 H^3 D$  & Class $6$    &$\psi^4H$ \\\hline
$\mathcal{O}_{\nu L1}^{(7)}$& $\epsilon_{ij}(\nu^T_R C\gamma_\mu L^i) (iD^\mu H^j)(H^\dagger H)$ & $\mathcal{O}_{Q\nu QLH2}^{(7)}$&$\epsilon_{ij}(\bar{Q}\nu_R)(Q^iCL^j)H$\\\cline{1-2}
Class $4$& $\psi^2 H^2 X $ & $\mathcal{O}_{dL\nu uH}^{(7)}$& $\epsilon_{ij}(\bar{d}L^i)(\nu^T_RCu)\tilde{H}^j$\\\cline{1-2} 
$\mathcal{O}_{\nu eW}^{(7)}$&$(\epsilon\tau^I)_{ij}(\nu^T_RC\sigma^{\mu\nu}e)(H^iH^j)W^I_{\mu\nu}$ & $\mathcal{O}_{dQ\nu eH}^{(7)}$&$\epsilon_{ij}(\bar{d}Q^i)(\nu^T_R Ce)H^j $ \\
 & & $\mathcal{O}_{Qu\nu eH}^{(7)}$&$(\bar{Q}u)(\nu^T_R Ce)H$ \\
  & & $\mathcal{O}_{Qe\nu uH}^{(7)}$&$(\bar{Q}e)(\nu^T_R Cu)H$ \\
\hline
\end{tabular}
\caption{LNV \textoverline{dim-7} operators \cite{Liao:2016qyd} involving a sterile neutrino that affect \NLDBD\ at tree level.  
} \label{tab:O7R}
\end{table}}

\subsection{Evolution to the electroweak scale}
To evolve the higher-dimensional operators from $\mu = \Lambda$ to the electroweak scale, $\mu\simeq m_W$, we briefly discuss the required renormalization group equations (RGEs). Although for several classes of operators the complete set of RGEs have been derived, in particular for  the \textoverline{dim-6} \cite{Jenkins:2013wua,Jenkins:2013zja,Alonso:2013hga} and \textoverline{dim-7}  \cite{Liao:2016hru} operators without right-handed neutrinos, here we only consider the RGEs due to one-loop QCD effects.
Most of the operators in Tables~\ref{tab:O6L}-\ref{tab:O7R} do not undergo QCD renormalization or consist of a quark bilinear and evolve either like a scalar or tensor operator. For these operators the RGEs are rather simple
\bea\label{eq:STrge}
\frac{d C_{S,T}}{d\ln\mu} &=& \frac{\al_s}{4\pi}\gamma_{S,T}C_{S,T} \,,\qquad \gamma_S = -6 C_F\,, \qquad \g_T = 2C_F\,,\\
C_S&\in&\{C_{LeQu1}^{(6)}\,, C_{LedQ}^{(6)},\, C_{LLQuH}^{(7)} \}\,,\qquad
C_T\in\{C_{uW}^{(6)}\,, C_{dW}^{(6)},\, C_{LeQu3}^{(6)} \}\,,\nn
\eea
where $C_F = (N_c^2-1)/(2N_c)$ for $N_c=3$, the number of colors. 
In addition, there are several cases for which only combinations of couplings follow a simple RGE
\bea
C_{S,T}^{(1)} &=& \mp\frac{1}{2}\left[C_{LLQdH1}^{(7)}\pm \left(C_{LLQdH1}^{(7)}\right)^T\right]\,,\nn\\
C_{S,T}^{(2)} &=& \mp\frac{1}{4}\left[C_{LLQdH1}^{(7)}\mp \left(C_{LLQdH1}^{(7)}\right)^T\right]\mp\frac{1}{2}\left[C_{LLQdH2}^{(7)}\mp \left(C_{LLQdH2}^{(7)}\right)^T\right]\,,\nn\\
C_{S}^{(3)} &=&-\frac{1}{2} C_{LdQ\nu}^{(6)}+C_{L\nu Qd}^{(6)}\,,\qquad C_{T}^{(3)} =-\frac{1}{8} C_{LdQ\nu}^{(6)}\,,
\eea
where $C_{S,T}^{(i)}$ follow the same RGEs as $C_{S,T}$ in Eq.\ \eqref{eq:STrge}.
For some of the operators involving $\nu_R$ the linear combinations that run like a scalar or a tensor current are more involved and lead to the following RGEs
\bea\label{eq:rge7}
\frac{d \vec C_1 }{d\ln\mu} &=& \frac{\al_s}{4\pi}
\bma \g_T &0\\
\frac{\ga_S-\ga_T}{2}&\g_S&\\
\ema \cdot \vec C_1
\,,\qquad \vec C_1^T =\left( C_{d\nu QLD}^{(7)}\,,C_{dQ\nu e H}^{(7)}/y_e\right)\,,\nn\\
\frac{d \vec C_2 }{d\ln\mu} &=& \frac{\al_s}{4\pi}
\bma \g_T &0&0&0&0 \\
\frac{\ga_T-\ga_S}{2}&\g_S&0&0&0 \\
\g_S+\g_T &0&-\g_S&0&0 \\
0 &0&0&\g_T&0 \\
\frac{\g_S-\g_T}{2} &0&0&\frac{\g_T-\g_S}{2}&\g_S \ema \cdot \vec C_2
\,,\nn\\
\vec C_2^T &=&\left( C_{QL\nu u D}^{(7)}\,,\left( \bar M_R^*\right)^{-1} C_{Qu\nu L}^{(6)}\,,  C_{dL\nu u H}^{(7)}/y_d\,,C_{Qe\nu u H}^{(7)}/y_e\,,C_{Qu\nu e H}^{(7)}/y_e\right)\,,
\label{eq:uglyRG}\eea
where $y_f = \sqrt{2} m_f/v$. The $C_{QL\nu u D}^{(7)}$ and $C_{d\nu QLD}^{(7)}$ couplings induce additional operators that only contribute to neutral currents and are therefore not shown.

\subsection{Matching at the electroweak scale}\label{MatchLow}
After EWSB where the Higgs field takes its vacuum expectation value and after integrating out SM particles with masses of the order of the electroweak scale, the SMEFT operators can be matched to a new EFT. Operators in this EFT are only invariant under $SU(3)_C\times U(1)_{\mathrm{em}}$ gauge symmetries and only involve light quarks, charged leptons, neutrinos, photons, and gluons. For purposes of \NLDBD, operators containing first-generation quarks and no photons or gluons are the most interesting and we focus on this subset of operators. 

Below the electroweak scale, the Lagrangian in Eq.~\eqref{eq:smeft} can be matched to the following effective Lagrangian
\begin{eqnarray}\label{DL2lag}
\mathcal L_{} &=&  \mathcal L_{SM}-  \left[\frac{1}{2} \bar \nu^c_{L} \, M_L \nu_{L}   +  \frac{1}{2} \bar \nu^c_{R} \, M_R \nu_{R} +\bar \nu_L M_D\nu_R +\hc \right]\nn \\
&&+  \mathcal L^{(6)}_{\Delta L = 2}+  \mathcal L^{(6)}_{\Delta L = 0} +   \mathcal L^{(7)}_{\Delta L = 2} +   \mathcal L^{(7)}_{\Delta L =0 } + \mathcal L^{(9)}_{\Delta L = 2}\,,
\end{eqnarray}
where $\mathcal L_{\mathrm{SM}}$ now refers to interactions of dim-4 and lower of light SM fields, and $M_D = v Y_\nu/\sqrt{2}$. The relevant higher-dimensional operators are given by
\bea
\mathcal L^{(6)}_{\Delta L = 2}& =& \frac{2 G_F}{\sqrt{2}} \Bigg\{ 
  \bar u_L \gamma^\mu d_L \left[  \bar e_{R}  \gamma_\mu C^{(6)}_{\textrm{VL}} \,  \nu^c_{L}+ \bar e_{L}  \gamma_\mu \bar C^{(6)}_{\textrm{VL}} \,  \nu^c_{R} \right]+
  \bar u_R \gamma^\mu d_R \left[\bar e_{R}\,  \gamma_\mu  C^{(6)}_{\textrm{VR}} \,\nu_{L}^c+\bar e_{L}\,  \gamma_\mu  \bar C^{(6)}_{\textrm{VR}} \,\nu^c_{R}  \right]\nn \\
& & +
  \bar u_L  d_R \left[ \bar e_{L}\, C^{(6)}_{ \textrm{SR}}  \nu^c_{L} +\bar e_{R}\, \bar C^{(6)}_{ \textrm{SR}}  \nu^c_{R} \right]+ 
  \bar u_R  d_L \left[ \bar e_{L} \, C^{(6)}_{ \textrm{SL}}    \nu_{L}^c + \bar e_{R} \, \bar C^{(6)}_{ \textrm{SL}}    \nu_{R}^c \right]\nn \\
&&+  \bar u_L \sigma^{\mu\nu} d_R\,  \bar e_{L}  \sigma_{\mu\nu} C^{(6)}_{ \textrm{T}} \, \nu_{L}^c+  \bar u_R \sigma^{\mu\nu} d_L\,  \bar e_{R}  \sigma_{\mu\nu} \bar C^{(6)}_{ \textrm{T}} \, \nu_{R}^c
\Bigg\}  +{\rm h.c.}\label{lowenergy6_l2}\eea  
\bea
\mathcal L^{(6)}_{\Delta L = 0}& =& \frac{2 G_F}{\sqrt{2}} \Bigg\{ 
  \bar u_L \gamma^\mu d_L \left[  \bar e_{L}  \gamma_\mu c^{(6)}_{\textrm{VL}} \,  \nu_{L}+ \bar e_{R}  \gamma_\mu \bar c^{(6)}_{\textrm{VL}} \,  \nu_{R} \right]+
  \bar u_R \gamma^\mu d_R \left[\bar e_{L}\,  \gamma_\mu  c^{(6)}_{\textrm{VR}} \,\nu_{L}+\bar e_{R}\,  \gamma_\mu  \bar c^{(6)}_{\textrm{VR}} \,\nu_{R}  \right]\nn \\
& & +
  \bar u_L  d_R \left[ \bar e_{R}\, c^{(6)}_{ \textrm{SR}}  \nu_{L} +\bar e_{L}\, \bar c^{(6)}_{ \textrm{SR}}  \nu_{R} \right]+ 
  \bar u_R  d_L \left[ \bar e_{R} \, c^{(6)}_{ \textrm{SL}}    \nu_{L} + \bar e_{L} \, \bar c^{(6)}_{ \textrm{SL}}    \nu_{R} \right]\nn \\
&&+  \bar u_R \sigma^{\mu\nu} d_L\,  \bar e_{R}  \sigma_{\mu\nu} c^{(6)}_{ \textrm{T}} \, \nu_{L}+  \bar u_L \sigma^{\mu\nu} d_R\,  \bar e_{L}  \sigma_{\mu\nu} \bar c^{(6)}_{ \textrm{T}} \, \nu_{R}
\Bigg\}  +{\rm h.c.} \label{lowenergy6_l0}
\eea
\bea
\mathcal L^{(7)}_{\Delta L = 2} &=& \frac{2 G_F}{\sqrt{2} v} \Bigg\{ 
  \bar u_L \gamma^\mu d_L \left[ \bar e_{L} \, C^{(7)}_{\textrm{VL}} \,  i \overleftrightarrow{D}_\mu  \nu_{L}^c+ \bar e_{R} \, \bar C^{(7)}_{\textrm{VL}} \,  i \overleftrightarrow{D}_\mu  \nu_{R}^c \right] \nn \\
  &&+
  \bar u_R \gamma^\mu d_R \left[ \bar e_{L} \, C^{(7)}_{\textrm{VR}}\, i \overleftrightarrow{D}_\mu  \nu^c_{L}+ \bar e_{R} \, \bar C^{(7)}_{\textrm{VR}}\, i \overleftrightarrow{D}_\mu  \nu^c_{R} \right] \nn \\
&&+  \bar u_L \sigma^{\mu\nu} d_R\, \bar e_L \bar C_{\rm TR}^{(7)} \overleftarrow \partial_\mu \gamma_\nu \nu_R^c+  \bar u_R \sigma^{\mu\nu} d_L\, \bar e_L  \bar C_{\rm TL}^{(7)}\gamma_\nu\partial_\mu \nu_R^c
    \Bigg\}  +{\rm h.c.}\label{lowenergy7}\eea
    \bea 
  \mathcal L^{(7)}_{\Delta L = 0} &=& \frac{2 G_F}{\sqrt{2} v} \Bigg\{ 
  \bar u_L \gamma^\mu d_L \left[ \bar e_{R} \, c^{(7)}_{\textrm{VL}} \,  i \overleftrightarrow{D}_\mu  \nu_{L} +\bar e_{L} \, \bar c^{(7)}_{\textrm{VL}} \,  i \overleftrightarrow{D}_\mu  \nu_{R} \right] \nn \\
  &&+
  \bar u_R \gamma^\mu d_R \left[ \bar e_{R} \, c^{(7)}_{\textrm{VR}}\, i \overleftrightarrow{D}_\mu  \nu_{L}   +\bar e_{L} \, \bar c^{(7)}_{\textrm{VR}}\, i \overleftrightarrow{D}_\mu  \nu_{R}\right]\nn \\
  &&+
  \bar u_L \sigma^{\mu\nu} d_R\, \partial_\mu\left(\bar e_L c_{\rm TR}^{(7)}   \gamma_\nu \nu_L\right)+  \bar u_R \sigma^{\mu\nu} d_L\, \partial_\mu\left(\bar e_L c_{\rm TL}^{(7)}   \gamma_\nu \nu_L\right)
    \Bigg\}  +{\rm h.c.}\label{lowenergy7b}
    \eea
where $\overleftrightarrow D_\mu = D_\mu - \overleftarrow D_\mu$. 

Apart from dim-6 and -7 operators, several dim-9 operators can be induced as well. Although only a small subset is induced at the electroweak scale, almost all can be populated if a right-handed neutrino with $\Lambda_\chi<m_\nu<m_W$ is integrated out. We therefore list the complete set  
\bea \label{eq:Lag9}
\vL^{(9)}_{\Delta L =2} = \frac{1}{v^5}\sum_i\bigg[\left( C^{(9)}_{i\, \rm R}\, \bar e_R C \bar e^T_{R} + C^{(9)}_{i\, \rm L}\, \bar e_L C \bar e^T_{L} \right)  \, O_i +  C^{(9)}_i\bar e\g_\mu\g_5  C \bar e^T\, O_i^\mu\bigg]\,,
\eea
where $O_i$ and  $O_i^\mu$ are four-quark operators that are Lorentz scalars and vectors, respectively. The scalar operators have been discussed in Refs.\ \cite{Graesser:2016bpz,Prezeau:2003xn} and can be written as
\bea\label{LagSca}
O_ 1  &=&  \bar{q}_L^\alpha  \gamma_\mu \tau^+ q_L^\alpha \ \bar{q}_L^\beta  \gamma^\mu \tau^+ q_L^\beta\,, \qquad O^\prime_ 1  =  \bar{q}_R^\alpha  \gamma_\mu \tau^+ q_R^\alpha \ \bar{q}_R^\beta  \gamma^\mu \tau^+ q_R^\beta     \nn
\,\,, 
\\
O_ 2  &=&  \bar{q}_R^\alpha  \tau^+ q_L^\alpha \  \bar{q}_R^\beta  \tau^+ q_L^\beta\,, \qquad \qquad     O^\prime_ 2  =  \bar{q}_L^\alpha  \tau^+ q_R^\alpha \  \bar{q}_L^\beta  \tau^+ q_R^\beta \nn
\,\,,\\
O_ 3  &=&  \bar{q}_R^\alpha  \tau^+ q_L^\beta \  \bar{q}_R^\beta  \tau^+ q_L^\alpha\,, \qquad \qquad    O^\prime_ 3  =  \bar{q}_L^\alpha  \tau^+ q_R^\beta \  \bar{q}_L^\beta  \tau^+ q_R^\alpha \nn
\,\,,\\
O_ 4  &=&  \bar{q}_L^\alpha  \gamma_\mu \tau^+ q_L^\alpha \  \bar{q}_R^\beta  \gamma^\mu \tau^+ q_R^\beta    \nn
\,\,,\\
O_ 5  &=&  \bar{q}_L^\alpha  \gamma_\mu \tau^+ q_L^\beta \  \bar{q}_R^\beta  \gamma^\mu \tau^+ q_R^\alpha\,, \label{dim9scalar}
\eea
where $\tau^\pm = (\tau_1\pm i\tau_2)/2$ with $\tau_i$ the Pauli matrices and $\al$, $\bt$ are color indices. The $O_i'$ operators are related to the $O_i$ by parity. The vector operators take the form \cite{Graesser:2016bpz}
\bea \label{LagVec}
O_{6}^{\mu} &=& \left(\bar q_L \tau^+\g^\mu q_L\right)\left(\bar q_L \tau^+ q_R\right)\,\,,\qquad\qquad 
O_{6}^{\mu\, \prime} = \left(\bar q_R \tau^+\g^\mu q_R\right)\left(\bar q_R \tau^+ q_L\right)\,\,,
\nn\\
O_{7}^{\mu} &=& \left(\bar q_L t^a\tau^+\g^\mu q_L\right)\left(\bar q_L t^a\tau^+ q_R\right)\,\,,\qquad  \,\,
O_{7}^{\mu\, \prime} = \left(\bar q_R t^a\tau^+\g^\mu q_R\right)\left(\bar q_R t^a\tau^+ q_L\right)\,\,
,\nn\\
O_{8}^{\mu} &=& \left(\bar q_L \tau^+\g^\mu q_L\right)\left(\bar q_R \tau^+ q_L\right)\,\,,\qquad \qquad 
O_{8}^{\mu\, \prime} = \left(\bar q_R \tau^+\g^\mu q_R\right)\left(\bar q_L \tau^+ q_R\right)\,\
,\nn\\
O_{9}^{\mu } &=& \left(\bar q_L t^a\tau^+\g^\mu q_L\right)\left(\bar q_Rt^a\tau^+ q_L\right)\,\,,\qquad \,\,
O_{9}^{\mu \, \prime} = \left(\bar q_R t^a\tau^+\g^\mu q_R\right)\left(\bar q_Lt^a\tau^+ q_R\right)\,\,,\label{dim9vector}
\eea
where the second column of operators is  related to the first column by a parity transformation. Together, Eqs.\ \eqref{LagSca} and \eqref{LagVec} provide a complete basis of four-quark two-electron operators. Without loss of generality, we work in a basis without operators with tensor structures that can be replaced through Fierz relations in terms of operators involving quark bilinears with uncontracted color indices, for example  $O_T=\bar q^\alpha_R \sigma_{\mu\nu} \tau^+ q^\alpha_L\,\bar q^\beta_R \sigma^{\mu\nu} \tau^+ q^\beta_L=-8O_3-4O_2$. 

 As mentioned, we only included operators with first-generation quarks and no photons. In principle, there appear dipole-type operators containing $F_{\mu\nu}$ and operators with heavier quarks. We have kept all generations of leptons for now. To derive the matching contributions, we applied  the equations of motion of the various fields
\bea
i\slashed \partial \nu_L &=& M_D \nu_R +M_L^\dagger \nu^c_L\,,\qquad i\slashed \partial \nu_R = M_D^\dagger \nu_L +M_R^\dagger \nu^c_R\,,\nn\\
i\slashed \partial e_L &=& M_e e_R\,, \qquad i\slashed \partial u_L = m_u u_R\,,\qquad  i\slashed \partial d_L = m_d d_R\,,\nn\\
W^+_\mu&\simeq & -\frac{g}{\sqrt{2}}\frac{1}{m_W^2}\left(\bar e_L\gamma_\mu \nu_L + \bar d_L \gamma_\mu u_L\right)\,.
\eea
Here $M_e = {\rm diag} (m_e,\, m_\mu,\, m_\tau)$ which appears in the Lagrangian as $\vL_{M_e} = -\bar e_LM_ee_R+{\rm h.c.}$, while $m_{u,d}$ are the masses of the up and down quarks.
 Before giving the explicit matching conditions, it is convenient to first rotate to the mass basis of the neutrino fields. 
 
 \subsubsection{Rotation to the neutrino mass basis}\label{sec:rot}
 After EWSB the mass terms can be written as 
  \bea
 \mathcal L_m = -\frac{1}{2} \bar N^c M_\nu N +{\rm h.c.}\,,\qquad M_\nu = \bma M_L &M_D^*\\M_D^\dagger&M_R^\dagger \ema \,,
 \eea
where $N = (\nu_L,\, \nu_R^c)^T$ and $M_\nu$ is a $N\times N$ symmetric matrix (since $M_L$ and $M_R$ are symmetric matrices), with $N=3+n$. The mass matrix can be diagonalized by a single $N\times N$ unitary matrix, $U$, 
\bea\label{Mdiag}
U^T M_\nu U =m_\nu = {\rm diag}(m_1,\dots , m_{3+n})\,, \qquad N = U N_m\,.
\eea
In the general case $U$ contains $N(N+1)/2$ phases and $N(N-1)/2$ rotation angles and the $m_1,\dots ,m_N$ are real and positive. The kinetic and mass terms of the neutrinos can be written as
\bea
\mathcal L_\nu = \frac{1}{2} \bar \nu i\slashed \partial \nu -\frac{1}{2} \bar \nu^{ } m_\nu \nu\,,
\eea
in terms of the Majorana mass eigenstates $\nu = N_m +N_m^c = \nu^c$. The rotation to the mass basis is given by 
\bea
\nu_L = P_L(P U) \nu \,,\qquad \nu_L^c =P_R (P U^*) \nu\,,\nn\\
\nu_R =P_R (P_s U^*) \nu \,,\qquad \nu_R^c = P_L(P_s U) \nu\,,
\eea
where $P$ and $P_s$ are $3\times N$ and $n \times N$ projector matrices
\be
P = \begin{pmatrix}\mathcal I_{3\times 3} & 0_{3 \times n}  \end{pmatrix}\,,\qquad
P_s = \begin{pmatrix} 0_{n\times 3} & \mathcal I_{n \times n}  \end{pmatrix}\, .
\ee

The above rotations lead to the following form for the SM charged and neutral currents,
\bea
\vL  = \frac{g}{\sqrt{2}}\bar e_L \gamma^\mu P U\nu\,W^-_\mu+\frac{g}{2c_w}\bar \nu \gamma^\mu P_L \, \left(U^\dagger P^TP U\right) \nu \,Z_\mu+\dots\,.
\eea
Both currents involve the combination $PU$, which is an $3\times N$ non-unitary matrix, implying that the neutral current is no longer necessarily diagonal or universal. In general, the matrix $PU$  contains $(N-n)(N+n+1)/2$ phases. In the absence of higher-dimensional operators $N-n=3$ of these phases can be absorbed by the charged-lepton fields, leading to $(N-n)(N+n-1)/2=3(n+1)$ phases and an equal number of angles \cite{Schechter:1980gr}. In the case of $N=3$ the resulting matrix is the usual PMNS matrix. In the presence of higher-dimensional operators the same re-phasings of the electron fields can still be performed, but will result in redefinitions of the Wilson coefficients of these operators.

After rotating to the neutrino mass basis the operators in $\mathcal L^{(6)}_{\Delta L = 0}$ and $\mathcal L^{(6)}_{\Delta L = 2}$, and $\mathcal L^{(7)}_{\Delta L = 0}$ and $\mathcal L^{(7)}_{\Delta L = 2}$ can be written in rather compact form. We combine the dim-6 operators into
\bea\label{6final}
\mathcal L^{(6)}& =& \frac{2 G_F}{\sqrt{2}} \Bigg\{ 
  \bar u_L \gamma^\mu d_L \left[  \bar e_{R}  \gamma_\mu C^{(6)}_{\textrm{VLR}} \,  \nu+ \bar e_{L}  \gamma_\mu  C^{(6)}_{\textrm{VLL}} \,  \nu \right]+
  \bar u_R \gamma^\mu d_R \left[\bar e_{R}\,  \gamma_\mu  C^{(6)}_{\textrm{VRR}} \,\nu+\bar e_{L}\,  \gamma_\mu   C^{(6)}_{\textrm{VRL}} \,\nu  \right]\nn\\
& & +
  \bar u_L  d_R \left[ \bar e_{L}\, C^{(6)}_{ \textrm{SRR}}  \nu +\bar e_{R}\,  C^{(6)}_{ \textrm{SRL}}  \nu \right]+ 
  \bar u_R  d_L \left[ \bar e_{L} \, C^{(6)}_{ \textrm{SLR}}    \nu + \bar e_{R} \,  C^{(6)}_{ \textrm{SLL}}    \nu \right]\nn\\
&&+  \bar u_L \sigma^{\mu\nu} d_R\,  \bar e_{L}  \sigma_{\mu\nu} C^{(6)}_{ \textrm{TRR}} \, \nu+  \bar u_R \sigma^{\mu\nu} d_L\,  \bar e_{R}  \sigma_{\mu\nu}  C^{(6)}_{ \textrm{TLL}} \, \nu
\Bigg\}  +{\rm h.c.}\eea
while for the dim-7 operators we obtain
 \bea\label{7final}
\mathcal L^{(7)}&=& \frac{2 G_F}{\sqrt{2} v} \Bigg\{ 
  \bar u_L \gamma^\mu d_L \left[ \bar e_{L} \, C^{(7)}_{\textrm{VLR}} \,  i \overleftrightarrow{D}_\mu  \nu+\bar e_{R} \,  C^{(7)}_{\textrm{VLL}} \,  i \overleftrightarrow{D}_\mu  \nu \right] \nn\\
  &&+
  \bar u_R \gamma^\mu d_R \left[ \bar e_{L} \, C^{(7)}_{\textrm{VRR}}\, i \overleftrightarrow{D}_\mu  \nu+ \bar e_{R} \,  C^{(7)}_{\textrm{VRL}}\, i \overleftrightarrow{D}_\mu  \nu \right] \nn\\
&&+  \bar u_L \sigma^{\mu\nu} d_R\, \bar e_L  C_{\rm TR1}^{(7)} \overleftarrow D_\mu \gamma_\nu \nu
  +  \bar u_R \sigma^{\mu\nu} d_L\, \bar e_L  C_{\rm TL1}^{(7)}\gamma_\nu \partial_\mu \nu \nn\\
&&+  \bar u_L \sigma^{\mu\nu} d_R\, D_\mu\left(\bar e_L C_{\rm TR2}^{(7)}   \gamma_\nu \nu\right)
  +  \bar u_R \sigma^{\mu\nu} d_L\,D_\mu\left(\bar e_L C_{\rm TL2}^{(7)}   \gamma_\nu \nu\right)
    \Bigg\}  +{\rm h.c.}
    \eea
   The dim-9 operators contain no neutrino fields and are unaffected. The dim-6 and -7 operators are now mixtures of LNC and LNV terms, as the $\nu$ fields do not have a definite lepton number. The Wilson coefficients of the dim-6 operators are given by
\begin{align}\label{redefC6}
C_{\rm VLR}^{(6)} &= 	  C_{\rm VL}^{(6)}PU^*	+\bar c_{\rm VL}^{(6)}P_s U^*\,,\qquad 	&C_{\rm VRR}^{(6)} &= C_{\rm VR}^{(6)}PU^*+\bar c_{\rm VR}^{(6)}P_s U^*\,,\nn\\
C_{\rm VLL}^{(6)} &= \bar C_{\rm VL}^{(6)}P_sU	+     c_{\rm VL}^{(6)}P U\,,	\qquad 	&C_{\rm VRL}^{(6)} &= \bar C_{\rm VR}^{(6)}P_sU+ c_{\rm VR}^{(6)}P U\,,\nn\\
C_{\rm SLR}^{(6)} &= 	  C_{\rm SL}^{(6)}PU^*	+\bar c_{\rm SL}^{(6)}P_s U^*\,,\qquad  &C_{\rm SRR}^{(6)} &= C_{\rm SR}^{(6)}PU^*+\bar c_{\rm SR}^{(6)}P_s U^*\,,\nn\\
C_{\rm SLL}^{(6)} &=  \bar C_{\rm SL}^{(6)}P_sU+ c_{\rm SL}^{(6)}P U\,,		\qquad  &C_{\rm SRL}^{(6)} &= \bar C_{\rm SR}^{(6)}P_sU+ c_{\rm SR}^{(6)}P U\,,\nn\\
C_{\rm TLL}^{(6)} &=  \bar C_{\rm T}^{(6)}P_sU+ c_{\rm T}^{(6)}P U\,,		\qquad 	&C_{\rm TRR}^{(6)} &=  C_{\rm T}^{(6)}PU^*+ \bar c_{\rm T}^{(6)}P_s U^*\,,
\end{align}
 and those of the dim-7 operators become
 \begin{align}\label{redefC7}
C_{\rm VLL}^{(7)} &= c_{\rm VL}^{(7)}PU+\bar C_{\rm VL}^{(7)}P_s U\,,\qquad &C_{\rm VRL}^{(7)} &= c_{\rm VR}^{(7)}PU+\bar C_{\rm VR}^{(7)}P_s U\,,\nn\\
C_{\rm VLR}^{(7)} &= C_{\rm VL}^{(7)}PU^*+\bar c_{\rm VL}^{(7)}P_s U^*\,,\qquad &C_{\rm VRR}^{(7)} &=  C_{\rm VR}^{(7)}PU^*+\bar c_{\rm VR}^{(7)}P_s U^*\,,\nn\\
C_{\rm TL1}^{(7)} &= \bar C_{\rm TL}^{(7)}P_s U\,,\qquad &C_{\rm TL2}^{(7)} &= c_{\rm TL}^{(7)}PU\,,\nn\\
C_{\rm TR1}^{(7)} &= \bar C_{\rm TR}^{(7)}P_sU\,,\qquad &C_{\rm TR2}^{(7)} &= c_{\rm TR}^{(7)}PU\,.
\end{align}
The operators involving $\nu_{L,R}^c$  and $\nu_{R,L}$ contribute to the same terms in the mass basis, the only difference results from the flavor indices that are summed over, i.e.\ whether $P$ or $P_s$ appears. This notation will help simplify the calculation of the \nnpp\ transition operators.

Before discussing the matching conditions, we note that although the dim-7 operators are in principle independent, this is no longer true  in the approximation that the charged leptons carry zero momenta. In this case, derivatives on the charged-lepton fields can be dropped which allows one to neglect $C_{\rm TR1}^{(7)}$. In addition, the derivatives in the vector-like operators can now be moved onto the quark bilinears, which, after using the equations of motion, gives rise to interactions that have the same form as the scalar dim-6 operators. As a result, the contributions of the dim-7 vector operators can be captured by the following shifts of the dim-6 scalar operators,
\bea
C_{\rm SLL}^{(6)}&\to& C_{\rm SLL}^{(6)}+\frac{m_u}{v}C_{\rm VLL}^{(7)}-\frac{m_d}{v}C_{\rm VRL}^{(7)}\,,\quad 
C_{\rm SLR}^{(6)}\to C_{\rm SLR}^{(6)}+\frac{m_u}{v}C_{\rm VLR}^{(7)}-\frac{m_d}{v}C_{\rm VRR}^{(7)}\,,\nn\\
C_{\rm SRR}^{(6)}&\to& C_{\rm SRR}^{(6)}+\frac{m_u}{v}C_{\rm VRR}^{(7)}-\frac{m_d}{v}C_{\rm VLR}^{(7)}\,,\quad 
C_{\rm SRL}^{(6)}\to C_{\rm SRL}^{(6)}+\frac{m_u}{v}C_{\rm VRL}^{(7)}-\frac{m_d}{v}C_{\rm VLL}^{(7)}\,.
\label{eq:dim7V}\eea
While working within this approximation, we will often employ the above shifts to obtain the contributions from the dim-7 vector operators, instead of writing them out explicitly.
\subsubsection{Matching contributions to the neutrino mass terms}
Finally we explicitly give the matching conditions for the various effective interactions. We only consider tree-level relations. Some one-loop matching results can be found in Ref.~\cite{Chala:2020vqp}. The mass terms are given by
\bea\label{Massmatch}
M_L &=& -v^2 C^{(5)}-\frac{v^4}{2} C_{LH}\,,\nn\\
M_R &=& \bar M_R + v^2 \bar M_R^{(5)}-\frac{v^4}{2} C_{\nu H}^{(7)}\,,\nn\\
M_D &=&\frac{v}{\sqrt{2}} \left[Y_\nu -\frac{v^2}{2}C_{L\nu H}^{(6)}\right]\,,
\eea 
such that the Majorana mass of active neutrinos gets \textoverline{dim-5} and \textoverline{dim-7} contributions. Additional \textoverline{dim-7} contributions are induced at the loop level and discussed in Ref.~\cite{Cirigliano:2017djv}. The Majorana mass of sterile neutrinos gets a direct \textoverline{dim-3} contribution and higher-dimensional corrections. The Dirac mass gets a direct \textoverline{dim-4} contribution and a \textoverline{dim-6} correction. In principle, one expects the lowest-dimensional contribution to each mass term to dominate the mass, but power-counting estimates can be violated by small dimensionless numbers such as Yukawa couplings.

\subsubsection{Matching conditions for operators involving active neutrinos}
We now turn to the dim-6 operators involving active neutrinos $\nu_L$. The Wilson coefficients of LNC operators are given by 
\bea\label{match6LNC}
c_{\rm VL}^{(6)} &=& -2V_{ud}\mathbb{1}+2v^2\left[C_{LQ\,3}^{(6)}-C_{HL\,3}^{(6)}-C_{HQ\,3}^{(6)}\, \mathbb{1}\right]-\frac{4\sqrt{2}v}{g }M_e \left(C^{(6)}_{eW}\right)^\dagger\nn\\
&&-\frac{4\sqrt{2}v}{g} C^{(6)}_{\nu W} M_D^\dagger+4v^2 \left(C_{LHW}^{(7)}\right)^\dagger M_L\,,\nn\\
c_{\rm VR}^{(6)} &=&-v^2C_{Hud}^{(6)}\, \mathbb{1},\nn\\
c_{\rm SR}^{(6)} &=& v^2 \left(C_{LedQ}^{(6)}\right)^\dagger\,,\nn\\
c_{\rm SL}^{(6)} &=& v^2\left(C_{LeQu\,1}^{(6)}\right)^\dagger\,,\nn\\
c_{\rm T}^{(6)} &=& v^2\left(C_{LeQu\,3}^{(6)}\right)^\dagger\,.
\eea
The first contribution to $c_{VL}^{(6)}$ is the SM contribution. The remaining contributions to $c_{VL}^{(6)}$ and the other couplings are from BSM interactions and can be probed in $\beta$-decay experiments \cite{Cirigliano:2013xha,Cirigliano:2012ab}. The matching conditions for LNC dim-7 operators are
\bea\label{match7LNC}
 c_{\rm VL}^{(7)} &=&\frac{4\sqrt{2}}{g}v^2 \left(C_{eW}^{(6)}\right)^\dagger\,,\nn\\
 c_{\rm VR}^{(7)} &=& 0\,,\nn\\
 c_{\rm TR}^{(7)} &=& - 4\frac{\sqrt{2} }{g}v^2C_{dW}^{(6)}\, \mathbb{1},\nn\\
  c_{\rm TL}^{(7)} &=& - 4\frac{\sqrt{2} }{g}v^2 C_{uW}^{(6)\,*}\, \mathbb{1}\,,
\eea
such that the right-handed dim-7 coupling is not generated. The tensor operators are generated by coupling to leptons through the SM charged current and are therefore diagonal in lepton flavor space. 

 The analogous conditions for LNV interactions involving $\nu_L$ are given by 
\begin{eqnarray}\label{match6LNV}
\frac{1}{v^{3}}\, C^{(6)}_{\textrm{VL},ij} & = & - \frac{i}{\sqrt{2}} V_{ud}   C_{LHDe,ji}^{(7)\,*} + 4  V_{ud} \frac{m_e}{v}  C_{LHW,ji}^{(7)\,*} -\frac{4\sqrt{2}}{gv^2}\left(M_L C_{eW}^{(6)}\right)^*_{ji} +\frac{8}{gv}\left(M_D C_{\nu e W}^{(7)\,*}\right)_{ji}
\,, \nonumber \\ 
\frac{1}{v^{3}}\,C^{(6)}_{\textrm{VR},ij} &=& \frac{1}{\sqrt{2}}   C_{Leu\bar dH,ji}^{(7)\,*} \,,  \nonumber \\
\frac{1}{v^{3}}\,C^{(6)}_{\textrm{SR},ij} & =& \frac{1}{2\sqrt{2}} \left( C^{(7)}_{LL Q \bar d H\,2,ij} - C^{(7)}_{LL Q \bar d H\,2,ji}+  C^{(7)}_{LL Q \bar{d} H\,1 ,ij} \right)^* \nn\\
&&+ \frac{V_{ud}}{2}\frac{m_d}{v} \left( C^{(7)}_{LHD\,1,ij}- C^{(7)}_{LHD\,1,ji}- C^{(7)}_{LHD\,2,ji} \right)^*
-\frac{i}{2}\frac{m_u}{v}\left( C^{(7)}_{LL \bar d u D\,1,ij}- C^{(7)}_{LL \bar d u D\,1,ji}\right)^*\,, \nonumber \\  
\frac{1}{v^{3}}\,C^{(6)}_{\textrm{SL},ij}  &=&  
 \frac{1}{\sqrt{2}}   C_{LL \bar Q u H,ij}^{(7)\,*}+\frac{1}{2v} \left[\left(C_{QL\nu uD}^{(7)}\right)^\dagger M_D^T\right]_{ij} \nn\\
 &&- \frac{V_{ud}}{2}\frac{m_u}{v} \left( C^{(7)}_{LHD\,1,ij}- C^{(7)}_{LHD\,1,ji}-    C^{(7)}_{LHD\,2,ji} \, \right)^*
+\frac{i}{2}\frac{m_d}{v}\left(  C^{(7)}_{LL \bar d u D\,1,ij}- C^{(7)}_{LL \bar d u D\,1,ji}\right)^*\,,  \nonumber \\
\frac{1}{v^{3}}\,C^{(6)}_{\textrm{T},ij } &= &  \frac{1}{8 \sqrt{2}} \left(C^{(7)}_{LL Q \bar d H\,2,ij}+ C^{(7)}_{LL Q \bar d H\,2,ji}+  C^{(7)}_{LL Q \bar d H\,1,ij}\right)^* \,,
\end{eqnarray}
for dim-6 operators and 
\bea\label{match7LNV}
\frac{1}{v^{3}}\,C^{(7)}_{\textrm{VL},ij} &=& -  \frac{V_{ud}}{2}\left(  \,C^{(7)}_{LHD\,1,ij}+ C^{(7)}_{LHD\,1,ji} +   C^{(7)}_{LHD\,2,ji} + 8  C_{LHW,ji}^{(7)} \right)^* \,, \nonumber \\ 
\frac{1}{v^{3}}\,C^{(7)}_{ \textrm{VR},ij} &=& - \frac{i}{2}  \, \left( C^{(7)}_{LL \bar d u D\,1,ij}+ C^{(7)}_{LL \bar d u D\,1,ji}\right)^*\, ,
\eea
for dim-7 operators. The expressions for dim-6 and -7 LNV operators were obtained earlier in Ref.~\cite{Cirigliano:2017djv}, with the exception of the \textoverline{dim-6} contribution proportional to $M^T_D$ and $M_L$. 
Note that the contribution proportional to $ M_L C_{eW}^{(6)}$ scales as $\Lambda^{-3}$ as $M_L = \mathcal O(v^2/\Lambda)$ as given in Eq.~\eqref{Massmatch}, so that all terms scale as  $\Lambda^{-3}$.

\subsubsection{Matching conditions for operators involving sterile neutrinos}
In analogous fashion we obtain the Wilson coefficient of the operators involving sterile neutrinos $\nu_R$. For the LNC dim-6 operators we find
\bea\label{match6LNCsterile}
\bar c_{\rm VL}^{(6)} &=& \left[-v^2C_{H\nu e}^{(6)}+\frac{8v^2}{g} M_R^\dagger C_{\nu eW}^{(7)} - \frac{4\sqrt{2} v}{g}   \left(C_{\nu W}^{(6)}\right)^\dagger M_e-\frac{4\sqrt{2}v}{g}M_D^\dagger C_{eW}^{(6)} \right]^\dagger\,,\nn\\
\bar c_{\rm VR}^{(6)} &=& v^2\left(C_{du\nu e}^{(6)}\right)^\dagger\,,\nn\\
\bar c_{\rm SR}^{(6)}&=& -v^2C_{L\nu Qd}^{(6)}+\frac{v^2}{2} C_{LdQ\nu }^{(6)}\,,\nn\\
\bar c_{\rm SL}^{(6)}&=& v^2\left(C_{Qu\nu L}^{(6)}\right)^\dagger+\frac{v^2}{2} \left(C_{QL\nu uD}^{(7)}\right)^\dagger M_R\,,\nn\\
\bar c_{\rm T}^{(6)} &=& \frac{v^2}{8} C_{LdQ\nu }^{(6)}\,,
\eea
and for the LNC dim-7 operators
\bea\label{match7LNCsterile}
\bar c_{\rm  VL}^{(7)} &=& \frac{4\sqrt{2} v^2}{g} C_{\nu W}^{(6)}\,,\nn\\
\qquad \bar c_{\rm  VR}^{(7)} &=& 0\,.
\eea
Analogous to Eq.~\eqref{match7LNC} the right-handed coupling is not induced, while dim-7 tensor couplings are not generated for sterile neutrinos either. In the case of the active neutrinos, such tensor couplings arise from the product of a \textoverline{dim-6} operator involving quark fields and the SM weak interaction, the  latter of which only couples to active neutrinos. 

The dim-6 LNV conditions are given by
\bea
\bar C_{\rm VL}^{(6)} &=&-\frac{4\sqrt{2} v}{g}C_{\nu W}^{(6)} M_R^\dagger+\frac{v^3}{\sqrt{2}} C_{\nu L1}^{(7)}+ 8 \frac{v^2}{g} \left(C_{\nu eW}^{(7)}\right)^\dagger M_e\nn\\
&&+\left(\frac{v}{\sqrt{2}}\right)^3 \left(C_{Q\nu QLH2}^{(7)}\right)^\dagger+4v^2 \left(C_{LHW}^{(7)}\right)^\dagger M_D^*\,,\nn\\
\bar C_{\rm VR}^{(6)} &=& -\frac{v^2}{2} m_d \left(C_{QL\nu uD}^{(7)}\right)^\dagger+\left(\frac{v}{\sqrt{2}}\right)^3 \left(C_{dL\nu uH}^{(7)}\right)^\dagger\,,\nn\\
\bar C_{\rm SR}^{(6)} &= & \left[\frac{v^3}{\sqrt{2}} C_{dQ\nu eH}^{(7)} +\frac{v^2}{2} M_e^\dagger C_{d\nu QLD}^{(7)} -\frac{v^2}{2} m_d C_{\nu eD}^{(7)}-\frac{v^2}{2} m_u C_{du\nu eD}^{(7)}\right]^\dagger\,,\nn\\
\bar C_{\rm SL}^{(6)} &=& \left[\frac{v^3}{\sqrt{2}} C_{Qu\nu eH}^{(7)}-\left(\frac{v}{\sqrt{2}}\right)^3 C_{Qe\nu uH}^{(7)} + \frac{v^2}{2} m_d C_{du\nu eD}^{(7)}+\frac{v^2}{2}  C_{QL\nu uD}^{(7)}M_e +\frac{v^2}{2} m_u C_{\nu eD}^{(7)}\right]^\dagger\,,\nn\\
\bar C_{\rm T}^{(6)} &=& -\frac{v^3}{8\sqrt{2}} \left(C_{Qe\nu uH}^{(7)}\right)^\dagger + \frac{v^2}{8} M_e^\dagger \left(C_{QL\nu uD}^{(7)}\right)^\dagger\,.
\eea
All contributions scale as $\Lambda^{-3}$ except for the first contribution to $\bar C_{\rm VL}^{(6)}$ which scales as $\Lambda^{-2}$ and is proportional to a LNC \textoverline{dim-6} coefficient, while the LNV source is the Majorana mass of the sterile neutrino. 

Finally the dim-7 LNV Wilson coefficients become
\bea
\frac{1}{v^3}\bar C_{\rm VL}^{(7)} &=&-\frac{1}{2} \left(C_{\nu eD}^{(7)}\right)^\dagger-\frac{8}{g}\left(C_{\nu eW}^{(7)}\right)^\dagger\,,\nn\\
\frac{1}{v^3}\bar C_{\rm VR}^{(7)} &=& \frac{1}{2} \left(C_{du\nu eD}^{(7)}\right)^\dagger\,,\nn\\
\frac{1}{v^3}\bar C_{\rm TL}^{(7)} &=& -\frac{1}{2} \left(C_{QL\nu uD}^{(7)}\right)^\dagger\,,\nn\\
 \frac{1}{v^3}\bar C_{\rm TR}^{(7)} &=& \frac{1}{2} \left(C_{d\nu QLD}^{(7)}\right)^\dagger\,.
\eea

\subsubsection{Matching conditions for dim-9 operators without neutrinos}
The matching conditions for the dim-9 operators can be taken from Ref.~\cite{Cirigliano:2017djv}
\bea\label{match9}
\frac{1}{v^{3}}\,C^{(9)}_{1,ij} &=& - 2 V_{ud}^2 \left(   C^{(7)}_{LHD\, 1,ij}+ 4 \mathcal C_{LHW,ij}\right)^*\,, \nn\\
\frac{1}{v^{3}}\,C^{(9)}_{4,ij} &=& - 2 i V_{ud} \,    C^{(7)*}_{LL \bar d u D,ij}\,,\nn\\
 \frac{1}{v^{3}}\,C^{(9)}_{5,ij}&=&0\, .
\eea
Other dim-9 operators are induced from \textoverline{dim-9} contributions as discussed in Ref.~\cite{Cirigliano:2018yza}. 

\subsection{Evolution to the QCD scale}
Below the electroweak scale we again considering the one-loop QCD running of the operators in Eqs.\ \eqref{6final}, \eqref{7final}, and \eqref{eq:Lag9}. The dimension-six and -seven couplings evolve like scalar or tensor currents, as in Eq.\ \eqref{eq:STrge}, with
\bea
C_S\in\{C_{\rm SRR,SRL,SLR,SLL}^{(6)}\}\,,\qquad C_T\in\{C_{\rm TRR,TLL}^{(6)},\, C^{(7)}_{\rm TR1,TR2,TL1,TL2}\}\,.
\eea

For the scalar dim-9 couplings we have \cite{Cirigliano:2018yza,Buras:2000if,Buras:2001ra}
\bea\label{RGE9scalar}
\frac{d}{d\ln \mu} C^{(9)}_{1} &=& 6\left(1-\frac{1}{N_c}\right)\, \frac{\al_s}{4\pi}  C^{(9)}_{1}\,,\nn\\
\frac{d}{d\ln \mu}  \bma C^{(9)}_{2}\\C^{(9)}_{3}\ema  &=&  \frac{\al_s}{4\pi}\,\bma 8 + \frac{2}{N_c} - 6 N_c & -4 - \frac{8}{N_c} + 4 N_c \\4 - \frac{8}{N_c} &4 + \frac{2}{N_c} + 2 N_c \ema \bma C^{(9)}_{2}\\C^{(9)}_{3}\ema\, \,,\nn\\
\frac{d}{d\ln \mu}  \bma C^{(9)}_{4}\\C^{(9)}_{5}\ema  &=&  \frac{\al_s}{4\pi}\,\bma 6/N_c&0\\-6&-12 C_F\ema \bma C^{(9)}_{4}\\C^{(9)}_{5}\ema\,.
\eea
The RGEs do not depend on the lepton chirality, and we therefore omitted the subscripts $L$, $R$ in Eq.\ \eqref{RGE9scalar}.
The equations for the $C^{(9)\prime}_{1,2,3}$ coefficients are equivalent to those in Eq.~\eqref{RGE9scalar}, while the RGEs for the vector operators are given by
\bea\label{RGE9vector}
\frac{d}{d\ln \mu}  \bma C^{(9)}_{6}\\C^{(9)}_{7}\ema  &=&  \frac{\al_s}{4\pi}\,\bma - 2 C_F \frac{3 N_c -4}{N_c} &  2C_F\frac{(N_c+2)(N_c-1)}{N_c\sq} \\ 
4 \frac{N_c -2}{N_c}&  \frac{4 -  N_c + 2 N_c^2 + N_c^3}{N_c^2} \ema \bma C^{(9)}_{6}\\C^{(9)}_{7}\ema\, \,.
\eea

\subsubsection{Integrating out sterile neutrinos with $\Lambda_\chi  < m_{\nu} \leq v$}\label{GeVneutrinos}
In case one or more neutrinos have masses in the range $\Lambda_\chi  < m_{\nu } \leq v$, we should integrate them out before matching onto chiral perturbation theory. We can do so by writing the Lagrangian involving the heavy neutrinos as 
\bea
\vL_H=\sum_{i=N-n_H+1}^N \bigg[\frac{1}{2} \bar \nu_i i\slashed\partial \nu_i -\frac{1}{2}\bar \nu_i m_{\nu_i} \nu_i +\mathcal J_i \nu_i\bigg]\,,
\eea
where $i$ is a neutrino mass eigenstate index that runs over the heavy neutrinos, i.e.\ $\Lambda_\chi  < m_{\nu_i } \leq v$ for $i\in \{N-n_H+1,\, N\}$, with $n_H$ the number of heavy neutrinos. Furthermore, $\mathcal J_i$ incorporates the interactions of the $i$-th neutrino that are present in $\vL^{(6,7)}$~\footnote{Note that the hermitian conjugate terms in Eqs.\ \eqref{6final} and \eqref{7final} can also be written in terms of the $\nu_i$ fields instead of $\bar \nu_i$ fields, since $\bar \nu\Gamma e = \bar \nu^c\Gamma e=  \bar e^cC \Gamma^T C^{-1} \nu$, where $\Gamma $ denotes the Dirac structure.}. When integrating out the heavy neutrinos, combinations of the interactions in $\vL^{(6,7)}$ will give rise to dimension-nine operators. These can be derived by making use of the equations of motion,
\bea
\nu_i^T C \simeq \left(\mathcal J\bar m_\nu^{-1}\right)_i\,, \qquad i\in\{N-n_H+1,\, N\}\,,
\eea
where $\bar m_\nu^{-1}$ is a diagonal $N\times N$ mass matrix for the heavy neutrinos, $\bar m_\nu^{-1} = {\rm diag} (0,\dots 0,\,$ $ m_{\nu_{N-n_H+1}}^{-1},\dots m_{\nu_N}^{-1})$, and we neglected the kinetic term of the heavy neutrinos, which produces terms that are suppressed by $q/m_{\nu_i}$. Making the same approximation for the interactions in $\mathcal J$ allows us to drop the dim-7 terms. Appendix \ref{app:mnu2} discusses corrections to this approximation. We obtain
\bea
\mathcal  J_i &=&  J_i+ \bar J_i\nn\\
  J_i &\simeq& \frac{1}{v^2} \bigg[
  \bar u_L \gamma^\mu d_L \left[  \bar e_{R}  \gamma_\mu C^{(6)}_{\textrm{VLR}} + \bar e_{L}  \gamma_\mu  C^{(6)}_{\textrm{VLL}} \right]+
  \bar u_R \gamma^\mu d_R \left[\bar e_{R}\,  \gamma_\mu  C^{(6)}_{\textrm{VRR}} +\bar e_{L}\,  \gamma_\mu   C^{(6)}_{\textrm{VRL}}   \right]\nn\\
& & +
  \bar u_L  d_R \left[ \bar e_{L}\, C^{(6)}_{ \textrm{SRR}}  +\bar e_{R}\,  C^{(6)}_{ \textrm{SRL}}\right]+ 
  \bar u_R  d_L \left[ \bar e_{L} \, C^{(6)}_{ \textrm{SLR}}  + \bar e_{R} \,  C^{(6)}_{ \textrm{SLL}}  \right]\nn\\
&&+  \bar u_L \sigma^{\mu\nu} d_R\,  \bar e_{L}  \sigma_{\mu\nu} C^{(6)}_{ \textrm{TRR}} +  \bar u_R \sigma^{\mu\nu} d_L\,  \bar e_{R}  \sigma_{\mu\nu}  C^{(6)}_{ \textrm{TLL}} \bigg]_i\,,
\eea
where $\bar J_i$ are the interactions that arise from the hermitian conjugate in $\vL^{(6,7)}$, i.e.\ terms involving $e_{L,R}$ instead of $\bar e_{L,R}$. This leads to the following effective Lagrangian
\bea
\vL_H\simeq \frac{1}{2} \mathcal J \bar m_\nu^{-1} C \mathcal J^T\,,
\eea
in which the terms of interest for $0\nu\bt^-\bt^-$, namely those that have $L = -2$ (not $L = 0, 2$), are contained in the $\sim J \bar m_\nu^{-1}CJ^T$ part of the above Lagrangian. Instead, the operators with $L=2$, which give rise to $0\nu\bt^+\bt^+$ decays are contained in the $\sim \bar J \bar m_\nu^{-1}C\bar J^T$ term and are given by  the hermitian conjugate of the $L=-2$ interactions.
This procedure leads to the following matching conditions for the scalar operators
\bea\label{dim91}
C^{(9)}_{1\,R} &=& -\frac{v}{2} C_{\rm VLR}^{(6)}\bar m_\nu^{-1} C_{\rm VLR}^{(6)\, T}\,,\qquad C^{(9)\prime }_{1\,R} = -\frac{v}{2} C_{\rm VRR}^{(6)}\bar m_\nu^{-1} C_{\rm VRR}^{(6)\, T}\,,\nn\\
C^{(9)}_{2\,R} &=& \frac{v}{2} C_{\rm SLL}^{(6)}\bar m_\nu^{-1} C_{\rm SLL}^{(6)\, T}+8v C_{\rm TLL}^{(6)}\bar m_\nu^{-1} C_{\rm TLL}^{(6)\, T}\,,\qquad  C^{(9)\prime }_{2\,R} = \frac{v}{2} C_{\rm SRL}^{(6)}\bar m_\nu^{-1} C_{\rm SRL}^{(6)\, T}\,,\nn\\
C^{(9)}_{3\,R} &=&16v C_{\rm TLL}^{(6)}\bar m_\nu^{-1} C_{\rm TLL}^{(6)\, T}\,,\qquad C^{(9)\prime }_{3\,R}=0\,,\nn\\
C^{(9)}_{4\,R} &=&-\frac{v}{2}\left[ C_{\rm VRR}^{(6)}\bar m_\nu^{-1} C_{\rm VLR}^{(6)\, T}+C_{\rm VLR}^{(6)}\bar m_\nu^{-1} C_{\rm VRR}^{(6)\, T}\right]\,,\nn\\
C^{(9)}_{5\,R} &=&-\frac{v}{4}\left[ C_{\rm SRL}^{(6)}\bar m_\nu^{-1} C_{\rm SLL}^{(6)\, T}+C_{\rm SLL}^{(6)}\bar m_\nu^{-1} C_{\rm SRL}^{(6)\, T}\right]\,.
\eea
Analogous matching contributions arise for the $C^{(9)}_{i\,L}$ operators, which can be obtained from the above by replacing
\bea\label{dim92}
C^{(9)}_{i\,R} & \to &C^{(9)\prime }_{i\,L}\,, \qquad C^{(9) \prime}_{i\,R} \to C^{(9) }_{i\,L}\,,\qquad C^{(9)}_{4,5\,R} \to C^{(9) }_{4,5\,L}\, \nn\\
C_{\rm ALL}^{(6)}&\leftrightarrow & C^{(6)}_{\rm ARR}\,,\qquad C_{\rm ARL}^{(6)}\leftrightarrow C^{(6)}_{\rm ALR}\,,\qquad A\in \{S,V,T\}\,.
\eea
The matching conditions for the vector operators are given by
\bea\label{dim93}
C_6^{(9)}&=&\frac{v}{2}\left[ \frac{C_{\rm VLR}^{(6)}\bar m_\nu^{-1} C_{\rm SRR}^{(6)\, T}+C_{\rm SRR}^{(6)}\bar m_\nu^{-1} C_{\rm VLR}^{(6)\, T}}{2}-\frac{C_{\rm VLL}^{(6)}\bar m_\nu^{-1} C_{\rm SRL}^{(6)\, T}+C_{\rm SRL}^{(6)}\bar m_\nu^{-1} C_{\rm VLL}^{(6)\, T}}{2}\right]\nn\\
&&+\frac{1}{4}\left(\frac{2}{N_c}+1\right)C_7^{(9)}\,,\nn\\
C_7^{(9)}&=&8 v \left[ \frac{C_{\rm VLR}^{(6)}\bar m_\nu^{-1} C_{\rm TRR}^{(6)\, T}+C_{\rm TRR}^{(6)}\bar m_\nu^{-1} C_{\rm VLR}^{(6)\, T}}{2}\right]\,,\nn\\
C_8^{(9)}&=&\frac{v}{2}\left[ \frac{C_{\rm VLR}^{(6)}\bar m_\nu^{-1} C_{\rm SLR}^{(6)\, T}+C_{\rm SLR}^{(6)}\bar m_\nu^{-1} C_{\rm VLR}^{(6)\, T}}{2}-\frac{C_{\rm VLL}^{(6)}\bar m_\nu^{-1} C_{\rm SLL}^{(6)\, T}+C_{\rm SLL}^{(6)}\bar m_\nu^{-1} C_{\rm VLL}^{(6)\, T}}{2}\right]\nn\\
&&+\frac{1}{4}\left(\frac{2}{N_c}+1\right)C_9^{(9)}\,,\nn\\
C_9^{(9)}&=&8 v \left[ \frac{C_{\rm VLL}^{(6)}\bar m_\nu^{-1} C_{\rm TLL}^{(6)\, T}+C_{\rm TLL}^{(6)}\bar m_\nu^{-1} C_{\rm VLL}^{(6)\, T}}{2}\right]\,,\nn\\
C_6^{(9)\prime}&=&\frac{v}{2}\left[ \frac{C_{\rm VRR}^{(6)}\bar m_\nu^{-1} C_{\rm SLR}^{(6)\, T}+C_{\rm SLR}^{(6)}\bar m_\nu^{-1} C_{\rm VRR}^{(6)\, T}}{2}-\frac{C_{\rm VRL}^{(6)}\bar m_\nu^{-1} C_{\rm SLL}^{(6)\, T}+C_{\rm SLL}^{(6)}\bar m_\nu^{-1} C_{\rm VRL}^{(6)\, T}}{2}\right]\nn\\
&&+\frac{1}{4}\left(\frac{2}{N_c}+1\right)C_7^{(9)\prime}\,,\nn\\
C_7^{(9)\prime}&=&-8 v \left[ \frac{C_{\rm VRL}^{(6)}\bar m_\nu^{-1} C_{\rm TLL}^{(6)\, T}+C_{\rm TLL}^{(6)}\bar m_\nu^{-1} C_{\rm VRL}^{(6)\, T}}{2}\right]\,,\nn\\
C_8^{(9)\prime}&=&\frac{v}{2}\left[ \frac{C_{\rm VRR}^{(6)}\bar m_\nu^{-1} C_{\rm SRR}^{(6)\, T}+C_{\rm SRR}^{(6)}\bar m_\nu^{-1} C_{\rm VRR}^{(6)\, T}}{2}-\frac{C_{\rm VRL}^{(6)}\bar m_\nu^{-1} C_{\rm SRL}^{(6)\, T}+C_{\rm SRL}^{(6)}\bar m_\nu^{-1} C_{\rm VRL}^{(6)\, T}}{2}\right]\nn\\
&&+\frac{1}{4}\left(\frac{2}{N_c}+1\right)C_9^{(9)\prime}\,,\nn\\
C_9^{(9)\prime}&=&-8 v \left[ \frac{C_{\rm VRR}^{(6)}\bar m_\nu^{-1} C_{\rm TRR}^{(6)\, T}+C_{\rm TRR}^{(6)}\bar m_\nu^{-1} C_{\rm VRR}^{(6)\, T}}{2}\right]\,.
\eea

Furthermore, integrating out a heavy neutrino in principle induces four-quark two-lepton operators with an additional derivative. We discuss such terms in Appendix \ref{app:mnu2}.

\section{Chiral perturbation theory with (sterile) neutrinos}\label{ChPTneut}
Below the GeV scale, a description in terms of quarks and gluons as degrees of freedom breaks down. We therefore match to an effective description in terms of pions and nucleons. To keep the connection to QCD and the higher-dimensional operators we apply the framework of chiral perturbation theory ($\chi$PT) \cite{Weinberg:1978kz,Gasser:1983yg,Jenkins:1990jv,Bernard:1995dp}. $\chi$PT is the low-energy EFT of QCD and the $\chi$PT Lagrangian consists of all interaction among the effective low-energy degrees of freedom consistent with the chiral and space-time symmetry properties of the underlying microscopic theory. We apply two-flavored $\chi$PT in which pions appear as pseudo-Goldstone bosons of the approximate chiral symmetry of QCD. Up to small chiral-symmetry-breaking corrections, pionic interactions involve space-time derivatives. This feature allows for a perturbative expansion in $\epc=p/\Lambda_\chi$ where $p$ is the momentum scale of a process. For $p\sim m_\pi$ only a finite number of interactions need to be considered. Each interaction is proportional to a coupling constant, often called a low-energy constant (LEC), whose value cannot be obtained from symmetry considerations alone. The LECs can be fitted to experimental data, calculated using nonperturbative QCD methods such as lattice QCD, or estimated based on the power-counting scheme with naive dimensional analysis (NDA) \cite{Manohar:1983md}. The application of $\chi$PT to neutrinoless double beta decay was developed in Refs.~\cite{Prezeau:2003xn,Graesser:2016bpz,Cirigliano:2017djv,Cirigliano:2017tvr,Cirigliano:2018yza}. 

The extension of $\chi$PT to systems with more than one nucleon, as required for our purposes, is often called chiral EFT ($\chi$EFT) \cite{Hammer:2019poc} and has a more complicated power counting. The nuclear scale $p^2/2m_N$ becomes relevant in diagrams in which the intermediate state consists purely of propagating nucleons, which are enhanced with respect to the $\chi$PT counting. 
This leads to the need to resum certain classes of diagrams to all orders, which manifests in the appearance of bound states: atomic nuclei. The need to resum certain nuclear interactions also has important consequences for external currents \cite{Valderrama:2014vra}. Currents that are sandwiched between nuclear interactions that must be resummed  can appear at lower order in the power counting than expected based on NDA. For example, the exchange of a light Majorana neutrino between two nucleons leads to a  \nnpp\ transition operator whose matrix element between nuclear wave functions diverges \cite{Cirigliano:2018hja,Cirigliano:2019vdj}. This implies that a counterterm must be present, in the form a short-range $nnppee$ operator, to absorb the associated divergence. In this work, we determine the scaling of nucleon-nucleon currents by explicitly enforcing that the \nnpp\ amplitude is renormalized.

The discussion of $0\nu\beta\beta$ mediated by light neutrinos requires the consideration of two more scales: the energy of the outgoing electrons and the neutrino mass.
The electron energy is determined by the $Q$-value of the \NLDBD\ reactions. All isotopes of experimental interest have $Q$-values at the MeV scale $Q\sim \mathcal O(\rm MeV)$ which is small compared to the typical momentum exchange between nucleons $q \sim k_F \sim m_\pi$ where $k_F$ is the Fermi momentum of a nucleus. \nnpp\ transition operators that explicitly depend on the lepton momenta therefore give rise to suppressed \NLDBD\ amplitudes. To explicitly consider this suppression in the power-counting scheme we assign the counting rule $Q\sim m_e \sim m_\pi \epc^2$ \cite{Cirigliano:2018yza}. 
The electron energy is of similar size as the excitation energy of the nuclear intermediate states, which are related to violation of the so called ``closure approximation''. 
In chiral EFT, corrections to closure are associated to the propagation of ultrasoft (usoft) neutrinos, with $q^0 \sim |\vec q| \sim Q$, and, in the standard mechanism \footnote{Throughout this work, we refer to the standard mechanism as \NLDBD\ induced by three very light, $m_i \ll Q$, Majorana neutrinos.}, are suppressed by $Q/(4\pi k_F) \sim \epc^3$ \cite{Cirigliano:2017tvr} \footnote{Ref.\ \cite{Cirigliano:2017tvr} adopted the counting $Q\sim p^2/m_N \sim m_\pi \epc$, leading to the usoft contribution to be suppressed by $\epc^2$ rather than 
$\epc^3$. Since the usoft contribution to the $0\nu\beta\beta$ half-life is given explicitly in terms of the energy spectrum of the initial, intermediate, and final nuclear states
and of the zero-momentum matrix elements of the weak currents, it would be interesting to evaluate its size for $0\nu\beta\beta$ emitters of experimental interest, and assess which counting is more accurate.}. 

The mass of sterile neutrinos is a varying parameter, which can go from  $m_i \ll Q$, similar to the standard mechanism, all the way to 
$m_i \gg \Lambda_\chi$.
For almost massless neutrinos, $m_i \lesssim Q$, the $0\nu\beta\beta$ amplitude receives contributions from ``potential'' neutrinos, with $(q^0, \vec q) \sim (0, k_F)$,
from hard neutrinos, with $q_0 \sim \vec q \sim \Lambda_\chi$, 
and from the usoft regime discussed above. The usoft region is suppressed, unless the LO potential contribution cancels, as happens 
when the active neutrinos have no Majorana mass, $M_L=0$, all sterile neutrinos are light, and higher-dimensional operators are turned off. In this case,
the potential region is suppressed by $m^2_i/k_F^2$ \cite{Li:2011ss,deGouvea:2011zz}, while the usoft region is comparatively less suppressed, by $m_i^2/Q^2$, and the two become similar.
In this paper, we do not include the contributions from the ultrasoft region, which are phenomenologically important only in the narrow region $m_i  \lesssim Q$,
and very small $M_L$. We will address the intricacies of the usoft region in a forthcoming study. In several cases, the hard region gives contributions
that are comparable to those from potential neutrinos. We will consider the corresponding chiral Lagrangian in Sect.\ \ref{sec:hardNu}.

If we increase the neutrino mass to $m_i \sim m_\pi$, the usoft region disappears. In this case, soft neutrinos and pions with $q_0 \sim \vec q \sim m_\pi$ 
are explicit degrees of freedom in the theory, but they correct the $0\nu\beta\beta$ transition operator only at loop level, implying
a suppression by factors of $\epc$  \cite{Cirigliano:2017tvr}. Instead, in the region $m_i\sim \Lambda_\chi$, such loop corrections become large, $\sim m_i/\Lambda_\chi$, making this the most complicated region to describe rigorously. Finally, if $m_i \gg \Lambda_\chi$, the sterile neutrinos can be integrated out in perturbation theory, as discussed in Sect.\ 
\ref{GeVneutrinos}.

Within the framework of $\chi$EFT, extended with the additional scale considerations mentioned above, the \nnpp\ transition operators have been derived for several sources of LNV. For example, Refs.~\cite{Cirigliano:2017tvr,Cirigliano:2018hja,Cirigliano:2019vdj} calculated the transition operator in the standard mechanism up to next-to-next-to-leading order in the $\epc$ expansion. Refs.~\cite{Cirigliano:2017djv, Cirigliano:2018yza} calculated the first non-vanishing contribution for, respectively, \textoverline{dim-7} and \textoverline{dim-9} LNV operators. In this work we extend these calculations to contributions from sterile neutrinos.

\subsection{Chiral building blocks}\label{ChiralBB}
The construction of the $\chi$PT Lagrangian is well documented \cite{Gasser:1983yg,Gasser:1984gg,Bernard:1995dp}, and the application to \NLDBD\ is spelled out in Ref.~\cite{Cirigliano:2017djv}. Here we just repeat the main steps. It is convenient to write the QCD Lagrangian supplemented by the operators in Eqs.~\eqref{6final} and \eqref{7final} as
\bea
\mathcal L_{qq} &=& \bar q i \Dslash \partial q+ \bar q\bigg\{ l^\mu\gamma_\mu P_L +r^\mu \gamma_\mu P_R \nn\\
&&- (M+s+ip)P_L - (M+s-ip)P_R + t^{\mu\nu}_L\sigma_{\mu\nu} P_L  + t^{\mu\nu}_R\sigma_{\mu\nu} P_R\big\}q\,,
\eea  
with $q = (u\,d)^T$ a doublet of quark fields, and $M = \mathrm{diag}(m_u,\,m_d)$ is a diagonal matrix of the real quark masses. The external sources $s$, $p$, $l^\mu$, $r^\mu$, $t^{\mu\nu}_L$, and $t^{\mu\nu}_R$ can be read from Eqs.~\eqref{6final} and \eqref{7final}. Neglecting the SM electromagnetic and neutral weak interaction, and focusing on terms that create electrons instead of positrons, we identify
\bea\label{sources}
s+ip &=& -\frac{2 G_F}{\sqrt{2}}\left\{\tau^+\left(\bar e_L C_{\rm SLR}^{(6)} \nu + \bar e_R C_{\rm SLL}^{(6)} \nu\right)+\left(\tau^+\right)^\dagger\left(\bar e_L C_{\rm SRR}^{(6)} \nu + \bar e_R C_{\rm SRL}^{(6)} \nu\right)^\dagger\right\}\,,\nn\\
s-ip &=& \left(s+ip \right)^\dagger\,,\nn\\
l^\mu &=& \frac{2 G_F}{\sqrt{2}v}\tau^+ \bigg\{v\,\bar e_R \gamma^\mu C_{\rm VLR}^{(6)} \nu+v\,\bar e_L \gamma^\mu C_{\rm VLL}^{(6)} \nu \nn\\
&&+ \bar e_{L} \, C^{(7)}_{\textrm{VLR}} \,  i \overleftrightarrow{\partial}^\mu  \nu+\bar e_{R} \,  C^{(7)}_{\textrm{VLL}} \,  i \overleftrightarrow{\partial}^\mu  \nu \bigg\}+{\rm h.c.}\,,\nn\\
r^\mu &=& \frac{2 G_F}{\sqrt{2}v}\tau^+ \bigg\{v\,\bar e_R \gamma^\mu C_{\rm VRR}^{(6)} \nu+v\,\bar e_L \gamma^\mu C_{\rm VRL}^{(6)} \nu \nn\\
&&+ \bar e_{L} \, C^{(7)}_{\textrm{VRR}} \,  i \overleftrightarrow{\partial}^\mu  \nu+\bar e_{R} \,  C^{(7)}_{\textrm{VRL}} \,  i \overleftrightarrow{\partial}^\mu  \nu \bigg\}+{\rm h.c.}\,,\nn\\
t^{\mu\nu}_L&=&\frac{2 G_F}{\sqrt{2}v}\bigg\{\tau^+\left[v\,\bar e_R \sigma^{\mu\nu}C^{(6)}_{ \textrm{TLL}} \, \nu +\bar e_L   C_{\rm TL1}^{(7)}\gamma^\nu\partial^\mu \nu +  \partial^\mu\left(\bar e_L C_{\rm TL2}^{(7)}   \gamma^\nu \nu\right)\right]\nn\\
&&+\left(\tau^+\right)^\dagger\left[v\,\bar e_L \sigma^{\mu\nu}C^{(6)}_{ \textrm{TRR}} \, \nu +  \bar e_L  C_{\rm TR1}^{(7)} \overleftarrow \partial^\mu \gamma^\nu \nu +  \partial^\mu\left(\bar e_L C_{\rm TR2}^{(7)}   \gamma^\nu \nu\right)\right]^\dagger\bigg\}\,,\nn\\
t^{\mu\nu}_R&=&\left(t^{\mu\nu}_L\right)^\dagger\,.
\eea
The quark-level Lagrangian is formally invariant under local $SU(2)_L\times SU(2)_R$ transformations, $q_L\to Lq_L$ and $q_R\to Rq_R$  with $L$ and $R$  general $SU(2)$ matrices, provided that the spurions transform as 
\bea
l_\mu&\to &Ll_\mu L^\dagger-i \left(\partial_\mu L\right) L^\dagger\,,\qquad r_\mu\to Rr_\mu R^\dagger-i \left(\partial_\mu R\right) R^\dagger\,,\nn\\
M+s+ip&\to& R (M+s+ip) L^\dagger\,,\qquad t_{L}^{\mu\nu} \to  Rt_{L}^{\mu\nu}L^\dagger\,,\qquad t_{R}^{\mu\nu} \to  Lt_{R}^{\mu\nu}R^\dagger\,.
\eea
The chiral Lagrangian that describes the exchange of potential neutrinos is then constructed by building the most general interactions that are  invariant under these transformations.
\subsubsection{The pion sector}\label{sec:piLag}
In $\chi$PT pions are described by 
\be
U = u^2 = \mathrm{exp} \left(\frac{i \boldpi\cdot \boldtau}{F_0}\right)\,,
\ee
in terms of the Pauli matrices $\boldtau$, the pion triplet $\boldpi$, and $F_0$ is the decay constant in the chiral limit. We use $F_\pi = 92.2$ MeV for the physical pion decay constant, and, since we work at lowest order in $\chi$PT, we will use $F_\pi = F_0$. 
Under $SU(2)_L\times SU(2)_R$ transformations the pion field transforms as $U \rightarrow L U R^\dagger$. It is convenient to define a covariant derivative that transforms in the same way $D_\mu U \rightarrow L (D_\mu U)R^\dagger$ under local transformations, where
\be 
D_\mu U = \partial_\mu U - i l_\mu U + i U r_\mu\,,
\ee
where $l_\mu$ and $r_\mu$ are the external source terms given above. Quark masses explicitly break chiral symmetry and their effects are included by the spurion $\chi$ that transforms as $\chi \rightarrow L \chi R^\dagger$, and explicitly
\be
\chi = 2 B (M + s -ip)\,,
 \ee
 where $B$ is a LEC, often called the quark condensate, related to the pion mass via $m_\pi^2 = B(m_u+m_d)$. 
The LO chiral Lagrangian consists of the Lorentz- and chiral-invariant terms with the lowest number of derivatives
\be
\mathcal L_{\pi} = \frac{F_0^2}{4} \mathrm{Tr}\left[(D_\mu U)^\dagger (D^\mu U)\right]+ \frac{F_0^2}{4} \mathrm{Tr}\left[U^\dagger \chi + U \chi^\dagger\right]\,.
\ee
By expanding the $U$ field, we can immediately read off the interactions between pions, neutrinos, and electrons, that are induced by effective operators that contribute to $l^\mu$, $r^\mu$, $s$, and $p$. Contributions from the tensor sources require two additional derivatives and only appear at higher order. 
 For those sources, interactions in the pion-nucleon sector are more relevant. Interactions with more derivatives or insertions of $\chi$ also appear at higher order, but will not be necessary for our purposes.

\subsubsection{The pion-nucleon sector} \label{sec:piNLag}
We work with non-relativistic heavy-baryon nucleon fields denoted by $N=(p,\,n)^T$ characterized by the nucleon velocity $v^\mu = (1,\vec 0)$ and spin $S^\mu = (0,\,\boldsigma/2)$. Under chiral symmetry the nucleon field transforms as $N \rightarrow K N$ with $K$ an $SU(2)$ matrix, belonging to the diagonal subgroup of $SU(2)_L\times SU(2)_R$. The same matrix appears in the transformation of $u = \sqrt{U} \rightarrow LuK^\dagger = K u R^\dagger$. A nucleon covariant derivative can be defined as 
\be
\mathcal D_\mu N = (\partial_\mu + \Gamma_\mu)N\,,\qquad \Gamma_\mu = \frac{1}{2}\left[u^\dagger(\partial_\mu -il_\mu)u+u(\partial_\mu - ir_\mu)u^\dagger\right]\,,
\ee
such that $\mathcal D_\mu N \rightarrow K \mathcal D_\mu N$. It is useful to introduce two more objects with convenient symmetry properties
\bea\label{eq:defs}
u_\mu &=& -i\left[u^\dagger(\partial_\mu -il_\mu)u-u(\partial_\mu - ir_\mu)u^\dagger\right]\,,\nn\\
\chi_{\pm} &=& u^\dagger \chi u^\dagger \pm u \chi^\dagger u\,,
\eea
that transform as $X\to K XK^\dagger$, with $X\in \{\chi_\pm,\, u_\mu\}$.
Operators relevant for \NLDBD\ with the lowest number of derivatives are given by 
\begin{equation}\label{eq:3.1}
\mathcal L^{(1)}_{\pi N} =  i\bar N  v \cdot \mathcal D N + g_A \bar N S\cdot u N 
+ c_5\,\bar N \hat \chi_+ N  -2g_T \ep_{\mu\nu\al\bt}v^\al\,\bar N  S^\bt \left(u^\dagger t_R^{\mu\nu} u^\dagger + u t_L^{\mu\nu} u\right) N\,,
\end{equation} 
where $\hat \chi_+ = \chi_+ - \mathrm{Tr}(\chi_+)/2$ and $c_5$ and $g_T$ are two LECs. $c_5$ is connected to the strong proton-neutron mass splitting $(m_n -m_p)^{\mathrm{str}}$
and to the scalar charge $g_S$ via $c_5 = (m_n -m_p)^{\mathrm{str}}/(4B(m_u-m_d)) = - g_S/(4B)$ \cite{Gonzalez-Alonso:2013ura}.  $g_T$ is nowadays known from lattice QCD calculations. The numerical values of all LECs are given in Table~\ref{Tab:LECs}. 

For certain LNV sources we also require the NLO corrections. Particularly important are the contributions from the nucleon isovector magnetic moment $g_M$
and the tensor form factor $g_T^\prime$
\bea\label{eq:nucleonNLO}
\vL^{(2)}_{\pi N} &=& 
-\frac{g_M}{4m_N}\ep^{\mu\nu\al\bt}v_\al\,\bar N S_\bt f^+_{\mu\nu}N -\frac{g_T'}{m_N} v_\mu\,\bar N \left[  (u^\dagger t_R^{\mu\nu}u^\dagger + u t_L^{\mu\nu}u),\,  \mathcal D_\nu\right] N\,.\nn
\eea
The $m_N$ in the definitions in Eq.\ \eqref{eq:nucleonNLO} is conventional, and do not indicate that the LECs $g_{M}$ and $g_T^\prime$ are determined by reparameterization invariance.
At the same order in the Lagrangian, there arise recoil corrections to the axial, vector and tensor form factors
\bea
\vL^{(2)}_{\pi N,\, \rm rec} &=& \frac{1}{2m_N}\left(v^\mu v^\nu-g^{\mu\nu}\right)\left(\bar N\mathcal D_\mu \mathcal D_\nu N\right)- \frac{i g_A}{2 m_N}\bar N \{S\cdot \mathcal D, v\cdot u\}N
\\
&&- \frac{g_T}{m_N} \ep_{\mu\nu\al\bt}\, \bar N S_\bt \lbrace (u^\dagger t_R^{\mu\nu}u^\dagger + u t_L^{\mu\nu}u),\, i \mathcal D_\al \rbrace N \,.\nn
\eea
Notice however that these terms contribute to the neutrino potentials only at N$^2$LO, and we disregard them in what follows. 
Before turning towards the nucleon-nucleon sector, we first discuss the single neutron $\beta$-decay transition operator, which plays an important role in the descriptions of \NLDBD\ induced by sterile neutrinos.

\subsection{The neutron $\beta$-decay transition operator}\label{neutronbeta}
The chiral Lagrangians of the pion and pion-nucleon sector can be used to derive the $\beta$-decay amplitude of a single neutron. This amplitude provides a building block towards deriving the \NLDBD\ transition operators. Not all contributions to \NLDBD\ can be captured in this way, since LNV interactions such as $\pi\pi ee$, $\pi N ee$, and $\bar NN \bar NN ee$ operators, contribute to the \NLDBD\ transition operator without the exchange of a neutrino. The long-distance contributions from sterile neutrinos with masses below $\Lambda_\chi$, however, can be captured by combining two neutron $\beta$-decay transition operators that are derived here. 

\begin{figure}
\center
\includegraphics[width=0.7\textwidth]{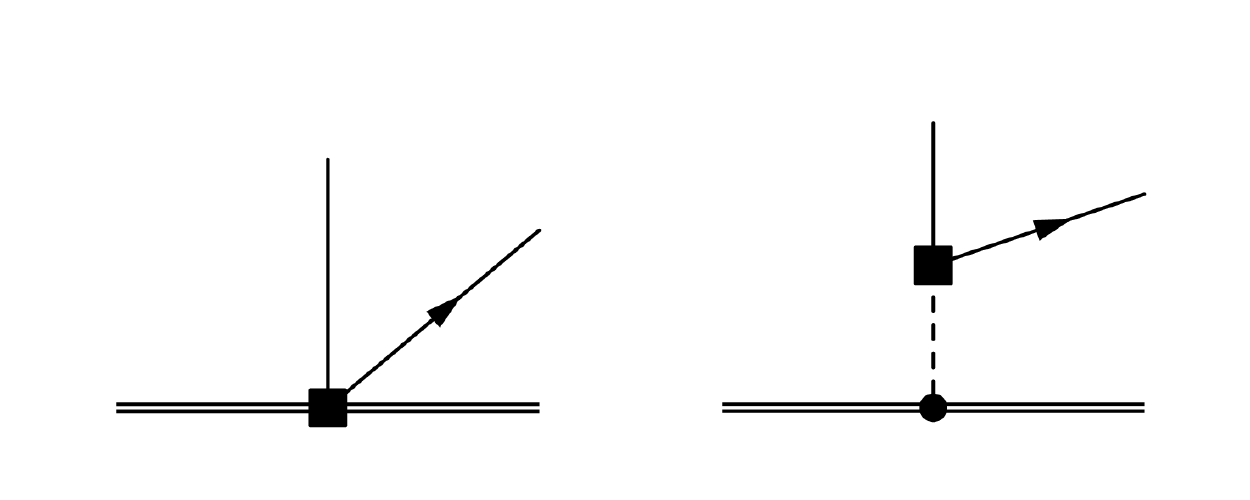}
\caption{Tree-level contributions to $n \rightarrow p e^- \nu$, in the presence of non-standard vector, axial, scalar, pseudoscalar, and tensor currents. Nucleons and pions are denoted 
by double and dashed lines, the electron by a single line with an arrow, while $\nu$, which is a Majorana mass eigenstate, by a single line (with no definite particle flow).
The insertion of the non-standard current is denoted by a square, while strong-interaction vertices by a circle. In the case of the vector, scalar, and tensor currents,  
only the first topology appears because parity forbids the couplings of the current to a single pion. Both diagrams contribute to the axial current at LO, while the pseudoscalar 
current is dominated by the pion pole in the second diagram.
}\label{neutroncurrent}
\end{figure}

At tree-level in $\chi$PT, there are two types of diagrams that contribute to $n\rightarrow p + e + \nu$ (where $\nu$ denotes a Majorana mass eigenstate) depicted in Fig.~\ref{neutroncurrent}. In the region $m_i\ll \Lambda_\chi$, loop corrections appear at next-to-next-to-leading order and will be neglected. They have been considered in Ref.~\cite{Cirigliano:2017tvr} for the case of the exchange of a light Majorana neutrino with SM couplings. The amplitude can be written in compact form 
 \bea\label{eq:currents0}
\mathcal A^{n\to pe^- \nu} = \bar N(\vec p')  \left[\frac{l_\mu+r_\mu}{2}J_V^\mu+ \frac{l_\mu-r_\mu}{2}J_A^\mu -s\,J_S +i p\, J_P +t_{R\, \mu\nu}\,J_{T_R}^{\mu\nu}+t_{L\, \mu\nu}\,J_{T_L}^{\mu\nu}\right] N(\vec p)\,,
\eea
where $N(\vec p)$ denotes a spinor of a non-relativistic nucleon field with three-momentum $\vec p$. 
The sources are given in Eq.\ \eqref{sources}, and include both LNC and LNV terms. Up to NLO in the chiral expansion we obtain 
\bea \label{eq:currents}
J^\mu_V  &=& g_V(\vec q^2) \left( v^\mu + \frac{p^\mu + p^{\prime \mu}}{2m_N} \right) + \frac{i  g_M(\vec q^2)}{m_N} \varepsilon^{\mu \nu \alpha \beta} v_\alpha S_\beta q_\nu  \,, \nn\\
J^\mu_A  &=& - g_A(\vec q^2)  \left( 2 S^\mu  - \frac{v^\mu}{2 m_N}\, 2 S \cdot (p + p^\prime) \right) + \frac{g_P(\vec q^2)}{2 m_N} 2 q^\mu \, S \cdot q\,, \nn\\
J_S &=&g_S(\vec q^2)\,, \nn\\
 J_P &=& B \frac{g_P(\vec q^2)}{m_N} S\cdot q\,,\nn\\
J^{\mu \nu}_{T_R} &=&  - 2 g_T(\vec q^2)  \varepsilon^{\mu \nu \alpha \beta} \left( v_{\alpha} + \frac{p_\alpha + p^\prime_\alpha}{2 m_N} \right) S_{\beta}  -i
  \frac{g_{T}^\prime(\vec q^2)}{2 m_N} (v^\mu q^\nu - v^\nu q^\mu) \,, \nn\\
  J^{\mu \nu}_{T_L} &=& J^{\mu \nu}_{T_R}\,,
\eea
where $p$ and $p'$ stand for the momenta of the incoming neutron and outgoing proton. $J^{\mu \nu}_{T_L}$ and $J^{\mu \nu}_{T_R}$ start to differ only at higher orders in the chiral expansion. We define  $q^\mu=(q^0,\, \vec q) = p^\mu-p^{\prime\,\mu}$, and $ \varepsilon^{\mu \nu \alpha \beta}$ is the totally antisymmetric tensor, with $\varepsilon^{0123}=+1$. We have written the currents in terms of form factors that depend on the momentum transfer $\vec q^2$. Up to the order we work, most form factors are constants with the important exception of $g_P(\vec q^2)$. 
Explicitly, we obtain
\bea \label{eq:FF}
g_V(\vec q^2) &=& g_V  \, ,\qquad g_A(\vec q^2) = g_A \, ,\qquad g_M(\vec q^2) = g_M\, ,\nn\\
g_S(\vec q^2) &=&g_S ,\qquad g_P(\vec q^2) =-\frac{2g_Am_N}{\vec q\sq+m_\pi\sq}\,,\nn\\
g_T(\vec q^2) &=& g_T \,,\qquad g_T'(\vec q^2) = g_T'\, ,
\eea
and the values of the (combinations of) LECs $g_V$, $g_A$, $g_M$, $g_S$, $g_T$, and $g_T'$ are given in Table~\ref{Tab:LECs}.  All form factors are $\mathcal O(1)$, except for $g_P(\vec q^2)$ that is enhanced by $m_N/m_\pi$. We stress that the form factors $g_P(\vec q^2)$, $g_M(\vec q^2)$, and $g'_T(\vec q^2)$ are associated with an inverted power of $m_N$ in the contributions to the hadronic currents in Eq.\ \eqref{eq:currents}. 
In practice, the NME calculations we use in Section \ref{sectionNME} include dipole form factors, that is they multiply Eq. \eqref{eq:FF} by  $(1 + \vec q^2/\Lambda^2)^{-2}$, with $\Lambda$ either the vector or axial mass, $\Lambda_V = 850$ MeV and $\Lambda_{A} = 1040$ MeV. These corrections appear at N$^{2}$LO in chiral EFT. In $0\nu\beta\beta$ candidates, these form factors shift the NME by 10\%-15\%, consistent with the chiral EFT expectation \cite{Menendez:2008jp}.

\subsection{Chiral Lagrangian induced by dimension-nine operators}\label{sec:ChiLag9}
The chiral Lagrangian induced by the dimension-nine operators in Eqs.\ \eqref{dim9scalar} and \eqref{dim9vector} was discussed in Refs.\ \cite{Prezeau:2003xn,Cirigliano:2018yza}.
These operators induce LNV couplings of  two pions, two nucleons and one pion, or four nucleons to two electrons. 
The $\pi\pi$ interactions only obtain significant contributions from the scalar operators.
Neglecting terms with more than two pions, which are only relevant at loop level or in multi-nucleon operators,   the pionic Lagrangian can be written as
\bea\label{eq:dim9pipi}
\vL_\pi 
&= &F_\pi^2\left[C_{L}^{\prime \pi\pi}\,  \partial_\mu \pi^- \partial^\mu \pi^-   
+ C_{L}^{\pi\pi}\pi^- \pi^- 
\right]
\frac{\bar e_L  C\bar e^T_L}{v^5}+(L\leftrightarrow R)\,.\eea
As we will see in the next subsection, these $\pi\pi$ interactions, as well as the $\pi N$ and $NN$ interactions discussed below, are not only induced by dim-9 operators, but also receive contributions from the exchange of hard neutrinos. In anticipation of these additional contributions, we will write the above couplings (as well as those to be introduced below) as 
\bea\label{eq:hardnu0}
C_{\bt}^{ \al} =  c_{\bt}^{\al}+\sum_{i=1}^{n_L}C_{i,\bt}^{\al}(m_i)\,,\qquad \al\in \{\prime \pi\pi,\, \pi\pi,\, \pi N,\, NN\}\,,\quad \bt \in \{L,R,V\}\,,
\eea
where we use  $ c_{\bt}^{\al}$ to denote the dim-9 contributions, while the remaining terms will be discussed in Sect.\ \ref{sec:hardNu}. The contributions from the dim-9 operators can then be written as
\bea\label{pipidim9}
c_{L}^{\prime \pi\pi} &=& \frac{5}{6} g_1^{\pi\pi}\left(C_{1L}^{(9)}+C_{1L}^{(9)\prime}\right)\,,\nn\\
c_{L}^{\pi\pi}&=&\frac{1}{2}\left[g_4^{\pi\pi}C_{4L}^{(9)}+g_5^{\pi\pi}C_{5L}^{(9)}-g_2^{\pi\pi}\left(C_{2L}^{(9)}+C_{2L}^{(9)\prime}\right)-g_3^{\pi\pi}\left(C_{3L}^{(9)}+C_{3L}^{(9)\prime}\right)\right]\,,
\eea
where the couplings with right-handed electron fields are obtained by the replacement $L\to R$.
The LECs, $g_i^{\pi\pi}$, were defined in \cite{Cirigliano:2018yza} and their sizes can be estimated using NDA
\begin{eqnarray}
g^{\pi\pi}_{1} = \mathcal O(1)\,,\qquad g_{2,3,4,5}^{\pi\pi} = \mathcal O(\Lambda_\chi^2)\,.
\end{eqnarray}
The LECs in Eq.\ \eqref{eq:dim9pipi} were computed in Ref.\ \cite{Nicholson:2018mwc}, and are found to be in agreement with these expectations. We report the values of the LECs in Table \ref{Tab:LECs}.

Pion-nucleon couplings are induced by both scalar and vector operators. For scalar operators, the $\pi N$ couplings are subleading, with the exception of the operator $O_1$.
For vector operators, they contribute to the LO $0\nu\beta\beta$ transition operator. Expanding in pion fields, the Lagrangian has the form
\bea\label{eq:dim9PiN}
\vL_{\pi N}^{}&=&\sqrt{2}g_A F_\pi \left[\bar p\, S\cdot (\partial \pi^-)n\right]
\Bigg\{\left[ C_L^{\pi N}\, \frac{\bar e_L  C\bar e^T_L}{v^5}+(L\leftrightarrow R)\right] 
  +C_V^{\pi N} v^\mu \frac{\bar e \g_\mu \g_5 C\bar e^T}{v^5} \Bigg\}\,,\nn\\
c_{L,R}^{\pi N} &=& g_1^{\pi N}\left(C_{1L,1R}^{(9)}+C_{1L,1R}^{(9)\prime}\right)\,, \qquad c_{V}^{\pi N} = g_V^{\pi N}C_V^{(9)}+\tilde g_V^{\pi N}\tilde C_V^{(9)}\,,
\eea
where $C_V^{(9)}\equiv C_{6}^{(9)}+C_{8}^{(9)}+C_{6}^{(9)\, \prime}+C_{8}^{(9)\, \prime}$,  
$\tilde{C}_V^{(9)} \equiv C_{7}^{(9)}+C_{9}^{(9)}+C_{7}^{(9)\, \prime}+C_{9}^{(9)\, \prime}$.
The LECs $g_{1,V}^{\pi N}$ and $\tilde{g}_{V}^{\pi N}$ were defined in Ref.\ \cite{Cirigliano:2018yza},
and they are $\mathcal O(1)$.

Finally, both scalar and vector operators induce nucleon-nucleon interactions. Following the definitions of Ref.\ \cite{Cirigliano:2018yza} and again expanding in pion fields, we have
\bea\label{eq:dim9NN}
\vL_{NN}^{} 
& =&  \left(\bar pn\right )\,\left(\bar pn\right ) \Bigg\{ 
\Bigg[
C_L^{NN}\, \frac{\bar e_L  C\bar e^T_L}{v^5}  +(L\leftrightarrow R) \Bigg]  +C_V^{NN}\,v^\mu \frac{\bar e \g_\mu \g_5 C\bar e^T}{v^5} \Bigg\}\,,\nn\\
c_{L,R}^{NN} &=& g_1^{NN}\left(C_{1L,1R}^{(9)}+C_{1L,1R}^{(9)\prime}\right)+g_2^{NN} \left(C_{2L,2R}^{(9)}+C_{2L,2R}^{(9)\prime}\right)+g_3^{NN}\left(C_{3L,3R}^{(9)}+C_{3L,3R}^{(9)\prime}\right)\nn\\
&&+g_4^{NN} C_{4L,4R}^{(9)}+g_5^{NN} C_{5L,5R}^{(9)}\,,\nn\\
c_V^{NN} &=&  g_6^{N N}C_V^{(9)}+ g_7^{N N}\tilde C_V^{(9)}\,.
\eea
The scaling of the nucleon-nucleon couplings follows the NDA expectation for $g^{NN}_{1,6,7}$, while $g^{NN}_{2,3,4,5}$ need to be enhanced with respect to NDA in order to renormalize the $nn\to pp\,ee$ amplitude \cite{Cirigliano:2018yza}. Explicitly, we have
\begin{eqnarray}
g^{NN}_{1,6,7} = \mathcal O(1), \qquad g^{NN}_{2,3,4,5} = \mathcal O\left((4\pi)^2\right).
\end{eqnarray}
Currently,  only NDA estimates are available for the $\pi N$ and $NN$ LECs.

\subsection{Chiral Lagrangian from the exchange of hard neutrinos}\label{sec:hardNu}
In addition to the long-range contributions originating from the exchange of potential neutrinos
between nucleons, mediated by the currents in Eq.\ \eqref{eq:currents0},
the $0\nu\beta\beta$ half-lives receive corrections from short-range operators, induced by the 
insertions of two currents connected by the exchange of hard, virtual neutrinos.
The origin of these contributions can be understood by considering the effective action induced by two insertions of the interactions in Eqs.\ \eqref{6final} and \eqref{7final}
\bea
iS_{\rm eff} = -\frac{1}{2!}\int d^4x\, d^4y \,T\left\{\left[ \vL^{(6)} (x)+\vL^{(7)} (x)\right]\left[ \vL^{(6)} (y)+\vL^{(7)} (y)\right]\right\}\,. \label{eq:Seff}
\eea
In terms of Eq.\ \eqref{eq:Seff},  the long-distance potential, derived in Sects.\ \ref{sec:piLag} and \ref{sec:piNLag}, arises from the region  $|\vec x-\vec y|\gtrsim 1/k_F$ where factorizing the two interactions is a good approximation. These long-distance contributions do not necessarily capture the region where $|x-y|\lesssim 1/\Lambda_\chi$. In fact, as we will argue below, NDA and renormalization imply that this region contributes at leading order in several cases. In order to correctly describe \NLDBD, the constructed chiral Lagrangian should be able to reproduce the amplitudes that result from inserting $S_{\rm eff}$ between initial and final states. In cases where the $|x-y|\lesssim 1/\Lambda_\chi$ region is important, this implies that additional short-distance interactions, of the same form as those induced by the dim-9 operators, are needed at LO in the chiral Lagrangian.

\subsubsection{Double insertions involving the vector couplings $C^{(6)}_{\rm VLR,VRR}$}
Before discussing these contributions in  generality, let us consider the amplitude $\langle h_f e_1 e_2 | S_{\rm eff}|h_i\rangle $, where  $h_{i,f}$ are hadronic states, for the example of the insertion of two vector operators. Since we are interested in amplitudes without initial- or final-state neutrinos, the neutrino fields in $\vL^{(6,7)}$ will be contracted among each other. Using this fact, and neglecting electron momenta, the Dirac algebra for the leptonic part can be performed, leading to
\bea
&&\langle h_f e_1 e_2 | S_{\rm eff}|h_i\rangle  =  \sum_i\frac{m_i}{2v^4}\int d^4\, x d^4 y \int \frac{d^4q}{(2\pi)^4} \frac{e^{-i q\cdot (x-y)}}{q^2-m_i^2+i\epsilon}\langle e_1 e_2| \bar e_R(x) e_R^c(x)|0\rangle \nn\\
&&\times \langle h_f|T\Bigg\{\left( C^{(6)}_{\rm VLR}\right)_{ei}^2  \bar u_L \gamma_\mu d_L(x) \, \bar u_L \gamma^\mu d_L(y)+ \left( C^{(6)}_{\rm VRR}\right)_{ei}^2 \bar u_R \gamma_\mu d_R (x) \,\bar u_R \gamma^\mu d_R(y)\nn\\
&&+2 \left( C^{(6)}_{\rm VRR}\right)_{ei} \left( C^{(6)}_{\rm VLR} \right)_{ei} \bar u_R \gamma_\mu d_R (x)\, \bar u_L \gamma^\mu d_L(y)
\Bigg\}|h_i\rangle +(L\leftrightarrow R)+\dots \label{eq:hardnu}
\eea
where the dots stand for terms proportional to other Wilson coefficients, as well as terms that arise from the $\slashed q$ term in the propagator. In this example, we will focus on the terms $\propto m_i$.

As mentioned above, Eq.\ \eqref{eq:hardnu} will induce operators of the same form as those induced by the dimension-nine operators. In particular, the $\left(C_{\rm VLR}^{(6)}\right)^2$, $\left(C_{\rm VRR}^{(6)}\right)^2$, and  $C_{\rm VLR}^{(6)}C_{\rm VRR}^{(6)}$ terms transform as the $O_1$, $O_1'$, and $O_4$ operators under chiral transformations. As a result, chiral symmetry allows  the following non-derivative pionic Lagrangian
\begin{eqnarray}\label{Lpipishort}
\mathcal L_{\pi\pi}  &=& 2 G_F^2 F_\pi^2 \sum_i m_i \,
g^{\pi\pi}_{\rm LR}(m_i) \, \textrm{Tr} [ \mathcal Q_L \mathcal Q_R]\,\bar e_R  C^{(6)}_{\rm VLR} \left(C^{(6)}_{\rm VRR} \right)^T e_R^c+(L\leftrightarrow R)\,,
\end{eqnarray}
where we introduced $\mathcal Q_L=u^{\dagger} \tau^+ u^{}, \,  \mathcal Q_R = u^{} \tau^+ u^{\dagger}$. By NDA the LEC $g^{\pi\pi}_{\rm LR}(m_i) $ is of order $\mathcal O(F_\pi^2)$, and we have explicitly given it a dependence on $m_i$. With this scaling, $g^{\pi\pi}_{\rm LR}(m_i) $ contributes at LO to \NLDBD, meaning that  the $|x-y|\lesssim 1/\Lambda_\chi$ region in  Eq.\ \eqref{eq:Seff} significantly contributes.

Very similar short-distance LECs are generated by the insertions of two electromagnetic currents, where hard virtual photons are exchanged instead of neutrinos. As explained in Refs.\ \cite{Cirigliano:2017tvr,Cirigliano:2018hja,Cirigliano:2019vdj}, this analogy can be made precise in the limit $m_i \rightarrow 0$, which allows for a relation between $g^{\pi\pi}_{\rm LR}(0)$ and the pion mass splitting, 
\begin{equation}\label{pionmasssplitting}
g^{\pi\pi}_{\rm LR}(m_i=0) = \frac{m_{\pi^\pm}^2 - m^2_{\pi^0}}{2 e^2 } \simeq 0.8 F_\pi^2\,,
\end{equation}
explicitly confirming the NDA expectations.
The $\left(C_{\rm VLR,VRR}^{(6)}\right)^2$ terms in principle give rise to pionic operators involving derivatives, which however induce subleading corrections to the long-distance neutrino potentials. None of the terms in Eq.\ \eqref{eq:hardnu} induce $\pi N$ couplings at leading order and we neglect them here. 

Additional interactions appear in the nucleon-nucleon sector. All combinations of couplings in Eq.\ \eqref{eq:Seff} give rise to short-distance nucleon-nucleon couplings, which are expected at N$^2$LO by NDA. However, as discussed in Refs.\ \cite{Cirigliano:2018hja,Cirigliano:2018yza,Cirigliano:2019vdj},
in the case of the standard mechanism and several dim-9 operators  they must appear at LO to guarantee that \nnpp\ amplitudes are properly renormalized and regulator independent.  The chiral Lagrangian is given by
\begin{eqnarray}\label{LNNshort}
\mathcal L_{NN} &=& \frac{1}{2} G_F^2  \,\bar e_R   e_R^c\sum_i m_i\nn\\
&&\times\Bigg\{   \left(C^{(6)}_{\rm VLR} \right)_{ei}^2  g_{\nu}^{\rm NN}(m_i)   \bar N \mathcal Q_L N \, \bar N \mathcal Q_L N + \left( C^{(6)}_{\rm VRR} \right)_{ei}^2 g_{\nu}^{\rm NN}(m_i)   \bar N \mathcal Q_R N \, \bar N \mathcal Q_R N \nn \\
& &  + 2 \left(C^{(6)}_{\rm VRR} \right)_{ei}\left(C^{(6)}_{\rm VRL}\right)_{ei} g_{\rm LR}^{\rm NN}(m_i) \, \left(
\bar N \mathcal Q_L N \, \bar N \mathcal Q_R N - \frac{1}{6} \textrm{Tr} (\mathcal Q_L \mathcal Q_R) \bar N \boldtau N \cdot \bar N \boldtau N \right)\Bigg\}\nn\\
&&+\left(L\leftrightarrow R\right)\,,\end{eqnarray}
where the $\left(C_{\rm VLR,VRR}^{(6)}\right)^2$ terms are related by parity and therefore come with the same LEC. In addition, we omitted traces that vanish for the form of $\mathcal Q_{L,R}$ relevant for \NLDBD, but in principle could be non-zero for other isospin components. 
From NDA, one finds $g^{\rm NN}_{i}(m_i) \sim \Lambda_\chi^{-2}$ which  implies the short-range operators contribute at N$^2$LO.  To absorb divergences in the scattering amplitudes, however,  
the scaling needs to be modified into $g^{\rm NN}_{i} \sim F_\pi^{-2}$, so that the $g_{\rm LR,\nu}^{\rm NN}$ operators in  $\mathcal L_{NN}$
contribute at LO.
The coupling $g_{\nu}^{\rm NN}$ was already encountered in Refs.\ \cite{Cirigliano:2018hja,Cirigliano:2019vdj},
since it also appears in the standard mechanism.

As was the case of the $\pi\pi$ interactions, the LECs that appear in the $NN$ sector can be related to LECs that appear due to the insertion of two electromagnetic currents. In this case,  it is the sum of $g_\nu^{\rm NN}$ and $g_{\rm LR}^{\rm NN}$ that is related to electromagnetic LECs, which affect isospin-breaking observables in nucleon-nucleon scattering. As a result, this combination of couplings can be obtained from the charge-independence breaking combination of scattering lengths, $\sim a_{nn}+a_{pp}-2a_{np}$, as detailed in Ref.\  \cite{Cirigliano:2019vdj}. Within pionful chiral EFT, at $m_i=0$ and in the \textoverline{MS} scheme, this leads to
\begin{equation}\label{eq:C12}
\frac{\tilde g_{\nu}^{\rm NN}(0) + \tilde g_{\rm LR}^{\rm NN}(0)}{2} = \frac{\tilde{\mathcal C_1} +\tilde{  \mathcal C_2 } }{2}=2.5-1.8\ln (m_\pi/\mu)\,,
\end{equation}
where we introduced $\tilde g_i = \left(\frac{4\pi}{m_N C}\right)^2 g_i$ and $\tilde{  \mathcal C_i } = \left(\frac{4\pi}{m_N C}\right)^2   \mathcal C_i$, while the electromagnetic couplings $\mathcal C_{1,2}$ were defined in Ref.\ \cite{Cirigliano:2019vdj}. Furthermore, $ C(\mu)=\Or(F_\pi^{-2})$ is the nucleon-nucleon contact interaction that appears at LO in the $^1S_0$ channel within chiral EFT. This coupling can be obtained by fitting to the isospin conserving nucleon-nucleon scattering lengths, and, within pionful EFT and using the \textoverline{MS} scheme, one has \cite{Cirigliano:2019vdj},
\bea
\frac{1}{ C(\mu)} = -0.24 \,{\rm fm}^{-2}-\frac{g_A^2 m_\pi^2 }{4 F_\pi^2}\left(\frac{m_N}{4\pi}\right)^2\ln\frac{\mu^2}{m_\pi^2}\,.
\eea
The above equations imply $\tilde g_\nu^{\rm NN}(0)+\tilde g^{\rm NN}_{\rm LR}(0) = \Or(1)$, or $ g_\nu^{\rm NN}(0)+ g^{\rm NN}_{\rm LR}(0) = \Or(F_\pi^{-2})$. This example explicitly confirms the arguments below Eq.~\eqref{LNNshort}.

\subsubsection{The general case}\label{sec:hardNuGen}
We now discuss
the general chiral  Lagrangian induced by hard neutrino exchange. This involves other combinations of Wilson coefficients, as well as the terms induced by the $\propto \slashed q$ term in the neutrino propagator, both can be constructed along similar lines. The induced interactions will have the form of the $\pi\pi$, $\pi N$, and $NN$ Lagrangians of Sect.\ \ref{sec:ChiLag9} such that all of these effects can be captured by the couplings defined in that section. For the non-derivative pion couplings of Eq.\ \eqref{eq:dim9pipi} we have
\bea
C_{i\, L,R}^{\pi\pi}&=& c_{i\,L,R}^{\pi\pi}+\frac{m_i v}{F_\pi^2}c_{i\,L,R}^{\nu\pi\pi}\,,\nn\\
c_{i\, L}^{\nu\pi\pi} &=&2g_{\rm LR}^{\rm \pi\pi}(m_i)\left(C_{\rm VLL}^{(6)}\right)_{ei}\left(C_{\rm VRL}^{(6)}\right)_{ei}-2g_{\rm S1}^{\pi\pi}(m_i)\left[\left(C_{\rm SLR}^{(6)}\right)^2_{ei}+\left(C_{\rm SRR}^{(6)}\right)^2_{ei}\right]\nn\\
&&+4g_{\rm S2}^{\pi\pi}(m_i)\left(C_{\rm SLR}^{(6)}\right)_{ei}\left(C_{\rm SRR}^{(6)}\right)_{ei}-2g_{\rm TT}^{\pi\pi}(m_i)\left(C_{\rm TRR}^{(6)}\right)^2_{ei}+c_{i\, L}^{\nu\pi\pi\,7} \,,
\nn\\
c_{i\, L}^{\pi\pi} &=& c_{i\, L}^{\pi\pi\,7} ,\qquad c_{i\, R}^{\pi\pi} = 0 \,.
\eea
Contributions from dim-7 operators are captured by $c_{i\, L}^{(\nu)\pi\pi\,7} $, which are discussed in Appendix \ref{app:hardNu7}, and all LECs scale as $g^{\pi\pi}_i=\Or(F_\pi^2)$. The contributions that are explicitly proportional to $m_i$ arise from choosing the $\sim m_i$ part of the neutrino propagator when performing the lepton contractions, as in Eq.\ \eqref{eq:hardnu}. The remaining terms arise from the $\slashed{q}$ part of the propagator, but only contribute at the dim-7 level. The right-handed coupling $c_{i\, R}^{\nu\pi\pi} $ can be obtained from $c_{i\, L}^{\nu\pi\pi}$  by interchanging the $L,R$ labels on the Wilson coefficients, $L\leftrightarrow R$, while leaving those on the LECs unchanged, and dropping  $c_{i\, L}^{\nu\pi\pi\,7} $. 

The derivative $\pi\pi$ couplings are given by
\bea
C_{i\,L,R}^{\prime \pi\pi}&=& \frac{v}{\Lambda_\chi}c_{i\,L,R}^{\prime \pi\pi}+\frac{m_i }{\Lambda_\chi}c_{i\,L,R}^{\prime \nu\pi\pi}\,,\nn\\
c_{i\,L}^{\prime \pi\pi}&=&g_{\rm S,VLL}^{\pi\pi}(m_i)\left[\left(C_{\rm SRR}^{(6)}\right)_{ei}\left(C_{\rm VLL}^{(6)}\right)_{ei}+\left(C_{\rm SLR}^{(6)}\right)_{ei}\left(C_{\rm VRL}^{(6)}\right)_{ei}\right]\nn\\
&&-\frac{g_{\rm T,VLL}^{\pi\pi}(m_i)}{4}\left(C_{\rm TRR}^{(6)}\right)_{ei}\left(C_{\rm VLL}^{(6)}\right)_{ei}+\left({\rm VLL}\leftrightarrow {\rm VRL}\right)  \,,\nn\\
c_{i\,L}^{\prime \nu\pi\pi}&=&c_{i\,L}^{\prime \nu\pi\pi\,7}\,,\qquad c_{i\,R}^{\prime \nu\pi\pi}=0\,,
\eea
where $c_{i\,R}^{\prime \pi\pi}$ can again be obtained from $c_{i\,L}^{\prime \pi\pi}$ with the interchange $L\leftrightarrow R$ and $c_{i\, L}^{\prime \nu\pi\pi\,7} $ is given in Appendix \ref{app:hardNu7}. The LECs related to these derivative couplings scale as $\Or(1)$, one of which was already encountered in Ref.\cite{Cirigliano:2018yza} where it was called $g_T^{\pi\pi}= \frac{m_N}{\Lambda_\chi}g_{\rm T,VLL}^{\pi\pi}(0)$. It should be noted that the terms proportional to $g_{\rm S,VLL}^{\pi\pi}$, $g_{\rm S,VRL}^{\pi\pi}$, and $g_{\rm S1,S2}^{\pi\pi}$ are generally suppressed by $F_\pi^2/\Lambda_\chi^2$ compared to the long-distance amplitudes in the limit $m_i\to 0$. In this limit these pieces only significantly contribute if the pseudo-scalar and axial couplings, that induce the long-distance contribution, are suppressed compared to the scalar and vector couplings.

The pion-nucleon couplings of \eqref{eq:dim9PiN}  can be written as,
\bea
C_{i\, L,R,V}^{ \pi N}&=& c_{i\,L,R,V}^{\pi N}+\frac{m_i }{\Lambda_\chi} c_{i\,L,R,V}^{\nu\pi N}\,,\nn\\
c_{i\,L}^{\pi N} &=&\frac{v}{\Lambda_\chi }\Bigg\{g_{\rm S,VLL}^{\pi N}(m_i)\left[\left(C_{\rm SRR}^{(6)}\right)_{ei}\left(C_{\rm VLL}^{(6)}\right)_{ei}+\left(C_{\rm SLR}^{(6)}\right)_{ei}\left(C_{\rm VRL}^{(6)}\right)_{ei}\right]\nn\\
&&-\frac{g_{\rm T,VLL}^{\pi N}(m_i)}{2}\left(C_{\rm TRR}^{(6)}\right)_{ei}\left(C_{\rm VLL}^{(6)}\right)_{ei}\Bigg\} +\left({\rm VLL}\leftrightarrow {\rm VRL}\right) \,,\nn\\
c_{i\,L}^{\nu\pi N} &=&c_{i\,L}^{\nu\pi N\,7}\,,\qquad c_{i\,R}^{\nu\pi N} =0\,,\nn\\
c_{i\,V}^{\pi N}&=&-\frac{1}{2}\frac{v}{\Lambda_\chi}g_{\rm VLL,VLR}^{\pi N}(m_i)\left[\left(C_{\rm VLL}^{(6)}\right)_{ei}\left(C_{\rm VLR}^{(6)}\right)_{ei}-\left(L\leftrightarrow R\right)\right]\,,\nn\\
c_{i\,V}^{\nu\pi N}&=&c_{i\,V}^{\nu\pi N\,7}\,,
\eea
where $g^{ \pi N}_\al=\Or(1)$, the right-handed coupling $c_{i\,R}^{\pi N}$ is given by $c_{i\,L}^{\pi N}$ with $L\leftrightarrow R$, and the dimension-7 contributions are again relegated to Appendix \ref{app:hardNu7}. Several of the above LECs are connected to those of Ref.\ \cite{Cirigliano:2018yza}, for which we have  $g_{\rm VLL,VLR}^{\pi N}(0) = g_{\rm VL}^{\pi N}\frac{\Lambda_\chi}{m_N}$ and $g_{\rm T,VLL}^{\pi N}(0) = g_{\rm T }^{\pi N}\frac{\Lambda_\chi}{m_N}$. 

Finally,  the contributions to the nucleon-nucleon couplings in Eq.\ \eqref{eq:dim9NN}  are given by,
\bea
C_{i\, L,R,V}^{ \rm N N}&=&c_{i\,L,R,V}^{\rm N N}+\frac{m_i }{\Lambda_\chi}c_{i\,L,R,V}^{\nu {\rm N N}}\,,\nn\\
c_{i\, L}^{NN}&=&\frac{v}{\Lambda_\chi }
\Bigg\{g_{\rm S,VLL}^{\rm NN}(m_i)\left[\left(C_{\rm SRR}^{(6)}\right)_{ei}\left(C_{\rm VLL}^{(6)}\right)_{ei}+\left(C_{\rm SLR}^{(6)}\right)_{ei}\left(C_{\rm VRL}^{(6)}\right)_{ei}\right]\nn\\
&&
-\frac{g_{\rm T,VLL}^{\rm NN}(m_i)}{2}\left(C_{\rm TRR}^{(6)}\right)_{ei}\left(C_{\rm VLL}^{(6)}\right)_{ei}\Bigg\}+\left({\rm VLL}\leftrightarrow {\rm VRL}\right) +c_{i\, L}^{{\rm NN\, 7}}\,,\nn\\
\frac{c_{i\, L}^{\nu NN}}{v\Lambda_\chi}&=&
\frac{g_{\nu}^{\rm NN}(m_i)}{4}\left[\left(C_{\rm VLL}^{(6)}\right)_{ei}^2+\left(C_{\rm VRL}^{(6)}\right)_{ei}^2\right]
+\frac{g_{\rm LR}^{\rm NN}(m_i)}{2}\left(C_{\rm VLL}^{(6)}\right)_{ei}\left(C_{\rm VRL}^{(6)}\right)_{ei}\nn\\
&&+\frac{g_{\rm S1}^{\rm NN}(m_i)}{4}\left[\left(C_{\rm SRR}^{(6)}\right)^2_{ei}+\left(C_{\rm SLR}^{(6)}\right)^2_{ei}\right]-\frac{g_{\rm S2}^{\rm NN}(m_i)}{2}\left(C_{\rm SRR}^{(6)}\right)_{ei}\left(C_{\rm SLR}^{(6)}\right)_{ei}\nn\\
&&+\frac{g_{\rm TT}^{\rm NN}(m_i)}{4}\left(C_{\rm TRR}^{(6)}\right)^2_{ei}+\frac{c_{i\, L}^{\nu NN\,7}}{v\Lambda_\chi}\,,\nn\\
c_{i\, V}^{NN}&=&-\frac{v}{\Lambda_\chi}\frac{g_{\rm VLL,VLR}^{\rm NN}(m_i)}{2}\left[\left(C_{\rm VLL}^{(6)}\right)_{ei}\left(C_{\rm VLR}^{(6)}\right)_{ei}-\left(L\leftrightarrow R\right)\right]\nn\\
&&+\frac{v}{\Lambda_\chi}g_{\rm T,SRL}^{\rm N N}(m_i)\left[\left(C_{\rm TRR}^{(6)}\right)_{ei}\left(C_{\rm SRL}^{(6)}\right)_{ei}-\left(L\leftrightarrow R\right)\right]\nn\\
&&+\frac{v}{\Lambda_\chi}g_{\rm T,SLL}^{\rm N N}(m_i)\left[\left(C_{\rm TRR}^{(6)}\right)_{ei}\left(C_{\rm SLL}^{(6)}\right)_{ei}-\left(L\leftrightarrow R\right)\right]+c_{i\, V}^{NN\,7}\,,\nn\\
\frac{c_{i\, V}^{\nu {\rm NN}}}{v\Lambda_\chi}&=&g_{\rm SLL,VLL}^{\rm NN}(m_i)\left[\left(C_{\rm SLL}^{(6)}+C_{\rm SRL}^{(6)}\right)_{ei}\left(C_{\rm VLL}^{(6)}+C_{\rm VRL}^{(6)}\right)_{ei}-\left(L\leftrightarrow R \right)\right]\nn\\
&&+g_{\rm TLL,VLL}^{\rm NN}(m_i)\left[\left(C_{\rm TLL}^{(6)}\right)_{ei}\left(C_{\rm VLL}^{(6)}-C_{\rm VRL}^{(6)}\right)_{ei}-\left(L\leftrightarrow R\right)\right]+\frac{c_{i\, V}^{\nu {\rm NN\, 7}}}{v\Lambda_\chi}\,.
\eea
The right-handed couplings $c_{i\, R}^{\rm NN, \nu{\rm NN}} $ can be obtained from $c_{i\, L}^{\rm NN, \nu{\rm NN}} $  by interchanging the $L,R$ labels on the Wilson coefficients, $L\leftrightarrow R$, while leaving those on the LECs unchanged, and dropping $c_{i\, L}^{\nu NN\,7}$. By NDA, the LECs related to the $c_{i}^{ \nu{\rm NN}} $ couplings scale as $g_i^{\rm NN}=\Or(\Lambda_\chi^{-2})$ while those contributing to the  $c_{i}^{ {\rm NN}} $ couplings follow the scaling $g^{\rm NN}_i=\Or(1)$. However, apart from the terms proportional to $g_{\rm VLL,VLR}^{\rm NN}$, $g_{\rm S,VLL}^{\rm NN}$, $g_{\rm S,VRL}^{\rm NN}$, $g_{\rm T,VLL}^{\rm NN}$, and $g_{\rm T,VRL}^{\rm NN}$, one has to enhance the scaling of all NN LECs by $\Lambda_\chi^2/F_\pi^2$ in order to obtain renormalized amplitudes. We report the RGEs for the enhanced LECs in Appendix \ref{app:RGE}. Finally, two of the above LECs are related to those discussed in  Ref.\ \cite{Cirigliano:2018yza}, namely  $g_{\rm VLL,VLR}^{\rm N N}(0) = g_{\rm VL}^{NN}\frac{\Lambda_\chi}{m_N}$ and $g_{\rm T,VLL}^{\rm N N}(0) = g_{\rm T}^{N N}\frac{\Lambda_\chi}{m_N}$.

\subsection{Summary}\label{sec:3summary}

The LECs needed to construct the neutrino potential at LO, and their current determinations, are summarized in Table \ref{Tab:LECs}.
The LECs that enter the neutron $\beta$ decay operators discussed in  Sect.\ \ref{neutronbeta}  are well determined, either from experiment, 
as in the case of $g_{A,M}$, which  appear in SM currents, or from lattice QCD, in the case of $g_S$, $g_T$ and $B$. The one exception is $g_T^\prime$, which contributes to the tensor current at recoil order and is not very important in $\beta$ decays. The evaluation of this LEC could be pursued  
with the same methods discussed in Refs.\ \cite{Bhattacharya:2016zcn,Gupta:2018qil,Aoki:2019cca}. The $\pi\pi$ couplings induced by dim-9 operators have been computed in Lattice QCD 
\cite{Nicholson:2018mwc}, with uncertainty better than 10\%. 
The $\pi\pi$, $\pi N$, and $NN$ couplings induced by dim-6 and dim-7 operators are functions of the neutrino mass. In the case of $g^{\pi\pi}_{\rm LR}$, both the small- and large-$m_i$
behavior are known, allowing us to obtain a reliable interpolation formula, as we will discuss in  Sect.\ \ref{interLEC}. In several other cases, only the large $m_i$ behavior is known. The calculation of these couplings as a function of $m_i$ could use techniques similar to the Hubbard-Stratanovich transformation proposed in Ref.\ \cite{Monge-Camacho:2019nby} for the $g_{1,\ldots,5}^{\pi\pi}$ couplings,
with the difference that the scalar particle $\sigma$ introduced in Ref.\ \cite{Monge-Camacho:2019nby} is kept light.
The determination of the pion-nucleon and nucleon-nucleon couplings, induced by dim-6, -7 and -9 operators, is much more uncertain. At the moment, only the combination 
$g_{\nu}^{\rm NN}(0) + g^{\rm NN}_{\rm LR}(0)$ is known, in a variety of renormalization schemes \cite{Cirigliano:2018yza,Cirigliano:2019vdj}, via its relation to charge-independence breaking
in nucleon-nucleon scattering. All other couplings require dedicated Lattice QCD calculations of LNV nucleon-nucleon scattering amplitudes.
In the literature, the LECs in Table \ref{Tab:LECs} are often estimated using uncontrolled assumptions such as ``factorization'' of the product of two weak currents. While this might be unavoidable at the moment, 
we will show that varying the LECs in a range suggested by their NDA scaling  introduces uncertainties in the $0\nu\beta\beta$ half-lives that are as large as those in the nuclear matrix elements,
and should not be neglected.

As argued above, the exchange of hard neutrinos within chiral EFT leads to counterterms that are expected to induce $\Or(1)$ effects in many cases. However, as we will discuss in Sect.\ \ref{sectionNME}, the nuclear matrix elements for isotopes of experimental interest are all calculated using various many-body methods, for which it is a priori unclear how the conclusions of Chiral EFT carry over. Thus, although the extraction of the counterterm from $NN$ scattering in Eq.\ \eqref{eq:C12},  as well as \textit{ab initio} calculations in light nuclei \cite{Pastore:2017ofx,Cirigliano:2019vdj}, suggest that hard-neutrino exchange has an $\Or(1)$ impact on the \NLDBD\ half-life, we cannot say with certainty to what extent this is true in many-body calculations for the heavy nuclei of experimental interest. This implies that it is in principle possible that the effects of hard neutrinos, which are $\Or(1)$ in the Chiral approach, turn out to be smaller in the calculation of NMEs of larger nuclei, such as those of Refs.\ \cite{Hyvarinen:2015bda,Menendez:2017fdf,Barea:2015kwa,Barea} (depicted in  Table \ref{tab:comparison}). To deal with these issues when deriving \NLDBD\ constraints in Sect.\ \ref{pheno}, we will conservatively employ the above mentioned NMEs and their uncertainties, while estimating the theoretical error due to the unknown hard-neutrino LECs by using their NDA values.

\begin{table}[t!]
\renewcommand{\arraystretch}{1.2}
\center\small
\begin{tabular}{|c|cc||c|cc|}
\hline
 \multicolumn{3}{|c||}{ $n\rightarrow pe\nu$, $\pi \rightarrow e \nu$ } &  \multicolumn{3}{c|}{$\pi \pi \rightarrow e e:\quad \Or^{(9)}$} \\
 \hline
 $g_A$ & $1.271\pm 0.002$ & \cite{Olive:2016xmw}      & $g^{\pi\pi}_{1}$   		& $  0.36 \pm 0.02 $             & \cite{Nicholson:2018mwc}  \\
 $g_S$ & $1.02\pm 0.10$ & \cite{Gupta:2018qil,Aoki:2019cca} & $g^{\pi\pi}_{2}$   		& $  2.0  \pm 0.2 $  \, GeV$^2$  & \cite{Nicholson:2018mwc}  \\
 $g_M$ &  $ 4.7$ &  \cite{Olive:2016xmw}              & $g^{\pi\pi}_{3}$ 	        & $ -0.62 \pm 0.06$  \, GeV$^2$  & \cite{Nicholson:2018mwc}  \\
 $g_T$ & $0.99\pm 0.03$ & \cite{Gupta:2018qil,Aoki:2019cca} & $g^{\pi\pi}_{4}$   		& $ -1.9  \pm 0.2$   \, GeV$^2$  & \cite{Nicholson:2018mwc}\\  
 $|g'_T|$ & $\mathcal O(1)$  	&		      & $g^{\pi\pi}_{5}$ 		& $ -8.0  \pm 0.6$   \, GeV$^2$  & \cite{Nicholson:2018mwc}  \\
 $B$      &    $2.7$~GeV       &     &  &&\\\hline
  \multicolumn{3}{|c||}{$n \rightarrow p\pi ee:\, \Or^{(9)},\, \Or^{(6,7)}\otimes  \Or^{(6,7)}$} & \multicolumn{3}{c|}{$\pi \pi \rightarrow e e:\quad \Or^{(6,7)}\otimes  \Or^{(6,7)}$} \\\hline
   $|g^{\pi N}_{i} |$       & $\mathcal{O}(1)$ &&  $|g^{\pi\pi}_{\rm T,VLL}|,\, |g^{\pi\pi}_{\rm S,VLL}|,|g^{\pi\pi}_{\rm T,VRL}|,\, |g^{\pi\pi}_{\rm S,VRL}|\,$     & $\mathcal O(1)$  & \\ 
 &&& $|g^{\pi\pi}_{\rm LR}|,\, |g^{\pi\pi}_{\rm S1,S2}|$& $ \mathcal{O}(F_\pi^2) $ & \\
  &&& $|g^{\pi\pi}_{\rm TT}|,\, |g^{\pi\pi}_{\rm  TL}|,\, |g^{\pi\pi}_{\rm TL,TR}|$& $ \mathcal{O}( F_\pi^2) $ & \\
 \hline
  \multicolumn{3}{|c||}{$nn\rightarrow pp\, ee:\quad \Or^{(9)}$} & \multicolumn{3}{c|}{$nn\rightarrow pp\, ee:\quad \Or^{(6,7)}\otimes  \Or^{(6,7)}$} \\\hline
$|g^{N N}_{1,6,7}|$    & $\mathcal{O}(1)$ &    & $|g^{N N}_{\nu}|,\,|g^{N N}_{\rm LR}|,\,|g^{N N}_{\rm S1}|$         & $ \mathcal{O}(1/F_\pi^2) $&  \\  
 $ |g^{NN}_{2,3,4,5}|$ & $ \mathcal{O}((4\pi)^2) $		  &    & $|g^{N N}_{\rm S2}|,\,|g^{N N}_{\rm TT}|,\,|g^{N N}_{\rm SLL,VLL}|$ &  $ \mathcal{O}(1/F_\pi^2) $  & \\ 
&  &    & $|g^{N N}_{\rm TLL,VLL}|,\,|g^{N N}_{\rm TL}|,\,|g^{N N}_{\rm TL,TR}|$ & $ \mathcal{O}(1/F_\pi^2) $ &   \\
  &  &    & $|g^{N N}_{\rm TL,T}|,\,|g^{N N}_{\rm TR,T}|$ & $ \mathcal{O}(1/\Lambda_\chi^2) $ &     \\
    &  &    & $|g^{N N}_{\rm S,VLL}|,\,|g^{N N}_{\rm T,VLL}|\,|g^{N N}_{\rm VLL,VLR}|$ & $ \mathcal{O}(1) $ &     \\
        &  &    & $|g^{N N}_{\rm S,VRL}|,\,|g^{N N}_{\rm T,VRL}|$ & $ \mathcal{O}(1) $ &     \\
        &  &    & $|g^{N N}_{\rm T,SRL}|,|g^{N N}_{\rm T,SLL}|,|g^{N N}_{\rm TL,V}|,|g^{N N}_{\rm TR,V}|$ & $ \mathcal{O}((4\pi)^2) $ &     
\\\hline
\end{tabular}
\caption{The low-energy constants relevant for the dim-3, dim-6, dim-7, and dim-9 operators. The headings show the type of long-distance ($n\rightarrow pe\nu$, $\pi \rightarrow e \nu$) or short-distance processes the LECs induce, while the labels $\Or^{(9)}$ and  $ \Or^{(6,7)}\otimes  \Or^{(6,7)}$ indicate whether the corresponding LECs are induced by dim-9 operators or by the insertion of two dim-6(-7) interactions.
Whenever known,  we quote the values of the LECs at $\mu=2$ GeV in the $\overline{\rm MS}$ scheme. }  
\label{Tab:LECs}
\end{table}

\section{The \nnpp\ transition operator including sterile neutrinos}\label{potentials}
\begin{figure}
\includegraphics[width=\textwidth]{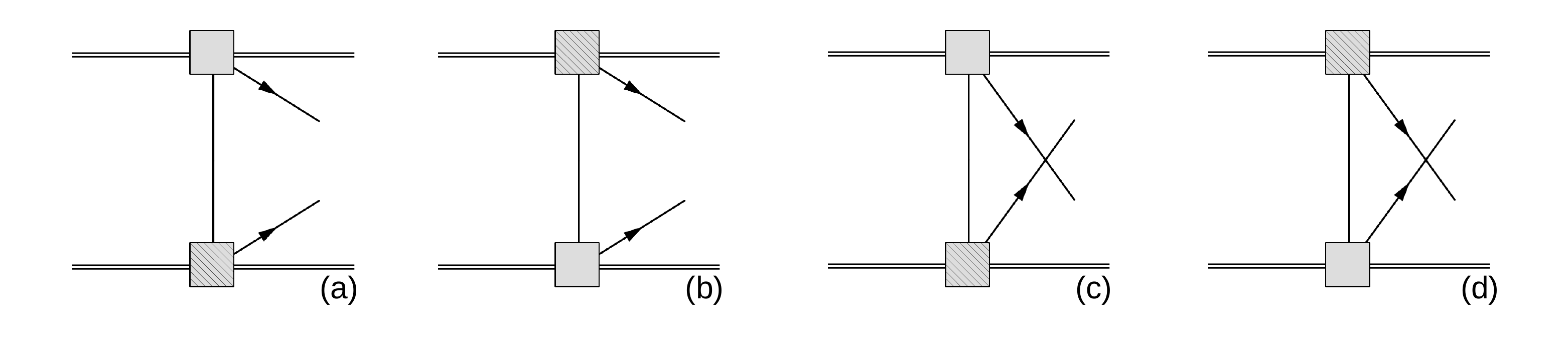}
\caption{
Tree-level contributions to the $0\nu\beta\beta$ transition operator arising from the exchange of a light neutrino. The notation for nucleons, electrons, and neutrinos is as in Fig.\ \ref{neutroncurrent}.
The squares denote the nucleon vector, axial, scalar, pseudoscalar, and tensor form factors, which, at LO in chiral EFT, include one or both diagrams in Fig.\ \ref{neutroncurrent}. The currents acting on the two nucleons can be different, which we denoted by hatching one of the two squares. LNV arises from the mass of the neutrinos, which in general are Majorana eigenstates,
or from the couplings of the neutrinos to the nucleons, which receive LNV contributions at \textoverline{dim-7}.}\label{Fig2}
\end{figure}

We now turn to the main part of this work: the derivation of the \nnpp\ transition operator. This transition operator will be inserted between nuclear wave functions and is sometimes called the ``neutrino potential''. The transition operator is not necessarily due to the exchange of a neutrino as other mechanisms exist, for instance via the contact $\pi\pi e e$, $n p \pi ee$ and $nnppee$ interactions 
discussed in  Sect.\ \ref{sec:ChiLag9} and \ref{sec:hardNu}. 
Such mechanisms have been discussed in detail in Ref.~\cite{Cirigliano:2018yza} and the derivation of the potential in the presence of sterile neutrinos
amounts to generalizing the couplings $C^{(\prime)\pi\pi}$, $C^{\pi N}$ and $C^{NN}$ as in Eq.\ \eqref{eq:hardnu0}, to include the contributions of hard-neutrino exchange.
We therefore focus here on the neutrino potential arising from the exchange of a light neutrino, with mass below the chiral-breaking scale $\Lambda_\chi$. In general the induced neutrino potential arises from the four diagrams in Fig.~\ref{Fig2}, where any combination of hadronic currents can be used. The top (bottom) incoming and outgoing nucleons have momenta $ p_1$ ($ p'_1$) and $ p_2$ ($ p'_2$), respectively, and we define $q_{1,2} = p_{1,2}-p'_{1,2}$. The top (bottom) electron has outgoing four-momenta $k_1$ ($k_2$). In diagrams $(a)$ and $(b)$ the neutrino then carries momentum $q_{11} = q_1 - k_1 = -q_2 + k_2$. In diagrams $(c)$ and $(d)$ the neutrino carries momentum $q_{12} = q_1-k _2 = - q_2 + k_1$. In most cases, we can neglect the electron momenta in the neutrino propagators and hadronic currents. In those cases, $q_{11} = q_{12} = q_1 = - q_2 \equiv q$. Finally, we define the notation $J_x(i)$, where $x=\{V,A,S,P,T_R,T_L\}$ and $i=\{1,2\}$, that implies that the expression in Eq.~\eqref{eq:currents} should be evaluated for nucleon $i$ using the momenta $p_i$, $p'_i$, and $q_i$. 

We begin by studying the so-called standard mechanism of \NLDBD\ which is the exchange of a light Majorana neutrino. We review how to derive the well-known form of the neutrino potential appearing in this scenario and how it is affected by the presence of additional sterile neutrinos that interact via left-handed currents. This warm-up calculation provides a useful guide towards obtaining the neutrino potential arising from other interactions. The calculation of the remaining terms is in principle straightforward, however, as it is rather lengthy we have checked our results by use of  the \textsc{Mathematica} package \textsc{FeynCalc}~\cite{Mertig:1990an,Shtabovenko:2016sxi}. 

\subsection{The standard mechanism with sterile neutrinos}
We start by considering the \nnpp\ transition operator arising from potential neutrinos that interact via the  $C_{\rm VLL}^{(6)}$  term in Eq.~\eqref{6final}. This term includes the SM weak interaction as can be seen from Eqs.~\eqref{redefC6} and \eqref{match6LNC}. We use the same vertex for the top and bottom nucleon propagators such that diagrams $(a)$ and $(b)$ add coherently and the resulting factor $2$ is cancelled by the $1/2!$ from using the same vertex twice. Diagrams $(a)$ and $(b)$ then sum to
\bea
V^{(a)+(b)}_\nu &=& i (\tau_1^+ \tau_2^+)\sum_{i=1}^{n_L}\left(\frac{+2i G_F}{\sqrt{2}}C_{\rm VLL}^{(6)} \right)_{ei}\left(\frac{-2 iG_F}{\sqrt{2}}(C_{\rm VLL}^{(6)})^T\right)_{ie}\times \frac{1}{2}\left[J^\mu_V(1) +J^\mu_A(1)\right]\nn\\
&& \times\frac{1}{2}\left[J^\nu_V(2) +J^\nu_A(2)\right]\times \bar u(k_1)\gamma_\mu P_L \frac{i(\dslash q_{11}+m_i)} {q^2_{11}-m_i^2} \gamma_\nu P_R u^c(k_2)\,,
\eea
where $u(k_i)$ denotes an electron spinor with momenta $k_i$, and we introduced $n_L = N-n_H$, such that the sum runs over all neutrino eigenstates with masses below $\Lambda_\chi$. The potential is related to the amplitude via $\mathcal A=-V$. This expression can be simplified into
\bea
V^{(a)+(b)}_\nu &=& (\tau_1^+ \tau_2^+) \frac{G_F^2}{2} \sum_{i=1}^{n_L}\left(C_{\rm VLL}^{(6)} \right)_{ei}\left(C_{\rm VLL}^{(6)} \right)^T_{ie}\left[J^\mu_V(1) +J^\mu_A(1)\right] \left[J^\nu_V(2) +J^\nu_A(2)\right]\nn\\
&&\times \frac{m_i}{\vec q^2 + m_i^2} \bar u(k_1) \gamma_\mu \gamma_\nu P_R u^c(k_2) + \dots\,,
\eea
where the dots denote corrections proportional to the lepton momenta or nucleon energy which are suppressed by additional powers of $\epc$. Similarly, the remaining two diagrams sum to
\bea
V^{(c)+(d)}_\nu &=&- (\tau_1^+ \tau_2^+)\frac{G_F^2}{2}\sum_{i=1}^{n_L}  \left(C_{\rm VLL}^{(6)} \right)_{ei}\left(C_{\rm VLL}^{(6)} \right)^T_{ie}\left[J^\mu_V(1) +J^\mu_A(1)\right] \left[J^\nu_V(2) +J^\nu_A(2)\right]\nn\\
&&\times \frac{m_i}{\vec q^2 + m_i^2} \bar u(k_2) \gamma_\mu \gamma_\nu P_R u^c(k_1) + \dots\,,
\eea
where the overall sign difference is from exchanging the two electrons. Summing all diagrams then gives
\bea
V_\nu &=& (\tau_1^+ \tau_2^+)G_F^2 \sum_{i=1}^{n_L} \left(C_{\rm VLL}^{(6)} \right)_{ei}\left(C_{\rm VLL}^{(6)} \right)^T_{ie}\left[J_V(1) +J_A(1)\right]\cdot \left[J_V(2) +J_A(2)\right]\nn\\
&&\times \frac{m_i}{\vec q^2 + m_i^2} \bar u(k_1)  P_R u^c(k_2) + \dots\,
\eea

The product of hadronic currents can be explicitly calculated from Eq.~\eqref{eq:currents} and contains parity-even and parity-odd components. As the remaining part of $V_\nu$ is an even function of $\vec q$, only the parity-even parts contribute to the $0^+ \rightarrow 0^+$ transitions of experimental interest. The relevant hadronic currents are therefore
\bea
J_V(1)\cdot J_V(2)  &=& g_V^2(\vec q^2) - \frac{g_M^2(\vec q) \vec q^2}{6 m_N^2}\left(\boldsigma_1 \cdot \boldsigma_2+ \frac{1}{2}S^{(12)}\right)\,,\nn\\
J_A(1)\cdot J_A(2)  &=&  -g^2_A\bigg\{ \boldsigma_1 \cdot \boldsigma_2\, \left(\frac{g^2_A(\vec q^2)}{g_A^2} + \frac{g_P(\vec q^2)g_A(\vec q^2) \vec q^2}{3 g^2_A m_N}+\frac{g_P^2(\vec q^2) \vec q^4}{12 g_A^2 m_N^2}\right)\nn\\
&& -S^{(12)}\left( \frac{g_P(\vec q^2)g_A(\vec q^2) \vec q^2}{3 g^2_A m_N}+\frac{g_P^2(\vec q^2) \vec q^4}{12 g_A^2 m_N^2}\right)\bigg\}\,,
\eea
where we have defined the tensor operator 
\be
S^{(12)} =  \boldsigma_1 \cdot \boldsigma_2 - 3\, \boldsigma_1 \cdot \hat  {\vec q}\,\boldsigma_2 \cdot \hat  {\vec q}\,.
\ee
It is useful to split the Fermi (F), Gamow-Teller (GT), and Tensor (T) operators into their separate contributions arising from vector, axial, pseudoscalar, and magnetic currents, as the corresponding nuclear matrix elements are reported in the literature. We define the combinations
\begin{eqnarray}\label{eq:hK(q)}
h_{GT}(\vec q^2) &=&  h^{AA}_{GT}(\vec q^2) + h^{AP}_{GT}(\vec q^2) + h^{PP}_{GT}(\vec q^2) + h^{MM}_{GT}(\vec q^2)\,,\nn\\
h_{T}(\vec q^2)  &=&  h^{AP}_{T}(\vec q^2) + h^{PP}_{T}(\vec q^2) + h^{MM}_{T}(\vec q^2)\,.
\end{eqnarray}
For the F, GT, and T functions, we have
\begin{eqnarray}\label{smff}
h_F(\vec q^2) &=& \frac{g_V^2(\vec q^2)}{g_V^2}\,\qquad
h^{AA}_{GT,T}(\vec q^2) =\frac{g_A^2(\vec q^2)}{g_A\sq}\,, \quad \, h_{GT}^{AP}(\vec q^2) = \frac{g_P(\vec q^2)g_A(\vec q^2) \vec q^2}{3 g^2_A m_N}\,,\nn\\
h_{GT}^{PP}(\vec q^2) &=& \frac{ g^2_P(\vec q^2)\vec q^4}{12 g_A^2 m_N^2}\,, \qquad
h^{MM}_{GT}(\vec q^2) = \frac{ g_M^2(\vec q^2) \vec q^2}{6g_A\sq m_N^2}\,,
\end{eqnarray}
and $h^{AP}_T(\vec q^2) = -h^{AP}_{GT}(\vec q^2)$, $h^{PP}_T(\vec q^2) = - h^{PP}_{GT}(\vec q^2)$, and  $h^{MM}_{T}(\vec q^2) = h^{MM}_{GT}(\vec q^2)/2$. 

We then obtain for the neutrino potential 
\bea
V_\nu &=& -(\tau_1^+ \tau_2^+)g_A^2 G_F^2 \sum_{i=1}^{n_L} \left(C_{\rm VLL}^{(6)} \right)^2_{ei}\,\left(\frac{m_i}{\vec q^2 + m_i^2}\right)  \nn\\
&&\times \left[ - \frac{g_V^2}{g_A^2}h_F(\vec q^2) +  \boldsigma_1 \cdot \boldsigma_2\, h_{GT}(\vec q^2) + S^{(12)}\,h_{T}(\vec q^2)\right]\times \bar u(k_1)  P_R u^c(k_2)\,.
\eea

This expression reduces to the familiar expression for the neutrino potential for the case of 3 light Majorana neutrinos. We set $n=0$ and we turn off all higher-dimensional operators except for the active Majorana mass (which is formally a \textoverline{dim-5} operator). In this limiting case, $C_{\rm VLL}^{(6)} = -2 V_{ud}\, P U$ and
\be
\sum_{i=1}^{n_L} \left(C_{\rm VLL}^{(6)} \right)^2_{ei}\,\left(\frac{m_i}{\vec q^2 + m_i^2}\right) \simeq \frac{4 V_{ud}^2}{\vec q^2} \left[ P U m_\nu U^T P^T \right]_{ee} =  \frac{4 V_{ud}^2}{\vec q^2} (M_L^*)_{ee} \,,
\ee
where we used Eq.~\eqref{Mdiag} and used $m_i \ll \vec q$. The neutrino potential becomes proportional to the Majorana mass of the active neutrinos and agrees with the usual result for the standard mechanism

\subsection{The general neutrino transition operator with sterile neutrinos}
The neutrino potentials arising from the other interactions in Eqs.~\eqref{6final} and \eqref{7final} can be obtained in analogous fashion to the calculation in the previous subsection. We give here the results for all combinations of interactions that lead to a non-vanishing potential when the electron mass and momenta are neglected. 
The limited cases in which the first non-vanishing contribution to the transition operator involves lepton momenta are discussed in Ref.\ \cite{Cirigliano:2017djv}.

The neutrino potential can be divided into three separate leptonic structures 
   \bea
   V_{\rm } &=& -(\tau_1^+ \tau_2^+) g_A^2 G_F^2 \sum_{i=1}^{n_L} \frac{1}{\vec q^2+m_i^2}\,\bar u(k_1) \left\{ V_L\,  P_R + V_R\,  P_L  + V_M\, \gamma^0 \gamma^5 \right\}u^c(k_2) \,.
   \eea
We separate the structures 
\be
V_{L,R,M} = V^{(6)}_{L,R,M}+ \frac{m_\pi}{v}V^{(7)}_{L,R,M}+\frac{\vec q^2+m_i^2}{m_\pi^2}V_{L,R,M}^{(sd)}\,,
\ee 
into three parts. The part with superscript $(6)$ denotes contributions from the dim-6 operators in Eq.~\eqref{6final} and is given by
   \bea\label{V6}
   V^{(6)}_L &=& m_i \left(C_{\rm VLL}^{(6)} +C_{\rm VRL}^{(6)} \right)^2_{ei}\left[-\frac{g_V^2}{g_A^2} h_F + \boldsigma_1 \cdot \boldsigma_2 \,h_{GT}^{MM} + S^{(12)}\,h^{MM}_{T} \right] \nn \\
    &&+ m_i\left(C_{\rm VLL}^{(6)} - C_{\rm VRL}^{(6)} \right)^2_{ei}\left[\boldsigma_1 \cdot \boldsigma_2\,(h_{GT}^{AA}+h_{GT}^{AP}+h_{GT}^{PP})+S^{(12)}\,(h_{T}^{AP}+h_{T}^{PP})
    \right] \nn \\
        &&+ m_i \left(C_{\rm SRR}^{(6)} +C_{\rm SLR}^{(6)} \right)^2_{ei} \frac{g^2_S}{g^2_A}h_F \nn\\
    &&+ m_i \left(C_{\rm SRR}^{(6)} -C_{\rm SLR}^{(6)} \right)^2_{ei} \frac{B^2}{2m_\pi^2}\left[\boldsigma_1 \cdot \boldsigma_2\,(h_{GT}^{AP}+2 h_{GT}^{PP})+S^{(12)}\,(h_{T}^{AP}+2 h_{T}^{PP})\right]\nn\\
    &&- m_i \left(C_{\rm TRR}^{(6)}\right)^2_{ei}   \boldsigma_1 \cdot \boldsigma_2\,\frac{16 g_T^2}{g_A^2} h_{GT}^{AA}\nn\\
    &&- B \left(C_{\rm VLL}^{(6)} - C_{\rm VRL}^{(6)} \right)_{ei} \left(C_{\rm SRR}^{(6)} -C_{\rm SLR}^{(6)} \right)_{ei}\left[\boldsigma_1 \cdot \boldsigma_2\,(h_{GT}^{AP}+2 h_{GT}^{PP})+S^{(12)}\,(h_{T}^{AP}+2 h_{T}^{PP})\right]\nn\\
    &&+ m_N\left(C_{\rm VLL}^{(6)} + C_{\rm VRL}^{(6)} \right)_{ei} \left(C_{\rm TRR}^{(6)} \right)_{ei}\Bigg[\frac{16 g_T}{g_M} \left(\boldsigma_1 \cdot \boldsigma_2 \,h_{GT}^{MM} + S^{(12)}\,h^{MM}_{T} \right)\nn\\
    &&-4    \frac{g^\prime_T  g_V }{g_A^2} \frac{\vec q^2}{m_N^2} h_F\Bigg]\,,\\
    V^{(6)}_R &=& V^{(6)}_L \big{\rvert}_{L \leftrightarrow R}\,,
    \eea
    \bea 
    V^{(6)}_M &=& m_N\left(C_{\rm VLL}^{(6)}  C_{\rm VLR}^{(6)}  - L \leftrightarrow R \right)_{ei}\frac{4g_A}{g_M}\left[ \boldsigma_1 \cdot \boldsigma_2 \,h_{GT}^{MM} + S^{(12)}\,h^{MM}_{T}  \right]\nn\\
    &&-m_i \left[\left(C_{\rm VLL}^{(6)} + C_{\rm VRL}^{(6)} \right)_{ei}\left(C_{\rm SLL}^{(6)} +C_{\rm SRL}^{(6)} \right)_{ei} - L \leftrightarrow R \right]\frac{g_S g_V}{g_A^2}h_F\nn\\
    &&-m_i\left[\left(C_{\rm VLL}^{(6)}  -   C_{\rm VRL}^{(6)} \right)_{ei}\left(C_{\rm TLL}^{(6)}\right)_{ei} - L \leftrightarrow R \right] \frac{2 g_T}{g_A} \left[\boldsigma_1 \cdot \boldsigma_2\,(2 h_{GT}^{AA}+h_{GT}^{AP})+S^{(12)}\,h_{T}^{AP}\right]\nn\\
    &&+B\,\, \left[ \left(C_{\rm SRR}^{(6)} -C_{\rm SLR}^{(6)} \right)_{ei} \left(C_{\rm TLL}^{(6)}\right)_{ei}- L \leftrightarrow R \right] \frac{2 g_T}{g_A} \left[\boldsigma_1 \cdot \boldsigma_2\, h_{GT}^{AP}+S^{(12)}\,h_{T}^{AP}\right]\,.
   \eea
Instead, $V^{(7)}_{L,R,M}$ arise from the dim-7 operators in Eq.~\eqref{7final}. As these terms are parametrically suppressed by a power of $m_\pi/v$ or $\Lambda_\chi/v$, we relegate the explicit expressions to Appendix \ref{Appdump}.   

Finally, the short-distance part of the potentials, $V_{L,R,M}^{(sd)}$, are induced by dimension-nine operators as well as the exchange of hard neutrinos, see Sects.\ \ref{sec:ChiLag9} and \ref{sec:hardNu}, they are given by,
\bea
V_L^{(sd)} &=& -4\frac{m_\pi^2}{v}\Bigg\{
\left(\frac{C_L^{\pi\pi}}{m_\pi^2}+C_L^{\prime \pi\pi}\right)\left[\left(\frac{h_{GT}^{AP}}{2}+h_{GT}^{PP}\right)\boldsigma_1 \cdot \boldsigma_2+\left(\frac{h_{T}^{AP}}{2}+h_{T}^{PP}\right)S^{(12)}\right]\nn\\
&&+\frac{C_L^{\pi N}-C_L^{\prime \pi\pi}}{2}\left(h_{GT}^{AP}\,\boldsigma_1 \cdot \boldsigma_2+h_T^{AP}S^{(12)}\right)-\frac{2}{g_A^2}C_L^{NN}h_F
\Bigg\}\,,\nn\\
    V^{(sd)}_R &=& V^{(sd)}_L \big{\rvert}_{L \leftrightarrow R}\,,\nn\\
    V_M^{(sd)}&=&
     -4\frac{m_\pi^2}{v}\Bigg\{
\frac{C_V^{\pi N}}{2}\left(h_{GT}^{AP}\,\boldsigma_1 \cdot \boldsigma_2+h_T^{AP}S^{(12)}\right)-\frac{2}{g_A^2}C_V^{NN}h_F
\Bigg\}\,.
\eea

\section{The neutrinoless double beta decay master formula including sterile neutrinos }\label{MasterFormula}

 Armed with the neutrino potentials we define the \nnpp\ amplitude by
 \be \label{eq:ampV}
 \mathcal A=\langle 0^+ | \sum_{m,n}\int \frac{d^3\vec q}{(2\pi)^3} e^{i \vec q \cdot \vec r} V(\vec q^2)| 0^+ \rangle\,,
 \ee
  where $V(\vec q^2)$ are the neutrino potentials from the previous section, and the sum extends over all the nucleons in the nucleus. $\vec r = \vec r_n - \vec r_m$ is the distance between nucleons $m$ and $n$ and $|\vec r| =r$, and the potentials are inserted between the $0^+$ initial- and final-state nuclei of experimental interest. The leptonic part of the neutrino potentials can be taken outside of the nuclear wave functions and we define
 \bea\label{amplitude}
 \mathcal A = \frac{g_A^2 G_F^2 m_e}{\pi R_A}&\times&\Bigg\{ \left[\sum_{i=1}^{n_L} \mathcal A_L(m_i) + \sum_{i=n_L+1}^{N} \mathcal A^{(9)}_L(m_i)\right]\,\bar u(k_1)  P_R u^c(k_2) \nn\\
 && +\left[\sum_{i=1}^{n_L} \mathcal A_R(m_i) + \sum_{i=n_L+1}^{N} \mathcal A^{(9)}_R(m_i)\right]\, \bar u(k_1)  P_L u^c(k_2)\nn\\
 &&+ \left[\sum_{i=1}^{n_L} \mathcal A_M(m_i) + \sum_{i=n_L+1}^{N} \mathcal A^{(9)}_M(m_i)\right]\,  \bar u(k_1)  \gamma^0 \gamma^5 u^c(k_2) \Bigg\}\nn\\
 \equiv \frac{g_A^2 G_F^2 m_e}{\pi R_A} &\times& \left[ \mathcal A_L\,\bar u(k_1)  P_R u^c(k_2)  + \mathcal A_R\,\bar u(k_1)  P_L u^c(k_2)+\mathcal A_M\,  \bar u(k_1)  \gamma^0 \gamma^5 u^c(k_2)\right]\,,
 \eea
where $R_A = 1.2\, A^{1/3}$ fm is the nuclear radius in terms of the atomic number $A$ and $\bar u(k_{1,2})$ denote the spinors of the outgoing electrons.
  This factor is introduced to align the definitions of the NMEs to those in the literature. The subamplitudes $\mathcal A_{L,R,M}(m_i)$ and $\mathcal A^{(9)}_{L,R,M}(m_i)$ depend on nuclear and hadronic matrix elements, the neutrino masses, and the Wilson coefficients of the higher-dimensional operators. They are discussed in detail below. We have explicitly separated contributions from light neutrinos (with masses $m_i < \Lambda_\chi$) and heavy neutrinos ($m_i > \Lambda_\chi$). If corrections due to electron masses and momenta are kept, additional terms appear, see e.g. Ref.~\cite{Cirigliano:2017djv,Cirigliano:2018yza}. 
 
 \begin{table}
\center
\begin{tabular}{|c|cccc|}
\hline
\hline
\cite{Horoi:2017gmj}	    & $^{76}$Ge & $^{82}$Se & $^{130}$Te & $^{136}$Xe \\ 

\hline
$G_{01}$    & 0.22 & 1. & 1.4 & 1.5 \\
$G_{04}$    & 0.19 & 0.86 & 1.1 & 1.2 \\
$G_{06}$    & 0.33 & 1.1 & 1.7 & 1.8 \\
$G_{09}$    & 0.48 & 2. & 2.8 & 2.8 \\\hline
\hline
$Q/{\rm MeV} $ \cite{Stoica:2013lka} & 2.04& 3.0&2.5 & 2.5 \\
\hline\hline
\end{tabular}
\caption{Phase space factors in units of $10^{-14}$ yr$^{-1}$ obtained in Ref.~\cite{Horoi:2017gmj}. The last row shows the $Q$ value of \NLDBD\ for various isotopes, where $Q = M_i - M_f -2m_e$.}
\label{Tab:phasespace}
\end{table}
 
With the definitions of the amplitudes in Eqs.\ (\ref{eq:ampV}) and (\ref{amplitude}), we express the inverted half-life for $0^+\to 0^+$ transitions as
\bea\label{eq:T1/2}
\left(T^{0\nu}_{1/2}\right)^{-1} &=& g_A^4 \bigg\{ G_{01} \, \left( |\mathcal A_{L}|\sq + |\mathcal A_{R}|\sq \right)
- 2 (G_{01} - G_{04}) \textrm{Re} \mathcal A_{L}^* \mathcal A_{R} 
\nn\\
&&+ G_{09}\, |\mathcal A_{M}|\sq + G_{06}\, {\rm Re}\left[ (\mathcal A_{L} - \mathcal A_{R} )\mathcal A_{M}^*\right] \bigg\}\,. 
\eea
For a derivation of this formula we refer to Refs.~\cite{Doi:1985dx,Cirigliano:2017djv}. Here $G_{0j}$ are electronic phase-space factors given in Table~\ref{Tab:phasespace} that have been calculated in the literature \cite{Kotila:2012zza,Stefanik:2015twa,Horoi:2017gmj}. 

The subamplitudes depend on a product of Wilson coefficients and hadronic and nuclear matrix elements. To keep the expressions somewhat compact, we list here only the contributions from the standard mechanism and dim-6 interactions. Contributions from dim-7 interactions are given in Appendix \ref{Appdump}. The amplitude $\mathcal A_L$, which includes the standard mechanism, is given by  
\begin{eqnarray}\label{Anu}
\mathcal A_L(m_i) &=&  -\frac{1}{4 m_e} \Bigg\{ m_i\mathcal M_V(m_i) \left( C^{(6)}_{\rm VLL} + C^{(6)}_{\rm VRL} \right)^2_{ei} 
					+ m_i \mathcal M_A(m_i) \left( C^{(6)}_{\rm VLL} - C^{(6)}_{\rm VRL} \right)^2_{ei}  \nn \\
					& & 
 + \mathcal M_{PS}(m_i) \bigg[  m_i \frac{B^2}{m_\pi^2}  \left( C^{(6)}_{\rm SRR} -C^{(6)}_{\rm SLR} \right)_{ei}
  - 2  B \left(C^{(6)}_{\rm VLL} - C^{(6)}_{\rm VRL}\right)_{ei}   
\bigg] \left(C^{(6)}_{\rm SRR} - C^{(6)}_{\rm SLR}\right)_{ei} \nn\\
 & & + m_i \mathcal M_S(m_i) \left( C^{(6)}_{\rm SRR} + C^{(6)}_{\rm SLR} \right)^2_{ei} - m_i \mathcal M_T(m_i) \left( C^{(6)}_{\rm TRR} \right)^2_{ei} 
 \nn \\
& & +  m_N \mathcal M_{TV}(m_i) \left(C^{(6)}_{\rm VLL} + C^{(6)}_{\rm VRL}\right)_{ei} \left(C^{(6)}_{\rm TRR}\right)_{ei} 
  \Bigg\}+\mathcal A_L^{(\nu)}(m_i)\,,    
\end{eqnarray}
where $\mathcal M_i(m_i)$ are combinations of LECs and NMEs defined below and $B =\frac{m_\pi^2}{m_u+m_d}$ is an LEC introduced in Sect.\ \ref{sec:piLag}, also see Table \ref{Tab:LECs}.
Most of the terms above describe the long-distance contributions, while $\mathcal A_L^{(\nu)}$ is due to the exchange of hard neutrinos and given by
\bea
\mathcal A_L^{(\nu)}(m_i) &=& \frac{m_\pi^2}{m_e v} \Bigg[\left(\frac{C_{i\,L}^{\pi\pi}}{m_\pi^2}+ C_{i\,L}^{\prime\pi\pi}\right) \mathcal M_{PS,sd} + \frac{C_{i\,L}^{\pi N}-C_{i\,L}^{\prime\pi\pi}}{2}  \left( M^{AP}_{GT, \, sd}+M^{AP}_{T, \, sd} \right) \nn\\
&&- \frac{2}{g_A^2}  C^{NN}_{i\,L} M_{F,\, sd} \Bigg]\,,\label{eq:ALhardNu}
\eea
where the subscript `sd' on the NMEs refers to their short-distance nature and the combinations of couplings, $C^{\alpha}_{i,\beta}$, are defined in  Sect.\ \ref{sec:hardNu}.
The subamplitude for right-handed electrons, which does not appear for the standard mechanism, has a very similar structure. At the dim-6 level, this  amplitude can be obtained by exchanging $L \leftrightarrow R$
\bea
\mathcal A_R(m_i)  &=& \mathcal A_L(m_i)  \big{\rvert}_{L \leftrightarrow R}\,,
\eea
where $\mathcal A_R^{(\nu)}(m_i)$ can be obtained by replacing $C_{i\, L}^{\al}\to C_{i\, R}^{\al}$ in Eq.\ \eqref{eq:ALhardNu}.
Once dim-7 operators are included there appear differences between $\mathcal A_L(m_i) $ and $\mathcal A_R(m_i)$ due to the dim-7 tensor operators. The explicit formulae are given in Appendix \ref{Appdump}.

The ``magnetic'' subamplitude $\mathcal A_M$ is given by~\footnote{We dub this amplitude ``magnetic'' since in left-right symmetric models it is dominated by $\mathcal M_{VA}$ \cite{Doi:1985dx},
which is proportional to the nucleon magnetic moment $g_M$. Since these models inspired most of the early literature on non-standard contributions to $0\nu\beta\beta$, we retain the denomination magnetic.  
}
\begin{eqnarray}\label{AM}
\mathcal A_M(m_i) &=& 
\frac{1}{2 m_e} \Bigg\{  -m_N \mathcal M_{VA}(m_i) \left( C^{(6)}_{\rm VLL}\right)_{ei}\left( C^{(6)}_{\rm VLR}\right)_{ei} \nn \\
& &  +\frac{1}{2} m_i \mathcal M_S(m_i) \frac{g_V}{g_S} \left( C^{(6)}_{\rm VLL} + C^{(6)}_{\rm VRL} \right)_{ei} \left( C^{(6)}_{\rm SLL} + C^{(6)}_{\rm SRL} \right)_{ei}  \nn \\
& & +m_i \mathcal M_{T A}(m_i) 
 \left( C^{(6)}_{\rm VLL} - C^{(6)}_{\rm VRL} \right)_{ei} \left( C^{(6)}_{\rm TLL} \right)_{ei}  \nn \\
& & + \mathcal M_{T P}(m_i)
  B \left( C^{(6)}_{\rm SLL} - C^{(6)}_{\rm SRL} \right)_{ei} \left( C^{(6)}_{\rm TRR} \right)_{ei}- \left(L \leftrightarrow R\right) \bigg\}  +\mathcal A_M^{(\nu)}(m_i)\,.
\end{eqnarray}
Here $\mathcal A_M^{(\nu)}(m_i)$ again describes the contributions from hard neutrinos,
\bea
\mathcal A_M^{(\nu)} &=& \frac{m_\pi^2}{m_e v}  \bigg[  - \frac{2}{g_A^2}C^{NN}_{i\,V}   \, M_{F,\, sd} 
+ \frac{1}{2}C_{i\,V}^{\pi N}
\left( M^{AP}_{GT, \, sd}  + M^{AP}_{T, \, sd} \right) \bigg]\,.
\eea

Finally, we have the subamplitudes related to heavy neutrinos. Since they are induced by the same $\pi\pi$, $\pi N$, and $NN$ interactions as those arising from hard-neutrino exchange, the resulting amplitudes are very similar to those mentioned above. In particular, one can obtain the dim-9 amplitudes using the following replacements,
\bea\label{dim9}
\mathcal A_{L,R,V}^{(9)} =\mathcal A_{L,R,V}^{(\nu)}(m_i)\Bigg|_{C^\bt_{i\, L,R,V}\to c_{L,R,V}^\bt}\,.
\eea
The combinations of couplings $c_{\al}^{\bt}$ are defined in Sect.\ \ref{sec:ChiLag9}. 

In the above expressions we have defined the combinations of NMEs and LECs
\begin{eqnarray}\label{m1}
\mathcal M_{V}(m_i) &=&  -\frac{g_V^2}{g_A^2} M_F(m_i) + M^{MM}_{GT}(m_i) + M^{MM}_{T}(m_i)\,, \nn \\ 
\mathcal M_{A}(m_i) &=&  M^{AA}_{GT}(m_i) + M^{AP}_{GT}(m_i)  +M^{PP}_{GT}(m_i)  +  M^{AP}_{T}(m_i) + M^{PP}_{T}(m_i)\,, \nn\\
\mathcal M_{PS}(m_i)     &=& \frac{1}{2} M^{AP}_{GT}(m_i) + M^{PP}_{GT}(m_i)  + \frac{1}{2} M^{AP}_{T}(m_i) + M^{PP}_{T}(m_i)\, ,\nn \\
\mathcal M_{PS, sd} &=& \frac{1}{2} M^{AP}_{GT,\,sd}(0) + M^{PP}_{GT,\, sd}(0)  + \frac{1}{2} M^{AP}_{T,\,sd}(0) + M^{PP}_{T,\, sd}(0)\, ,\nn \\
\mathcal M_{S}(m_i) &=& \frac{g^2_S}{g^2_A} M_F(m_i) 
\, ,\nn \\
\mathcal M_{T}(m_i) &=& 16\frac{g^2_T}{g^2_A} M^{AA}_{GT}(m_i) \, , 
\end{eqnarray}
which arise from the insertions of the same currents on the nucleon lines, and the combinations
\begin{eqnarray}\label{m2}
  \mathcal M_{T\, V}(m_i) &=& -4\frac{g^\prime_T  g_V
  }{g_A^2} \frac{m^2_\pi}{m_N^2} M_{F,\, sd}(m_i)  + \frac{16 g_T}{ g_M} \left[ M_{GT}^{MM}(m_i) + M_T^{MM}(m_i)\right]\,, \nn \\
\mathcal M_{V\,A}(m_i) &=& 2\frac{g_A}{g_M}  \left[M_{GT}^{MM}(m_i) + M_{T}^{MM}(m_i) \right]\,, \nn \\
\mathcal M_{V\, S}(m_i)&=& \frac{g_V g_S}{g^2_A} M_F(m_i)
\, ,\nn \\
\mathcal M_{T\,A}(m_i) &=& \frac{g_T}{g_A}  \left[2M_{GT}^{AA}(m_i) + M_{GT}^{AP}(m_i) + M_T^{AP}(m_i) \right]\,, \nn \\
\mathcal M_{T\,P}(m_i) &=& \frac{g_T}{g_A}  \left[M_{GT}^{AP}(m_i) + M_T^{AP}(m_i) \right]\,,
\end{eqnarray}
which appear when two different currents interfere. We  explicitly denoted the dependence on the neutrino mass in Eqs.\ \eqref{m1} and \eqref{m2}. 

\section{Nuclear matrix elements}\label{sectionNME}
While the expressions for the subamplitudes given in the previous section look complex, they actually only depend on a relatively small set of structures. It is useful to introduce the Fourier-transformed functions in $r$-space 
\bea\label{hr}
h^{ab}_K(r, m_i)  &=& \frac{2}{\pi}R_A\int_0^{\infty} d|\vec q| \frac{\vec q^2}{\vec q^2 + m_i^2}\,h_K^{ab}(\vec q^2)\,j_\lambda(|\vec q|r)\, ,\nn\\
h^{ab}_{K,\, sd}(r, m_i)  &=& \frac{2}{\pi}\frac{R_A}{m_\pi^2}\int_0^{\infty} d|\vec q| \frac{\vec q^4}{\vec q^2 + m_i^2}\,h_K^{ab}(\vec q^2)\,j_\lambda(|\vec q|r)\,,
\eea
where $j_\lambda(|\vec q|r)$ are spherical Bessel functions, and the functions $h_K^{ab}(\vec q^2)$ are defined in Eq.~\eqref{smff} for 
$ K = \{F,\,GT,\,T\}$, and $ab = \{AA,AP,PP,MM\}$ for $K=GT$, $ab =\{AP,PP,MM\}$ for $K=T$, while for $K=F$ the $ab$ superscript should be ignored. Finally,  $\lambda=0$ for $K=\{F,\,GT\}$ and $\lambda =2$ for $K=T$. The factor of $R_A$ in Eq.~\eqref{hr} cancels against the $1/R_A$ in Eq.~\eqref{amplitude}. Note that the $h^{ij}_{K,\, sd}(r,m_i)$  are normalized using a factor of $m^{-2}_\pi$ instead of $(m_N m_e)^{-1}$ as was done in Ref.\ \cite{Hyvarinen:2015bda}. Apart from this rescaling, these definitions agree with the literature once we neglect the energy of the intermediate states, which is a subleading correction in chiral EFT.

We define the nuclear matrix elements (NMEs) from these functions via
\bea\label{NME}
M_{F,\,(sd)}(m_i) &=& \langle 0^+| \sum_{m,n}h_{F,\,(sd)}(r,m_i)\tau^{+(m)}\tau^{+(n)}|0^+\rangle\,,\nn\\
M^{ab}_{GT,\,(sd)}(m_i) &=& \langle 0^+| \sum_{m,n}h^{ab}_{GT,\,(sd)}(r,m_i)\,\boldsigma^{(m)}\cdot \boldsigma^{(n)}\,\tau^{+(m)}\tau^{+(n)}|0^+\rangle\,,\nn\\
M^{ab}_{T,\,(sd)}(m_i) &=& \langle 0^+| \sum_{m,n}h^{ab}_{T,\,(sd)}(r,m_i)\,S^{(mn)}(\hat {\vec r})\,\tau^{+(m)}\tau^{+(n)}|0^+\rangle\,,
\eea
where the tensor in coordinate space is defined as
\be
S^{(mn)}(\hat {\vec r}) = - \left(\boldsigma^{(m)}\cdot \boldsigma^{(n)}-3\,\boldsigma^{(m)}\cdot \hat{\vec r}\,\boldsigma^{(n)}\cdot \hat{\vec r}\right)\,.
\ee
The set of NMEs have been calculated with various nuclear many-body methods and for different isotopes in the limit $m_i \rightarrow 0$. We use calculations in the quasi-particle random phase approximation (QRPA) \cite{Hyvarinen:2015bda}, the Shell Model \cite{Menendez:2017fdf}, and the interacting boson model  \cite{Barea:2015kwa,Barea} and their values are given in Table~\ref{tab:comparison}. We focus on these particular calculations because, in those works, the results were presented in terms of the different components (i.e. $AA$, $AP$, $PP$, $MM$) of the $F$, $GT$, and $T$ long- and short-distance matrix elements. At leading order in Chiral EFT, not all NMEs are independent. The momentum dependence of $g_V(\vec q^2)$ and $g_A(\vec q^2)$ is a higher-order effect in $\chi$PT. Neglecting this dependence gives leading-order relations such as
\be
M^{AA}_{GT,\,sd}(0) = - 3 M_{F,\,sd}(0)\,.
\ee
This and other relations were obtained in Ref.~\cite{Cirigliano:2017djv}.

\begin{table}
\center
$\renewcommand{\arraystretch}{1.5}
\begin{array}{l||rrr|rr|rr |rr}
 \text{NMEs} & \multicolumn{3}{c|}{\text{}^{76} \text{Ge}} & \multicolumn{2}{c|}{\text{}^{82} \text{Se}} & \multicolumn{2}{c|}{ \text{}^{130} \text{Te}} & \multicolumn{2}{c}{  \text{}^{136} \text{Xe}}  \\
& \text{\cite{Hyvarinen:2015bda}} &   \text{\cite{Menendez:2017fdf}}  
& \text{\cite{Barea:2015kwa,Barea}} & \text{\cite{Hyvarinen:2015bda}} &    \text{\cite{Menendez:2017fdf}} & \text{\cite{Hyvarinen:2015bda}} &   \text{\cite{Menendez:2017fdf}} & \text{\cite{Hyvarinen:2015bda}} &   \text{\cite{Menendez:2017fdf}} \\
 \hline
 M_F 			   & $-$1.74    &  $-$0.59 	& $-$0.68 	& $-$1.29 	&  $-$0.55	& $-$1.52    	&   $-$0.67	& $-$0.89  	&   $-$0.54 \\
 M_{GT}^{AA} 		   & 5.48       &  3.15	    	& 5.06 		& 3.87 		&  2.97		& 4.28    	&   2.97	& 3.16   	&   2.45 \\
 M_{GT}^{AP} 		   & $-$2.02    & $-$0.94	& $-$0.92 	& $-$1.46    	& $-$0.89   	& $-$1.74 	&  $-$0.97	& $-$1.19   	&  $-$0.79  \\
 M_{GT}^{PP} 		   & 0.66  	&  0.30		& 0.24 		& 0.48          &  0.28 	& 0.59   	&   0.31   	& 0.39   	&   0.25\\
 M_{GT}^{MM} 		   & 0.51 	& 0.22		& 0.17 		& 0.37       	& 0.20 		& 0.45 		&  0.23 	& 0.31 		&  0.19 \\
 M_T^{AA} 	   	   &  -    	& - 		& - 		&  -         	& -		&  -     	&  -      	&  -     	&   -	\\
 M_T^{AP} 	   	   & $-$0.35 	& $-$0.01	& $-$0.31 	& $-$0.27    	& $-$0.01 	& $-$0.50	&     0.01	& $-$0.28 	&     0.01	\\
 M_T^{PP} 	 	   & 0.10     	&    0.00	& 0.09  	& 0.08       	& 0.00 		& 0.16 		&  $-$0.01 	& 0.09 		&  $-$0.01 	 \\
 M_T^{MM}           	   & $-$0.04	&   0.00	& $-$0.04	& $-$0.03    	&   0.00 	& $-$0.06 	&    0.00	& $-$0.03 	&    0.00  \\\hline
 M_{F,\, sd}  	   	   & $-$3.46 	& $-$1.46	& $-$1.1  	& $-$2.53   	& $-$1.37 	& $-$2.97 	&  $-$1.61 	& $-$1.53   	&  $-$1.28	\\
 M^{AA}_{GT,\, sd}  	   &    11.1 	& 4.87		& 3.62		& 7.98    	& 4.54 		& 	10.1 	&  5.31 	&    5.71    	&  4.25  \\
M^{AP}_{GT,\, sd}	   & $-$5.35 	& $-$2.26 	& $-$1.37 	& $-$3.82    	& $-$2.09 	& $-$4.94 	&  $-$2.51 	& $-$2.80  	&  $-$1.99  \\
M^{PP}_{GT,\, sd}	   & 1.99 	& 0.82		& 0.42		& 1.42      	& 0.77 		& 1.86 		&  0.92 	& 1.06  	&   0.74\\
M^{AP}_{T,\, sd}	   & $-$0.85 	&  $-$0.05	&  $-$0.97	& $-$0.65    	&  $-$0.05	& $-$1.50 	&   0.07	& $-$0.92  	&   0.05		\\  
M^{PP}_{T,\, sd} 	   & 0.32 	&  0.02		&  0.38		& 0.24       	&  0.02		& 0.58 		&   $-$0.02	& 0.36  	&   $-$0.02 \\
\end{array}$
\caption{
Comparison of NMEs computed in the quasi-particle random phase approximation  \cite{Hyvarinen:2015bda}, shell model  \cite{Menendez:2017fdf}, and interacting boson model  \cite{Barea:2015kwa,Barea}
for several nuclei  of experimental interest.  All NMEs are evaluated at $m_i =0$. The NMEs are defined in Eq.\ \eqref{NME}.}
\label{tab:comparison}
\end{table}

\subsection{Interpolation formulae}\label{interpolation}

To calculate \NLDBD\ decay rates when sterile neutrinos are present, we require an understanding of the $m_i$ dependence of the NMEs. For certain linear combinations of the NMEs given above, this mass dependence has been explicitly calculated \cite{Blennow:2010th, Barea:2015zfa}, but the results are often not split up as in Table \ref{tab:comparison}. Furthermore, in certain cases we also require the derivative of the NMEs with respect to the neutrino masses. Inspired by Ref.~\cite{Barea:2015zfa}, we therefore construct an  interpolation formula for the $m_i$ dependence of the NMEs. For most NMEs, we know the behavior in the small and large neutrino-mass limits. For instance, for $M_F(m_i)$ we have
\bea\label{NMElimits}
\lim_{m_i\rightarrow 0} M_{F}(m_i) &=& M_F\,,\qquad
\lim_{m_i\rightarrow \infty} M_{F}(m_i) = \frac{m_\pi^2}{m_i^2} M_{F,\,sd}\,.
\eea
Note that when we give an NME without an $(m_i)$ dependence, it is implied that it is an NME given in Table~\ref{tab:comparison} corresponding to $m_i=0$. 
We stress that the meaning of an NME becomes ambiguous for $m_i \gtrsim \Lambda_\chi$. For example, in the standard mechanism, the contributions are proportional to the NME $\mathcal M_V+\mathcal M_A$. However, for large neutrino masses, $m_i \gtrsim \Lambda_\chi$, the neutrino mass eigenstate should be integrated out before matching onto $\chi$EFT, which would contribute via the dim-9 operators in Eqs.~\eqref{dim91}-\eqref{dim93}, and its effects are no longer captured by $\mathcal M_V+\mathcal M_A$ alone. This implies that the correct large-$m_i$ limit is not necessarily equivalent to naively taking the $m_i\to\infty$ limit in the NMEs. We discuss this issue in detail in the next section. 

To nevertheless capture the $m_i$ dependence of the NMEs in the $m_i\lesssim \Lambda_\chi$ region, we can construct a simple Pad\'e approximation of order $(0,1)$ that interpolates between the limits in Eq.~\eqref{NMElimits}
\be\label{intsimple}
M_{F\,\mathrm{int}}(m_i) = M_{F,\, sd}\frac{m_\pi^2}{m_i^2 + m_\pi^2 \frac{M_{F,\, sd}}{M_F}}\,.
\ee
We can do the same for $M^{AA}_{GT,T\,\mathrm{int}}(m_i)$, $M^{AP}_{GT,T\,\mathrm{int}}(m_i)$, and $M^{PP}_{GT,T\,\mathrm{int}}(m_i)$. The formulae for the tensor NMEs are less reliable due to smallness and large model dependence of the tensor NMEs. The functional form in Eq.~\eqref{intsimple} was used in Ref.~\cite{Barea:2015zfa} and was shown to agree well with the explicit $m_i$ dependence calculated in the interacting boson model. 

\begin{figure}
\begin{center}
\includegraphics[scale =.7 ]{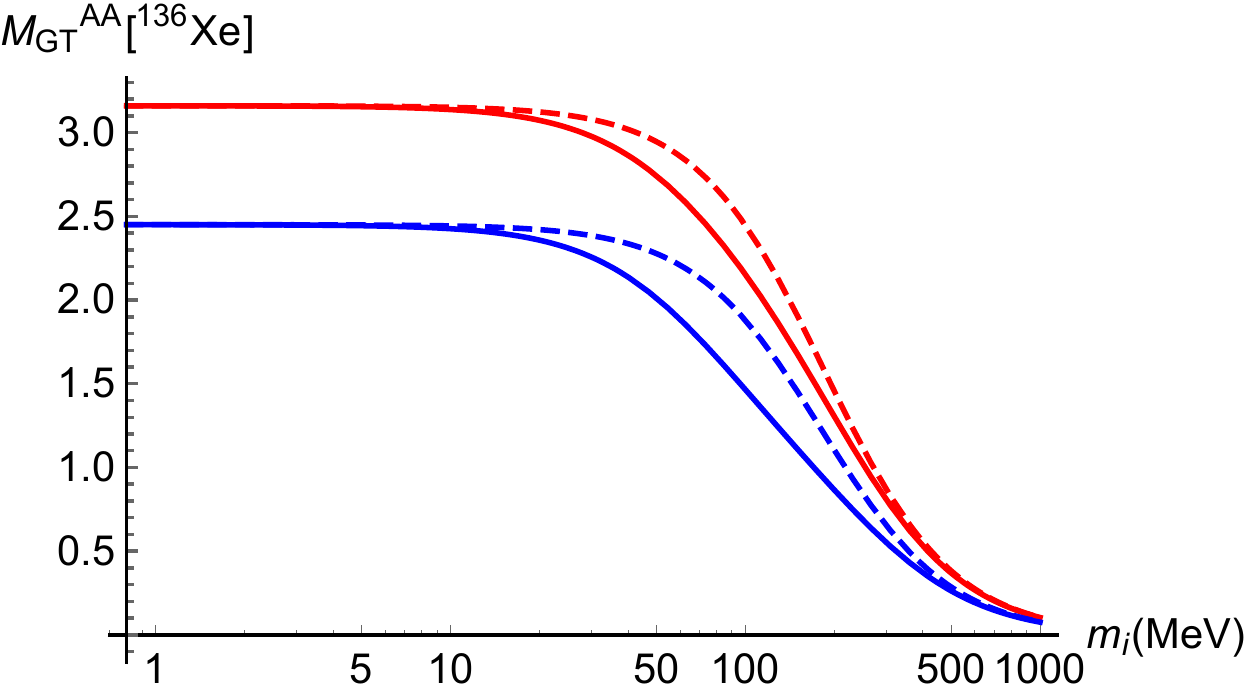}
\caption{The NME $M^{AA}_{GT}(m_i)$ for ${}^{136}$Xe from the interpolation formula in Eq.~\eqref{intsimple} (dashed) and Eq.~\eqref{intcomplex} (solid) using the quasi-particle random phase approximation 
\cite{Hyvarinen:2015bda} (red) and the Shell Model \cite{Menendez:2017fdf} (blue).}\label{MGTAA_interpolation}
\end{center}
\end{figure}

In case of $M^{AA}_{GT}(m_i)$ we can use additional information to further constrain the interpolation formula
\bea\label{GTtest}
M^{AA}_{GT}(m_i=m_\pi) & =&  -\frac{3}{2}M^{AP}_{GT}\,,\nn\\
\frac{\partial}{\partial m_i^2} M^{AP}_{GT}(m_i) \bigg|_{m_i=m_\pi} &=& \frac{3}{m_\pi^2}\left(M^{PP}_{GT} + \frac{1}{2} M^{AP}_{GT}\right)\,.
\eea
With additional constraints, we can construct an order $(1,2)$  Pad\'e  approximation 
\be\label{intcomplex}
M_{\mathrm{int2}}(m_i) = \frac{a_0 + a_1 m_i^2/m_\pi^2 }{1 + b_1 m_i^2/m_\pi^2+ b_2 m_i^4/m_\pi^4}\,.
\ee
For an NME, $M$, with the large $m_i$ behavior of Eq.\ \eqref{NMElimits}, the coefficients are given by
\bea
a_0&=& M(0)\,, \qquad a_1 = M_{sd}\frac{M(m_\pi)^2-M(0)\left(M(m_\pi)+M'(m_\pi)\right)}{M(m_\pi)^2+M'(m_\pi)M_{sd}}\,,\nn\\
b_1 &=& \frac{a_0+a_1}{M(m_\pi)}-(b_2+1)\,,\qquad b_2 = a_1/M_{sd}\,,
\eea
where $M'(m_\pi) \equiv \frac{\partial M(m_i)}{\partial \ln m_i^2}\Big|_{m_i=m_\pi}$.

In Fig.~\ref{MGTAA_interpolation} we plot $M^{AA}_{GT\,\mathrm{int}}(m_i)$ and $M^{AA}_{GT\,\mathrm{int2}}(m_i)$ for neutrino masses between 1 MeV and 1 GeV for ${}^{136}$Xe based on two nuclear many body methods. The two interpolation formulae agree within $25\%$ over the whole range of neutrino masses, and the associated spread is smaller than the spread between different many-body methods. We therefore use Eq.~\eqref{intsimple} for the NMEs where we do not have sufficient information to construct the more accurate approximation in Eq.~\eqref{intcomplex}. 

Armed with these interpolation formulae it is straightforward to obtain the $m_i$ dependence of the remaining NMEs in Table~\ref{tab:comparison}. For the magnetic GT NME we use
\bea
M^{MM}_{GT}(m_i) &=& \frac{g_M^2}{6 g_A^2} \left[\frac{m_\pi^2}{m_N^2} M^{AA}_{GT,\,sd}  - \frac{m_i^2}{m_N^2}M^{AA}_{GT}(m_i)\right]\,,
\eea
while for the short-distance NMEs we obtain
\bea
M_{F,\,sd}(m_i) &=& M_{F,\,sd}- \frac{m_i^2}{m_\pi^2}M_{F}(m_i)\,,\nn\\
M^{ab}_{GT,\,sd}(m_i) &=& M^{ab}_{GT,\,sd}- \frac{m_i^2}{m_\pi^2}M^{ab}_{GT}(m_i)\,,\nn\\
M^{ab}_{T,\,sd}(m_i) &=& M^{ab}_{T,\,sd}- \frac{m_i^2}{m_\pi^2}M^{ab}_{T}(m_i)\,.
\eea

\subsubsection{$\mathcal O(m_i^2)$ corrections in the small neutrino mass limit}
\label{mi2_correction}
From the functional form of Eqs.~\eqref{intsimple} and \eqref{intcomplex} it is obvious that the NMEs quickly saturate for small neutrino masses, $m_i \ll m_\pi$, and become constant. However, in certain interesting cases it is important to understand how fast the functions become constant. For instance, as observed in Ref.~\cite{Blennow:2010th}, in scenarios with  light, $m_i \ll m_\pi$, sterile neutrinos and no additional higher-dimensional operators, i.e.\ $M_L=0$, the leading \nnpp\ transition operator is proportional to 
\be\label{vanish}
\sum_{i=1}^{N} \left(C_{\rm VLL}^{(6)} \right)^2_{ei}\,\left(\frac{m_i}{\vec q^2 + m_i^2}\right) \simeq  \frac{4 V_{ud}^2}{\vec q^2} \left[ P U m_\nu U^T P^T \right]_{ee} =  \frac{4 V_{ud}^2}{\vec q^2} (M_L^*)_{ee} = 0 \,,
\ee
and vanishes. In this case, the $m_i^2/\vec q^2$ correction is necessary to get a non-vanishing result. This correction can in principle be estimated from expanding the interpolation formulae in the small $m_i^2$ limit. We write for $m_i \ll m_\pi$
\be\label{NMEprime}
M^{ab}_{\{F,GT,T\}}(m_i) = M^{ab}_{\{F,GT,T\}} + M^{\prime\,ab}_{\{F,GT,T\}}\frac{m_i^2}{m_\pi^2}\,.
\ee
Using the interpolation formula in Eq.~\eqref{intsimple}, we can directly calculate $M^{\prime\,ab}_{\{F,GT,T\}}$ and we give the results in Table~\ref{tab:comparisonderivative} for ${}^{136}$Xe. We stress that the results for the derivatives for the NMEs in the small $m_i$ regime are associated with significant uncertainties even beyond those from the dependence on the nuclear many-body method. By using the interpolation formulae in Eq.~\eqref{intcomplex} for $M_{GT}^{AA}(m_i)$ instead of Eq.~\eqref{intsimple} leads to $\mathcal O(100\%)$ corrections in $M_{GT}^{\prime\,AA}$. More importantly, for neutrino potentials scaling as $1/\vec q^4$, contributions from ultrasoft neutrinos can be as important as those from potential neutrinos that are considered here. We leave these corrections to future work, but stress that our results for $M'$ should be taken as order-of-magnitude estimates. 
\begin{table}
\center
$\renewcommand{\arraystretch}{1.5}
\begin{array}{l|rr}
 \text{NMEs} ( \text{}^{136} \text{Xe}) 
  & \text{\cite{Hyvarinen:2015bda}} &   \text{\cite{Menendez:2017fdf}} \\
 \hline
 M'_F 			& 0.52  	&   0.23 \\
 M_{GT}^{\prime\,AA} 		  & $-$1.75   	&   $-$1.41 \\
 M_{GT}^{\prime\,AP} 			& 0.51   	&  0.31  \\
 M_{GT}^{\prime\,PP} 		    	& $-$0.14   	&  $-$0.09\\
 M_{GT}^{\prime\,MM} 		 	& $-$0.15 		&  $-$0.11
\end{array}$
\caption{
Comparison of the derivative of $\text{}^{136} \text{Xe}$ NMEs with respect to $m_i^2$ in the quasi-particle random phase approximation  \cite{Hyvarinen:2015bda} and shell model  \cite{Menendez:2017fdf}. The NMEs are defined in Eq.\ \eqref{NMEprime}.}
\label{tab:comparisonderivative}
\end{table}

\section{Neutrino mass dependence of subamplitudes}\label{interLEC}

The master formula in Eq.\ \eqref{amplitude}, combined with the results presented in Ref.\ \cite{Cirigliano:2018yza}, describes all possible contributions to $0\nu\beta\beta$
from sterile and active neutrinos, capturing both the regime of heavy sterile neutrinos, $m_{i} \gg \Lambda_\chi$,
and light sterile neutrinos $m_{i} \ll \Lambda_\chi$. In the former regime, the heavy neutrino is integrated out at the quark level, while in the latter regime it has to be kept as a degree of freedom in chiral EFT. In the region $m_i \sim \Lambda_\chi$, however, both  approaches are questionable.
Within chiral EFT, diagrams arising at the $n$-loop level give corrections $\sim \left(m_i/\Lambda_\chi\right)^{2n}$ and the loop expansion breaks down. Instead, when integrating out the heavy neutrino at the quark level, operators involving additional derivatives, $\sim \left(\frac{\partial}{m_i}\right)^n\times \Or^{(9)}$, cannot be neglected as they induce corrections $\sim \left(\Lambda_\chi\right/m_i)^{n}$. This implies  the $m_i \sim \Lambda_\chi$ regime is beyond the reach of chiral EFT and 
of perturbative QCD methods, and it is therefore hard to treat rigorously. In this section we discuss the $m_i$ dependence of the amplitudes in more detail and employ what is known of the amplitudes in the two regimes
to suggest approximate interpolation formulae 
that link the low- and high-mass regions.

Before discussing the interpolation it is useful to  consider the two regimes in an example involving  one neutrino mass eigenstate with mass $m_i$ which couples to left-handed electrons, and to right- and left-handed quark vector currents.
The low-energy amplitudes depend on the neutrino masses in two ways, explicitly through the light neutrino propagator, and implicitly via the mass dependence of the low-energy constants 
in Eqs.\ \eqref{Lpipishort} and \eqref{LNNshort}. 
The resulting $0\nu\beta\beta$ amplitude, valid in the region $m_i\ll \Lambda_\chi$, can be written as
\begin{eqnarray}\label{limit0}
\mathcal A_L &=& - \frac{m_i}{4 m_e}  \bigg\{  
2 C^{(6)}_{\rm VLL} C^{(6)}_{\rm VRL}  \left(\bar{\mathcal{ M}}_V(m_i) -\bar{ \mathcal {M}}_A(m_i)\right) \nn\\
&&\qquad+ 
\left[ \left(C^{(6)}_{\rm VLL}\right)^2 + \left(C^{(6)}_{\rm VRL}\right)^2 \right]  \left(\bar{\mathcal{ M}}_V(m_i) + \bar{\mathcal {M}}_A(m_i)\right)
\bigg\},
\end{eqnarray}
where we include the contributions from the hard neutrino exchange amplitudes, $\mathcal A_L^{(\nu)}(m_i)$,  in $\bar{ \mathcal{M}}_{V,A}(m_i)$.
In the limit of large neutrino masses, $\Lambda_\chi \ll m_i$, we can integrate out the neutrino in perturbation theory, as discussed in Sec.\ \ref{GeVneutrinos},
and consider the hadronization of the four-fermion operators with coefficients $C^{(9)}_{1\rm L}$, $C^{(9)\prime}_{1\rm L}$ and $C^{(9)}_{4\rm L}$. Using Eq.\ \eqref{dim9}, we find
\begin{eqnarray}\label{limit1}
\mathcal A_L^{(9)} &=&  
-\frac{1}{2 m_e } \left( C^{(6)}_{\rm VLL} m_\nu^{-1} C^{(6)}_{\rm VLL} + C^{(6)}_{\rm VRL} m_\nu^{-1} C^{(6)}_{\rm VRL}   \right) \\
& & \times 
\left\{  \frac{5}{6} m_\pi^2 g_1^{\pi\pi} \left( M^{PP}_{GT, \, sd} + 
M^{PP}_{T, \, sd} \right) + \frac{m_\pi^2}{2}  g^{\pi N}_{1}  \left( M^{AP}_{GT, \, sd} + 
M^{AP}_{T, \, sd} \right) - \frac{2}{g_A^2} m_\pi^2 g^{\rm NN}_1 M_{F,\, sd}  \right\} \nn \\
& & - \frac{1}{2m_e}\left( C^{(6)}_{\rm VRL} m_\nu^{-1} C^{(6)}_{\rm VLL} + C^{(6)}_{\rm VLL} m_\nu^{-1} C^{(6)}_{\rm VRL}\right)  \left\{  \frac{1}{2} g_4^{\pi\pi} \mathcal M_{PS,sd} 
- \frac{2}{g_A^2} m_\pi^2 g^{\rm NN}_4 M_{F,\, sd} \right\}\,,\nn
\end{eqnarray}
where $g_4^{\pi\pi} = \mathcal O(\Lambda_\chi^2)$ and $g^{\rm NN}_4 = \mathcal O(\Lambda_\chi^2/F_\pi^2)$, while $g^{\pi\pi,\,\pi \rm N,\, NN}_{1} = \mathcal O(1)$.

Although the amplitudes in Eqs.\ \eqref{limit0} and \eqref{limit1} look rather different from one another, one can see that they take similar forms in the large-$m_i$ limit. To naively take this limit for  the long-range contributions, as discussed in  Sect.\ \ref{interpolation}, we use
\begin{eqnarray}\label{eq:NMElimits}
\lim_{m_i \rightarrow \infty} M_K^{(ab)} &=& \frac{m^2_\pi}{m^2_i} M_{K\, sd}^{(ab)} , \quad  \lim_{m_i \rightarrow \infty} M^{AA}_{GT} = - 3 \frac{m^2_\pi}{m^2_i} M_{F,\, sd}\,,
\end{eqnarray}
and neglect the magnetic contributions which lead to short-range derivative operators, subleading in the power counting. 
Using the above expressions, we obtain naive estimates of the LECs $g_{1}^{\pi\pi}$, $g^{\pi N}_{1}$, and $g^{\rm NN}_{1}$ in Eq.\ \eqref{limit1}. By matching terms that depend on the same NMEs in Eqs.~\eqref{limit0} and \eqref{limit1}, which is equivalent to matching the $\pi\pi\to ee$ and $nn\to pp\,ee$ amplitudes, we obtain  
\begin{equation}\label{naive}
g_1^{\pi\pi} = \frac{3}{5},\quad  g_1^{\rm \pi N} = 1, \quad g_{1}^{\rm NN} =  \frac{1}{4} \left( 1 + 3 g_A^2 - 2 m_i^2 g_{\nu}^{\rm NN}(m_i) \bigg|_{m_i \geq \Lambda_\chi}\right)\,.
\end{equation}
These equations were obtained by setting $ \mathcal A_L^{(9)}=\mathcal A_L$ in the regime $m_i\gg \Lambda_\chi$ where the left-hand side should be reliable, while the right-hand side receives large contributions from loop diagrams that were neglected. This implies that Eq.\ \eqref{naive} can only give an order-of-magnitude estimate. Nevertheless,
neglecting the $g_{\nu}^{\rm NN}(m_i)$ contribution, we see that these estimates are consistent with the NDA expectation and coincide with the ``factorization'' approximation.
The value of $g_{1}^{\pi\pi}$ extracted from the lattice, $g_1^{\pi\pi}(\mu=2 \, {\rm GeV})=0.36$, is about 40\% smaller than Eq.\ \eqref{naive}.

While these estimates are not very accurate, they at least give the right scaling. This is not so clear for the LECs in the third line of Eq.\ \eqref{limit1}. For instance, we can obtain 
\be\label{g4NNmatch}
g_{4}^{\rm NN} =  \frac{1}{4}\left(  1 -3 g_A^2 - 2m_i^2 g^{\rm NN}_{\rm LR}(m_i)\bigg|_{m_i \geq \Lambda_\chi}\right)\,,
\ee
where the first two terms on the right-hand side are $\mathcal O(1)$, whereas Table ~\ref{Tab:LECs} tells us that $g_4^{\rm NN} = \mathcal O ((4\pi)^2)$. In similar fashion, we obtain
\be\label{g4match}
g_4^{\pi\pi} =  -4\frac{m_i^2 g^{\pi\pi}_{\rm LR}(m_i) }{F_\pi^2}\bigg|_{m_i \geq \Lambda_\chi} + \mathcal O(m_\pi^2)\,.
\ee
$g_4^{\pi\pi} =\mathcal O(\Lambda_\chi^2)$ which is not clear from the right-hand side. Similarly, it is not obvious that the $g_\nu^{\rm NN}$ term in Eq.\ \eqref{naive} scales the same as the left-hand side which is  $\Or(1)$. In all of these cases, the comparison of the naive limit of Eq.\ \eqref{limit0} with Eq.\ \eqref{limit1}  suggests  that the hard-neutrino LECs should have a non-trivial $m_i$ dependence.
As we will argue below for the case of $g_{\rm LR}^{\pi\pi}$, it is indeed the  $m_i$-dependence of these LECs which ensures that the matching relations are restored.

\subsection{A dispersive representation}\label{dispersive}

In the case of the $g^{\pi\pi}_{\rm LR}$ we can investigate its $m_i$ dependence by exploiting the isospin relation to the pion mass splitting. Modifying the dispersive representation derived in Ref.\ \cite{Knecht:1998sp}
to account for a massive neutrino, we find~\footnote{Here we  used $g^{\pi\pi}_{\rm LR}(m_i)$ as an `effective' LEC that captures both the hard-neutrino exchange contributions as well as the loop diagrams $\sim \left(m_i/\Lambda_\chi\right)^n$ which become non-negligible for $m_i\sim \Lambda_\chi$. Explicitly, we have $\mathcal A(\pi(q)\pi(-q)\to e_Le_L)|_{q=0}=8C^{(6)}_{\rm VLL}C^{(6)}_{\rm VRL}\frac{m_i }{v^4}\frac{g^{\pi\pi}_{\rm LR}(m_i)|_{\rm eff}}{F_\pi^2} \bar u_Lu_L^c $, such that $g^{\pi\pi}_{\rm LR}(m_i)|_{\rm eff} =g^{\pi\pi}_{\rm LR}(0)$ in the limit $m_i\to 0$. Here and in what follows we use the notation $g^{\pi\pi}_{\rm LR}(m_i)|_{\rm eff} \to g^{\pi\pi}_{\rm LR}(m_i) $.}
\begin{eqnarray}\label{gpipi}
g_{\rm LR}^{\pi\pi}(m_i) = \frac{1}{F_\pi^2} \frac{3}{32 \pi^2} \int_0^{\infty} d Q^2 \frac{Q^2}{Q^2+m_i^2} (- Q^2 \Pi_{\rm LR}(Q^2))\,,
 \end{eqnarray}
where $\Pi_{\rm LR}$ is the vacuum matrix element of the time-ordered product of a left-handed and right-handed current, see e.g. Ref.~\cite{Knecht:1998sp}. The correlator $\Pi_{\rm LR}$ is exactly zero in perturbation theory and in the chiral limit, making it an order parameter for spontaneous symmetry breaking. As such, 
the correlator falls off rapidly, $Q^4 \Pi_{\rm LR}(Q^2)\to 0$, as $Q^2 \rightarrow \infty$ and the integral in Eq.\ \eqref{gpipi} is finite. This behavior leads to the Weinberg sum rules, which are discussed for example in Ref.\ \cite{Knecht:1998sp}.
In the large-$N_c$ limit, the correlator can be modeled by an infinite sum of narrow axial and vector resonance contributions, subjected to the Weinberg sum rules
\begin{eqnarray}
&&- Q^2 \Pi_{\rm LR}(Q^2) = F_\pi^2  + \sum_A f_A^2 m_A^2 \frac{Q^2}{m_A^2 + Q^2} - \sum_V f_V^2 m_V^2 \frac{Q^2}{m_V^2 + Q^2}\,, \\
&&F_\pi^2= \sum_V f_V^2 m_V^2 -\sum_A f_A^2 m_A^2 \,,   \qquad   \sum_V f_V^2 m_V^4 -\sum_A f_A^2 m_A^4=0\,. \label{WSR1}
\end{eqnarray}
In this parametrization the integral in Eq.\ \eqref{gpipi} can be done explicitly and, after imposing the Weinberg sum rules, we obtain
\bea
g_{\rm LR}^{\pi\pi}(m_i) =\frac{3}{32\pi^2F_\pi^2}\left[ \sum_V \frac{f_V^2 m_V^6}{m_i^2-m_V^2}\log\frac{m_V^2}{m_i^2}-\sum_A \frac{f_A^2 m_A^6}{m_i^2-m_A^2}\log\frac{m_A^2}{m_i^2}\right]\,,
\eea
Considering a two-resonance model with one axial and one vector resonance, the Weinberg sum rules allow us to solve for $f_{V,A}$ in terms of the resonance masses and $F_\pi$.
Using $m_V = m_\rho = 770$ MeV and $m_A = m_{a_1}= 1.24$ GeV, and taking the massless neutrino limit, we obtain
\begin{eqnarray}\label{gpipi2}
g_{\rm LR}^{\pi\pi}(0)\bigg |_{2\mathrm{-res}} =  \frac{3}{32 \pi^2} \frac{m_A^2 m_V^2}{m_A^2 - m_V^2}  \log\left(\frac{m_A^2}{m_V^2}\right)  \simeq 1.02 \, F_\pi^2\,,
\end{eqnarray}
which is in reasonable agreement with the determination from the pion mass splitting, $g_{\rm LR}^{\pi\pi}(0) \simeq 0.8 \, F_\pi^2$, see Eq.~\eqref{pionmasssplitting}. 
Considering the large-$m_i$ limit one instead obtains
\begin{eqnarray}\label{gpipi3}
g_{\rm LR}^{\pi\pi}(m_i)\bigg |_{2\mathrm{-res}} \rightarrow  \frac{3}{32 \pi^2} \frac{m_A^2 m_V^2}{(m_A^2- m_V^2) m_i^2} \left( m_A^2 \log\left(\frac{m_i^2}{m_A^2}\right) - m_V^2 \log \left( \frac{m_i^2}{m_V^2}\right)  \right)\,.\end{eqnarray}
Using $m_A \sim m_V \sim \mathcal O(\Lambda_\chi)$ and $(4\pi)^2 = \Lambda_\chi^2/f_\pi^2$, for large neutrino masses the LEC scales as
\begin{eqnarray}\label{gpipi4a}
g_{\rm LR}^{\pi\pi}(m_i) \rightarrow \frac{\Lambda^2_\chi}{m_i^2} \, F_\pi^2\,,
\end{eqnarray}
and the left- and right-hand sides of Eq.~\eqref{g4match} are of the same size. 

We can be more precise  by taking into account additional constraints on $g_{\rm LR}^{\pi\pi}$. Firstly, we can consider the asymptotic limit of $\Pi_{\rm LR}$ for $Q^2 \rightarrow \infty$. In the chiral limit, the correlator can be obtained through the operator product expansion and it is dominated by the matrix elements of dimension-six operators
\cite{Braaten:1991qm,Knecht:1998sp}
\begin{eqnarray}\label{newsum}
\lim_{Q^2 \rightarrow \infty} Q^6 \, \Pi_{\rm LR}(Q^2) &=& 8 \pi^2 \frac{\alpha_s}{\pi}  \left\{
\langle 0 | \bar q_L^\alpha \tau^+ \gamma^\mu q_L^{\beta} \,  \bar q_R^\alpha \tau^- \gamma_\mu q_R^{\beta}| 0  \rangle 
- \frac{1}{N_c} \langle 0 |  \bar q_L \tau^+ \gamma^\mu q_L \,  \bar q_R \tau^- \gamma_\mu q_R | 0 \rangle \right\} \nn \\ 
&=& 8 \pi^2 \frac{\alpha_s}{4\pi} F_\pi^4 \left( g^{\pi\pi}_{5} - \frac{1}{N_c} g_{4}^{\pi\pi} \right)
= \sum_{V,A} \left( f_V^2 m_V^6 - f_A^2 m_A^6\right)\,.
\end{eqnarray}
Using this, we can rewrite Eq.\ \eqref{gpipi} for $m_i \gg \Lambda_\chi$ as
\begin{eqnarray}\label{gpipi4}
g_{\rm LR}^{\pi\pi}(m_i \gg \Lambda_\chi) &=& \frac{1}{F_\pi^2} \frac{3}{32 \pi^2} \int_0^{\Lambda^2} d Q^2 \frac{Q^2}{Q^2+m_i^2} (- Q^2 \Pi_{\rm LR}(Q^2))\nn\\
& &+ \frac{1}{F_\pi^2} \frac{3}{32 \pi^2} \int_{\Lambda^2}^{\infty} d Q^2 \frac{Q^2}{Q^2+m_i^2} (- Q^2 \Pi_{\rm LR}(Q^2))\,,
\end{eqnarray}
in the first term, $m_i \gg Q$, and we can drop $Q^2$ in the denominator, while in the second term we use the asymptotic expression in Eq.\ \eqref{newsum} for the correlator. This gives
\begin{eqnarray}\label{gpipi5}
g_{\rm LR}^{\pi\pi}(m_i \gg \Lambda_\chi) &=& -\frac{3}{16} F_\pi^2 \frac{\alpha_s}{\pi} \left( g_5^{\pi\pi} - \frac{1}{N_c} g_{4}^{\pi\pi} \right) \frac{1}{m^2_i} \log \frac{m_i^2}{\mu^2} \nn\\
& & - \frac{3}{32 \pi^2 F_\pi^2} \frac{1}{m_i^2} \sum_{V,A}  \left( f_A^2 m_A^6 \log \frac{m_A^2}{\mu^2} - f_V^2 m_V^6 \log \frac{m_V^2}{\mu^2}\right)\,.
\end{eqnarray}
The dependence on $\Lambda$ and $\mu$ cancels after applying the sum rule in Eq.~\eqref{newsum}.

Secondly, at large values of $m_i$, the expression for $g_{\rm LR}^{\pi\pi}$ has to reproduce the amplitude obtained from integrating out the neutrino at the quark level as done in  Sect.\ \ref{GeVneutrinos}. In this case, matching the $\pi\pi\to ee$  amplitudes gives, see Eq.~\eqref{g4match}, 
\begin{eqnarray}\label{gpipi6}
g_{\rm LR}^{\pi\pi}(m_i \gg \Lambda_\chi) &=& -\frac{F_\pi^2}{4 m_i^2} g^{\pi\pi}_{4}(m_i)\,.
\end{eqnarray}
Together with Eq.~\eqref{gpipi5} this gives an expression for $g^{\pi\pi}_{4}(m_i)$, which can be written in terms of $g^{\pi\pi}_{4}(\mu)$ using the RGE of this LEC, see Eq.~\eqref{RGE9scalar},
\bea
\frac{d}{d\ln \mu}  \bma g^{\pi\pi}_{4}\\g^{\pi\pi}_{5}\ema  &=& - \frac{\al_s}{4\pi}\,\bma 6/N_c&0\\-6&-12 C_F\ema^T \bma g^{\pi\pi}_{4}\\g^{\pi\pi}_{5}\ema\,,
\eea
and its perturbative solution
\begin{equation}\label{gpipi7}
g_{4}^{\pi\pi}(m_i) = g_4^{\pi\pi}(\mu) + \frac{3}{4} \frac{\alpha_s}{\pi} \left( g_5^{\pi\pi} - \frac{1}{N_c} g_{4}^{\pi\pi}\right) \log \frac{m_i^2}{\mu^2}\,.
\end{equation} 
By combining Eqs.~\eqref{gpipi5}-\eqref{gpipi7} we can derive an expression for the LEC $g_4^{\pi\pi}(\mu)$, 
\begin{eqnarray}\label{gpipi8}
g_{4}^{\pi\pi}(\mu) =  \frac{3}{8 \pi^2 F_\pi^4} \sum_{V,A} \left( f_A^2 m_A^6 \log \frac{m_A^2}{\mu^2} - f_V^2 m_V^6 \log \frac{m_V^2}{\mu^2}\right)\,.
\end{eqnarray}
Note that the $\log m_i^2$ dependence has dropped out in the above expression, as one would expect. By using the  two-resonance approximation and the corresponding Weinberg sum rules, the  above already allows for an estimate of the LEC,
\begin{eqnarray}\label{gpipi9}
g_{4}^{\pi\pi}(\mu)\bigg |_{2\mathrm{-res}} =  \frac{3}{8 \pi^2 F_\pi^2} \frac{m_V^2 m_A^2}{m_A^2-m_V^2} \left( m_A^2 \log \frac{m_A^2}{\mu^2} - m_V^2 \log \frac{m_V^2}{\mu^2}\right) 
\simeq  - 1.5 \, {\rm GeV}^2,
\end{eqnarray}
for $\mu=2$ GeV. This is in reasonable agreement with the direct lattice-QCD calculation $g_{4}^{\pi\pi}(\mu=2\,{\rm GeV}) = -1.9$ GeV$^2$ \cite{Nicholson:2018mwc}.

\begin{figure}[t!]
\center
\includegraphics[width=0.6\textwidth]{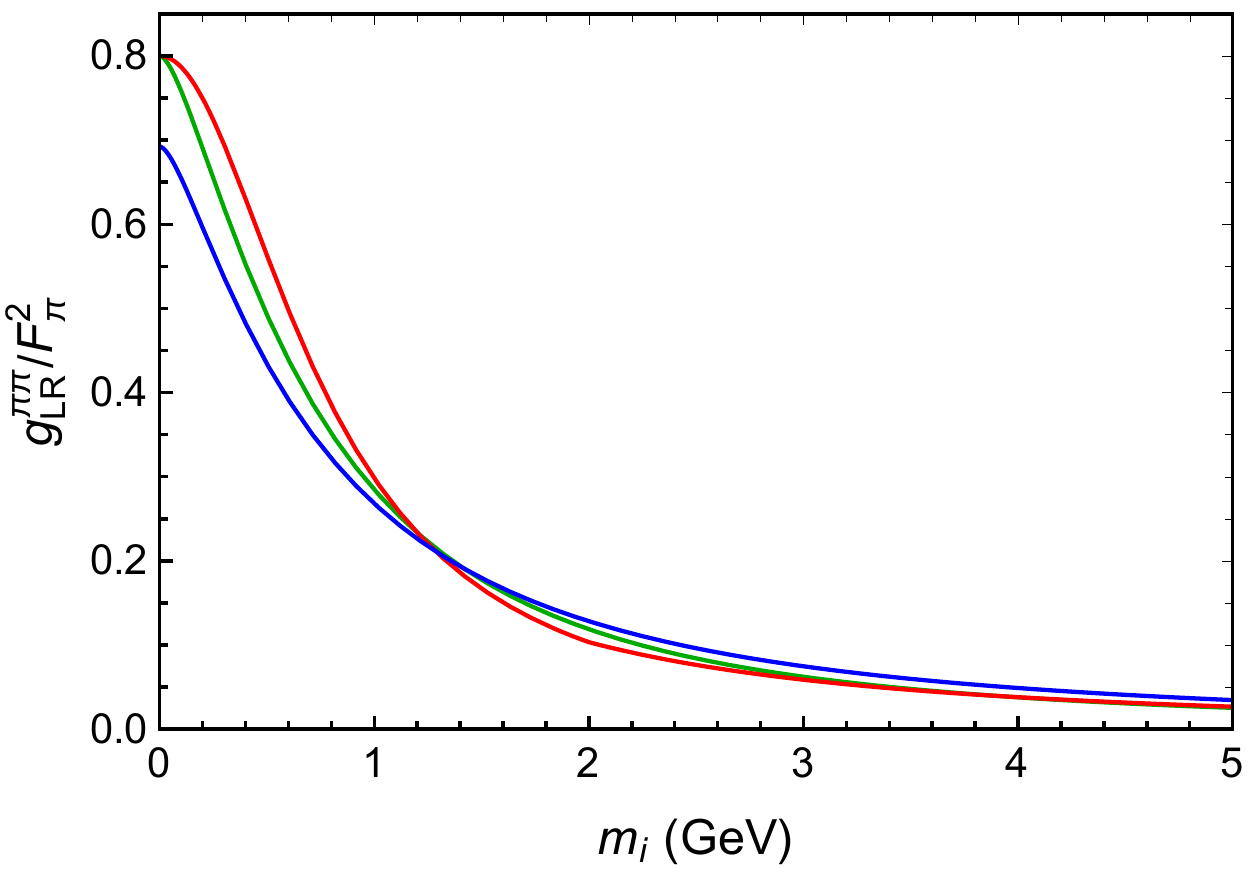}
\caption{The LEC $g^{\pi\pi}_{\rm LR}$ in units of $F_\pi^2$ as a function of the neutrino mass $m_i$. The blue and green lines denote the three- and five-resonance models, while the red line denotes the naive interpolation formula in Eq.\ \eqref{naiveinter0}. 
}\label{hadronicinterpolation}
\end{figure}

The final condition that can be imposed on $g_{\rm LR}^{\pi\pi}$ is the known behavior in the limit $m_i\to0$, where we have, $g^{\pi\pi}_{\rm LR}(m_i=0)  \simeq 0.8 F_\pi^2$, see Eq.\ \eqref{pionmasssplitting}. All in all, we then have five constraints, the two Weinberg sum rules Eq.\ \eqref{WSR1}, knowledge from the large $Q^2$ behavior of the correlator in Eq.\ \eqref{newsum}, and the high- and low-mass limits of $g^{\pi\pi}_{\rm LR}$, Eqs.\ \eqref{gpipi6} and \eqref{pionmasssplitting}. Clearly, not all of these constraints can be satisfied in the  two-resonance approximation, even though the predictions for the high- and low-mass limits,  Eqs.\ \eqref{gpipi9} and \eqref{gpipi2} are not very far off. To incorporate all constraints we can take into account five resonances, namely the $\rho$, $\rho(1450)$,  $\rho(1700)$, $a_1(1260)$, and $a_1(1640)$ \cite{Tanabashi:2018oca}, and we use the physical masses while fixing their couplings through the five constraints. We show the resulting  $g^{\pi\pi}_{\rm LR}$ in Fig.\ \ref{hadronicinterpolation} in green, while a three-resonance approximation, for which we do not include  $\rho(1700)$ and $a_1(1640)$ and do not impose Eqs.\ \eqref{pionmasssplitting} and \eqref{gpipi2}, is shown in blue. As is clear from the figure, the three- and five-resonance case approximations agree very well.

\subsection{A naive interpolation formula}\label{naiveIP}
It will prove useful to construct a simpler interpolation formula that can be applied to LECs for which we have less information or where the resonance model is not applicable. We follow a similar strategy as in the start of this section and impose
\bea\label{eq:intAmp}
\mathcal A_{A\, \rm int}(m_i)\big|_{m_i\gg \Lambda_\chi} &=& \mathcal{A}_{A}^{(9 )}(m_i)\,,
\eea
where $\mathcal A_{A\, \rm int}(m_i)=\mathcal A_{A}^{(ld)}(m_i) +\bar{\mathcal {A}}_{A\, \rm int}^{(\nu)}(m_i)$ with $A\in\{L,R,V\}$ and  $\mathcal A_{A}^{(ld)}(m_i)=\mathcal A_{A}^{}(m_i)-\mathcal A_{A}^{(\nu)}(m_i)$ is the purely long-distance part of the amplitude constructed in Sect.\ \ref{MasterFormula}. Instead, $\bar{\mathcal {A}}_{A\, \rm int}^{(\nu)}=\mathcal {A}_{A}^{(\nu)}\big|_{g_\al\to \bar g_\al}$  involves `effective' LECs, $\bar g_\al (m_i)$, that capture both the hard-neutrino exchange contributions as well as the loop corrections that become large in the $m_i\sim \Lambda_\chi$ regime. In the limit of zero neutrino mass, we have $\bar g_\al(0) = g_\al(0)$. In this way the interpolated amplitude has the correct limiting behavior in both the low- and high-mass regions.

The condition Eq.\ \eqref{eq:intAmp} can be used to obtain expressions for these `effective' hard-neutrino LECs at a scale $m_i\gg\Lambda_\chi$ in terms of LECs arising from dim-nine operators and possible long-distance contributions. To do so, we employ the large-$m_i$ limits of the NMEs, see Eq.\ \eqref{eq:NMElimits}, and demand that the contributions proportional to each NME in Eq.\ \eqref{eq:intAmp} match (this is equivalent to demanding that Eq.\ \eqref{eq:intAmp} not only holds for the \nnpp\ amplitude, but also for the $\pi\pi\to ee$ and $n\to p\pi \,ee$ subamplitudes). We then construct an interpolation formula
\bea\label{eq:intLECs}
 g_\al (m_i)\big|_{\rm naive} =  \frac{g_\al (0)}{1+g_\al (0)\left[\theta(m_0-m_i)\frac{m_0^2}{m_i^2}\bar g_\al (m_0)+\theta(m_i-m_0)\bar g_\al (m_i)\right]^{-1}}\,,
\eea
where $m_0\simeq 2$ GeV is a scale at which the procedure of integrating out the heavy neutrino becomes reliable and we use Eq.\ \eqref{eq:intAmp} to obtain expressions for $\bar g_\al (m_{0,i})$. The `effective' LECs in the large-$m_i$ region scale as $m_i^{-2}$, due to the $m_i^2$ in the denominator of the neutrino propagator. This scaling ensure that $ g_\al (m_i)\big|_{\rm naive}$   reduces to  $\bar g_\al (m_i)$  for $m_i\to \infty$, while it reproduces $g_\al (0)$ in the opposite, $m_i\to 0$, limit. Using $ g_\al (m_i)\big|_{\rm naive}$ in the interpolated amplitude, $\mathcal A_{A\, \rm int}(m_i)$, then allows us to obtain amplitudes for any value of $m_i$. 

Applying this procedure to the case of $g^{\pi \pi}_{\rm LR}$, we again obtain $\bar g^{\pi \pi}_{\rm LR}(m_i)|_{m_i\gg \Lambda_\chi} = -\frac{F_\pi^2}{4 m_i^2} g^{\pi\pi}_{4}(m_i)$ from Eq.\ \eqref{eq:intAmp}, which, in combination with Eq.\ \eqref{eq:intLECs}, leads to
\begin{eqnarray}\label{naiveinter0}
g^{\pi \pi}_{\rm LR} (m_i)\bigg |_{\rm naive} &=& \frac{g_{\rm LR}^{\rm \pi \pi}(0) }{1  -  m_i^2 \frac{4 g_{\rm LR}^{\rm \pi \pi}(0)}{F_\pi^2}\left[\theta(m_0-m_i)g^{\pi\pi}_{4}(m_0) +\theta(m_i-m_0)g^{\pi\pi}_{4}(m_i)\right]^{-1}   }\,,
\end{eqnarray}
where on the right-hand-side we use the observed value for $g_{\rm LR}^{\rm \pi \pi}(0)=0.8 F_\pi^2$ (see Eq.~\eqref{pionmasssplitting}). At $m_i = m_0 = 2$ GeV this function
approximates Eq.~\eqref{gpipi6} and it has the correct logarithmic dependence on $m_i$ in the perturbative QCD regime due to the $\theta(m_i-m_0)g^{\pi\pi}_{4}(m_i)$ term. This naive interpolation formulae is shown in red in Fig.~\ref{hadronicinterpolation}, where it is compared to the results from the three- and five-resonance models. The  formulae agree over the whole range of $m_i$ within $20\%$. Based on this success, we will use similar naive interpolation formulae for the other LECs.

It remains to understand the relations in Eqs.~\eqref{naive} and \eqref{g4NNmatch} and the $m_i$ dependence of $g_{\nu}^{\rm NN}(m_i)$ and $g^{\rm NN}_{\rm LR}(m_i)$.
These equations imply that $g_{\rm LR}^{\rm NN}$ is enhanced with respect to $g_{\nu}^{\rm NN}$ in the $m_i\geq \Lambda_\chi$ region. This can be understood from the different RGEs of these couplings. For $m_i \ll \Lambda_\chi$, the RGE for $g_{\rm LR}^{\rm NN}$ receives contributions from both light neutrino exchange and the $\pi \pi \rightarrow e^- e^-$ coupling, while $g_{\nu}^{\rm NN}$ only from the former
\begin{eqnarray}
\frac{d}{d \log \mu} \tilde g_{\nu}^{\rm NN}    &=& \frac{1}{2}(1 + 2 g_A^2)\,, \nn \\
\frac{d}{d \log \mu} \tilde g_{\rm LR}^{\rm NN} &=& \frac{1}{2}\left(1 - 2 g_A^2 + 2 g_A^2 \frac{g^{\pi\pi}_{\rm LR}(m_i)}{F_\pi^2} \right),\label{gnurge}
\end{eqnarray}
where $g_i^{\rm NN} = (m_N C/4\pi)^2 \tilde g_i^{\rm NN} \sim \tilde g_i^{\rm NN}/F_\pi^2$. While the first term is independent of $m_i$, the second term in the RGE of  $g_{\rm LR}^{\rm NN}$ scales as $\Lambda_\chi^2/m_i^2$ for large $m_i$ as can be seen explicitly from Eq.~\eqref{naiveinter0}. It is this behavior which ensures that the left- and right-hand side of Eq.~\eqref{g4NNmatch} match. 

We can now construct interpolation formulae for $g_{\nu}^{\rm NN}(m_i)$ and $g_{\rm LR}^{\rm NN}(m_i)$ similar to Eq.~\eqref{naiveinter0}, by using the matching conditions at large neutrino masses
\begin{eqnarray}\label{gNNmatch}
\bar g_{\nu}^{\rm NN}(m_i \gg \Lambda_\chi) &=& -\frac{2}{m_i^2}  \hat{g}^{\rm NN}_1(m_i) = -\frac{2}{m_i^2} \left( g^{\rm NN}_{1}(m_i) - \frac{1}{4} (1 + 3 g_A^2) \right)\,,\nn\\
\bar g_{\rm LR}^{\rm NN}(m_i \gg \Lambda_\chi) &=& -\frac{2}{m_i^2}  \hat{g}^{\rm NN}_4(m_i) = -\frac{2}{m_i^2} \left( g^{\rm NN}_{4}(m_i) - \frac{1}{4} (1 - 3 g_A^2) \right)\,,
\end{eqnarray}
to obtain
\begin{eqnarray}\label{naiveinterNN}
g^{\rm NN}_{\nu} (m_i)\bigg |_{\rm naive}  \!\!\!\!\!\!&=& \frac{g^{\rm NN}_{\nu}(0) }{1  -  m_i^2 \frac{g^{\rm NN}_{\nu}(0)}{2}\left[\theta(m_0-m_i)\hat g_{1}^{\rm NN}(m_0) +\theta(m_i-m_0)\hat g_{1}^{\rm NN}(m_i)\right]^{-1}   }\,,\nn\\
g^{\rm NN}_{\rm LR} (m_i)\bigg |_{\rm naive}   \!\!\!\!\!\!&=&  \frac{g^{\rm NN}_{\rm LR}(0)}{1  -  m_i^2 \frac{g^{\rm NN}_{\rm LR}(0)}{2}\left[\theta(m_0-m_i)\hat g_{4}^{\rm NN}(m_0) +\theta(m_i-m_0)\hat g_{4}^{\rm NN}(m_i)\right]^{-1}   }\,,
\end{eqnarray}
where we use the following RGEs to express  $g^{\rm NN}_{1,4}(m_i)$ in terms of $g_{1,4}^{\rm NN}(\mu \simeq 2\,\rm GeV)$
\begin{eqnarray}\label{gNNRGE}
g_{1}^{\rm NN}(m_i) &=& g_{1}^{\rm NN}(\mu) - \frac{3}{4} \frac{\alpha_s}{\pi} \left( 1 - \frac{1}{N_c}\right)g_{1}^{\rm  NN} \log \frac{m_i^2}{\mu^2}\,,\nn\\
g_{4}^{\rm NN}(m_i) &=& g_4^{\rm NN}(\mu) + \frac{3}{4} \frac{\alpha_s}{\pi} \left( g_5^{\rm NN} - \frac{1}{N_c} g_{4}^{\rm NN}\right) \log \frac{m_i^2}{\mu^2}\,.
\end{eqnarray}

\begin{figure}[t!]
\center
\includegraphics[width=0.45\textwidth]{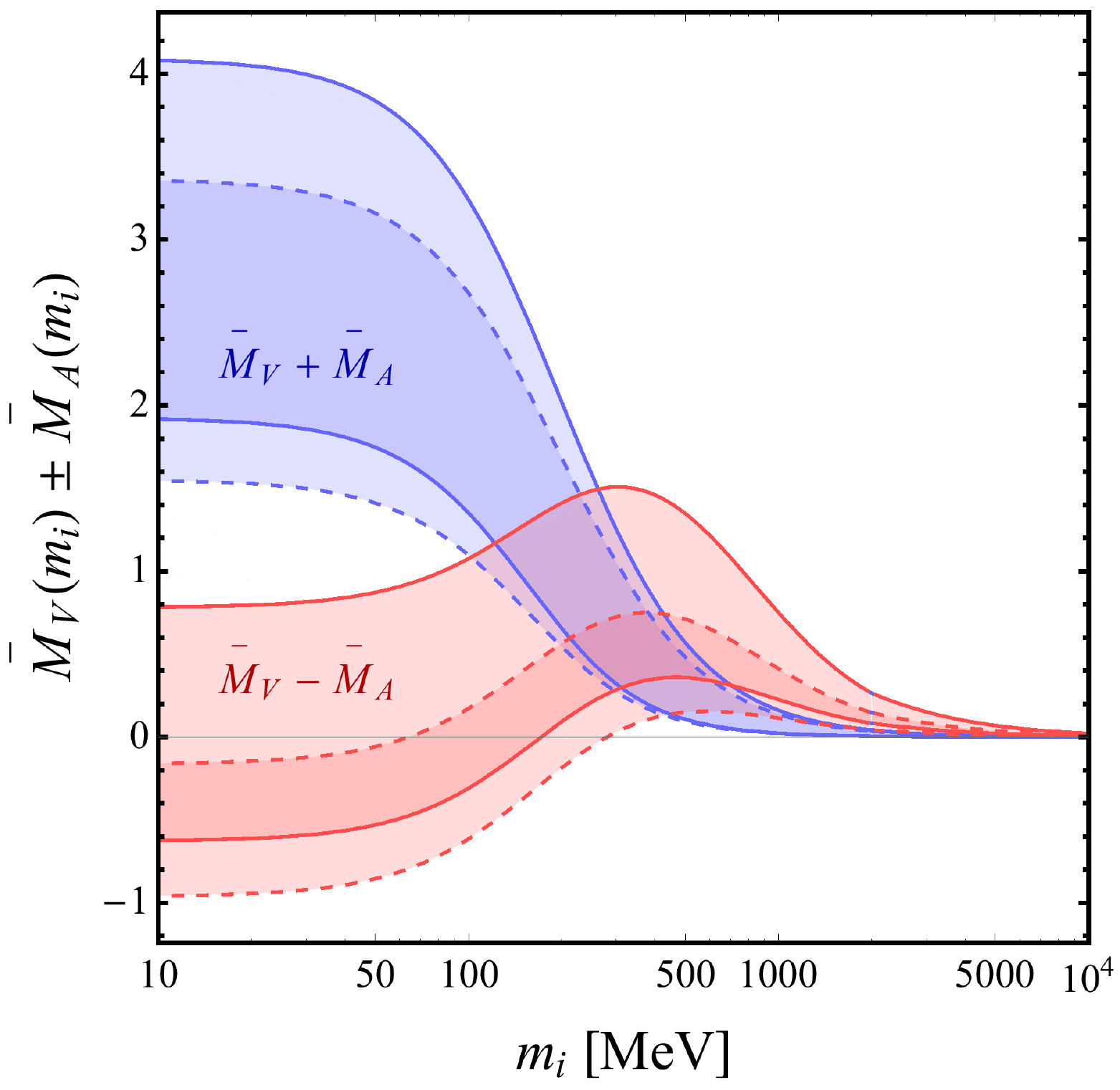}
\includegraphics[width=0.48\textwidth]{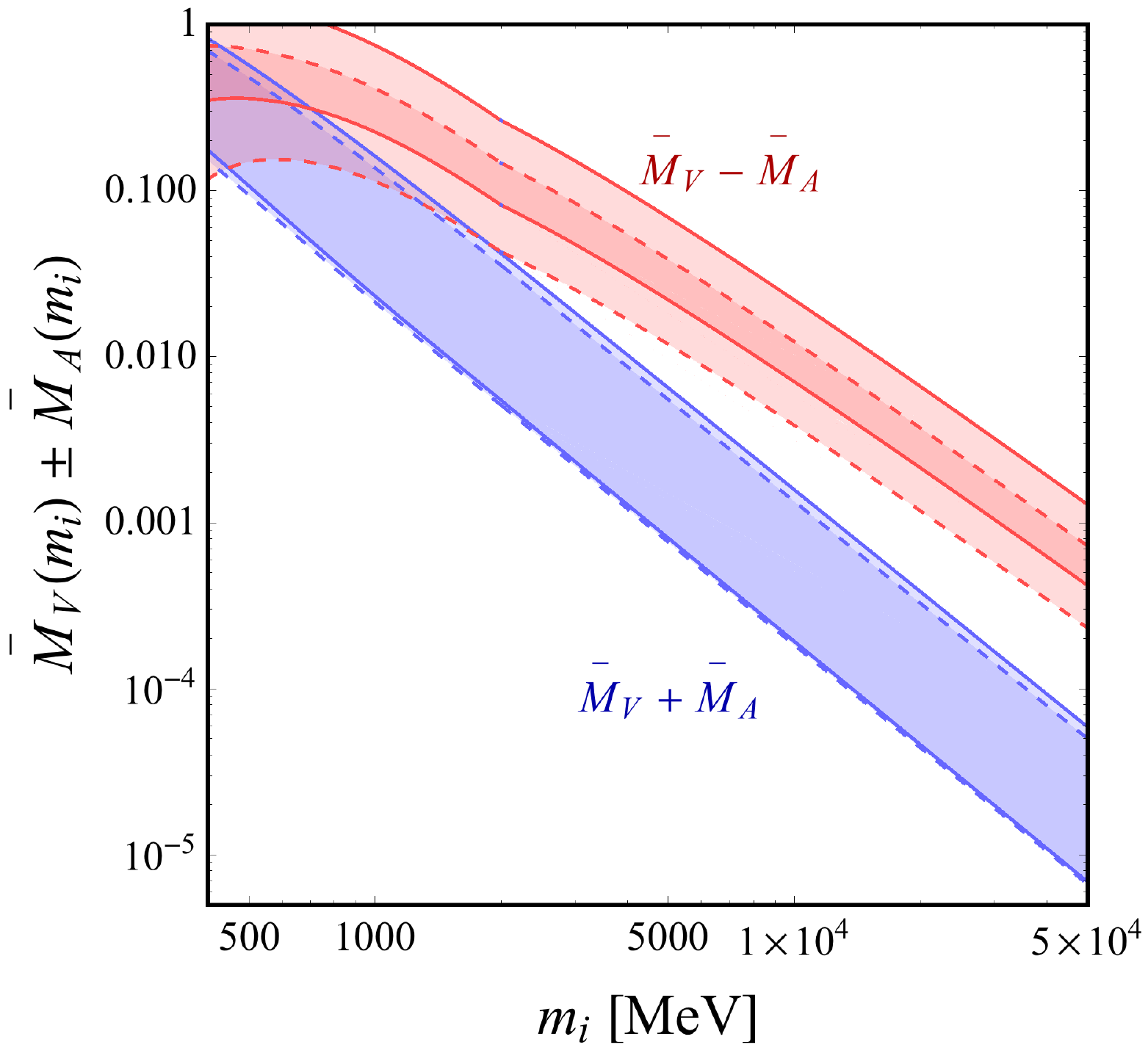}
\includegraphics[width=0.47\textwidth]{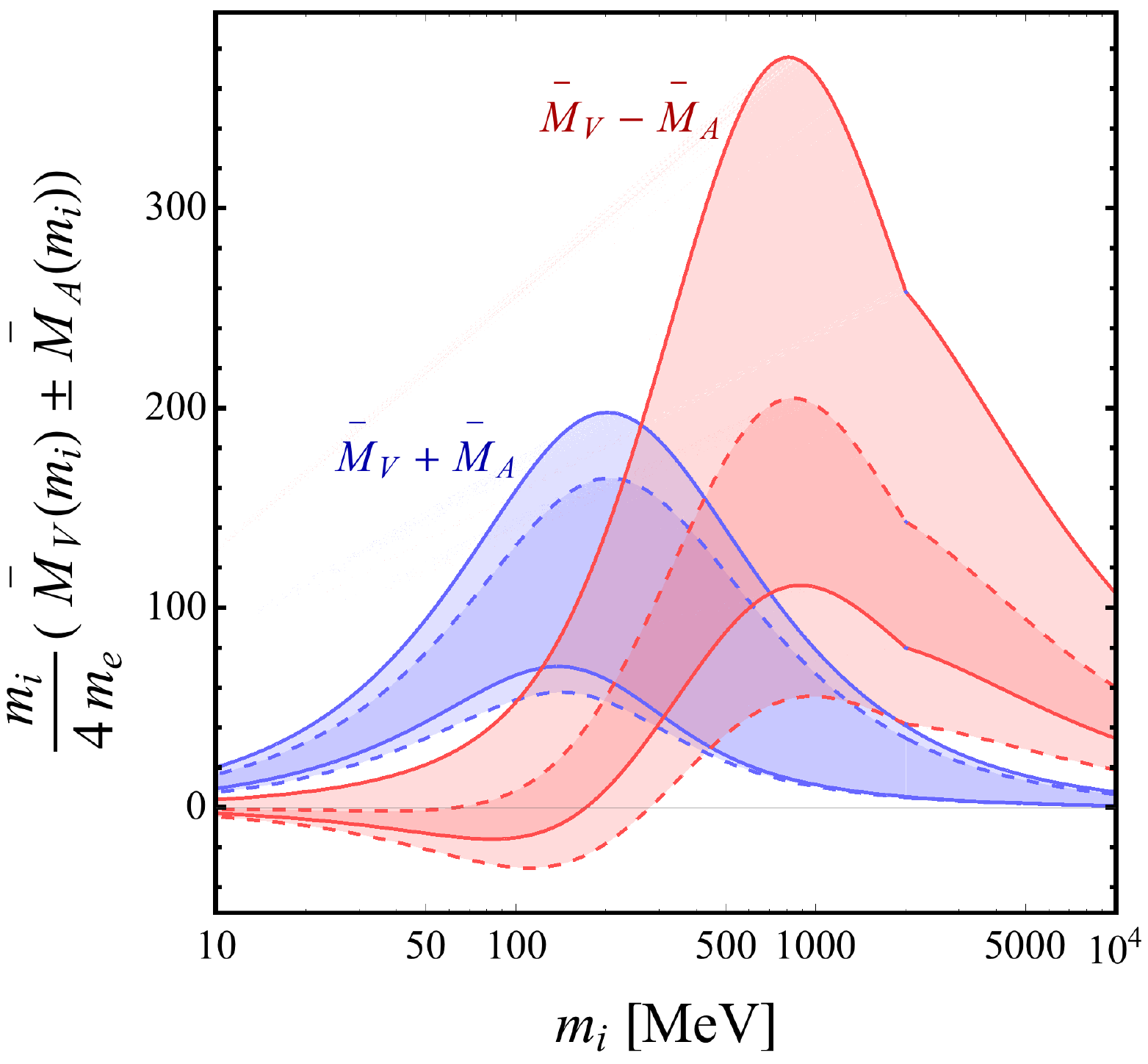}
\includegraphics[width=0.49\textwidth]{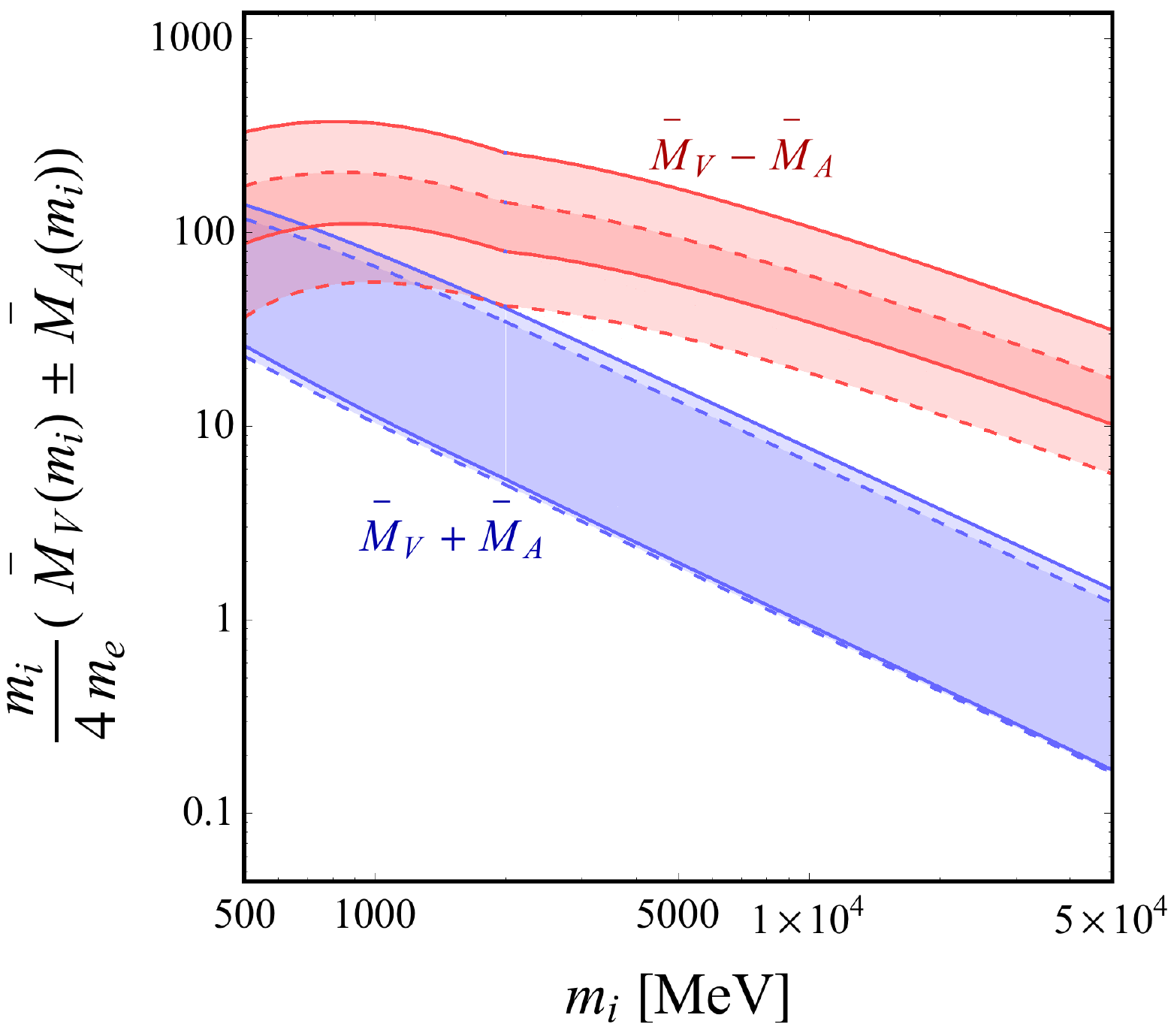}
\caption{ Top-left panel: the dependence of $\bar{\mathcal{M}}_V +\bar{\mathcal{M}}_A$ (blue) and $\bar{\mathcal{M}}_V -\bar{\mathcal{M}}_A$ (red) as a function of $m_i$ for ${}^{136}$Xe NMEs obtained with QRPA (solid) or the Shell Model (dashed). The bands reflect the uncertainty due to LECs associated with hard-neutrino exchange. Top-right panel: same as top-left panel but for larger neutrino masses. The relative enhancement of $\bar{\mathcal{M}}_V - \bar{\mathcal{M}}_A$ at large neutrino masses is due to the behavior of the LEC $g_{\rm LR}^{\pi\pi}(m_i)$. Bottom-left panel: the dependence of $(m_i/4m_e)(\bar{\mathcal{M}}_V \pm\bar{\mathcal{M}}_A)$ (blue) as a function of $m_i$ for ${}^{136}$Xe NMEs. Bottom-right panel: same as bottom-left panel but larger $m_i$ region.}
\label{AVpmA}
\end{figure}

Armed with the interpolation formulae we are now ready to calculate the $0\nu\beta\beta$ amplitude in Eq.~\eqref{limit0}, starting with the combination of NMEs and LECs $\bar{\mathcal{M}}_V \pm \bar{\mathcal{M}}_A$. The combination $\bar{\mathcal{M}}_V + \bar{\mathcal{M}}_A$, relevant for scenarios without higher-dimensional operators, depends on NMEs such as $M_F(m_i)$ for which we use interpolation formulae of the form in Eq.~\eqref{intsimple}. In addition, there is a dependence on $\bar g_\nu^{\rm NN}(m_i)$ for which we use Eq.~\eqref{naiveinterNN}. Unfortunately, the latter formula depends on two LECs, $g_\nu^{\rm NN}(0)$ and $g_1^{\rm NN}(m_0)$, that have not been determined with nonperturbative methods. 
For a discussion of the required lattice QCD calculation, as well as recent steps towards such a determination, we refer to Refs.~\cite{Cirigliano:2019jig,Drischler:2019xuo,Feng:2018pdq,Tuo:2019bue,Detmold:2018zan}. We use two reasonable choices for the LECs to assess the associated uncertainty. We use the NDA estimates 
\be
g_\nu^{\rm NN}(0) = \mp \frac{1}{(2 F_\pi)^2}\,,\qquad g_1^{\rm NN}(m_0) = \frac{1}{4}(1+3 g_A^2)  \pm 1\,,
\ee
where the latter choice is guided by the factorization approximation in Eq.~\eqref{naive}. 
The choice for $g_\nu^{\rm NN}(0)$ is dictated by the NDA expectation $g_\nu^{\rm NN}(0)\sim g_{\rm LR}^{\rm NN}(0)$,
and by the extraction of $g_{\nu}^{\rm NN}(0) + g_{\rm LR}^{\rm NN}(0)$ from isospin breaking in nucleon-nucleon scattering \cite{Cirigliano:2019vdj}.
As discussed in Ref.\ \cite{Cirigliano:2019vdj}, the value of  $(2 F_\pi)^2(g_{\nu}^{\rm NN}(0) + g_{\rm LR}^{\rm NN}(0) )$ 
varies between $0.2$ and $\sim 2.5$ in various high-quality chiral interactions, depending on the form and value of the ultraviolet regulators.

For the left-right combination $\bar{\mathcal{M}}_V -\bar{\mathcal{M}}_A$ we require, in addition to the usual NMEs, the `effective' LECs $\bar g_{\rm LR}^{\pi\pi}(m_i)$ and $\bar g_{\rm LR}^{\rm NN}(m_i)$ for which we use the interpolations in Eqs.\ \eqref{naiveinter0} and \eqref{gNNmatch}. The former is reasonably well understood, see Fig.~\ref{hadronicinterpolation},  while the latter is as uncertain as $g_\nu^{\rm NN}(0)$. In this case, we include this uncertainty by assigning a $50\%$ error on the contribution from $g_{\rm LR}^{\pi\pi}(m_i)$, as both effects are expected to appear at the same order. 
The dependence of $\bar{\mathcal{M}}_V \pm\bar{\mathcal{M}}_A$ on $m_i$ is depicted in the top panels of Fig.~\ref{AVpmA}. The blue and red bands correspond, respectively, to $\bar{\mathcal{M}}_V \pm\bar{\mathcal{M}}_A$ for ${}^{136}$Xe NMEs obtained with QRPA (solid) or the Shell Model (dashed). The bands are obtained by varying the LECs as discussed above and it is clear that these LECs, and not the NMEs, provide the dominant uncertainty. At small neutrino masses, $\bar{\mathcal{M}}_V + \bar{\mathcal{M}}_A$ and $\bar{\mathcal{M}}_V -\bar{\mathcal{M}}_A$ are of similar size with the former being a bit larger, but this behavior changes drastically once $m_i$ increases. Around a few hundred MeV, $\bar{\mathcal{M}}_V - \bar{\mathcal{M}}_A$ becomes larger due to $m_i$ dependence of $g_{\rm LR}^{\pi\pi}(m_i)$. 

In the bottom panels of Fig.~\ref{AVpmA} we plot the combinations $(m_i/ 4m_e)(\bar{\mathcal{M}}_V \pm \bar{\mathcal{M}}_A)$ that appear in the subamplitude $\mathcal A_L$. The amplitudes peak in the hundreds of MeV to GeV range, but the exact location depends on the underlying operators. The uncertainties are sizable, at the order-of-magnitude level, and dominated by uncertainties in the LECs. The amplitudes show a non-trivial behavior on $m_i$ which is in part due to the $m_i$ dependence of the `effective' hard-neutrino LECs. The contributions from these LECs are not included in interpolating formulae in
the literature, which are purely based on the $m_i$-dependence of the NMEs \cite{Barea:2015zfa}. This leads to significant differences for the case of $\bar{\mathcal{M}}_V-\bar{\mathcal{M}}_A$ where $g_{\rm LR}^{\pi\pi}$ dominates the $m_i\gtrsim \Lambda_\chi$ region. 

Similar interpolation formulae can be derived for the other LECs introduced in  Sect.\ \ref{sec:hardNuGen} by matching via Eq.\ \eqref{eq:intAmp} and employing the interpolation formula in Eq.\ \eqref{eq:intLECs}. This procedure allows us to smoothly interpolate between the $m_i \ll \Lambda_\chi$ and $m_i \gg \Lambda_\chi$ limits. In Appendix \ref{AppLQ} we discuss several cases that we require in Sect.~\ref{lepto}.

\section{Phenomenology}\label{pheno}
\begin{table}[t]
\begin{center}
\begin{tabular}{ c  c c c c c}
\hline\hline
Isotope & Experiment & \multicolumn{2}{c}{Current limit $(\times 10^{25} {\rm yr}) $} & \multicolumn{2}{c}{ Future sensitivity $(\times 10^{25} {\rm yr}) $}\\
\hline
$^{48}$Ca 	& ELEGANT$-$IV 		& $5.8\times 10^{-3}$ 	&\cite{Umehara:2008ru}   	& $-$ 		&\\
		& CANDLES 		& $6.2\times 10^{-3}$ 	&\cite{CANDLES_TAUP2019} 	& $ 10^{-2}$ 	&\cite{Iida:2016vfi}\\
		& NEMO$-3$ 		& $2.0\times10^{-3}$ 	&\cite{Arnold:2016ezh} 		& 		&\\
$^{76}$Ge 	& MAJORANA DEMONSTRATOR	& $2.7$ 		& \cite{Alvis:2019sil} 		& $-$		&\\
		& GERDA 		& $9.0$			&\cite{Agostini:2019hzm} 	&  $-$		&\\
		& LEGEND 		& $-$ 			&				& $10^3$  	&\cite{Abgrall:2017syy}\\
$^{82}$Se	& CUPID 		& $3.5\times10^{-1}$ 	&\cite{Azzolini:2019tta}  	& 		&\\
		& NEMO$-3$ 		& $2.5\times 10^{-2}$ 	&\cite{Arnold:2018tmo} 		&		& \\
		& SuperNEMO 		& $-$ 			&				& $10$		& \cite{Patrick:2017eso}\\
$^{96}$Zr 	& NEMO$-3$ 		& $9.2\times10^{-4}$    &\cite{Barabash:2010bd} 	&		&\\
$^{100}$Mo 	& NEMO$-3$ 		& $1.1\times10^{-1}$    &\cite{Arnold:2015wpy} 		& 		&\\
		& CUPID$-$1T 		& $-$ 			& 				&$9.2\times10^{2}$ &\cite{CUPIDInterestGroup:2019inu}\\ 
		& AMoRE 		& $9.5\times 10^{-3}$ 	&\cite{Alenkov:2019jis} 	& $5.0\times 10$ &\cite{Salvio:2019agg}\\
$^{116}$Cd 	& NEMO$-3$ 		& $1.0\times10^{-2}$ 	&\cite{Arnold:2016bed} 		&		&\\
$^{128}$Te 	& $-$ 			& $1.1\times10^{-2}$  	&\cite{Arnaboldi:2002te} 	& $-$ 		& \\
$^{130}$Te 	& CUORE 		& $3.2$ 		&\cite{Adams:2019jhp} 		& $9.0$ 	&\cite{Adams:2018nek}\\
		& SNO$+$ 		& $-$ 			& 				& $1.0\times10^2$ 	&\cite{Paton:2019kgy}\\
$^{136}$Xe	& KamLAND-Zen 		& $10.7$ 		&\cite{KamLAND-Zen:2016pfg} 	& $2.0\times10^2$ 	&\\
		& EXO$-200$ 		& $3.5$ 		&\cite{Anton:2019wmi} 		& $10^3$		& \cite{Albert:2017hjq}\\
		& NEXT 			& $-$ 			&				& $2.0\times 10^{2}$ 	&\cite{Gomez-Cadenas:2019sfa}\\
		& PandaX 		& $-$ 			&				& $1.0\times 10^2$ 	&\cite{Han:2017fol}\\
$^{150}$Nd 	& NEMO$-3$  		& $2.0\times10^{-3}$ 	&\cite{Arnold:2016qyg}  	& 		& \\
\hline
\end{tabular}
\caption{Current and future experimental limits on $T^{0\nu}_{1/2}$ at  $90\%$ C.L.}
\label{table:constraints}
\end{center}
\end{table}
We now turn to applications of the master formula in Eq.~\eqref{eq:T1/2} by investigating several scenarios involving sterile neutrinos. We emphasize that our purpose is not to find phenomenologically viable models of neutrino masses, but mainly to illustrate the use of the framework developed in this work. The search for sterile neutrinos is a very rich field with searches in a wide range of experiments, see e.g. Refs.~\cite{Barry:2011wb,Abazajian:2012ys,Helo:2018qej,Chrzaszcz:2019inj,Bolton:2019pcu,Bryman:2019bjg}, of which \NLDBD\ is only a small, but crucial, part. The framework presented here can be used directly in future global analyses of sterile neutrinos. 

In what follows we study several relatively simple scenarios. We start by considering minimal scenarios in which we extend the SM by one or two sterile neutrinos that are gauge-singlets and do not interact via higher-dimensional interactions. In these so-called $ 3+1$ and $3+2$ models, \NLDBD\ arises solely from the Majorana masses of the sterile neutrinos. We begin by studying whether \NLDBD\ can be measured in these minimal models and discuss the $m_i$ dependence of the resulting decay rates. After considering these cases, we turn on several higher-dimensional operators 
that are induced in BSM scenarios involving leptoquarks and determine the impact of such interactions on the \NLDBD\ predictions.

The current experimental bounds on the half-lives of various isotopes are summarized in Table~\ref{table:constraints}, where the expected future sensitivities are also shown.  In our numerical analyses, we use the limit on $T^{0\nu}_{1/2} (^{136}{\rm Xe})$ obtained by KamLAND-Zen, which is the strongest one at present, and take into account the following future prospects 
\begin{align}
T^{0\nu}_{1/2} \left(^{136}{\rm Xe}\right)&>2.0\times 10^{27}~[{\rm yr}]\hspace{0.2cm}~({\rm KamLAND2}-{\rm Zen})\,, \label{nEXO} \\
T^{0\nu}_{1/2} \left(^{136}{\rm Xe}\right)&>1.0\times 10^{28}~[{\rm yr}]\hspace{0.8cm}~({\rm nEXO})\,. \label{KamLAND2}
\end{align}
The prospects for the LEGEND experiment are of high interest as well, with an expected sensitivity of $T^{0\nu}_{1/2}\left(^{76}{\rm Ge} \right)\sim 10^{28}~$yr. 

\begin{figure}[t!]
\center
\includegraphics[scale=.45]{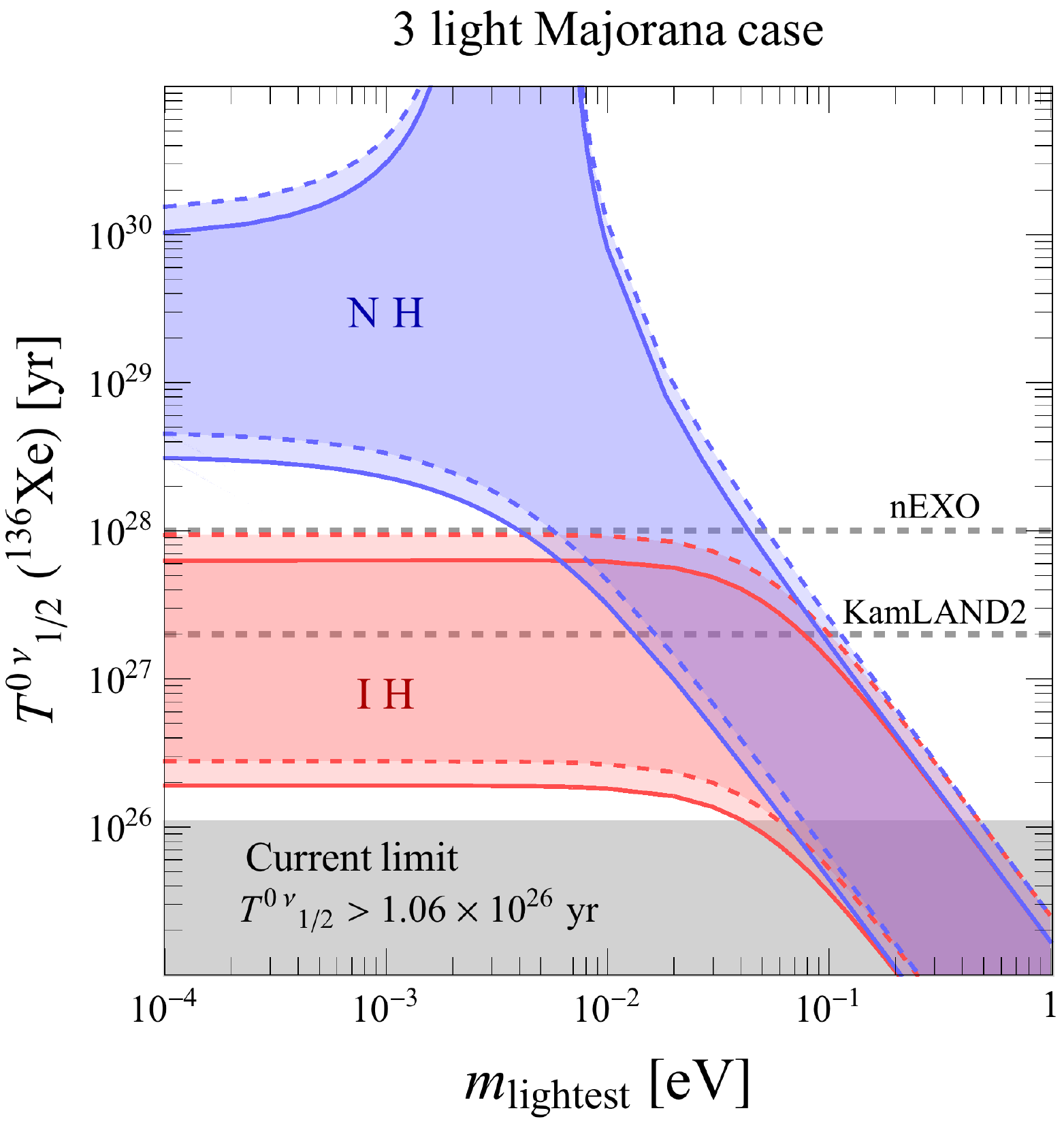}
\includegraphics[scale=.45]{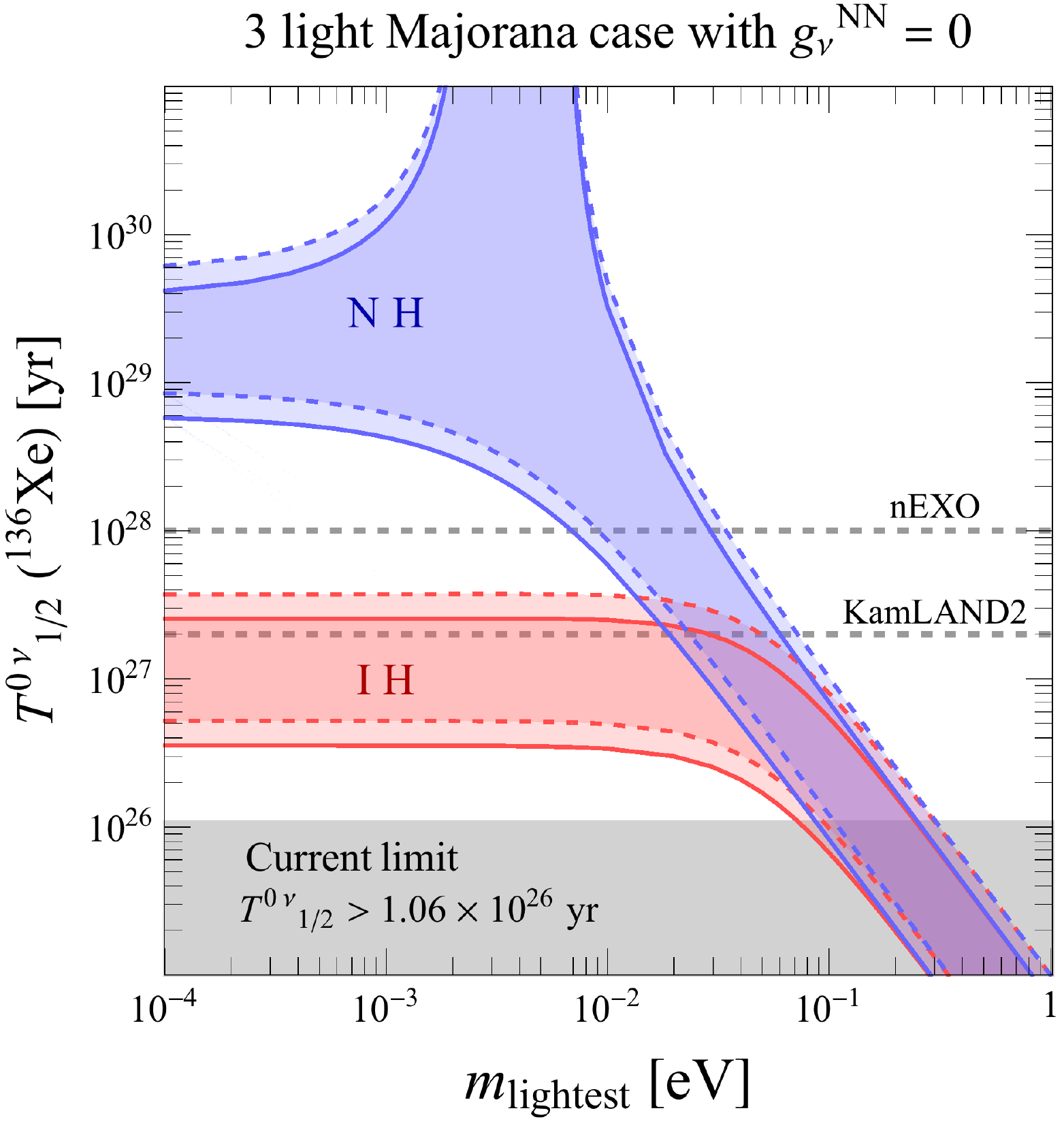}
\caption{$0\nu\beta\beta$ half-life of $^{136}$Xe as a function of the lightest neutrino mass in the scenario of 3 light Majorana neutrinos in case of the normal hierarchy (blue) and inverted hierarchy (red). We show results obtained with QRPA (solid) or the Shell Model (dashed) NMEs. In the left panel, the 
bands reflect the uncertainty due to LECs associated to hard-neutrino exchange and the unknown Majorana phases. In the right panel, we ignore the contributions from hard-neutrino exchange as typically done in the literature and the bands arise solely from varying the unknown Majorana phases. The shaded regions in both panels correspond to the present experimental limit, and expected future limits are depicted by the two dashed lines. }
\label{fig:Lifetime_3light}
\end{figure}

Before we begin analyzing the scenarios with sterile neutrinos mentioned above, we first briefly discuss the scenario of 3 light Majorana eigenstates, that is, the standard mechanism. This case corresponds to only turning on the LNV operators that generate Majorana masses for the left-handed neutrinos, $M_L$ in Eq.\ \eqref{DL2lag}. We use the standard parametrization and write 
\bea\label{eq:Uparam}
U = R^{23}W^{13}R^{12}{\rm diag}(1,e^{i\la_1},e^{i\la_2})\,,
\eea
 in terms of the rotation matrices $\left[W^{ab}(\theta_{ab},\dt_{ab})\right]_{ij} =\dt_{ij}+ (\dt_{ia}\dt_{jb}e^{i\dt_{ab}}-\dt_{ib}\dt_{ja}e^{-i\dt_{ab}})s_{ab}+ (\dt_{ia}\dt_{ja}+\dt_{ib}\dt_{jb})(c_{ab}-1)$ and $R^{ab}(\theta_{ab}) =W^{ab} (\theta_{ab},0)$, so that
\be
m_{\bt\bt}=\sum_{i=1}^3 m_i U_{ei}^2 = m_{1} c_{12}\sq c_{13}\sq +m_{2} e^{2i \lambda_1}s_{12}\sq c_{13}\sq +m_{3} e^{2i (\lambda_2-\dt_{13})}s_{13}\sq \,,
\ee
in terms of the sines (cosines) of the neutrino mixing angles, $s_{ij}$ ($c_{ij}$), the Dirac phase $\dt_{13}$, and the Majorana phases, $\lambda_{1,2}$. We set the mixing angles to their central values \cite{Tanabashi:2018oca} (see Table~\ref{table:BenchMark_angle}). The relevant subamplitude is $\mathcal A_L(m_i)$ that depends on the combination $\bar{\mathcal{M}}_V(m_i) + \bar{\mathcal{M}}_A(m_i)$, but since all mass eigenstates are at the eV scale or below, we actually only require $\bar{\mathcal{M}}_V(0) + \bar{\mathcal{M}}_A(0)$. This combination depends on the unknown LEC $g_\nu^{\rm NN}(0)$ associated with the exchange of hard neutrinos. This contribution is usually not considered in the literature, but as demonstrated in Refs.~\cite{Cirigliano:2018hja,Cirigliano:2019vdj} appears at the same order as the exchange of potential neutrinos. 
To calculate decay rates we marginalize over the Majorana phases and vary $g_\nu^{\rm NN}(0)$ between $\pm (2 F_\pi)^{-2}$ as discussed below Eq.~\eqref{naiveinterNN}. 

The resulting ${}^{136}$Xe half life is depicted in the left panel of Fig.~\ref{fig:Lifetime_3light} for the normal hierarchy (NH) in blue and inverted hierarchy (IH) in red as a function of the lightest neutrino mass, $m$. We used two sets of NMEs obtained with the QRPA (solid) and Shell Model (dashed). Around $m = 5 \cdot 10^{-3}$ eV, the usual `funnel' appears for the NH due to a possible cancellation in $m_{\bt\bt}$. For smaller $m$, the uncertainty on the half life is roughly two orders of magnitude for both the NH and IH. This uncertainty arises roughly in equal parts from the uncertainties in the LECs and Majorana phases, and in smaller amount from the change in NMEs between the QRPA and Shell Model. This can be seen more clearly by comparing to the right panel of Fig.~\ref{fig:Lifetime_3light} where we have set $g_\nu^{\rm NN}(0) =0$ and thus ignored contributions from hard-neutrino exchange as usually done in the literature. While the $\Or(1)$ contribution from the hard-neutrino LEC is consistent with the chiral expectations discussed in Sect.\ \ref{sec:hardNu}, it is possible that this effect turns out to be smaller when consistently evaluated in the many-body methods used in Refs.\ \cite{Hyvarinen:2015bda,Menendez:2017fdf,Barea:2015kwa,Barea} (see the discussion at the end of Sect.\ \ref{sec:3summary}). If this turns out to be the case, the uncertainty shown in the right panel of Fig.~\ref{fig:Lifetime_3light} is more appropriate. The fact that the bands in both the NH and IH are significantly smaller in this panel highlights the importance of pinning down the value of $g_\nu^{\rm NN}(0)$ with nonperturbative methods.

\subsection{3+1 model}

The simplest scenario we investigate is the $3+1$ model where the SM is extended by one gauge-singlet neutrino. That is $n=1$ and $N=4$. This model leads to two massless neutrinos and is thus ruled out by the combined atmospheric, $\Delta m^2_{\rm ATM}\simeq 2.4 \cdot 10^{-3}$ eV$^2$ and solar, $\Delta m^2_{\rm SOL}\simeq 7.5\cdot 10^{-5}$ eV$^2$, squared mass differences. 
Nevertheless, due to its simplicity, the scenario provides a useful toy model. In the flavor basis, we write the neutrino mass matrix as
\begin{align}
M_\nu = \bma 0 & 0 & 0 & M^*_{D,1}\\ 
		      0 & 0 & 0 & M^*_{D,2}\\
		      0 & 0 & 0 & M^*_{D,3}\\
		      M^*_{D,1} & M^*_{D,2}  & M^*_{D,3} & M_R		      
		    \ema\,,
\end{align}
and we set $M^*_{D,1} = M^*_{D,2} = M^*_{D,3} \equiv M^*_{D}$. The neutrino mass matrix is then described by just two parameters $M^*_D$ and $M_R$.  
Note that in the absence of higher-dimensional operators, $(M_\nu)_{ij}=0$ for $i,j=1,2,3$ is required by gauge invariance. 
This simple setup predicts two massless neutrinos, $m_1=m_2=0$, and two massive neutrinos described by
\begin{align}
m_3 = \frac{1}{2}\left[\sqrt{|M_R|^2 +12 |M_{D}|^2}-|M_R| \right]\,,\hspace{0.5cm}
m_4 = \frac{1}{2}\left[ \sqrt{|M_R|^2 +12 |M_{D}|^{2}}+|M_R| \right] \label{3p1_m4}\,,
\end{align}
where we assumed $m_3<m_4$, while the inverted relations are
\begin{align}
|M_R| =m_4 - m_3\,,\hspace{0.5cm} |M_D| = \frac{1}{\sqrt{3}} \sqrt{m_3 m_4}\,.
\end{align}
It is straightforward to diagonalize the mass matrix to obtain the PMNS matrix, which can be parametrized as \cite{Giunti:2019aiy}
\bea
U &=&D_L R^{34}R^{24}R^{23}R^{14}R^{13}W^{12}D_R\,,\nn\\
 D_{L}&=&e^{i(\al_D+\al_R/2)}{\rm diag}(1,1,1,e^{-i(\al_R+\al_D)})\,\qquad D_{R}={\rm diag}(1,1,i,1)\,,
\eea 
where $W^{ij}$ and $R^{ij}$ are defined below Eq.\ \eqref{eq:Uparam} and $\al_{D,R} = {\rm Arg}\,M_{D,R}$. In this fairly simple case nearly all the mixing angles can be expressed in terms of the neutrino masses
\bea
s_{23}=c_{34}/\sqrt{2},\,\quad s_{34}=t_{24},\,\quad s_{24}=s_{13}\sqrt{m_3/m_4}=t_{14}=\sqrt{\frac{m_3}{2m_3+3m_4}}\,,
\eea
where $t_{ij}=s_{ij}/c_{ij}$, while  $s_{12}$ is unconstrained due to the two vanishing neutrino masses. One useful result of the above is 
\bea
U_{e3}^2=-\frac{m_4}{m_3}U_{e4}^2 =- \frac{1}{3}\frac{m_4}{m_3+m_4}e^{i(2\al_D+\al_R)}\,.
\label{eq:3plus1}
\eea

As mentioned, this model of neutrino masses is too restrictive to reproduce the oscillation data. To nevertheless investigate the effect a sterile neutrino would have on \NLDBD\ in this scenario we  approximate $\Delta m^2_{\rm ATM} \simeq 0$, and set $m_3 =\sqrt{\Delta m^2_{31}} \simeq 0.05$ eV. 
For $m_4 < \Lambda_\chi$ the \NLDBD\ half life is then simply given by (Eq.~\eqref{eq:T1/2})
\begin{align}
\left(T^{0\nu}_{1/2} \right)^{-1}=g_A^4G_{01}\left|\sum_{i=1}^{4}{\cal A}_L(m_i)\right|^2\,,
\end{align}
where (Eqs.~\eqref{Anu}, \eqref{limit0})
\be
{\cal A}_L(m_i) = -\frac{m_i}{4m_e}  \left[\bar{\mathcal{M}}_V(m_i)+\bar{\mathcal{M}}_A(m_i)\right] (C_{\rm VLL}^{(6)})^2_{ei}\,,
\ee
and  (Eqs.~\eqref{redefC6} and \eqref{match6LNC})
\be
C_{\rm VLL}^{(6)} = -2 V_{ud} (P U)_{ei} =  -2 V_{ud} U_{ei} \,.
\ee
All other Wilson coefficients vanish. The combinations of NMEs and LECs $\bar{\mathcal{ M}}_{V,A}(m_i)$ are defined in Eq.~\eqref{m1}, with the inclusion of the short-range pieces in Eq.\ \eqref{eq:ALhardNu},
and their dependence on the neutrino masses is discussed in Sects.~\ref{interpolation} and \ref{interLEC}. For $m_4 > \Lambda_\chi$, instead, the half life becomes 
\begin{align}\label{m4heavy}
\left(T^{0\nu}_{1/2} \right)^{-1}=g_A^4G_{01}\left|\sum_{i=1}^{3}{\cal A}_L(m_i) + \mathcal A^{(9)}_L(m_4)\right|^2\,
\end{align}
as the fourth mass eigenstate is integrated out on the quark level. In this case $\mathcal A^{(9)}_L(m_4)$ is defined in Eqs.~\eqref{dim9}  and the neutrino mass dependence enters via
\be C_{1L}^{(9)} = -\frac{v}{2 m_4} (C_{\rm VLL}^{(6)})^2 = -\frac{2 v V_{ud}^2}{m_4}U^2_{e4}\,. \ee
The interpolation formulae described in Sect.\ \ref{naiveIP} ensure that the $m_4 \leq \Lambda_\chi$ and $m_4 \geq \Lambda_\chi$ limits smoothly match. 

\begin{figure}[t!]
\center
\includegraphics[scale =.5 ]{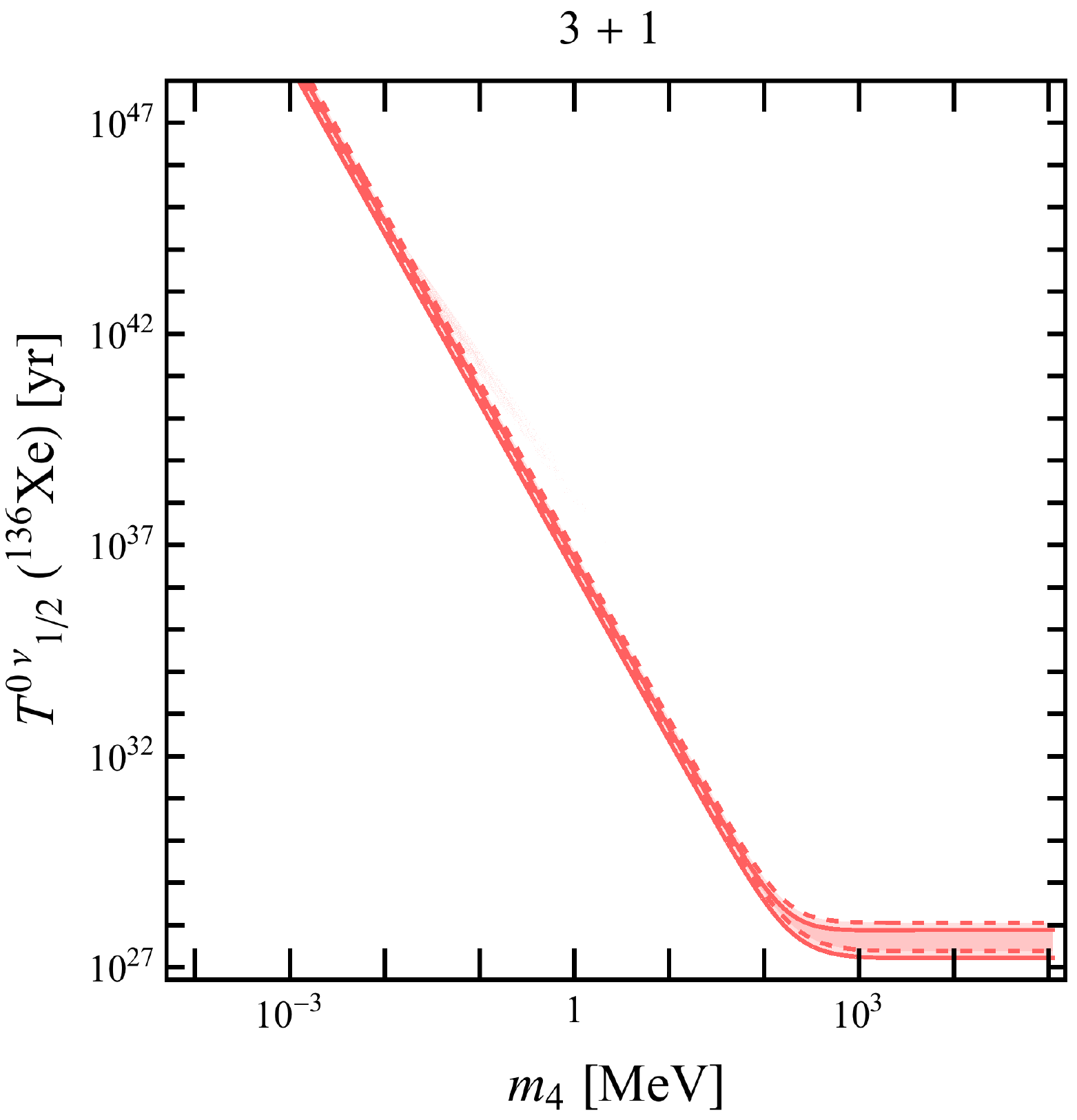}
\includegraphics[scale =.5 ]{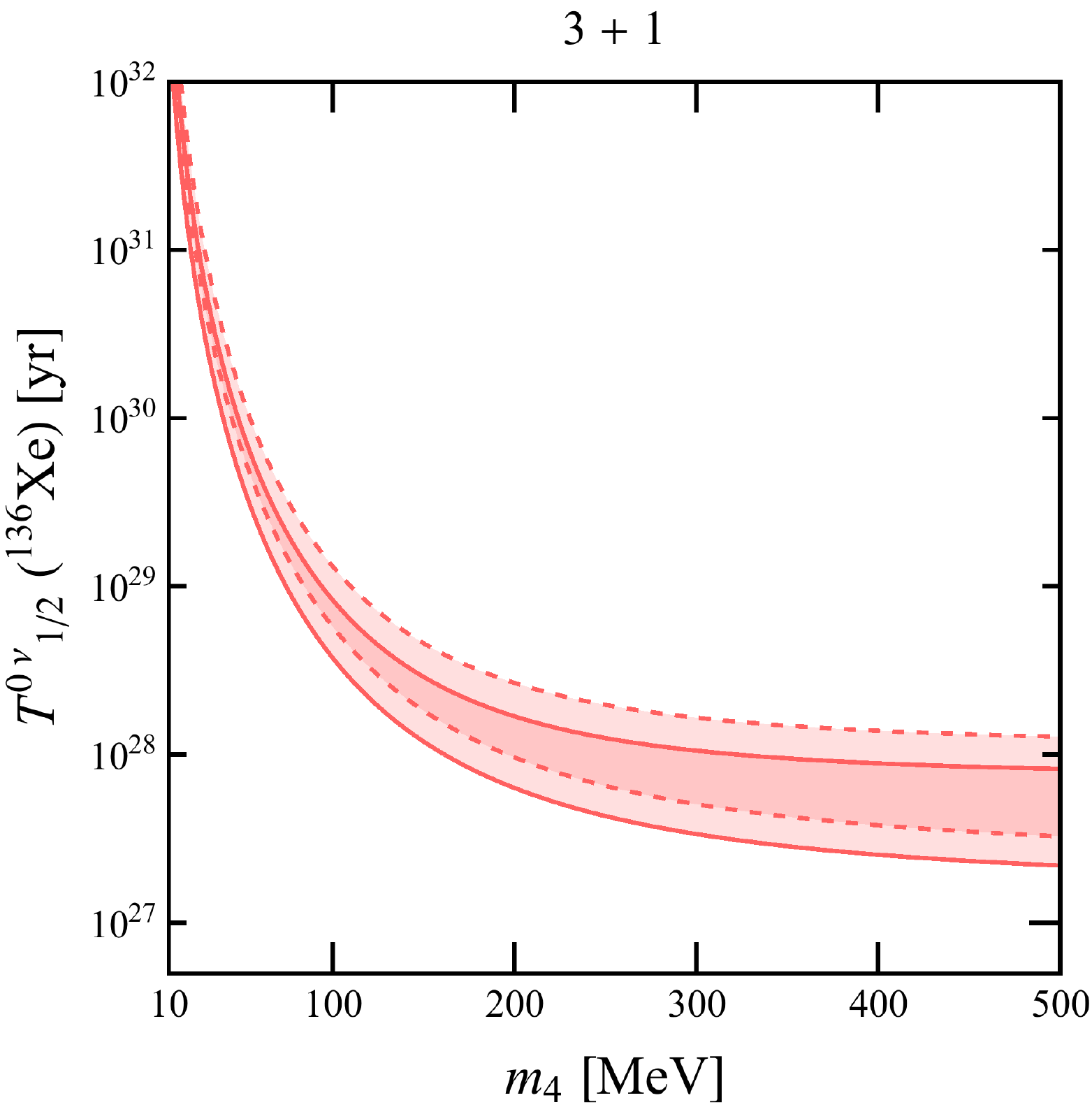}
\caption{$0\nu\beta\beta$  half-life of $^{136}$Xe as a function of $m_4$ in the $3+1$ model. We show results obtained with the QRPA (solid) or Shell Model (dashed) NMEs. The bands reflect the uncertainty due to LECs associated with hard-neutrino exchange. Right-panel: same as left panel but for neutrino masses around the pion mass.}
\label{fig:3p1SM_halflife_jordy}
\end{figure}

The result for the \NLDBD\ ${}^{136}$Xe rate is depicted in the left panel of Fig.~\ref{fig:3p1SM_halflife_jordy} as a function of $m_4$. It can be divided into three regions. For $m_4 \ll m_\pi$ the life time increases as $m_4^4$ for decreasing neutrino masses. For $m_4 \geq \Lambda_\chi$ the lifetime becomes independent of the neutrino mass. The intermediate region is more complicated and shown in more detail in the right panel of Fig.~\ref{fig:3p1SM_halflife_jordy}. The behavior in the three regions can be understood from the neutrino mass dependence of $\bar{\mathcal{M}}_V(m_i)+\bar{\mathcal{M}}_A(m_i)$ shown in Fig.~\ref{hadronicinterpolation}. For small neutrino masses $m_4 \ll m_\pi \sim k_F$, the NMEs become almost independent of the neutrino mass. The dominant contribution, however, is proportional to 
\be
\sum_{i=1}^4 m_i U_{ei}^2 = 0\,,
\ee
as can be seen explicitly from Eq.\ \eqref{eq:3plus1}, see also Eq.~\eqref{vanish} and Ref.~\cite{Blennow:2010th}. The first non-vanishing contributions in this regime are suppressed by $m_i^2/m_\pi^2$ as discussed in Sect.~\ref{mi2_correction} and the half life is proportional to
\be\label{3p1SM_AL}
\left|\sum_{i=1}^4 m^3_i U_{ei}^2\right|^2 = \left|M^{*\,2}_D M_R\right|^2\sim m^2_3 m_4^4\,,
\ee
in the regime $m_4 \gg m_3$. 
For large neutrino masses, $m_4 > \Lambda_\chi$, we need to compare the two terms in Eq.~\eqref{m4heavy} which are both non-zero. The first term involves the sum over light neutrinos, but only the $m_3$ contribution is nonzero, and depends on $U_{e3} \simeq -1/\sqrt{3}$ which is constant in the $3+1$ model up to $m_3/m_4$ corrections, see Eq.\ \eqref{eq:3plus1}. Similarly, we can see that $m_4 U_{e4}^2$ is roughly independent of $m_4$ from Eq.\ \eqref{eq:3plus1}, but in this case the amplitude scales as $\mathcal A_L^{(9)}(m_4)\sim C_{1L}^{(9)}\sim  U_{e4}^2/m_4\sim m_4^{-2}$ and thus quickly drops off. In fact, already for $m_4 = 1$ GeV the fourth neutrino only contributes at the $10\%$ level. 

From Fig.~\ref{fig:3p1SM_halflife_jordy} it is clear that for small neutrino masses, the $3+1$ toy model predicts extremely slow \NLDBD\ rates, orders of magnitude away from present or projected sensitivities. For $m_4 \geq 1$ GeV, the half lives range from roughly $1.7 \cdot 10^{27}$ yr to $1.5 \cdot 10^{28}$ yr depending on the choice of NMEs and LECs. The uncertainty in the half life is at the order-of-magnitude level and mainly due to our poor knowledge of the short-distance LEC $g_\nu^{\rm NN}(m_i)$. Although the uncertainty in the small-$m_4$ region looks small this is probably unrealistic. In this regime, the amplitudes depend on the derivative of the NMEs with respect to $m_i$, which we estimated by expanding the interpolation formula. However, as discussed in Sect.~\ref{mi2_correction}, this expression might not be accurate in this region and a realistic uncertainty would likely be at the order-of-magnitude level as well. Since the decay rates are immeasurably small, this uncertainty is not too relevant.

\subsection{3+2 model}\label{sect3p2}
We now consider a minimal $3+2$ model, where we extend the SM with two sterile neutrinos ($n=2$). This model is more realistic as it can readily accommodate the measured neutrino mass splittings, mixing angles, and CP phase. We closely follow the analysis of Ref.~\cite{Donini:2012tt} to conveniently parametrize the $5\times5$ mixing matrix in terms of neutrino masses and PMNS parameters. The original $5\times5$ mass matrix is given by
\begin{align}
M_{\nu}=
\begin{pmatrix}
0 & 0 & 0 & M_{D_{14}}^* & M_{D_{15}}^* \\
0 & 0 & 0 & M_{D_{24}}^* & M_{D_{25}}^*\\
0 & 0 & 0 & M_{D_{34}}^* & M_{D_{35}}^*\\
M_{D_{14}}^* & M_{D_{24}}^* & M_{D_{34}}^* & M_{R_{44}} & M_{R_{45}}\\
M_{D_{15}}^* & M_{D_{25}}^* & M_{D_{35}}^* & M_{R_{45}} & M_{R_{55}}
\end{pmatrix} \label{mass3p2_org}
\end{align}
This mass matrix leads to one massless and four massive neutrinos and can be diagonalized by a $5\times5$ unitary matrix $U$ that consists of physical parameters \cite{Donini:2012tt}
\begin{align}\label{eq:paramDiag}
U^TM_\nu U={\rm diag}(m_1,m_2,m_3,m_4,m_5)\,,\qquad U=
\begin{pmatrix}
U_{aa} & U_{as} \\
U_{sa} & U_{ss}
\end{pmatrix}\,,
\end{align} 
where, in the normal hierarchy, we have
\begin{align}
U_{aa}&=U_{\rm PMNS}
\begin{pmatrix}
1 & 0\\
0 & H
\end{pmatrix},\hspace{1cm}
U_{as}=iU_{\rm PMNS}
\begin{pmatrix}
0\\
Hm_{l}^{\frac{1}{2}}R^{\dagger}M_h^{-\frac{1}{2}}
\end{pmatrix}\,,\\
U_{sa}&=i
\begin{pmatrix}
0 & \overline{H}M_h^{-\frac{1}{2}}Rm_l^{1/2} 
\end{pmatrix},\hspace{1.3cm}
U_{ss}=\overline{H}\,.
\end{align}
Here, $m_l$ and $M_h$ are $2\times2$ mass matrices
\begin{align}
m_l=
\begin{pmatrix}
m_2 & 0\\
0 & m_3
\end{pmatrix}, \hspace{0.5cm}
M_h=
\begin{pmatrix}
m_4 & 0\\
0 & m_5
\end{pmatrix}\,,
\end{align}
and $U_{\rm PMNS}$ and $R$ are $3\times 3$ and $2\times 2$  matrices
\begin{align}
U_{\rm PMNS}&=
R^{23}W^{13}R^{12}\,{\rm diag}(1,1,e^{i\alpha})\,,
\end{align}
\begin{align}
R=
\begin{pmatrix}
\cos\left(\theta_{45}+i\gamma_{45} \right) & \sin\left(\theta_{45}+i\gamma_{45} \right)\\
-\sin\left(\theta_{45}+i\gamma_{45} \right) & \cos\left(\theta_{45}+i\gamma_{45} \right)
\end{pmatrix}\,,
\end{align}
where $\theta_{45}\in [0,\pi/2]$ and $\gamma_{45}\in (-\infty, \infty)$. As the name implies, $U_{PMNS}$ is the usual PMNS matrix consisting of 3 mixing angles, 1 Dirac phase, and 1 Majorana phase (there is only 1 Majorana phase because $m_1=0$). The matrices $H$ and $\overline{H}$ are composed of the above mass and rotation matrices
\begin{align}\label{eq:paramH}
H&=\left[I+m_l^{\frac{1}{2}}R^{\dagger}M_h^{-1}Rm_l^{\frac{1}{2}} \right]^{-\frac{1}{2}},\hspace{0.5cm}
\overline{H}=\left[I+M_h^{-\frac{1}{2}}Rm_lR^{\dagger}M_h^{-\frac{1}{2}} \right]^{-\frac{1}{2}}.
\end{align}

The above form of $U$ assumes $m_1=0$ making it directly applicable to the case of the NH, while in the case of the IH we instead have $m_3=0$. To account for this change we can replace $m_{2,3}\to m_{1,2}$ in the above, leading to a solution of $U^TM_\nu U={\rm diag}(0,m_1,m_2,m_4,m_5)$, after which  the mass matrix can be brought into its usual ordering by rearranging the columns of $U$. Although this procedure leads to a perfectly adequate parametrization of $U$, it does not lead to the familiar identification of the mixing angles $s_{ij}$ with the solar and reactor angles. To ensure that the usual $U_{PMNS}$ appears as the upper left-hand  block in our parametrization of $U$ (in the limit $m_{4,5}\to \infty$) we simply follow the steps of the derivation in Ref.\ \cite{Donini:2012tt}, starting from  Eq.\ \eqref{eq:paramDiag} with $m_3=0$ instead of $m_1=0$. This leads to a form of $U$ which can again be written as in Eq.\ \eqref{eq:paramDiag}, but now 
\bea
U_{aa} &=& U_{PMNS}\hat H\,,\qquad U_{as} = iU_{PMNS}\hat HR_3\bma 0\\ m_l^{1/2}R^\dagger M_h^{-1/2}\ema\,,\nn\\
U_{sa} &=& i\bma 0 & \bar HM_h^{-1/2}Rm_l^{1/2}\ema R_3^T\,,\qquad U_{ss}=\bar H\,,
\eea
where $m_l={\rm diag}(m_1,m_2)$, while $\hat H = R_3\bma 1&0\\0& H\ema R_3^T$, with $H$ as in Eq.\ \eqref{eq:paramH} but using $m_l={\rm diag}(m_1,m_2)$, and $R_3=W^{13}(-\pi/2,0)W^{23}(\pi/2,0)$. $M_h$ and $\bar H$ are left unchanged.
To obtain these expressions we used the relation 
\bea {\rm diag}(m_1,m_2,0,m_4,m_5)=\bma R_3&0\\0&1\ema {\rm  diag}(0,m_1,m_2,m_4,m_5) \bma R_3&0\\0&1\ema ^T \,,\eea 
to write Eq.\ \eqref{eq:paramDiag} in the form of Eq.\ (2.3) of Ref.\ \cite{Donini:2012tt}. The rest of the derivation then closely follows that of Ref.\ \cite{Donini:2012tt}, leading to a similar form of $U$ as in the NH case, but with additional factors of $R_3$.

\begin{table}[t!]
\begin{center}
\begin{tabular}{||c||c c c c || }
\hline
NH	&	&   $\Delta m^2_{21}$ [eV$^2]$  & $\Delta m^2_{32}$ [eV$^2$] & $\lambda_{1,2}$   \\ 
		\hline
 	&	&    $7.39 \cdot 10^{-5}$ &  $2.449 \cdot 10^{-3}$ & 0  \\ 
$3\sigma$ &  & $[6.79,~8.01]\cdot 10^{-5}$ & $[2.358,~2.544] \cdot 10^{-3}$ & $[0,~\pi]$   \\ 
\hline
	& $\sin^2\theta_{12}$ & $\sin^2\theta_{23}$ & $ \sin^2\theta_{13}$  & $\delta_{13}/\pi\,$ \\
\hline 	&  $3.10\cdot 10^{-1}$ & $5.58\cdot 10^{-1}$ & $2.241\cdot 10^{-2}$ & 1.23 \\
$3\sigma$ 	& $[2.75,~3.50]\cdot 10^{-1}$ & $[4.27,~6.09]\cdot10^{-1}$ & $[2.046,~2.440]\cdot 10^{-2}$ & [0.78,~2.06] \\
	\hline \hline 
IH &    & $\Delta m^2_{21}$ [eV$^2]$ & $\Delta m^2_{32}$ [eV$^2$]   & $\lambda_{1,2}$   \\
\hline
	&	&   $7.39 \cdot 10^{-5}$    & $-2.509 \cdot 10^{-3}$  & 0  \\ 
$3\sigma$ &   & $[6.79,~8.01]\cdot 10^{-5}$ & $[-2.603,~-2.416]\cdot 10^{-3}$  & $[0,\pi]$   \\ 
\hline
	& $\sin^2\theta_{12}$ & $\sin^2\theta_{23}$ & $ \sin^2\theta_{13}$  & $\delta_{13}/\pi\,$ \\
	\hline
 	&  $3.10\cdot10^{-1}$ & $5.63\cdot10^{-1}$ & $2.261\cdot10^{-2}$ & 1.58 \\
$3\sigma$ 	& $[2.75,~3.50]\cdot10^{-1}$ & $[4.30,~6.12]\cdot 10^{-1}$ & $[2.066,~2.461]\cdot 10^{-2}$ & [1.14,~1.97]\\
\hline \hline
3+2 	&  $\theta_{45}$ & $\gamma_{45}$ & & \\
\hline 
& $\pi/8$  & 0.5 & & \\
\hline
\end{tabular}
\caption{Input parameters used for the analysis of the standard three-light Majorana neutrino scenario, depicted in Fig.\ \ref{fig:Lifetime_3light},  and the $3+2$ model.
The values of the light neutrino mass splittings and the PMNS angles are taken from Ref.\ \cite{Tanabashi:2018oca}.}
\label{table:BenchMark_angle}
\end{center}
\end{table}


\begin{figure}[t]
\begin{center}
\includegraphics[scale =.5 ]{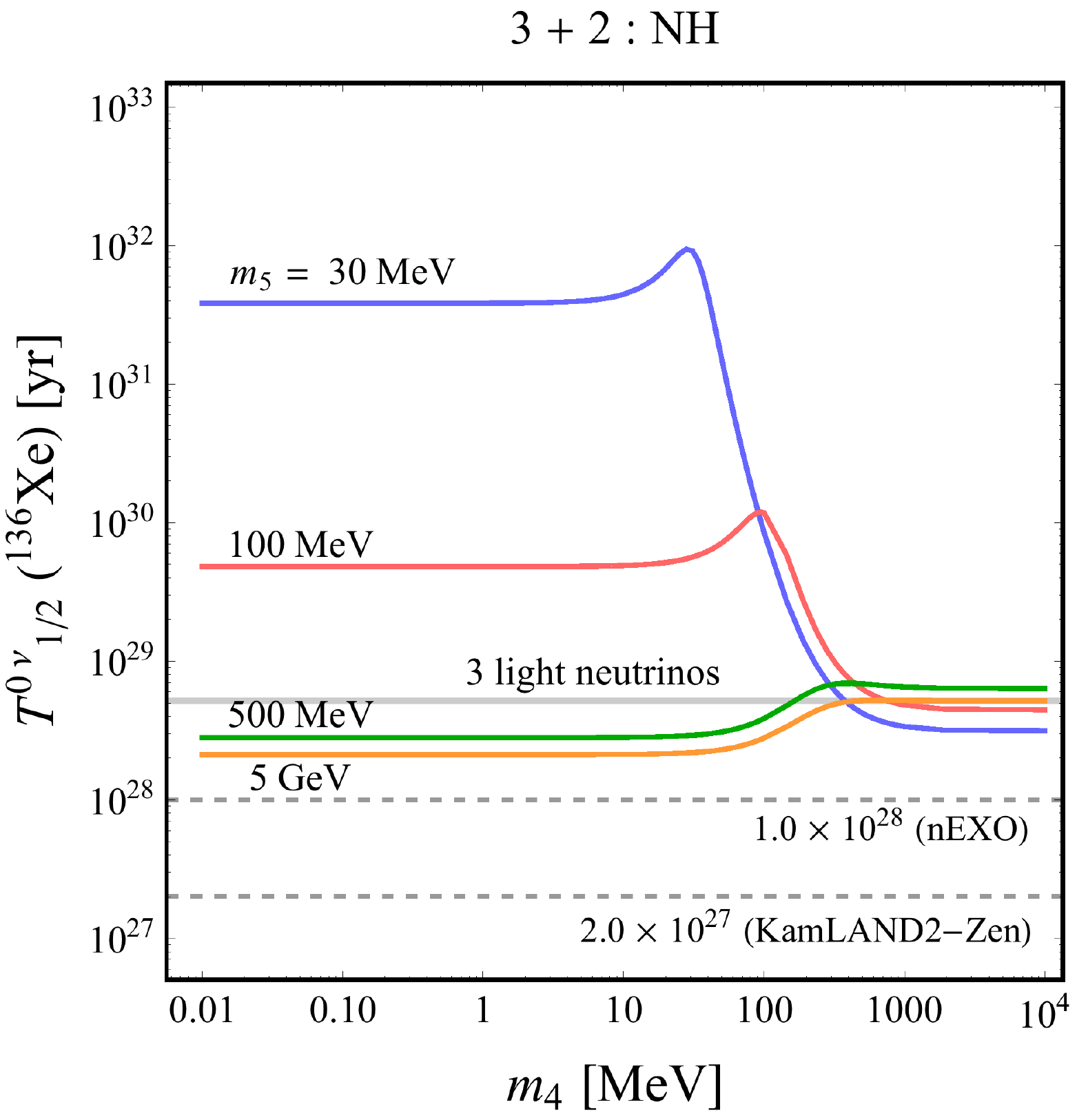}
\includegraphics[scale =.5 ]{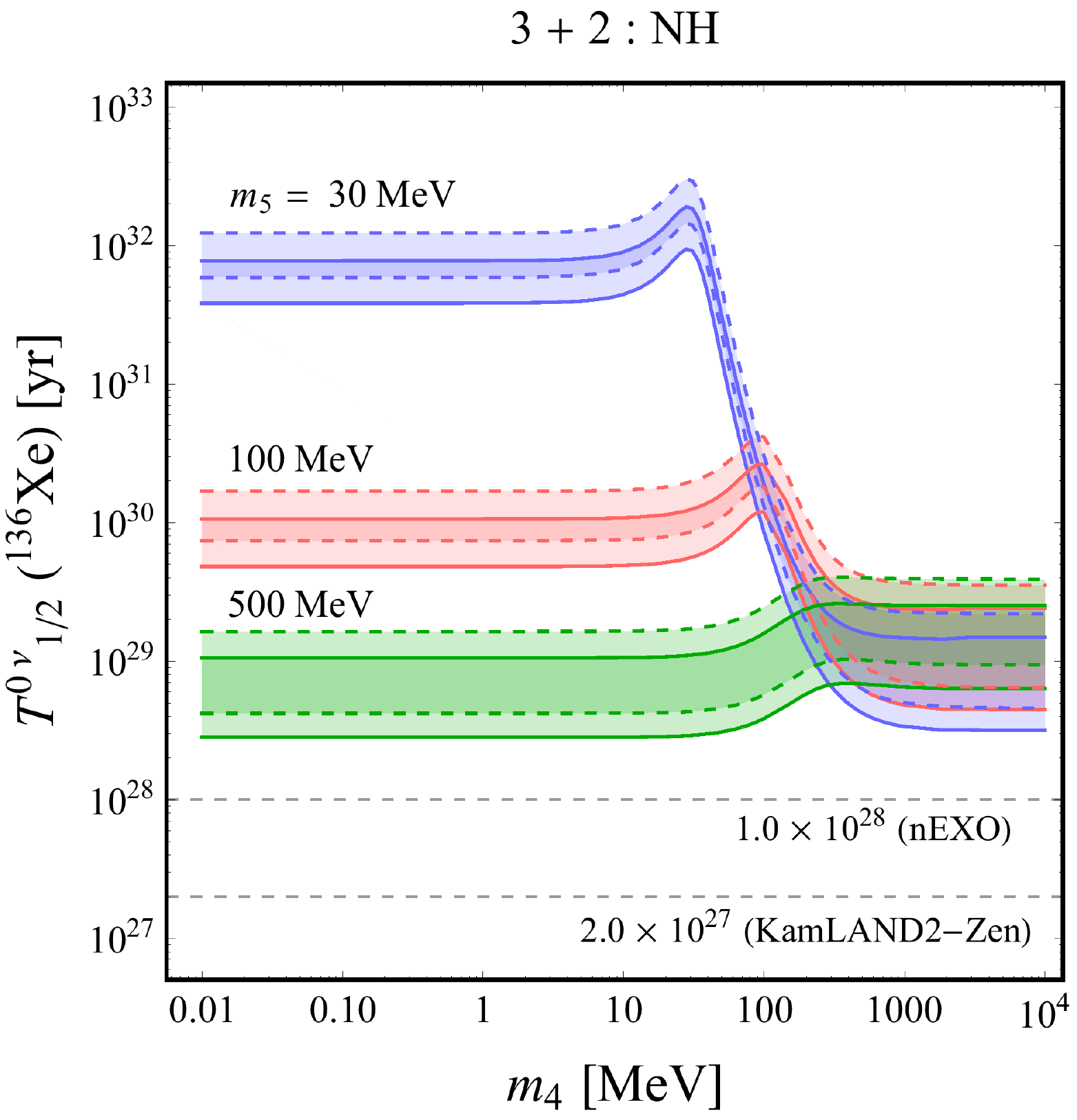}
\caption{Left panel: $0\nu\beta\beta$  half-life of $^{136}$Xe as a function of $m_4$ for $m_5=30$ MeV (blue), $m_5=100$ MeV (red), $m_5=500$ MeV (green), and $m_5=5000$ MeV (orange) in the NH. The gray horizontal line is the predicted half life for the standard mechanism for the same choices of neutrino parameters and LECs.
Right panel: Same as left panel but now we included uncertainties from NMEs and LECs. Bands correspond to $m_5=30$ MeV (blue), $m_5=100$ MeV (red), $m_5=500$ MeV (green). } 
\label{fig:3p2SM_halflife_Jordy}
\end{center}
\end{figure}

We do not wish to perform a fully general analysis of the parameters in the mixing matrix. Instead, we fix all parameters except for $m_4$ and $m_5$. We work in the normal hierarchy and set $m_2= \sqrt{\Delta m^2_{\rm SOL}} \simeq 8.58\cdot 10^{-3}$ eV and $m_3 = \sqrt{\Delta m^2_{\rm ATM}} \simeq 0.05$ eV. We pick the best-fit values for the PMNS mixing angles and Dirac phase \cite{Tanabashi:2018oca}. For simplicity we set the Majorana phase to zero, $\alpha=0$. While the choice of $\theta_{45}$ does not affect the unitary matrix drastically, $U_{e2}$ and $U_{e3}$ can deviate from the experimental values if $\gamma_{45}\gtrsim {\cal O}(1)$. Taking into account this restriction, we pick moderate values for $\theta_{45}$ and $\gamma_{45}$. All choices of parameters are given in Table \ref{table:BenchMark_angle}. 

We show the lifetime $T^{0\nu}_{1/2} (^{136}{\rm Xe})$ in the case of the NH as a function of $m_4$ for four different values of $m_5$ in the left-panel of Fig.~\ref{fig:3p2SM_halflife_Jordy}. We use the QRPA  NMEs of Ref.~\cite{Hyvarinen:2015bda} and a specific value of the short-distance LECs to not clutter the plots too much. We set
\be\label{guessLECs1}
g_\nu^{\rm NN}(0) = - \frac{1}{(2 F_\pi)^2}\,,\qquad g_1^{\rm NN}(m_0) = (5+3 g_A^2)/4\,,
\ee
as discussed in Sect.~\ref{naiveIP}. We can observe a few things. For small $m_5 < m_\pi$  and $m_4\ll m_5$ the half life becomes independent of $m_4$ and scales as $m_5^4$, similar to the behavior of the $3+1$ scenario for small $m_4$ (the left part of Fig.~\ref{fig:3p1SM_halflife_jordy}). This is the `cancellation regime', where the NMEs and LECs become $m_i$ independent and
\be
\mathcal A_L(m_i) \sim \sum_{i=1}^5 U_{ei}^2 m_i (\bar{\mathcal{M}}_V(m_i)+(\bar{\mathcal{M}}_A(m_i)) \sim \mathcal O(U_{ei}^2m^3_i)\propto m_5^2\,,
\ee
since $U_{e5}\sim \sqrt{m_l/m_5}$.
The scaling with $m_5^4$ breaks down for larger values of $m_5$, 
in fact, for $m_5\geq 500$ MeV and $m_4 \ll m_\pi$ the half life becomes essentially independent of both $m_4$ and $m_5$ as can be seen by comparing the left part of the green and orange lines. 

\begin{figure}[t]
\begin{center}
\includegraphics[scale =.5]{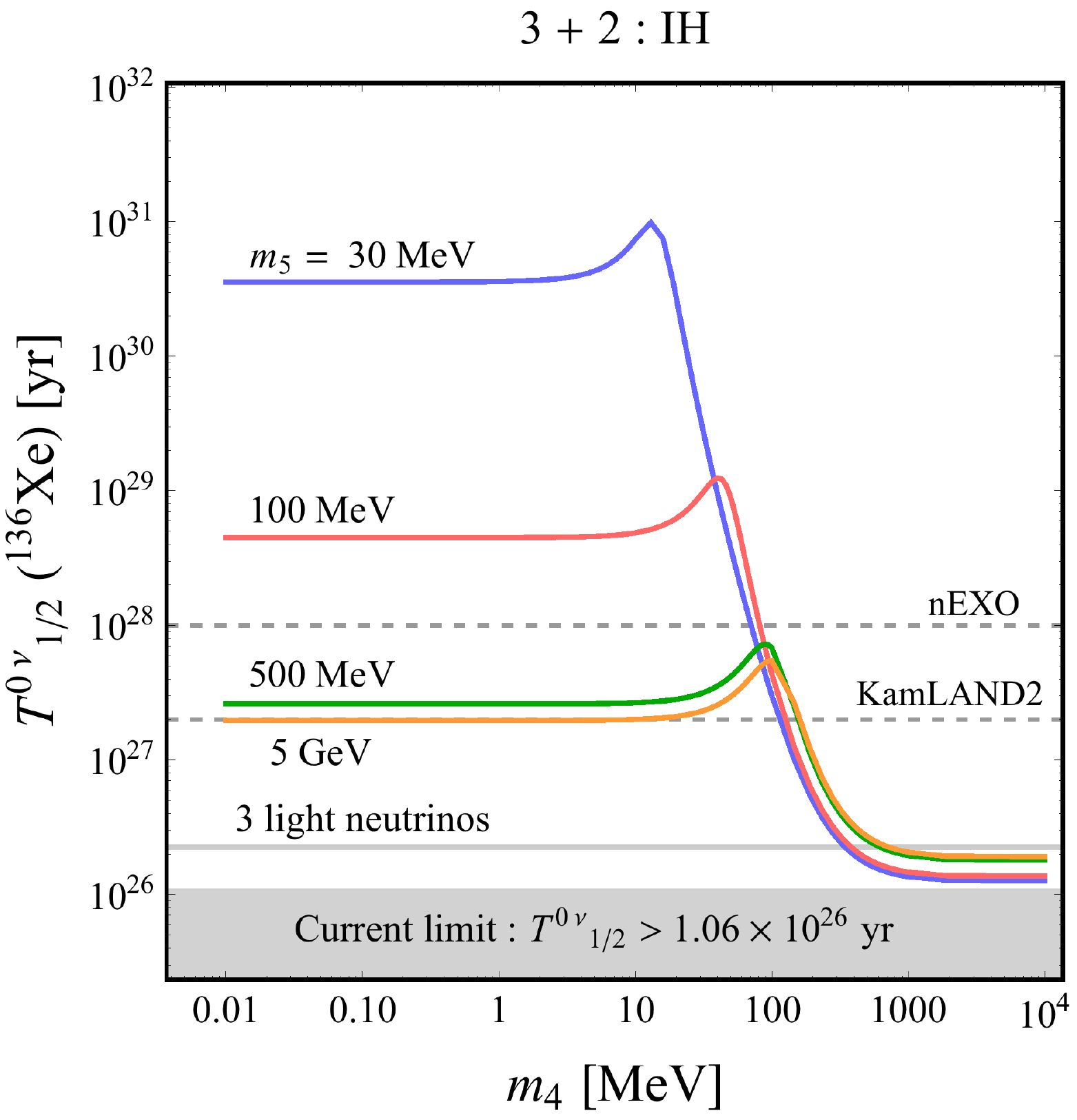}\hfill
\includegraphics[scale =.5 ]{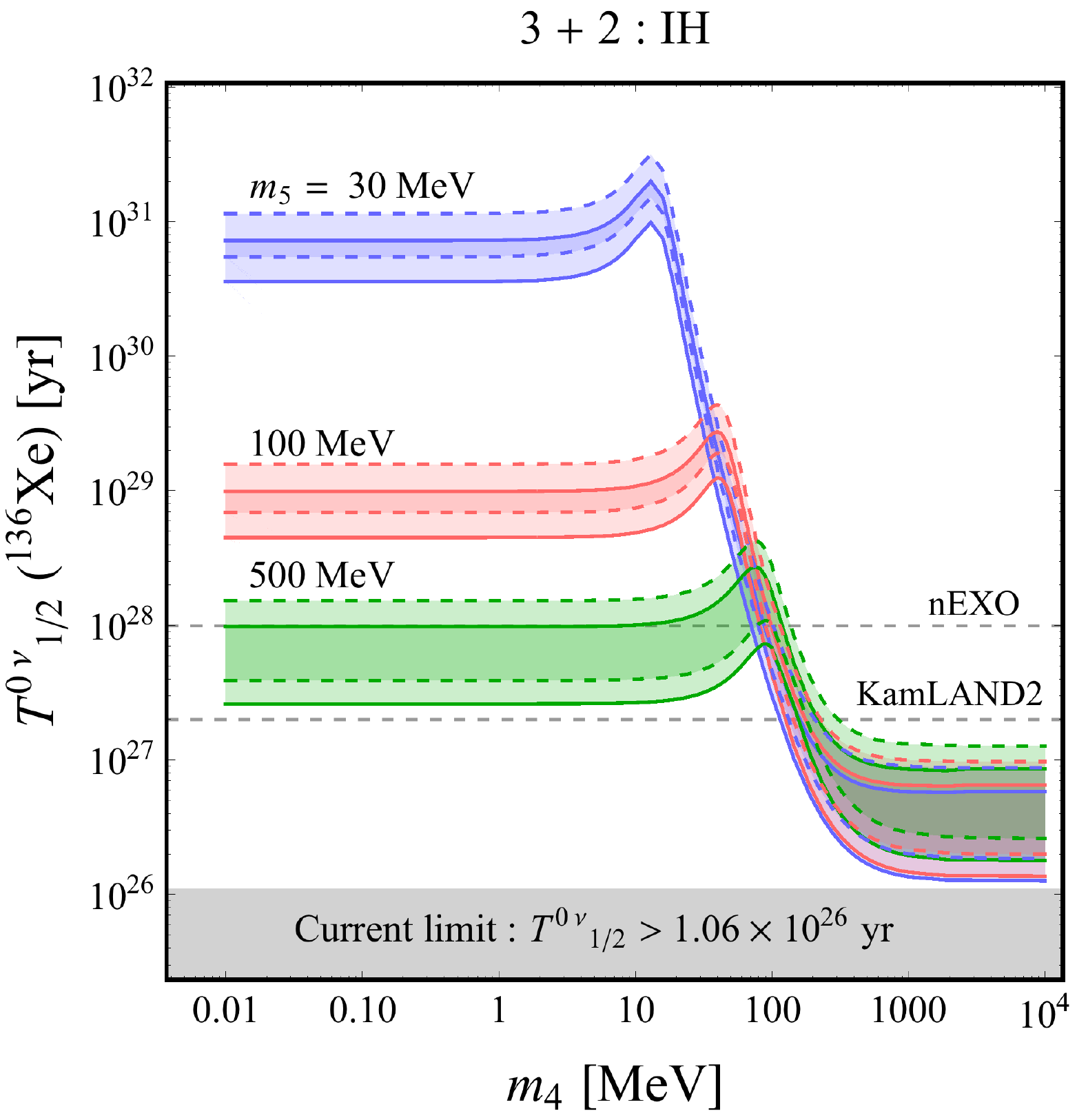}
\caption{Left panel: $0\nu\beta\beta$  half-life of $^{136}$Xe as a function of $m_4$ for $m_5=30$ MeV (blue), $m_5=100$ MeV (red), $m_5=500$ MeV (green), and $m_5=5000$ MeV (orange) in the IH. The gray horizontal line is the predicted half life for the standard mechanism for the same choices of neutrino parameters and LECs. Right panel: Same as left panel but now we included uncertainties from NMEs and LECs. Bands correspond to $m_5=30$ MeV (blue), $m_5=100$ MeV (red), $m_5=500$ MeV (green). } 
\label{fig:3p2IH}
\end{center}
\end{figure}

For $m_4 \gg m_\pi$ and $m_5 \geq \Lambda_\chi$ (the right part of the green and orange lines), the scenario becomes similar to a standard seesaw scenario with 3 light Majorana neutrinos and `decoupled' heavy states, see the horizontal gray line. The lifetime becomes roughly $10^{29}$ yr, the predicted lifetime in the NH for a massless lightest neutrino and vanishing Majorana phases. Shorter half lives are possible if one neutrino is heavy while the other is at or below the pion mass. The right sides of the blue and red curves correspond to such a scenario with a half life of roughly $2\cdot 10^{28}$ yr, not too far from the projected nEXO limits. Finally, for almost degenerate $m_4$ and $m_5$ there is a cancellation leading to a peak in the half life. In the right panel of Fig.~\ref{fig:3p2SM_halflife_Jordy}, we show the same results for three choices of $m_5$, now including the uncertainty from the short-distance LECs and we show results for two choices of NMEs. The uncertainties are at the order-of-magnitude level.

The case of the IH is shown in Fig.\ \ref{fig:3p2IH}, which shows a behavior that is very similar to that of the NH scenario. In particular, at large values of $m_5$ the half life becomes almost independent of $m_{4,5}$, while, in the small mass region, the scaling $\sim m_5^4$ reappears. We again see a cancellation when $m_4\sim m_5$, and the half life approximates that of a seesaw scenario with 3 light neutrinos in the large-$m_{4,5}$ region. In contrast, the absolute value of the half life does differ between the scenarios and is roughly an order of magnitude smaller in the IH than in the NH. In fact, nEXO will be sensitive to minimal $3+2$ scenarios in the IH for which at least one mass eigenstate has mass larger than roughly $500$ MeV. 

\begin{figure}[t]
\begin{center}
\includegraphics[width=0.47\textwidth ]{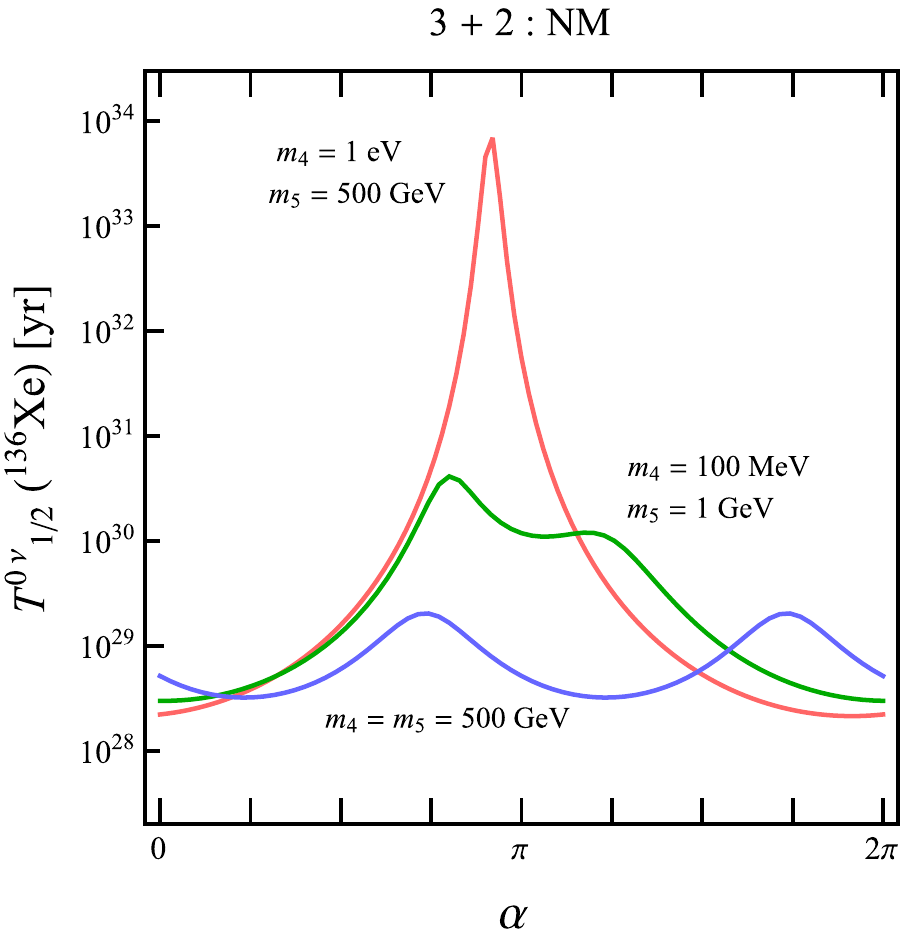}\hfill
\includegraphics[width=0.50\textwidth]{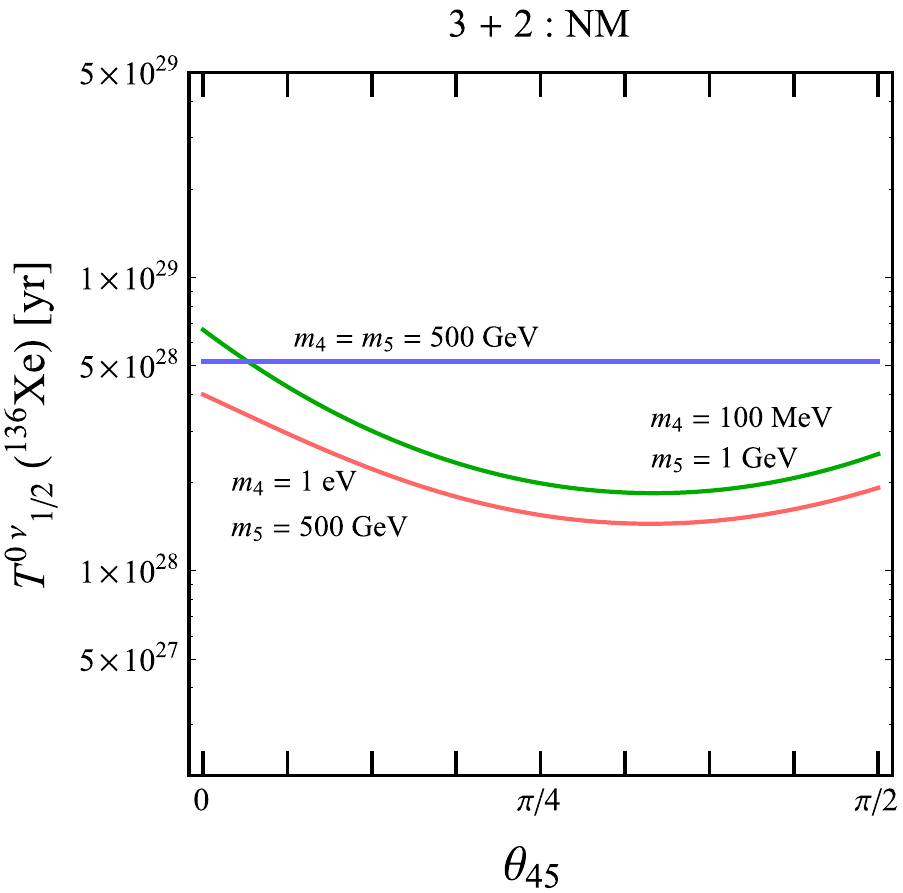}
\caption{Left panel: $0\nu\beta\beta$ decay half-life of $^{136}$Xe  in the NH  as a function of the Majorana phase, $\alpha$, for $m_{4,5} \gg 1$ GeV (blue), for $m_4 =1$ eV and $m_{5} \gg 1$ GeV (red), and $m_4 =100$ MeV and $m_5=1$ GeV. LECs and NMEs as in the left panel of Fig.~\ref{fig:3p2SM_halflife_Jordy}. Right panel: similar but now for $\alpha=0$ and we vary the mixing angle $\theta_{45}$.} 
\label{angledep}
\end{center}
\end{figure}

It should be mentioned that  Figs.\ \ref{fig:3p2SM_halflife_Jordy} and \ref{fig:3p2IH} depend on the choice of mixing angles and phases, for which we used $\alpha=0$ and the values in Table \ref{table:BenchMark_angle}. Varying the phases can lead to significantly larger half lives, especially when they are tuned to induce cancellations between the different contributions, while in other parts of parameter space (often milder) enhancements are possible. To get an idea of these effects we show the dependence on the Majorana phase $\alpha$ and the mixing angle $\theta_{45}$ in the NH in the left- and right-panel of Fig.~\ref{angledep} for three choices of $m_4$ and $m_5$. The blue lines correspond to decoupled mass eigenstates with $m_{4,5} \gg 1$ GeV. In this case, the Majorana phase is still relevant and the half life varies by roughly a factor 5 depending on $\alpha$, but only has a mild $\theta_{45}$ dependence. The red line corresponds to $m_4 = 1$ eV while $m_5=500$ GeV is decoupled. In this case, the dependence on $\alpha$ is significant and for $\alpha=\pi$ the half life increases by roughly six orders of magnitude compared to $\al=0$, while the dependence on $\theta_{45}$ is far less severe. Finally, the green line correspond to an intermediate scenario with $m_4 =100$ MeV and $m_5 =1$ GeV. 

\subsection{A leptoquark scenario}\label{lepto}
In this section we illustrate the use of the EFT framework by performing the matching in the case of an explicit model of BSM physics. Our main goal is to illustrate the framework and we consider a simple SM extension.
We extend the SM with right-handed neutrinos that interact with leptoquarks (LQs). Leptoquarks are hypothetical particles that convert quarks to leptons and vice versa. All possible representations of LQs are summarized in Ref.\ \cite{Dorsner:2016wpm}, and among them is a scalar LQ that transforms as an $SU(3)_c$ triplet, an $SU(2)_L$ doublet, and carries nonzero hypercharge$: \tilde R\left({\bf 3},~{\bf 2},~1/6\right)$. The LQ interaction with sterile neutrinos is given by 
\begin{align}\label{eq:LQlag}
{\cal L}_{\rm LQ}=-{y}^{RL}_{ab}\bar{d}_{Ra}\tilde R^i\epsilon^{ij}L_{Lb}^{j}+y^{\overline {LR}}_{ab}\bar{Q}^{i}_{La}\tilde R^i\nu_{Rb} +{\rm h.c.}\,,
\end{align}
where $a,b$ and $i,j$ are flavor and $SU(2)$ indices, respectively. 
In addition to these interactions, we include the right-handed neutrino Majorana mass terms and Yukawa interactions as in Eq.~\eqref{eq:smeft}. The LQ interactions are LNC so that LNV only arises from the Majorana mass of the sterile neutrinos. Integrating out ${\tilde R}$, one \textoverline{dim-6} operator  in Table \ref{tab:O6R} is generated
\begin{align}
{\cal L}^{(\bar{6})}_{\nu_R}=C^{({6})}_{\substack{LdQ\nu\\ abcd}}\left(\bar{L}^i_ad_b \right)\epsilon^{ij}\left(\bar{Q}^j_c\nu_{Rd} \right)+{\rm h.c.}\,,
\end{align}
where
\begin{align}
C^{(\bar{6})}_{\substack{LdQ\nu\\abcd}}=\frac{1}{m^2_{\rm LQ}}y^{\overline{LR}}_{cd}y^{RL*}_{ba}\,,
\end{align}
with $m_{\rm LQ}$ being the mass of the $\tilde R$ LQ. The above operator, as well as $\Or_{L\nu Qd}$ generated through the RGEs of Eq.\ \eqref{eq:STrge}, induces the dim-6 scalar and tensor operators in Eq.~\eqref{lowenergy6_l0} below the electroweak scale. We focus on operators involving the first-generation quarks and charged leptons
\begin{align}
{\cal L }^{(6)}_{\Delta L=0}=\frac{2G_F}{\sqrt{2}}\bigg[\bar{c}_{\substack{{\rm SR}\\ea}}^{(6)}~\bar{u}_Ld_R\bar{e}_L\nu_{Ra}+\bar{c}^{(6)}_{\substack{{\rm T}\\ea}}~\bar{u}_L \sigma^{\mu\nu}d_R\bar{e}_L\sigma^{\mu\nu}\nu_{Ra}\bigg]\,,
\end{align}
where the subscripts $e$ and $a$ denote the charged-lepton and neutrino flavor, and 
\begin{align}
\bar{c}^{(6)}_{\substack{{\rm SR}\\ea}}=4\bar{c}^{(6)}_{\substack{{\rm T}\\ea}}=\frac{v^2}{2m^2_{\rm LQ}}y^{\overline{LR}}_{1a}y^{RL*}_{1e}\,,
\end{align}
where $a$ runs from $1$ to $n$. 

Of course, we also have to include the SM weak interactions and together we obtain the following matching to the operators  in Eq.~\eqref{redefC6}
\begin{align}
\left(C^{(6)}_{\rm VLL}\right)_{ei}=-2 V_{ud} U_{ei}\,, \hspace{0.5cm}
\left(C^{(6)}_{\rm SRR}\right)_{ei}=4\left(C^{(6)}_{\rm TRR}\right)_{ei}=\sum^n_{a=1} \bar{c}^{(6)}_{\substack{{\rm SR}\\ ea}}U^*_{3+a,i}\,.
\end{align}
With these nonzero Wilson coefficients the $0\nu\beta\beta$ decay rate can be directly read from the master formula in Eq.\ \eqref{eq:T1/2}. We use the procedure outlined in Sect.\ \ref{naiveIP} to obtain matching relations for the relevant LECs and we explicitly give the resulting interpolation formulae in App.~\ref{AppLQ}.

\begin{figure}[t]
\begin{center}
\includegraphics[scale =.45 ]{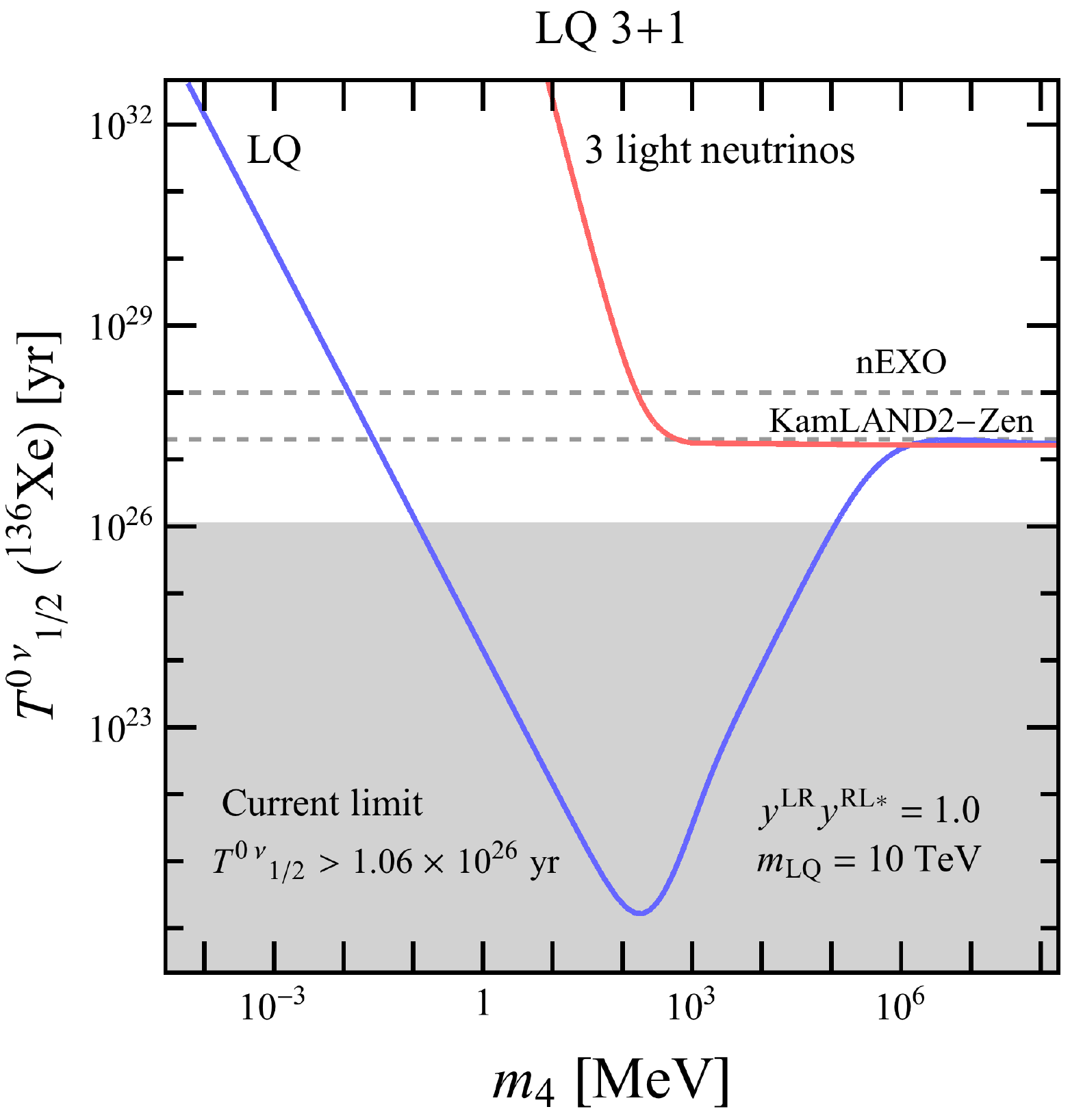}
\includegraphics[scale =.45]{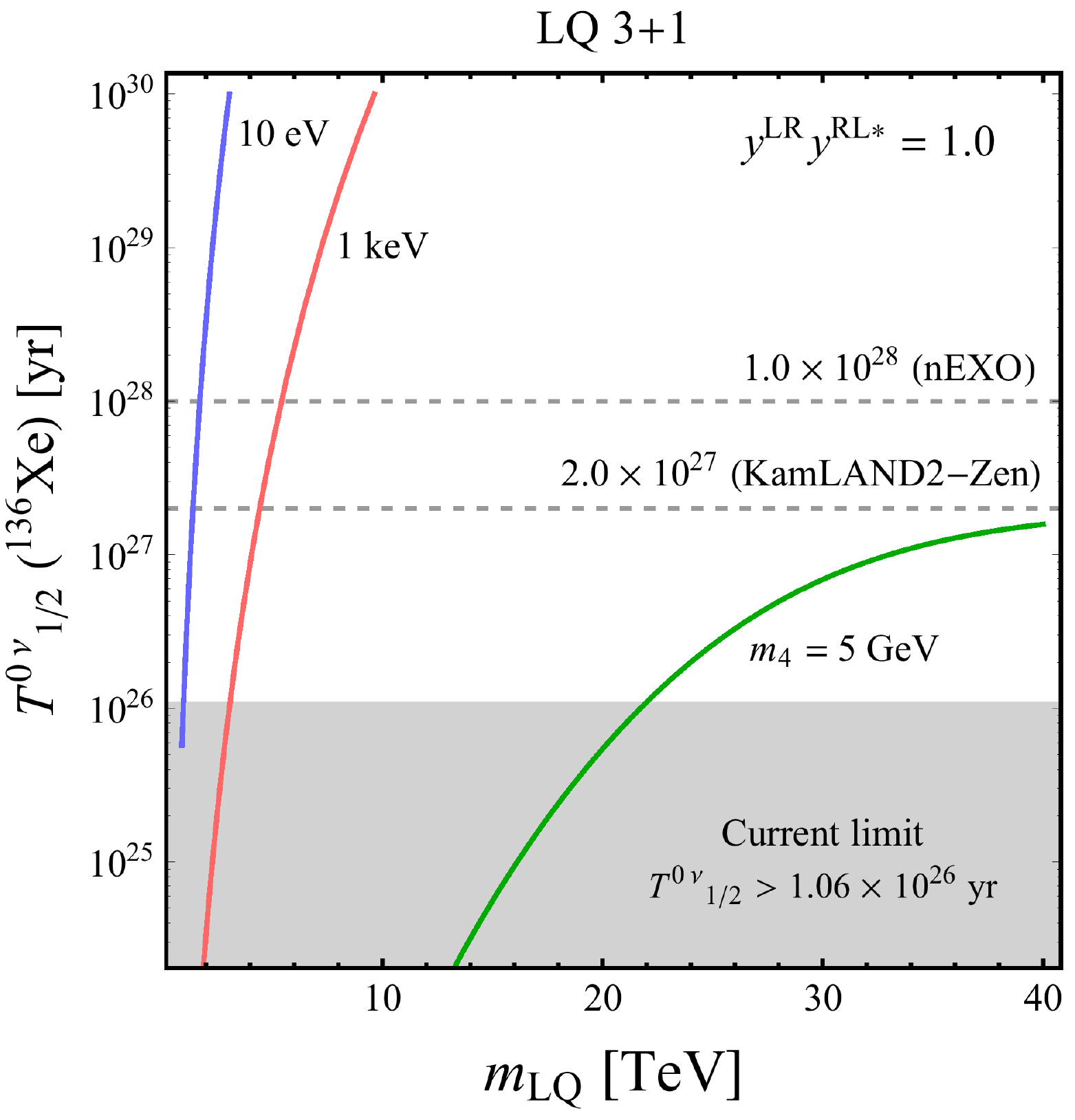}
\caption{ Left panel: $T^{0\nu}_{1/2} (^{136}{\rm Xe})$ as a function of $m_4$ in the LQ $3+1$ model. The blue line is the total half life whereas the red line depicts the contribution from just the standard mechanism (exchange of neutrinos interacting via left-handed currents).  We have set $y^{\overline{LR}}_{11}y^{RL*}_{12}=1.0$ and $m_{\rm LQ}=10~$TeV. Right panel: Similar but now we fixed $m_4 =10$ eV (blue), $m_4 = 1$ keV (red), and $m_4=5$ GeV (green) and vary $m_{\rm LQ}$. In both panels the Dirac and Majorana phases were set to zero, $m_3=\sqrt{\Delta m_{\rm ATM}^2}$, and the QRPA NMEs were used.}
\label{fig:3p1LQ_halflife}
\end{center}
\end{figure}
\subsubsection{A 3+1 scenario with leptoquark interactions}

We first consider the $3+1$ scenario, but now with the additional LQ interactions described above. 
This only provides a simple toy model in which to study the effects of the LQs, as it also leads to two massless neutrinos and cannot explain the observed oscillation data. A more realistic 3+2 scenario is studied in the next subsection, we note, however, that is also possible to explain the neutrino masses in LQ models without introducing sterile neutrinos, see e.g.\ \cite{Mahanta:1999xd,AristizabalSierra:2007nf,Babu:2019mfe}.
For simplicity we set the couplings $y^{\overline{LR}}_{11}y^{RL*}_{1e}=1$ as deviations can be absorbed in $m_{\rm LQ}$. For the contributions from the standard mechanism we set the unknown LECs as in Eq.~\eqref{guessLECs1}. 
The LQ interactions induce a large number of contributions to the subamplitude $\mathcal A_L$ as given in Eq.~\eqref{ALLQ}. Unfortunately most of these contributions are associated to LECs we do not control. To simplify the analysis somewhat we only include contributions from purely pionic operators, where  lattice results are available that can help constrain the LECs. The missing contributions from pion-nucleon and nucleon-nucleon interactions can be added once more information about the LECs is obtained. The missing contributions appear at the same order as the pionic contributions we do include, and thus correspond to a significant uncertainty. For the pionic operators we require the interpolation formulae discussed in Eq.~\eqref{naiveinterpipi} of App.\ \ref{AppLQ}. These depend on $g_{1,2,3,4,5}^{\pi\pi}$ that are known (see Table \ref{Tab:LECs})  and the unknown couplings $g^{\pi\pi}_{\rm{S1}}(0)$, $g^{\pi\pi}_{\rm{TT}}(0)$, $g^{\pi\pi}_{\rm{S,VLL}}(0)$, $g^{\pi\pi}_{\rm{T,VLL}}(0)$. In this section we use the NDA estimates
\bea
g^{\pi\pi}_{\rm{S1}}(0) = - g^{\pi\pi}_{\rm{TT}}(0) = \frac{1}{F_\pi^2}\,,\qquad g^{\pi\pi}_{\rm{S,VLL}}(0) = g^{\pi\pi}_{\rm{T,VLL}}(0) = -1\,,
\eea
where the signs were chosen in such a way that the LECs do not change sign when varying the neutrino mass.

We plot the resulting \NLDBD\ half-life of $^{136}$Xe in the left panel of Fig.~\ref{fig:3p1LQ_halflife} for $m_{\rm LQ} = 10$ TeV as a function of $m_4$. The red line denotes the limit of $m_{\rm LQ} \rightarrow \infty$ corresponding the $3+1$ scenario discussed above. We have set the Dirac and Majorana phases to zero, $m_3=\sqrt{\Delta m_{\rm ATM}^2}$, and used QRPA NMEs. The theoretical uncertainty due to NMEs and LECs is not shown, but is similar to the $3+1$ and $3+2$ scenarios and thus roughly one-to-two orders of magnitude on the half life. 
The plot shows that for a light fourth neutrino, $m_4 < 100$ GeV, the LQ interactions completely dominate over the standard mechanism. In fact, the current experiments rule out $10$ TeV interactions for $100\, {\rm keV} < m_4 <100$ GeV. In the LQ scenario, $0\nu\beta\beta$ experiments are most constraining at $m_4 \simeq 170$ MeV, where we find $m_{LQ} > 56$ TeV. Future experiments could push this towards $m_{LQ} > 150$ TeV. 

In the right panel of Fig.~\ref{fig:3p1LQ_halflife} we show the half life as a function of $m_{\rm LQ}$ for three specific values of $m_4$. We set $m_4 =10$ eV, with the eV-scale being motivated by anomalies in neutrino experiments \cite{Aguilar:2001ty, AguilarArevalo:2008rc, Aguilar-Arevalo:2013pmq, Aguilar-Arevalo:2018gpe} (but see Ref.~\cite{Dentler:2018sju} as well),  $m_4 =1$ keV, as motivated by models of sterile neutrino DM \cite{Abazajian:2001nj, Kusenko:2009up}, and $m_4 = 5$ GeV as motivated by studies of low-scale leptogenesis \cite{Canetti:2012vf}. For these cases, the limits on $m_{\rm LQ}$ are respectively $m_{LQ} > 1$ TeV, $m_{LQ} > 3$ TeV, and $m_{LQ} > 21$ TeV. While it is difficult for present and future \NLDBD\ experiments to probe scenarios with just light sterile neutrinos, these results illustrate that prospects are much better in scenarios where the sterile neutrinos have non-standard interactions.

\begin{figure}[h!]
\begin{center}
\includegraphics[scale =.36]{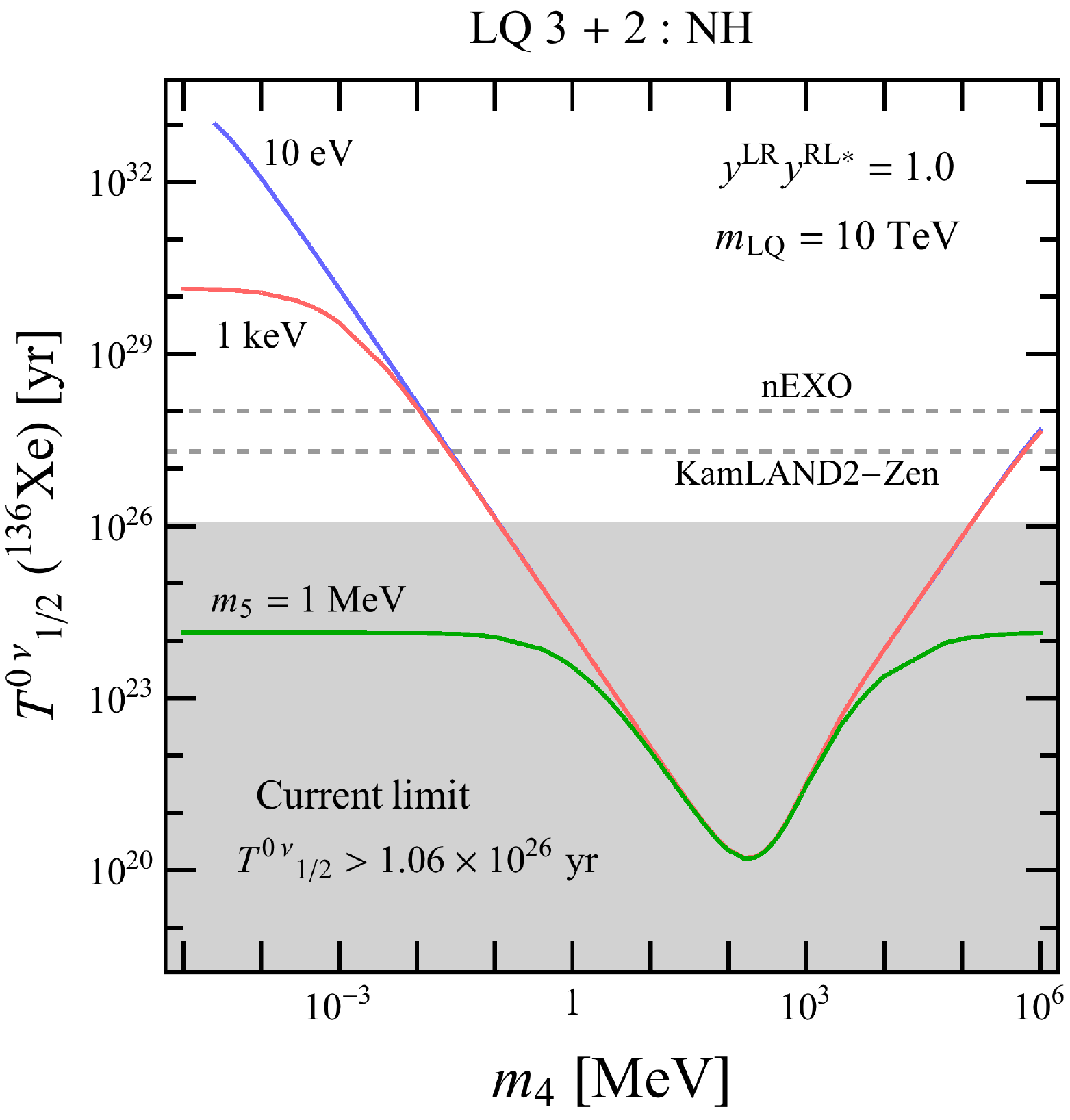}
\includegraphics[scale =.36]{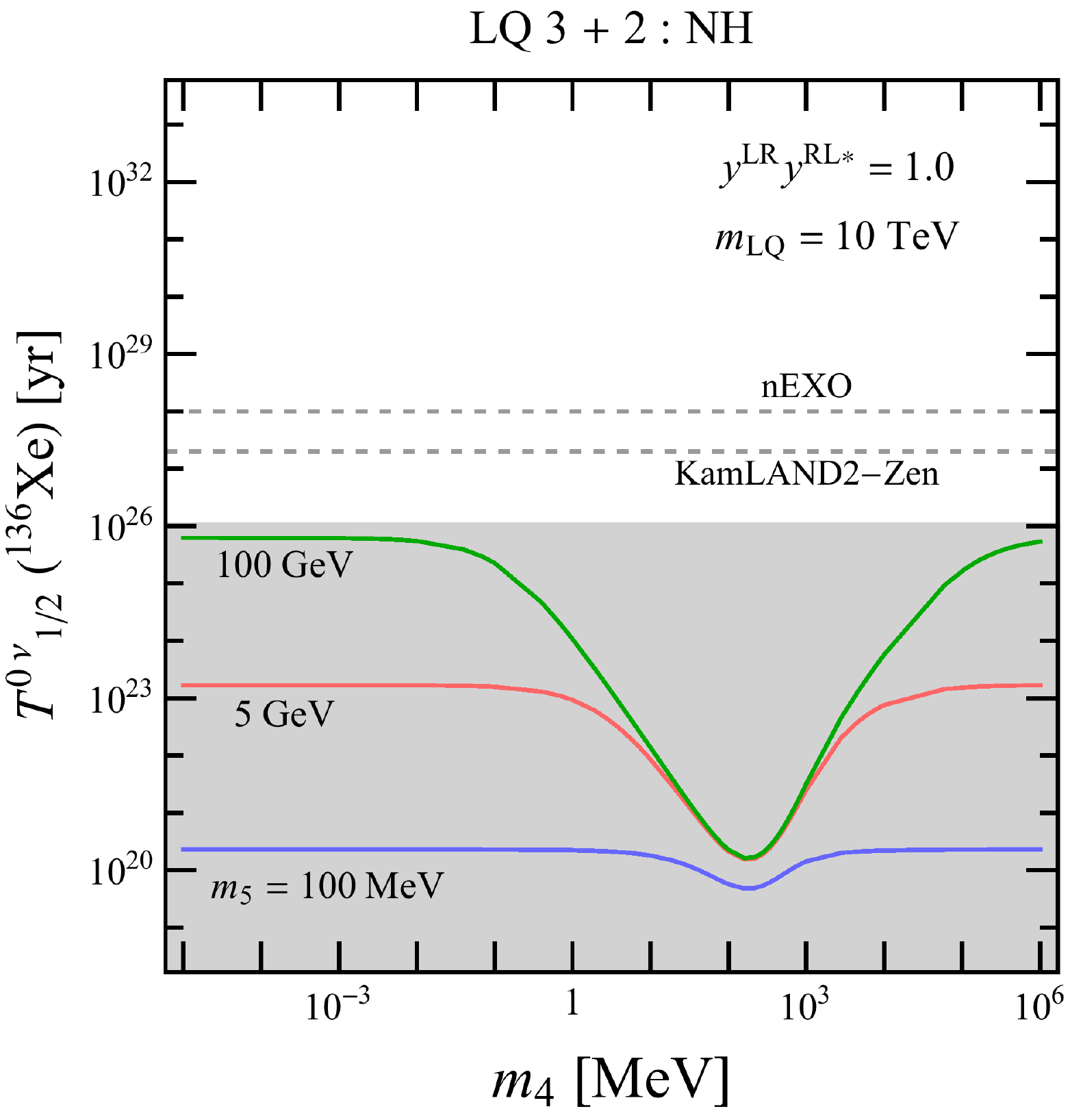}
\includegraphics[scale =.36]{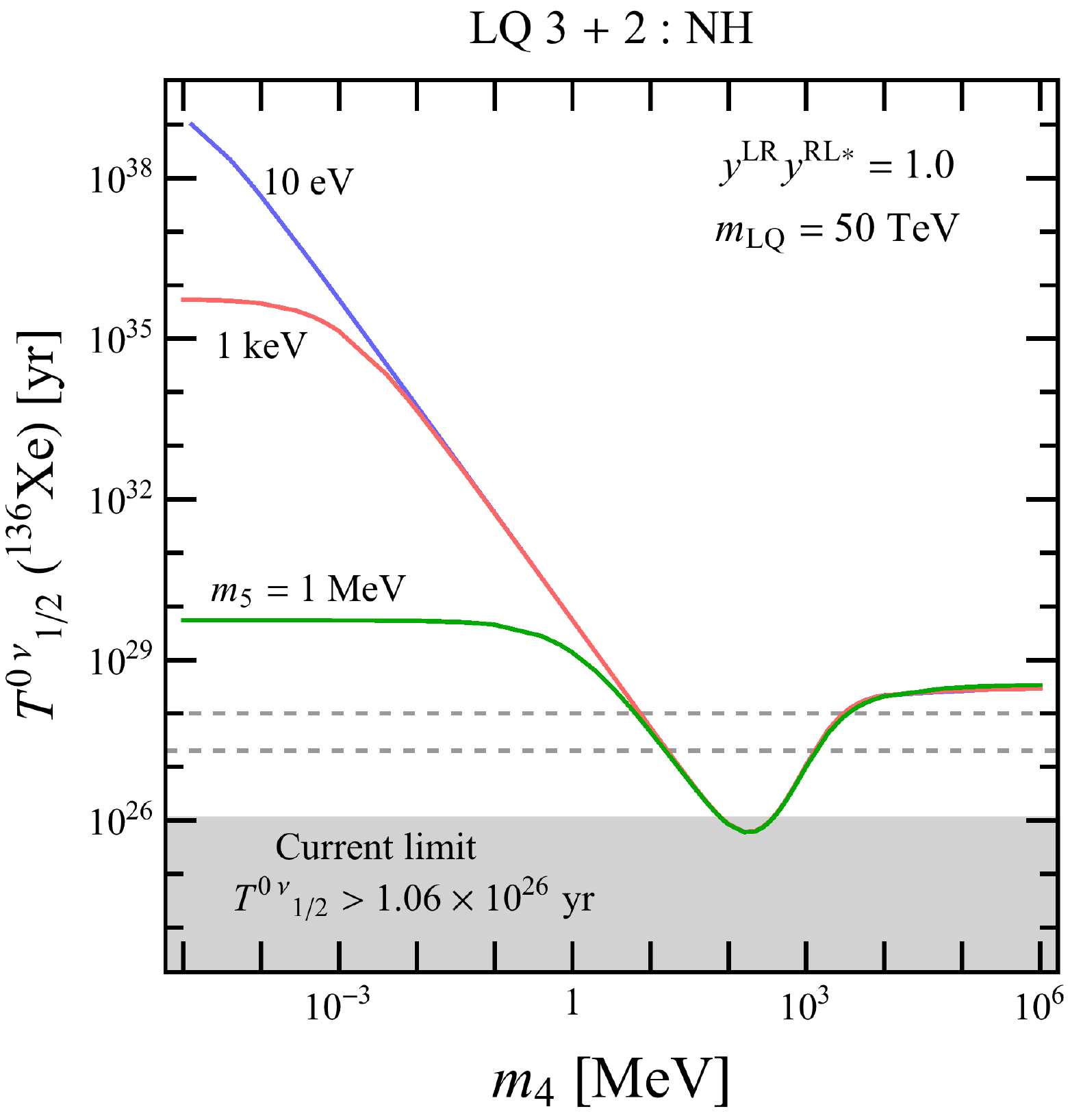}
\includegraphics[scale =.36]{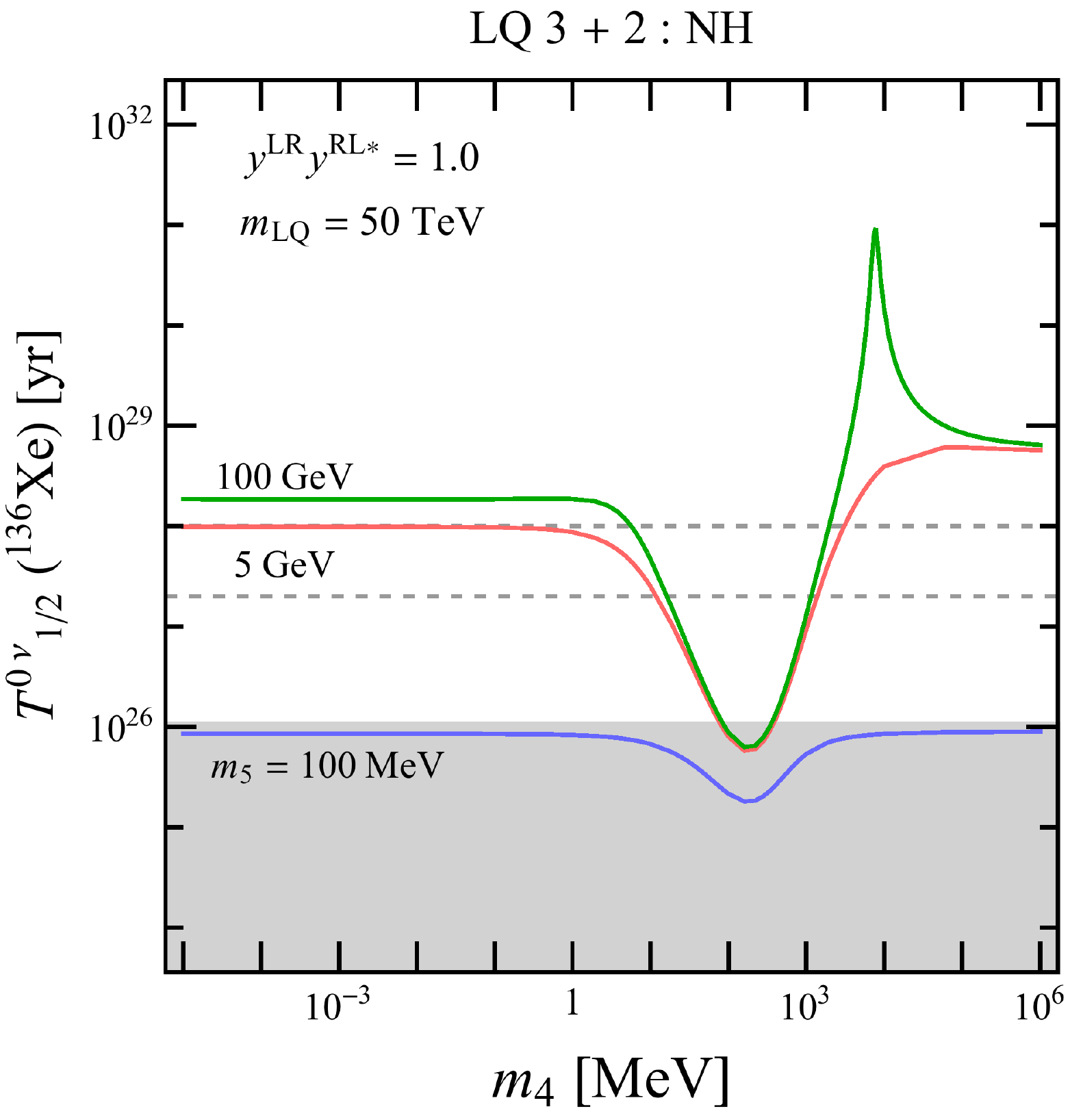}
\caption{ Top-left panel: $T^{0\nu}_{1/2} (^{136}{\rm Xe})$ as a function of $m_4$ in the LQ $3+2$ leptoquark model for different choices of $m_5=10$ eV (blue), $m_5=1$ keV (red), $m_5=1$ MeV (green). We consider a leptoquark mass of $10$ TeV. Top-right panel: same as top-left but now for $m_5=100$ MeV (blue), $m_5=5$ GeV (red), $m_5=100$ GeV (green). Bottom panels are the same as the top panels but now for a leptoquark mass of $50$ TeV. All panels use the same phases, mixing angles, and LECs as in Sect.~\ref{sect3p2}.}
\label{fig:3p2LQ_halflife}
\end{center}
\end{figure}

\begin{figure}[t]
\begin{center}
\includegraphics[scale =.45 ]{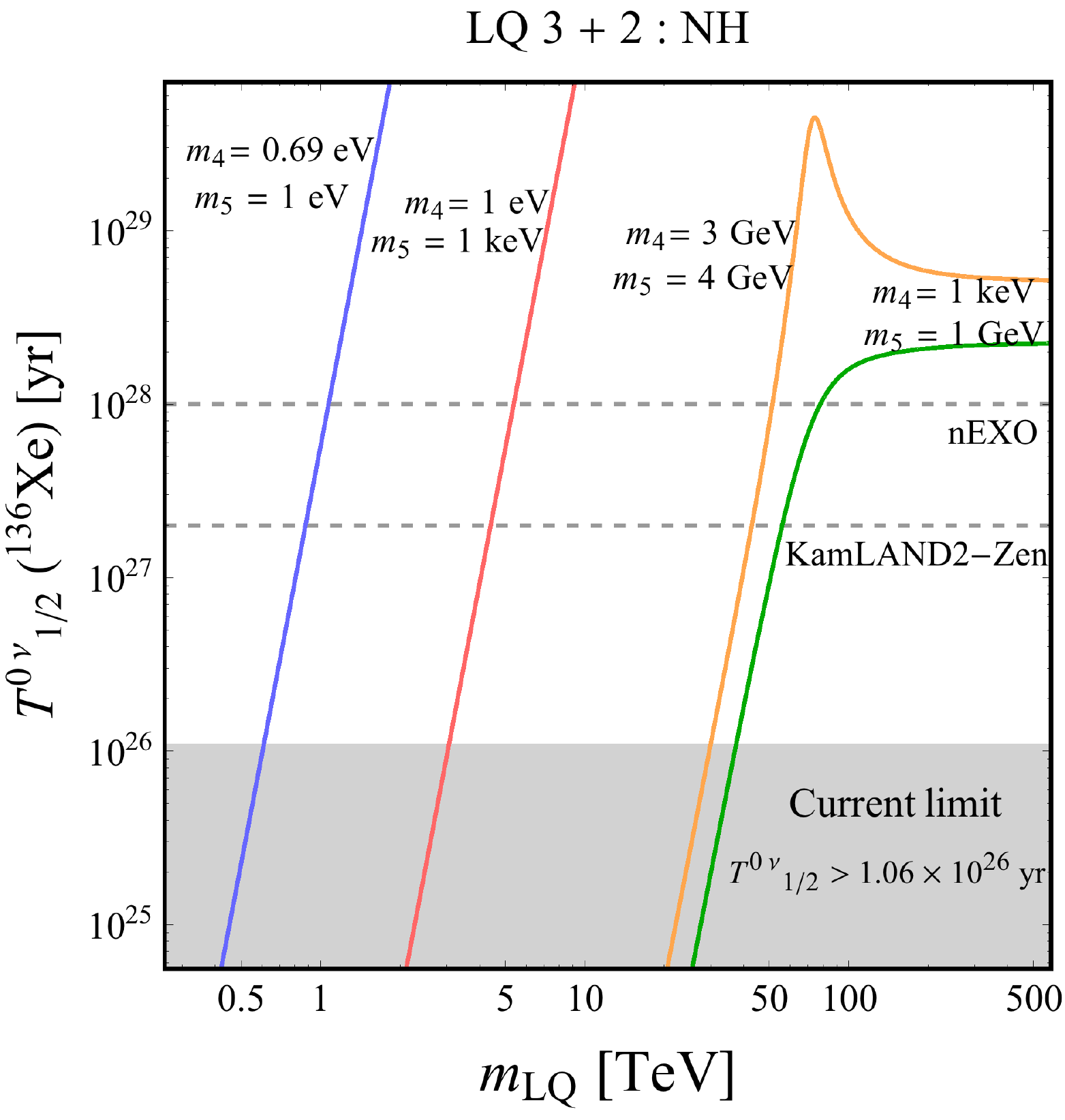}
\caption{ $T^{0\nu}_{1/2} (^{136}{\rm Xe})$ as a function of $m_{\rm LQ}$ for 4 choices of $m_4$ and $m_5$. The blue line corresponds to $m_4 =0.69$ eV and $m_5 =1$ eV, the red line to $m_4 =1$ eV and $m_5 =1$ keV, the green line to $m_4=1$ keV and $m_5= 1$ GeV, and the orange line to $m_4=3$ GeV and $m_5 =4$ GeV. The same phases, mixing angles, and LECs as in Sect.~\ref{sect3p2} were used.} 
\label{fig:3p2LQ_halflife2}
\end{center}
\end{figure}

\subsubsection{A 3+2 scenario with leptoquark interactions}
We now repeat the analysis in the $3+2$ model. In Sect.~\ref{sect3p2} we concluded that detection of $0\nu\beta\beta$ is not possible with the next generation of experiments in the pure $3+2$ model for the NH (see Fig.~\ref{fig:3p2SM_halflife_Jordy}). It is interesting to investigate what the scale of BSM physics should be to make a detection possible. We take the LQ interactions as an example of BSM physics in order to determine the relevant mass of $m_{\rm LQ}$ for different choices of $m_4$ and $m_5$. To reduce the number of parameters we set $y^{\overline{LR}}_{11}y^{RL*}_{1e}=y^{\overline{LR}}_{12}y^{RL*}_{1e}=1$ and use the same phases, angles and LECs as in Sect.~\ref{sect3p2}. We do not show the dependence on the LQ couplings explicitly, but note that they always appear in the combination $\frac{y^{\overline{LR}}_{1a}y^{RL*}_{1e} }{m^2_{\rm LQ}}$, so that a shift in the couplings of, for example, $y^{\overline{LR}}_{1a}y^{RL*}_{1e}\to 10^{-2}$, is equivalent to rescaling the LQ mass by a factor of $10$.

In the top-left panel of Fig.~\ref{fig:3p2LQ_halflife}, we set $m_{\rm LQ} = 10$ TeV and $m_5=10$ eV (blue), $m_5=1$ keV (red), $m_5=1$ MeV (green) and vary $m_4$. Current experiments already rule out such scenarios for a significant range of $m_4$. For $m_5$ in the MeV range, the resulting half lives are too short and excluded. For light $m_5 \leq$ keV, the scenario is excluded for $100\,\mathrm{keV} < m_4 < 100\,\mathrm{GeV}$. In the top-right panel, we set $m_5=100$ MeV (blue), $m_5=5$ GeV (red), $m_5=100$ GeV (green) and conclude that such values are ruled out independent of $m_4$. 
Of course, the results depend on the scale of BSM physics. In the bottom panels we repeat the analysis but now set $m_{\rm LQ} = 50$ TeV. The bottom-left panel, corresponding to light $m_5$, tells us that future experiments are sensitive if $m_4 > 10$ MeV, while scenarios where both $m_4$ and $m_5$ are light lead to decay rates that are too slow to observe. The right panel shows that current limits already exclude the scenario for $m_5 = 100$ MeV independent of $m_4$. For larger $m_5$, nEXO can observe \NLDBD\ events if $m_4 < 1$ GeV. 

Our main goal is not to exhaustively investigate the parameter space of this LQ model. Instead, we wish to highlight that future \NLDBD\ experiments are sensitive to a class of BSM models with light sterile neutrinos if they have weak non-standard interactions with SM fields. How weak these interactions can be depends on the exact value of the neutrino masses, but scales of $1$-$100$ TeV can be accessed. We illustrate this in Fig.~\ref{fig:3p2LQ_halflife2}. In scenarios with $m_4$ and $m_5$ around $1$ eV, the ${}^{136}$Xe half life from just the standard mechanism is roughly $10^{60}$ yr, but nEXO can make a detection for $m_{\rm LQ} \leq 1$ TeV. In this particular case, LHC experiments already strongly constrain such LQ masses \cite{Sirunyan:2018btu,Aaboud:2019jcc}, but for heavier neutrinos much higher LQ mass scales can be probed. For instance, for $m_4$ at the keV scale and $m_5$ at the GeV scale, next-generation experiments are sensitive to leptoquark masses up to $100$ TeV.

This discussion highlights the important role that $0\nu\beta\beta$ experiments can play in understanding the nature and interactions of sterile neutrinos. While a positive $0\nu\beta\beta$ signal alone cannot unambiguously be interpreted as evidence for sterile neutrinos, $0\nu\beta\beta$ experiments, in conjunction with  direct observations in laboratory experiments or astrophysical probes, can put severe constraints on the possible  interactions  of these particles that are very competitive with other low-energy probes and high-energy collider experiments.

\section{Conclusions}\label{conclusions}

Sterile neutrinos are  natural candidates to explain the origin of active neutrino masses via the type-I seesaw mechanism \cite{Minkowski:1977sc,GellMann:1980vs,Mohapatra:1980yp},
which, however, does not definitively point towards a specific mass scale. While the smallness of neutrino masses can be realized with very massive $\nu_R$, $m_{\nu_R}\sim 10^{15}$ GeV, with $\mathcal O(1)$ Yukawa couplings to
SM leptons, the sterile neutrino mass scale can be lowered to the TeV scale \cite{Mohapatra:1980yp}, inducing interesting signals at the Large Hadron Collider \cite{Aaboud:2018spl},
or even to the GeV or sub-GeV region \cite{Shaposhnikov:2006nn,Canetti:2012vf}. In such scenarios the sterile neutrinos could provide a dark matter candidate and induce the correct matter-antimatter asymmetry in the Universe
\cite{Shaposhnikov:2006nn,Canetti:2012vf,Boyarsky:2018tvu,Adhikari:2016bei}.  
While muon, meson, and nuclear $\beta$ decays constrain the interactions of sterile neutrinos with SM particles 
to be more feeble than the weak interaction, in generic models $\nu_R$ can be charged under new forces, mediated by bosons with masses much larger than the $W$ and $Z$ masses.
These interactions, while weak, can leave traces in high-precision
experiments, including searches for $0\nu\beta\beta$. 

In this paper, we studied the impact of light sterile neutrinos with mass smaller than the electroweak scale, $m_{\nu_R} < v$, on $0\nu\beta\beta$, in a systematic framework 
which relies on a tower of EFTs. The contribution of sterile neutrinos to $0\nu\beta\beta$ has been investigated extensively 
in the literature \cite{Blennow:2010th,Mitra:2011qr,deGouvea:2011zz,Barea:2015zfa,Giunti:2015kza}, in the minimal scenario in which $\nu_R$ has only renormalizable interactions, namely 
the Majorana mass term $M_R$ and a Yukawa interaction $Y_\nu$ in Eq.\ \eqref{eq:smeft}. Here we extend these works in several, significant directions 

\begin{itemize}
\item We extend the SM by adding a (family) of sterile neutrinos $\nu_R$, which are singlets under the SM gauge group.
At scales larger than the electroweak scale, we allow the $\nu_R$ to interact with SM degrees of freedom via a dimension-four Yukawa interaction, and 
the most general set of gauge-invariant higher-dimensional operators up to \textoverline{dim-7}, in the framework of the sterile-neutrino-extended SMEFT \cite{delAguila:2008ir,Cirigliano:2012ab,Liao:2016qyd}.
With the exclusion of the Majorana mass term in Eq.\ \eqref{eq:smeft}, we consider operators with at most one $\nu_R$, which are the most relevant for $0\nu\beta\beta$.
The \textoverline{dim-6} LNC operators with active and sterile neutrinos are listed in Tables \ref{tab:O6L} and \ref{tab:O6R}, respectively,
while the \textoverline{dim-7} LNV operators are given in Tables \ref{tab:O7L} and \ref{tab:O7R}. 
The renormalization-group equations describing the QCD evolution for the coefficients of the \textoverline{dim-6} and \textoverline{dim-7} operators are given 
in Eqs.\ \eqref{eq:STrge} and \eqref{eq:rge7}.

\item We match the SMEFT operators onto $SU(3) \times U(1)_{\rm em}$-invariant operators, by integrating out heavy SM degrees of freedom. 
For $2 \, {\rm GeV} \lesssim m_{\nu_R} < v$, the sterile neutrino is also integrated out in this step, and one is left with dim-9 operators  
in Eqs.\ \eqref{LagSca} and \eqref{LagVec}, with coefficients given by Eqs.\ \eqref{dim91}, \eqref{dim92}, and \eqref{dim93}.
If $\nu_R$ is lighter, it remains dynamical in chiral EFT. In this case, at the quark level, one finds the generalized LNC and LNV $\beta$-decay Lagrangians in Eqs.\ \eqref{lowenergy6_l2}, \eqref{lowenergy6_l0}, \eqref{lowenergy7}, and \eqref{lowenergy7b}, which involve, vector, axial, scalar, pseudoscalar, and tensor currents as well as derivative operators. 
Rotating these interactions to the neutrino mass basis gives rise to the operators in Eqs.\ \eqref{6final} and \eqref{7final}, whose coefficients can be expressed in terms of the gauge-invariant operators in  Sect.\ \ref{MatchLow}.

\item We systematically construct the chiral Lagrangian in the presence of light sterile neutrinos with non-standard interactions, starting from the  
quark-level  operators in Eqs.\ \eqref{6final} and \eqref{7final}. This Lagrangian includes terms with explicit neutrinos, that couple to pions and nucleons via the vector, axial, scalar, pseudoscalar, and tensor currents. These are constructed in  Sects.\ \ref{ChiralBB} and \ref{neutronbeta}, and the neutron $\beta$-decay transition operator is summarized in Eq.\ \eqref{eq:currents0}.
In addition, the Chiral Lagrangian contains LNV operators that couple pions and nucleons to two leptons, which are induced by the exchange of virtual light neutrinos.  
In  Sect.\ \ref{sec:hardNu} we construct, for the first time, these operators in full generality and we show that the $\pi\pi e e$ and $ n n p p e e $ couplings  contribute at LO for several higher-dimensional operators. We discuss the dependence of the LECs of these operators on the neutrino mass, and demonstrate that they are needed 
to guarantee that the $0\nu\beta\beta$ amplitude has a smooth dependence on the neutrino mass.

\item We identify all the chiral EFT low-energy constants required for the derivation of the $0\nu\beta\beta$ operator at LO. They are summarized in Table \ref{Tab:LECs}.
With the exception of the recoil-order LEC appearing in the tensor current, $g_T^{\prime}$, all LECs in the neutron $\beta$-decay operator are well known, either from experiment or Lattice QCD.
The $\pi\pi$ couplings induced by dim-9 operators are also well known \cite{Nicholson:2018mwc}, while $\pi\pi$ couplings induced by hard-neutrino exchange
and $\pi N$ and $NN$ couplings induced by dim-9 operators and hard-neutrino exchange are at the moment mostly undetermined. The ignorance of these LECs
causes a sizable hadronization uncertainty in the $0\nu\beta\beta$ half-lives, which, though usually neglected, is often as big as the error from nuclear matrix elements, see for example Figs.~\ref{AVpmA} and \ref{fig:Lifetime_3light}.

\item We derive the $0\nu\beta\beta$ transition operator in  Sect.\ \ref{potentials} and the
master formula for $0\nu\beta\beta$ in  Sect.\ \ref{MasterFormula}. In spite of the large number of operators, we find that the 
final form of the amplitude and the half-lives, Eqs.\ \eqref{amplitude} and \eqref{eq:T1/2}, are rather compact and involve structures that are very similar to those found
for the exchange of active neutrinos in Refs.\ \cite{Cirigliano:2017djv,Cirigliano:2018yza}.
In particular, they involve the same nuclear matrix elements, with the difference that, for $1 \, {\rm MeV}  < m_{\nu} < \Lambda_\chi$,  the NMEs acquire a non-trivial dependence on the neutrino mass.

\item We study the dependence of the $0\nu\beta\beta$ half-lives on the neutrino masses. For neutrino masses below $\Lambda_\chi$, the dependence arises in two ways, explicitly via the neutrino propagators
in the neutrino potentials  and implicitly via the LECs induced by hard-neutrino exchange. In contrast, if $m_{\nu_R} \gg \Lambda_\chi$, the mass dependence appears through the matching coefficients of dim-9 operators. 
We derive  interpolation formulae, grounded in QCD and $\chi$PT, for both the nuclear matrix elements, in  Sect.\ \ref{interpolation},
and the LECs, in  Sect.\ \ref{interLEC}. These formulae allow us to smoothly interpolate between the $m_{\nu_R} \ll \Lambda_\chi$ and the $m_{\nu_R} \gg \Lambda_\chi$ regimes, as shown in Figs.\ \ref{AVpmA} and \ref{fig:3p1LQ_halflife}, and to get the correct chiral EFT scaling of the amplitudes.
The interpolation formulae for the LECs are partially phenomenological, due to the difficulties to treat the intermediate region $m_\pi \ll m_{\nu_R} \lesssim \Lambda_{\chi}$
rigorously, but can be systematically improved by calculating pion, nucleon, and two-nucleon LNV matrix elements with nonperturbative methods for different neutrino masses.

\item As a consequence of the systematic construction of the chiral Lagrangian and the study of the mass dependence
of NMEs and LECs, we can address all sources of theoretical uncertainties on the $0\nu\beta\beta$ half-lives, induced by active and sterile neutrinos.
An important finding of this work is that the hadronization uncertainty is very significant. 
We estimate this uncertainty by conservatively varying the unknown LECs in the range expected on the basis of naive dimensional analysis 
and internal theoretical consistency. We recommend to include the hadronization uncertainty in future analyses of $0\nu\beta\beta$. As an example, in the left panel of Fig.~\ref{fig:Lifetime_3light} we show the usual inverted and normal hierarchy predictions for the ${}^{136}$Xe half life as a function of the lightest neutrino mass with errors bands that include the hadronization uncertainty. This can be compared with the right panel where this hadronic uncertainty has been neglected.
\end{itemize}

To illustrate the use of the developed EFT framework, we studied several scenarios with light sterile neutrinos. Scenarios with two additional sterile neutrinos can reproduce all neutrino oscillation data. In the normal hierarchy, a minimal model where sterile neutrinos only interact via mixing, leads to  ${}^{136}$Xe half lives that for all choices of $m_4$ and $m_5$ are (slightly) above $10^{28}$ yr, the prospected sensitivity of next-generation experiments. In the inverted hierarchy a detection will be possible  if at least one of the neutrinos has a mass above $\mathcal O(500)$ MeV. These conclusions change drastically in the presence of higher-dimensional interactions. For instance,  $0\nu\beta\beta$ experiments already exclude scalar interactions at a scale of $10$ TeV if one of the neutrinos has a mass around the keV scale and the other at the GeV scale. Such mass ranges appear in
scenarios where sterile neutrinos can account for both Dark Matter and the matter-antimatter asymmetry \cite{Boyarsky:2018tvu}, although to account for both at least three sterile neutrinos are necessary. Depending on the exact neutrino masses, next-generation experiments are sensitive to scales up to $\mathcal O(100)$ TeV. The framework developed here, and the master formula in \eqref{eq:T1/2} in particular, can be used directly to assess the impact of \NLDBD\ experiments on any BSM scenario with light sterile neutrinos and should  prove useful when comparing \NLDBD\ with other probes of sterile neutrinos.

\section*{Acknowledgments}

We acknowledge stimulating discussions with Vincenzo Cirigliano, Bhupal Dev, Michael Graesser, Javier Men{\'e}ndez, and Jiang-Hao Yu. 
This work was supported by  the  RHIC Physics Fellow Program of the RIKEN BNL Research Center (JdV) and the US DOE under grant
numbers DE-SC0009919 (WD). This research was supported in part by the LDRD program at Los Alamos National Laboratory (EM, KF), the DOE topical collaboration on ``Nuclear Theory for Double-Beta Decay and Fundamental Symmetries'' (EM), the US DOE, Office of Science, Office of Nuclear Physics under  award  number DE-AC52-06NA25396 (EM,KF).

\newpage

\appendix

\section{Long-distance potentials induced by dim-7 operators}\label{Appdump}
Here we collect the potentials induced by dim-7 interactions. As the contributions of dim-7 vector operators can be obtained from the dim-6 potentials through a shift of the scalar interactions, see Eq.\ \eqref{eq:dim7V}, we only explicitly list the terms due to $C^{(7)}_{\rm TL}\equiv C^{(7)}_{\rm TL1}+C^{(7)}_{\rm TL2}$ and $C^{(7)}_{\rm TR}\equiv C^{(7)}_{\rm TR2}$. These are given by
   \bea\label{V76}
  V^{(7)}_L &=& 
 m_i \left(C_{\rm VLL}^{(6)} + C_{\rm VRL}^{(6)} \right)_{ei}\left( C_{\rm TR}^{(7)}+ C_{\rm TL}^{(7)}\right)_{ei}
 \Bigg[\frac{4 g_Tm_N}{g_Mm_\pi} \left(\boldsigma_1 \cdot \boldsigma_2 \,h_{GT}^{MM} + S^{(12)}\,h^{MM}_{T} \right)\nn\\
    &&-    \frac{g^\prime_T  g_V }{g_A^2} \frac{\vec q^2}{m_N m_\pi} h_F\Bigg]
 +m_\pi \left(C_{\rm TRR}^{(6)}\right)_{ei}\left(C_{\rm TR}^{(7)}\right)_{ei}\frac{8 g_T^2}{g_A^2}\left[ \boldsigma_1 \cdot \boldsigma_2 \,h_{GT,\,sd}^{AA}\right]\nn\\
  &&+m_\pi \left(C_{\rm TRR}^{(6)}\right)_{ei}\left(C_{\rm TL}^{(7)}\right)_{ei}\frac{8 g_T^2}{3 g_A^2}\left[ \boldsigma_1 \cdot \boldsigma_2 \,h_{GT,\,sd}^{AA}+2 S^{(12)}\, h^{AA}_{T,\,sd}\right]\nn\\
 &&+m_i \frac{m_\pi}{v}\frac{g_T^2}{g_A^2}\left[\left(C_{\rm TL}^{(7)}\right)^2_{ei}+\left(C_{\rm TR}^{(7)}\right)_{ei}^2\right] \boldsigma_1 \cdot \boldsigma_2 \,h_{GT,\,sd}^{AA}\nn\\
&&+m_i \frac{m_\pi}{v}\frac{2g_T^2}{3g_A^2}\left(C_{\rm TL}^{(7)}\right)\left(C_{\rm TR}^{(7)}\right)_{ei}\left[ \boldsigma_1 \cdot \boldsigma_2 \,h_{GT,\,sd}^{AA}+2S^{(12)}\,h_{T,\,sd}^{AA}\right]
  +\dots\,,\nn\\
      V^{(7)}_R &=&  0+\dots\,,\nn\\
      V^{(7)}_M &=& 
 m_\pi \left(C_{\rm VRR}^{(6)} - C_{\rm VLR}^{(6)} \right)_{ei}\left( C_{\rm TR}^{(7)} \right)_{ei}\frac{g_T}{2 g_A}\left[\boldsigma_1 \cdot \boldsigma_2\,(3 h_{GT}^{AP}+2 h_{GT,\,sd}^{AP})-S^{(12)}\,h_{T,\,sd}^{AP}\right]\nn\\
      &&+ m_\pi \left(C_{\rm VRR}^{(6)} - C_{\rm VLR}^{(6)} \right)_{ei}\left( C_{\rm TL}^{(7)} \right)_{ei}\frac{g_T}{2 g_A}\left[\boldsigma_1 \cdot \boldsigma_2\,( h_{GT}^{AP}+2 h_{GT,\,sd}^{AP})-S^{(12)}\,(2 h_T^{AP}+h_{T,\,sd}^{AP})\right]\nn\\
      &&+ m_i\frac{g_T}{2g_A}\frac{B}{m_\pi} \left(C_{\rm SRL}^{(6)} -C_{\rm SLL}^{(6)} \right)_{ei} \left(C_{\rm TR}^{(7)} - C_{\rm TL}^{(7)} \right)_{ei}\left[\boldsigma_1 \cdot \boldsigma_2\, h_{GT}^{AP}+S^{(12)}\,h_T^{AP}\right]+\dots \,,\nn\\
  \eea
where $h^{ab}_{K,\, sd} = h_K^{ab}\vec q^2/m_\pi^2$ and 
 the dots indicate that the complete dimension-seven potentials involve contributions from the dim-7 vector operators.

The resulting left-handed amplitude can be written as,
\begin{eqnarray}\label{Anu67}
\mathcal A^{(7)}_L(m_i) &=& \frac{m_\pi^2}{4m_e v} \Bigg\{
\left[\left( C^{(7)}_{\rm TR} \right)_{ei}  \mathcal M_{T1,\, sd}(m_i)+ \left(C^{(7)}_{\rm TL} \right)_{ei} \mathcal M_{T2,\, sd}\right]  \left( C^{(6)}_{\rm TRR} \right)_{ei}  \nn \\
& & -\frac{m_i m_N}{4m_\pi^2} \mathcal M_{TV}(m_i) \left(C^{(6)}_{\rm VLL} + C^{(6)}_{\rm VRL}\right)_{ei}
\left( C^{(7)}_{\rm TR} + C^{(7)}_{\rm TL}\right)_{ei}\nn\\
&&-\frac{g_T^2}{g_A^2}\frac{m_i}{v}\Bigg[\left(C^{(7)}_{\rm TL} + C^{(7)}_{\rm TR}\right)_{ei}^2M_{GT,\,sd}^{AA}\nn\\
&&+\frac{4}{3}\left( C^{(7)}_{\rm TL}\right)_{ei}\left( C^{(7)}_{\rm TR}\right)_{ei}\left(M_{T,\,sd}^{AA}-M_{GT,\,sd}^{AA}\right)\Bigg]
  \Bigg\}\,.    
\end{eqnarray}
The amplitude for right-handed electrons does not obtain contributions from the dim-7 tensor operators, which implies 
\begin{eqnarray}
\mathcal A^{(7)}_R(m_i) &=& 0+\dots\,,
\end{eqnarray}
while the ``magnetic'' amplitude $\mathcal A_M$ is given by
\begin{eqnarray}
\mathcal A^{(7)}_M(m_i) &=& -\frac{m_\pi^2}{8m_ev} \Bigg\{  
 \mathcal M_{T P}(m_i) 
\frac{ m_i B}{m_\pi^2}\left( C^{(6)}_{\rm SRL} - C^{(6)}_{\rm SLL} \right)_{ei}
\left( C^{(7)}_{\rm TR} - C^{(7)}_{\rm TL}  \right)_{ei}
 \nn\\
& &+ \left(C_{\rm VRR}^{(6)} - C_{\rm VLR}^{(6)}\right)_{ei} \left[  \left(C^{(7)}_{\rm TR}\right)_{ei} \mathcal M_{TA\, 1}(m_i)
+ \left(C^{(7)}_{\rm TL}  \right)_{ei}\mathcal M_{TA\, 2}(m_i)
\right]
\Bigg\}\,.
\end{eqnarray}
The additional matrix elements that appear in the above potentials are given by
\begin{eqnarray}
\mathcal M_{T1, \, sd}(m_i) &=& -8\frac{g^2_T}{g^2_A} M^{AA}_{GT,\, sd}(m_i) \, , \nn \\
\mathcal M_{T2, \, sd}(m_i) &=&- \frac{8 g^2_T}{3 g^2_A} \left( M^{AA}_{GT,\, sd}(m_i) + 2M^{AA}_{T,\, sd}(m_i)\right)\, , \nn \\
\mathcal M_{TA1}(m_i) &=& \frac{g_T}{ g_A} \left( 3 M^{AP}_{GT}(m_i) + 2 M^{AP}_{GT,\, sd}(m_i) - M^{AP}_{T,\, sd}(m_i)\right)\, , \nn \\
\mathcal M_{TA2}(m_i) &=& \frac{ g_T}{ g_A} \left( M^{AP}_{GT}(m_i) + 2 M^{AP}_{GT,\, sd}(m_i) -2 M^{AP}_{T}(m_i) - M^{AP}_{T,\, sd}(m_i)\right)\, .
\end{eqnarray}

\section{Hard-neutrino exchange contributions from dim-7 operators}\label{app:hardNu7}
In this appendix we collect the contributions from dim-7 tensor operators, $ C_{\rm TL,TR}^{(7)}$, due to hard-neutrino exchange, mentioned in Sect.\ \ref{sec:hardNu}. As noted in Sect.\ \ref{sec:rot}, all contributions from the dim-7 vector operators can again be obtained from a shift of the dim-6 scalar couplings, see Eq.\ \eqref{eq:dim7V}, and we do not explicitly discuss them here.
The contributions in the $\pi\pi$ sector can be written as 
\bea
c_{i\, L}^{\nu\pi\pi\,7} &=&\frac{\Lambda_\chi^2}{v^2}\Bigg\{g_{\rm TL,TR}^{\pi\pi}(m_i)\left(C_{\rm TL}^{(7)}\right)_{ei}\left(C_{\rm TR}^{(7)}\right)_{ei}
+g_{\rm TL}^{\pi\pi}(m_i)\left[\left(C_{\rm TL}^{(7)}\right)^2_{ei}+\left(C_{\rm TR}^{(7)}\right)^2_{ei}\right]\Bigg\}\,,\nn\\
c_{i\, L}^{\pi\pi\,7}&=&\frac{\Lambda_\chi^2}{F_\pi^2}\left[4g_{\rm TL,TR}^{\pi\pi}(m_i)\left(C_{\rm TL}^{(7)}\right)_{ei}\left(C_{\rm TRR}^{(6)}\right)_{ei}+8g_{\rm TL}^{\pi\pi}(m_i)\left(C_{\rm TR}^{(7)}\right)_{ei}\left(C_{\rm TRR}^{(6)}\right)_{ei}\right]
\,,\\
c_{i\,L}^{\prime \nu\pi\pi\,7}&=&-\frac{1}{16}g_{\rm T,VLL}^{\pi\pi}(m_i)\left[\left(C_{\rm TL}^{(7)}\right)_{ei}\left(C_{\rm VRL}^{(6)}\right)_{ei}+\left(C_{\rm TR}^{(7)}\right)_{ei}\left(C_{\rm VLL}^{(6)}\right)_{ei}\right]+\left({\rm VLL}\leftrightarrow {\rm VRL}\right)\,,\nn
\eea
where  $g^{\pi \pi}_{\rm TL},g^{\pi \pi}_{\rm TL,TR}=\Or(F_\pi^2)$.
In the $\pi N$ sector we have
\bea
c_{i\,L}^{\nu\pi N\,7} &=&-\frac{1}{8}g_{\rm T,VLL}^{\pi N}(m_i)\left[\left(C_{\rm TL}^{(7)}\right)_{ei}\left(C_{\rm VRL}^{(6)}\right)_{ei}+\left(C_{\rm TR}^{(7)}\right)_{ei}\left(C_{\rm VLL}^{(6)}\right)_{ei}\right]+\left({\rm VLL}\leftrightarrow {\rm VRL}\right)\,,\nn\\
c_{i\,V}^{\pi N\,7}&=&
\left[g_{\rm TL,V}^{\pi N}(m_i)\left(C_{\rm TL}^{(7)}\right)_{ei}+g_{\rm TR,V}^{\pi N}(m_i)\left(C_{\rm TR}^{(7)}\right)_{ei}\right]\left(C_{\rm VLR}^{(6)}-C_{\rm VRR}^{(6)}\right)_{ei}\,,\nn\\
c_{i\,V}^{\nu\pi N\,7}&=&\left[g_{\rm TL,T}^{\pi N}(m_i)\left(C_{\rm TL}^{(7)}\right)_{ei}+g_{\rm TR,T}^{\pi N}(m_i)\left(C_{\rm TR}^{(7)}\right)_{ei}\right]\left(C_{\rm TLL}^{(6)}\right)_{ei}\,,
\eea
where all the LECs scale as $g^{\pi N}_\al=\Or(1)$. 

Finally, the dim-7 contributions in the $NN$ sector are given by,
\bea
\frac{c_{i\, L}^{\nu NN\,7}}{v\Lambda_\chi}&=&-\frac{1}{8}\frac{g_{\rm T,VLL}^{N N}(m_i)}{v\Lambda_\chi}\left[\left(C_{\rm TL}^{(7)}\right)_{ei}\left(C_{\rm VRL}^{(6)}\right)_{ei}+\left(C_{\rm TR}^{(7)}\right)_{ei}\left(C_{\rm VLL}^{(6)}\right)_{ei}\right]+\left({\rm VLL}\leftrightarrow {\rm VRL}\right)\,\nn\\
&&+\frac{\Lambda_\chi^2}{v^2}\Bigg\{g_{\rm TL,TR}^{\rm NN}(m_i)\left(C_{\rm TL}^{(7)}\right)_{ei}\left(C_{\rm TR}^{(7)}\right)_{ei}
+g_{\rm TL}^{\rm NN}(m_i)\left[\left(C_{\rm TL}^{(7)}\right)^2_{ei}+\left(C_{\rm TR}^{(7)}\right)^2_{ei}\right]\Bigg\}\,,
\nn\\
c_{i\, L}^{\rm NN\,7}&=&\Lambda_\chi^2\left[4g_{\rm TL,TR}^{\rm NN}(m_i)\left(C_{\rm TL}^{(7)}\right)_{ei}+8g_{\rm TL}^{\rm NN}(m_i)\left(C_{\rm TR}^{(7)}\right)_{ei}\right]\left(C_{\rm TRR}^{(6)}\right)_{ei}\,,\nn\\
c_{i\, V}^{NN\,7}&=&\left[g_{\rm TL,V}^{\rm NN}(m_i)\left(C_{\rm TL}^{(7)}\right)_{ei}+g_{\rm TR,V}^{\rm NN}(m_i)\left(C_{\rm TR}^{(7)}\right)_{ei}\right]\left(C_{\rm VLR}^{(6)}-C_{\rm VRR}^{(6)}\right)_{ei}\,,\nn\\
\frac{c_{i\, V}^{\nu {\rm NN\,7}}}{v\Lambda_\chi}&=&\frac{1}{4v\Lambda_\chi}g_{\rm T,SRL}^{\rm NN}(m_i)\left[\left(C_{\rm TL}^{(7)}\right)_{ei}\left(C_{\rm SLL}^{(6)}\right)_{ei}+\left(C_{\rm TR}^{(7)}\right)_{ei}\left(C_{\rm SRL}^{(6)}\right)_{ei}\right]\nn\\
&&+\frac{1}{4v\Lambda_\chi}g_{\rm T,SLL}^{\rm NN}(m_i)\left[\left(C_{\rm TL}^{(7)}\right)_{ei}\left(C_{\rm SRL}^{(6)}\right)_{ei}+\left(C_{\rm TR}^{(7)}\right)_{ei}\left(C_{\rm SLL}^{(6)}\right)_{ei}\right]\nn\\
&&+\frac{\Lambda_\chi}{v}\left[g_{\rm TL,T}^{\rm NN}(m_i)\left(C_{\rm TL}^{(7)}\right)_{ei}+g_{\rm TR,T}^{\rm NN}(m_i)\left(C_{\rm TR}^{(7)}\right)_{ei}\right]\left(C_{\rm TLL}^{(6)}\right)_{ei}\,.
\eea
Of the above LECs that were not encountered in Sect.\ \ref{sec:hardNuGen},  $g_{\rm TL,TR}^{\rm NN}$, $g_{\rm TL}^{\rm NN}$ and $g_{\rm TL,V}^{\rm NN}$, $g_{\rm TR,V}^{\rm NN}$  are enhanced with respect to the NDA expectation and scale as $\Or(1/F_\pi^2)$ and $\Or(\Lambda_\chi^2/F_\pi^2)$, respectively. The RGEs that signal the enhancement of these LECs are collected in Appendix \ref{app:RGE}.
The LECs $g_{\rm TL,T}^{\rm NN}$ and $g_{\rm TR,T}^{\rm NN}$ do not need to be enhanced and scale as $\Or(1/\Lambda_\chi^2)$. 

As one would expect, the above equations lead to contributions that scale as $v^{-5}$ ($v^{-6}$) for one (two) insertions of dim-7 operators. Interestingly, the terms proportional to $g^{\pi\pi}_{\rm TL,TR}$ and $g_{\rm TL}^{\pi\pi}$ give rise to contributions that are enhanced by $\Lambda^2_\chi/F_\pi^2$ over the corresponding long-distance contributions.
In addition, we note that  $C_{\rm TL,TR}^{(7)}$ combined with dim-6 vector or scalar operators induces the same LECs that were already encountered in the insertions of dim-6 operators only (see Sect.\ \ref{sec:hardNuGen}), namely, $g_{\rm T,VLL}^{\al}$, $g_{\rm T,VRL}^{\al}$, $g_{\rm T,SLL}^{\al}$, and $g_{\rm T,SRL}^{\al}$. The hadronic parts of $C^{(7)}_{\rm TL,TR}\times  C^{(6)}_{i}$ insertions are very similar to those of  $C^{(6)}_{\rm TRR, TLL}\times  C^{(6)}_{i}$, and, although the former involve an additional derivative this can be supplied by a factor of $\slashed q$ from the neutrino propagator in the latter case. This leads to a relation between the amplitudes of $C^{(7)}_{\rm TL,TR}\times  C^{(6)}_{i}$, resulting from the $\propto m_i$ term in the neutrino propagator, and those of $C^{(6)}_{\rm TRR, TLL}\times  C^{(6)}_{i}$, which result from the $\slashed q$ term instead.
After working out the leptonic Dirac algebra one finds that the $S_{\rm eff}$ in several cases differ only by an overall constant $\sim m_i/v$. As these are relations between the effective actions they are also respected by the corresponding long-distance amplitudes.
\section{Non-perturbative renormalization of LECs}
\label{app:RGE}
Here we collect the RGEs which give rise to the enhancement of LECs over their NDA expectations.
These follow from the requirement that the LO $nn\to ppee$ amplitude is finite and $\mu$ independent, see Refs.\ \cite{Cirigliano:2018hja,Cirigliano:2018yza} for details. The RGEs for the LECs related to two insertions of dim-6 operators are given by,
\bea
2  \frac{d}{d\ln\mu}  \tilde{g}_{\text{S1}}^{\text{NN}}&=&-8 g_A^2 \frac{g_{\text{S1}}^{\pi \pi }}{F_{\pi }^2}-g_S^2\,,\nn\\
2  \frac{d}{d\ln\mu} \tilde{g}_{\text{S2}}^{\text{NN}}&=&-8 g_A^2\frac{ g_{\text{S2}}^{\pi \pi }}{F_{\pi }^2}+g_S^2\,,\\\frac{1}{2} \frac{d}{d\ln\mu}    \tilde{g}_{\text{TT}}^{\text{NN}}&=&-2g_A^2 \frac{g_{\text{TT}}^{\pi \pi }}{F_{\pi }^2}-12 g_T^2\,,\nn\\
\frac{d}{d\ln\mu}   \tilde{g}_{\nu }^{\text{NN}}&=&\frac{2 g_A^2+g_V^2}{2}\,,\nn\\
\frac{d}{d\ln\mu} \tilde{g}_{\text{LR}}^{\text{NN}}&=&2g_A^2\frac{g_{\text{LR}}^{\pi \pi }}{F_{\pi }^2}-g_A^2+\frac{g_V^2}{2}\,,\nn\\
\frac{d}{d\ln\mu}  \tilde{g}_{\text{SLL},\text{VLL}}^{\text{NN}}&=&\frac{g_S g_V}{8}\,,\nn\\
\frac{d}{d\ln\mu} \left(8   \tilde{g}_{\rm T,\text{SLL}}^{\text{NN}}-4   g_A^2 \tilde{g}_{\rm T,\text{SLL}}^{\text{$\pi $N}}\right)&=&4 g_A g_T B\Lambda _{\chi }-4 g_A^2 m_{\pi }^2 g_{\rm T,\text{SLL}}^{\text{$\pi $N}}\,,\nn\\
\frac{d}{d\ln\mu} \left(8   \tilde{g}_{\rm T,\text{SRL}}^{\text{NN}}-4   g_A^2 \tilde{g}_{\rm T,\text{SRL}}^{\text{$\pi $N}}\right)&=&-4 g_A  g_T B\Lambda _{\chi }-4g_A m_{\pi }^2 g_{\rm T,\text{SRL}}^{\text{$\pi $N}}\,,\nn\\
\frac{d}{d\ln\mu}  \tilde{g}_{\text{TLL},\text{VLL}}^{\text{NN}}&=&g_A g_T\,,\nn\\
\frac{d}{d\ln\mu} \tilde g_{\rm SLL,VRR}&=&-g_A^2 m_{\pi }^2 \left(g_{\text{SLL},\text{VRR}}^{\text{$\pi $N}}-2 g_{\text{SLL},\text{VRR}}^{\text{$\pi \pi $}}\right)\,,\nn\\
\tilde g_{\rm SLL,VRR} &=& \left(2 \tilde{g}_{\text{SLL},\text{VRR}}^{\text{NN}}-g_A^2 \tilde{g}_{\text{SLL},\text{VRR}}^{\text{$\pi $N}}+g_A^2 \tilde{g}_{\text{SLL},\text{VRR}}^{\text{$\pi \pi $}}\right)\,,\nn\eea
where we introduced $\tilde g_i = \left(\frac{4\pi}{m_N C}\right)^2 g_i$. The RGEs for the dim-7 pieces can be written as,
\bea
\frac{d}{d\ln\mu} \left(-8 \tilde{g}_{\text{TL},\rm V}^{\text{NN}}+4 g_A^2 \tilde{g}_{\text{TL},\rm V}^{\text{$\pi $N}}-\frac{64 \pi ^2 g_A g_T}{\tilde{C}^2 m_N^2}\right)&=&g_A \left(4 g_A m_{\pi }^2 g_{\text{TL},\rm V}^{\text{$\pi $N}}+g_T \left(m_{\pi }^2+2 m_{\nu _i}^2\right)\right)\,,\nn\\
\frac{d}{d\ln\mu}  \left(-8 \tilde{g}_{\text{TR},\rm V}^{\text{NN}}+4 g_A^2 \tilde{g}_{\text{TR},\rm V}^{\text{$\pi $N}}-\frac{64 \pi ^2 g_A g_T}{\tilde{C}^2 m_N^2}\right)&=&g_A \left(4 g_A m_{\pi }^2 g_{\text{TR},\rm V}^{\text{$\pi $N}}-g_T \left(m_{\pi }^2-2 m_{\nu _i}^2\right)\right)\,,\nn\\
\frac{d}{d\ln\mu}   \left( \tilde{g}_{\text{TL}}^{\text{NN}}+\frac{12 \pi ^2 g_T^2}{\tilde{C}^2 m_N^2 \Lambda _{\chi }^2}\right)&=&-\frac{3}{8} g_T^2 \frac{m_{\nu _i}^2}{\Lambda _{\chi }^2}+\frac{ g_A^2 g_{\text{TL}}^{\pi \pi } }{2F_{\pi }^2}\,,\nn\\
\frac{d}{d\ln\mu}  \left( \tilde{g}_{\text{TL},\text{TR}}^{\text{NN}}+\frac{8 \pi ^2 g_T^2}{\tilde{C}^2 m_N^2\Lambda _{\chi }^2}\right)&=&-\frac{1}{4}g_T^2 \frac{m_{\nu _i}^2}{\Lambda _{\chi }^2}+\frac{ g_A^2  g_{\text{TL},\text{TR}}^{\pi \pi }}{2F_{\pi }^2}\,.
\eea

\section{Derivative contributions to four-quark two-lepton operators}\label{app:mnu2}
Integrating out a heavy neutrino induces operators involving four-quark fields, two-lepton fields, and a derivative, 
in addition to the contributions discussed in Sect.\ \ref{GeVneutrinos}. These terms can be generated through dim-6 operators, proportional to $\bar m_\nu^{-2}$, or through dim-7  interactions, $\propto \bar m_\nu^{-1}$. When such operators are matched onto Chiral Perturbation Theory, they will give rise to interactions that come with new, unknown, LECs. For the dimension-6 contributions one expects the derivative operators to be suppressed by $m_q/m_\nu$ compared to the operators of Sect.\ \ref{GeVneutrinos}~\footnote{This estimate holds when the interactions in Sect.\ \ref{GeVneutrinos} induce $\pi\pi$ couplings at LO, while the corresponding derivative terms do not. This is the case for both the scalar and tensor dimension-6 operators.}. However, there are several Wilson coefficients for which the contributions in Sect.\ \ref{GeVneutrinos} scale as  $\sim v^4/\Lambda^4$, while the derivative couplings can be induced by interfering with the SM coupling, $C_{\rm VLL}^{(6)}$. This implies that there is a regime, $1\lesssim \frac{\Lambda^2 m_q}{v^2m_\nu}$, where the derivative interactions are important.

Here we will refrain from constructing the full set of derivative interactions that are induced in this way. The reason being that these interactions are only relevant in fairly specific scenarios and regimes of parameter space. In addition, their construction would lead to a sizable number of new operators all of which will come with unknown LECs, meaning that we cannot do better than an order-of-magnitude estimates in this regime.
We therefore restrict ourselves to listing the contributions can be captured by the usual dim-9 interactions, i.e.\ the terms that, through the equations of motions, can be written as $m_q\times \Or^{(9)}$.

The $\sim m_q$ terms resulting from interference with the SM-like vector operators are given by
\bea\label{deltaC9}
\dt  C^{(9)}_{1\,L} &=& -4vm_d\, C_{\rm VLL}^{(6)}\bar m_\nu^{-2} C_{\rm TRR}^{(6)\, T}\,,\qquad \dt C^{(9)\prime }_{1\,L} =  -4vm_u \,C_{\rm VRL}^{(6)}\bar m_\nu^{-2} C_{\rm TRR}^{(6)\, T}\,,\nn\\
\dt C^{(9)}_{2\,L} &=&v \left(m_uC_{\rm VLL}^{(6)}-m_d C_{\rm VRL}^{(6)}\right)\bar m_\nu^{-2} C_{\rm SLR}^{(6)\, T}\,,\qquad  \dt C^{(9)\prime }_{2\,L} = v\left(m_uC_{\rm VRL}^{(6)}-m_d C_{\rm VLL}^{(6)}\right)\bar m_\nu^{-2} C_{\rm SRR}^{(6)\, T}\,,\nn\\
\dt C^{(9)}_{4\,L} &=&-4 v\left(m_d C_{\rm VRL}^{(6)}+m_u C_{\rm VLL}^{(6)}\right)\bar m_\nu^{-2} C_{\rm TRR}^{(6)\, T},\nn\\
\dt C^{(9)}_{5\,L} &=&-\frac{v}{2}\left[\left(m_uC_{\rm VLL}^{(6)}-m_d C_{\rm VRL}^{(6)}\right)\bar m_\nu^{-2} C_{\rm SRR}^{(6)\, T}+\left(m_uC_{\rm VRL}^{(6)}-m_d C_{\rm VLL}^{(6)}\right)\bar m_\nu^{-2} C_{\rm SLR}^{(6)\, T}\right]
\,.
\eea
These $\sim m_q$ terms are expected to contribute at the same order as the original derivative couplings. Although NDA  counts a derivative as a factor of $\sim \Lambda_\chi\gg m_q$ when matching onto ChiPT, the insertion of $m_q$ allows for a chirality flip which changes the chiral symmetry properties of the operators. In the above cases this allows one to trade derivative operators that do not induce $\pi\pi$ terms at LO, for interactions that do induce $\pi\pi$ terms proportional to $m_q$.  As a result, one expects the relative scaling of $\Lambda_\chi\sim  (4\pi)^2 m_q$ for the derivative and $ m_q$ terms, respectively, allowing us to use the latter as an order-of-magnitude estimate for the total.

We can in principle derive similar $\sim m_q$ terms for the interference of $C_{\rm VLR}^{(6)}$ with $C_{\rm VLL}^{(6)}$. However, in this case a chirality flip does not lead to chiral enhancement (the induced operators are dim-9 vector operators which do not induce $\pi\pi$ terms at LO), implying that the $m_q$ terms cannot be used to estimate the derivative pieces in this case. If such interference terms are important we would have to construct the derivative terms to obtain an order-of-magnitude estimate.

We take a similar approach for the derivative operators involving dim-7 interactions. Namely, we only list the pieces $\sim m_q$ that can be written in terms of the dim-9 operators discussed in Sect.\  \ref{GeVneutrinos}. These pieces can be written as
\bea
\dt  C^{(9)}_{1\,L} &=& -\left[ C_{\rm VLL}^{(6)}
+\left(\frac{m_u}{v}C_{\rm TL}^{(7)}+\frac{m_d}{v} C_{\rm TR}^{(7)}\right)\right]\bar m_\nu^{-1} \left(m_uC_{\rm TL}^{(7)}+m_d C_{\rm TR}^{(7)}\right)^T
\nn\\
\dt C^{(9)\prime }_{1\,L} &=& - \left[ C_{\rm VRL}^{(6)}+\left(\frac{m_d}{v}C_{\rm TL}^{(7)}+\frac{m_u}{v} C_{\rm TR}^{(7)}\right)\right]\bar m_\nu^{-1} \left(m_dC_{\rm TL}^{(7)}+m_u C_{\rm TR}^{(7)}\right)^T
\,,\nn\\
\dt C^{(9)}_{4\,L} &=& -\left[ C_{\rm VLL}^{(6)}\bar m_\nu^{-1} \left(m_dC_{\rm TL}^{(7)}+m_u C_{\rm TR}^{(7)}\right)^T+ C_{\rm VRL}^{(6)}\bar m_\nu^{-1} \left(m_uC_{\rm TL}^{(7)}+m_d C_{\rm TR}^{(7)}\right)^T\right]\nn\\
&&-\frac{2}{v} \left(m_uC_{\rm TL}^{(7)}+m_d C_{\rm TR}^{(7)}\right)\bar m_\nu^{-1} \left(m_dC_{\rm TL}^{(7)}+m_u C_{\rm TR}^{(7)}\right)^T\,,
\eea
while possible terms $\propto \bar m_\nu^{-2}$ do not appear for the combinations $C_{\rm VLL}^{(6)}\times C_{\rm TL,TR}^{(7)}$ and $\left( C_{\rm TL,TR}^{(7)}\right)^2$.

We again stress that in all these cases one should in principle consider the additional operators involving derivatives, along with the unknown LECs that result from them, and the results above can only serve as an order-of-magnitude estimate.

\section{Matching and interpolation formulae for the leptoquark scenario}\label{AppLQ}

As discussed in Sect.~\ref{lepto}, right below the electroweak scale we can match the leptoquark model of Eq.\ \eqref{eq:LQlag} to the dim-6 operators involving the first-generation quarks and charged leptons
\begin{align}
{\cal L }^{(6)}_{\Delta L=0}=\frac{2G_F}{\sqrt{2}}\bigg[\bar{c}_{\substack{{\rm SR}\\ea}}^{(6)}~\bar{u}_Ld_R\bar{e}_L\nu_{Ra}+\bar{c}^{(6)}_{\substack{{\rm T}\\ea}}~\bar{u}_L \sigma^{\mu\nu}d_R\bar{e}_L\sigma^{\mu\nu}\nu_{Ra}\bigg]\,,
\end{align}
where
\begin{align}
\bar{c}^{(6)}_{\substack{{\rm SR}\\ea}}=4\bar{c}^{(6)}_{\substack{{\rm T}\\ea}}=\frac{v^2}{2m^2_{\rm LQ}}y^{\overline{LR}}_{1a}y^{RL*}_{1e}\,,
\end{align}
and $a$ runs from $1$ to $n$. 
For simplicity we neglect the evolution to lower energies and simply give the matching to the operators  in Eq.~\eqref{redefC6}. We obtain
\begin{align}
\left(C^{(6)}_{\rm VLL}\right)_{ei}=-2 V_{ud} U_{ei}\,, \hspace{0.5cm}
\left(C^{(6)}_{\rm SRR}\right)_{ei}=4\left(C^{(6)}_{\rm TRR}\right)_{ei}=\sum^n_{a=1} \bar{c}^{(6)}_{\substack{{\rm SR}\\ ea}}U^*_{3+a,i}\,.
\end{align}
With these nonzero Wilson coefficients the $0\nu\beta\beta$ decay rate can be directly read from the master formula. For mass eigenstates $m_i<\Lambda_\chi$, the interactions only introduce contributions to the subamplitude ${\cal A}_L(m_i)$ which is given by
\begin{eqnarray}\label{ALLQ}
{\cal A}_L(m_i)&=&-\frac{m_i}{4m_e}\bigg\{\left[{\cal M}_V(m_i)+{\cal M}_A(m_i) \right]\left(C_{\rm VLL}^{(6)} \right)_{ei}^2\nn\\
&&\qquad+\left[{\cal M}_S(m_i)+{\cal M}_{PS}(m_i)\frac{B^2}{m^2_{\pi}} \right]\left(C_{\rm SRR}^{(6)} \right)_{ei}^2-{\cal M}_T(m_i)\left(C_{\rm TRR}^{(6)} \right)_{ei}^2\bigg\}\nonumber\\
&&
+\frac{1}{4m_e}\left[2B{\cal M}_{PS}(m_i)\left(C^{(6)}_{\rm SRR} \right)_{ei}-m_N{\cal M}_{TV}(m_i)\left(C^{(6)}_{\rm TRR}\right)_{ei} \right]\left(C^{(6)}_{\rm VLL}\right)_{ei}\nonumber\\
 &&+\frac{m^2_{\pi}}{m_ev}\bigg[\left(\frac{m_iv}{m^2_{\pi}F_{\pi}^2} c_{iL}^{\nu\pi\pi}+\frac{v}{\Lambda_{\chi}}c^{\prime\pi\pi}_{iL} \right){\cal M}_{PS, sd}\nonumber\\
&&
 +\frac{1}{2}\left(c_{iL}^{\pi N}-\frac{v}{\Lambda_{\chi}}c^{\prime \pi\pi}_{iL}\right)\left(M_{GT,sd}^{AP}+M_{T,sd}^{AP} \right)-\frac{2}{g_A^2}C_{iL}^{NN}M_{F,sd} \bigg]\,.
\end{eqnarray}
The first three lines arise from the exchange of potential neutrinos while the last two lines arise from hard-neutrino exchange. The couplings are given by
\begin{align}\label{LECsLQ}
c_{iL}^{\nu\pi\pi}&=-2g_{\rm S1}^{\pi\pi}(m_i)\left(C_{\rm SRR}^{(6)} \right)^2_{ei}-2g^{\pi\pi}_{\rm TT}(m_i)\left(C_{\rm TRR}^{(6)} \right)^2_{ei}\,,\nn\\
c_{iL}^{\prime\pi\pi}&=\left[g^{\pi\pi}_{\rm S,VLL}(m_i)\left(C_{\rm SRR}^{(6)} \right)_{ei}-\frac{1}{4}g^{\pi\pi}_{\rm T,VLL}(m_i)\left(C_{\rm TRR}^{(6)} \right)_{ei} \right]\left(C_{\rm VLL}^{(6)} \right)_{ei}\,\nn\\
c^{\pi N}_{iL}&=\frac{v}{\Lambda_{\chi}}\left[g^{\pi N}_{\rm S,VLL}(m_i)\left(C_{\rm SRR}^{(6)} \right)_{ei}-\frac{1}{2}g^{\pi N}_{\rm T,VLL}(m_i)\left(C_{\rm TRR}^{(6)} \right)_{ei} \right]\left(C_{\rm VLL}^{(6)} \right)_{ei}\,,\nn\\
C^{NN}_{iL}&=\frac{v}{\Lambda_{\chi}}\left[g_{\rm S,VLL}^{NN}(m_i)\left(C_{\rm SRR}^{(6)} \right)_{ei}-\frac{1}{2}g_{\rm T,VLL}^{NN}(m_i)\left(C_{\rm TRR}^{(6)} \right)_{ei} \right]\left(C_{\rm VLL}^{(6)} \right)_{ei}\nonumber\\
&+m_iv\left[\frac{1}{4}g_{\nu}^{NN}(m_i)\left(C_{\rm VLL}^{(6)} \right)_{ei}^2+\frac{1}{4}g_{\rm S1}^{NN}(m_i)\left(C_{\rm SRR}^{(6)} \right)_{ei}^2+\frac{1}{4}g_{\rm TT}^{NN}(m_i)\left(C_{\rm TRR}^{(6)} \right)_{ei}^2 \right]\,.
\end{align}

Mass eigenstates with  $m_i>\Lambda_{\chi}$ are integrated out at the quark level and induce dim-9 operators. From Eqs.~\eqref{dim91} and \eqref{deltaC9} we find the following contributions
\begin{align}
C_{1L}^{(9)}&=-\frac{v}{2m_i}\left(C_{\rm VLL}^{(6)} \right)_{ei}^2-\frac{4m_dv}{m^2_i}\left(C^{(6)}_{\rm VLL} \right)_{ei}\left(C_{\rm TRR}^{(6)} \right)_{ei} \\
C_{2L}^{(9)\prime}&=\frac{v}{m_i}\left[\frac{1}{2}\left(C_{\rm SRR}^{(6)}\right)^2_{ei}+8\left(C_{\rm TRR}^{(6)} \right)_{ei}^2\right]-\frac{m_dv}{m^2_i}\left(C_{\rm VLL}^{(6)}\right)_{ei}\left(C_{\rm SRR}^{(6)}\right)_{ei},\\
C_{3L}^{(9)\prime}&=\frac{16 v}{m_i}\left(C_{\rm TRR}^{(6)}\right)^2_{ei},\\
C_{4L}^{(9)}&=-\frac{4m_uv}{m^2_i}\left(C_{\rm VLL}^{(6)} \right)_{ei}\left(C_{\rm TRR}^{(6)} \right)_{ei},\\
C_{5L}^{(9)}&=-\frac{m_uv}{2m^2_{i}}\left(C_{\rm VLL}^{(6)}\right)_{ei}\left(C_{\rm SRR}^{(6)}\right)_{ei},
\end{align}
The resulting amplitude ${\cal A}^{(9)}_L$ is 
\begin{align}
{\cal A}^{(9)}_L=\frac{m^2_{\pi}}{m_ev}\left[\left(\frac{c_L^{\pi\pi}}{m^2_{\pi}}+c^{\prime\pi\pi}_L \right){\cal M}_{PS,sd}+\frac{1}{2}\left(c^{\pi N}_L-c^{\prime\pi\pi}_L \right)\left(M^{AP}_{GT,sd}+M^{AP}_{T,sd} \right)-\frac{2}{g^2_A}c_L^{NN}M_{F,sd} \right],\nn
\end{align}
with $c_L^{\pi\pi}$, $c^{\prime\pi\pi}_L$, $c^{\pi N}_L$, and $c_L^{NN}$ defined in Eqs.~\eqref{pipidim9}, \eqref{eq:dim9PiN}, and \eqref{eq:dim9NN}.

We introduce interpolation formulae for the LECs appearing in Eqs.~\eqref{LECsLQ} along the lines of Sect.~\ref{naiveIP}. For the pionic terms we use
\bea\label{naiveinterpipi}
g_{\rm S1}^{\pi\pi}(m_i) &=& g_{\rm S1}^{\pi\pi}(0) \frac{1}{1 +  8 \frac{m_i^2}{F_\pi^2} g_{\rm S1}^{\pi\pi}(0) \left[g_2^{\pi\pi}(m_0)  - B^2\right]^{-1}  }\,,\nn \\
g_{\rm TT}^{\pi\pi}(m_i) &=& g_{\rm TT}^{\pi\pi}(0) \frac{1}{1 +  \frac{m_i^2}{F_\pi^2} g_{\rm TT}^{\pi\pi}(0) \left[4 g_3^{\pi\pi}(m_0)  + 2 g_2^{\pi\pi}(m_0)\right]^{-1}}\,,\nn\\
g_{\rm{S,VLL}}^{\pi\pi}(m_i) &=&g_{\rm{S,VLL}}^{\pi\pi}(0)\frac{1}{1-\frac{2 m_i^2}{\Lambda_\chi}g_{\rm{S,VLL}}^{\pi\pi}(0)\left[B+\frac{1}{2m_\pi^2}\left[{-}2 m_d g_2^{\pi\pi}(m_0)+m_u g_5^{\pi\pi}(m_0)\right]\right]^{-1}}\,,\nn\\
g_{\rm{T,VLL}}^{\pi\pi}(m_i) &=&g_{\rm{T,VLL}}^{\pi\pi}(0)\frac{1}{1+\frac{ m_i^2m_\pi^2}{8\Lambda_\chi}g_{\rm{T,VLL}}^{\pi\pi}(0)\left[ m_u g_4^{\pi\pi}(m_0)+ {\frac{5}{3}}m_d m_\pi^2 g_1^{\pi\pi}(m_0)\right]^{-1}}\,.
\eea
The pion-nucleon interpolation formula is
\bea
g_{\rm{S,VLL}}^{\pi N}(m_i) &=&g_{\rm{S,VLL}}^{\pi \pi}(m_i) \,,\nn\\
g_{\rm{T,VLL}}^{\pi N}(m_i) &=&g_{\rm{T,VLL}}^{\pi N}(0)\frac{1}{1+\frac{ m_i^2m_\pi^2}{4\Lambda_\chi}g_{\rm{T,VLL}}^{\pi N}(0)\left[ m_u g_4^{\pi \pi}(m_0)+ 2m_d m_\pi^2 g_1^{\pi N}(m_0)\right]^{-1}}\,.
\eea
A similar formula can be written down for $g_{\rm{S,VLL}}^{\pi N}(m_i)$ once four-quark operators involving derivatives are explicitly included, see App.~\ref{app:mnu2}.  The nucleon-nucleon interpolation formulae become
\bea
g_{\rm S1}^{NN}(m_i) &=& g_{\rm S1}^{NN}(0) \frac{1}{1 +  m_i^2 g_{\rm S1}^{NN}(0) \left[2 g_2^{NN}(m_0)  -\frac{1}{2}g_S^2\right]^{-1}  }\,,\nn \\
g_{\rm TT}^{NN}(m_i) &=& g_{\rm TT}^{NN}(0) \frac{1}{1 +  m_i^2 g_{\rm TT}^{NN}(0) \left[32 g_2^{NN}(m_0) +64 g_3^{NN}(m_0)-24 g_T^2\right]^{-1}  }\,,\nn \\
g_{\rm {S,VLL}}^{NN}(m_i) &=& g_{\rm {S,VLL}}^{NN}(0) \frac{1}{1 - \frac{m_i^2}{\Lambda_\chi} g_{\rm {S,VLL}}^{NN}(0) \left[ m_d\, g_2^{NN}(m_0)  + \frac{1}{2}m_u \,g_5^{NN}(m_0)\right]^{-1}  }\,,\nn \\
g_{\rm {T,VLL}}^{NN}(m_i) &=& g_{\rm {T,VLL}}^{NN}(0) \frac{1}{1 {+}  \frac{m_i^2}{8\Lambda_\chi} g_{\rm {T,VLL}}^{NN}(0) \left[ m_d\, g_1^{NN}(m_0)  +m_u \,g_4^{NN}(m_0)\right]^{-1}  }\,.
\eea
In all the above relations we use the matching scale $\mu = m_0 = 2$ GeV. With some effort the relations can be RGE improved as we did for $g^{\pi\pi}_{\rm LR}(m_i)$, $g^{NN}_{\rm LR}(m_i)$, and $g^{NN}_{\nu}(m_i)$ in Sect.~\ref{naiveIP}. This is more complicated for scalar and tensor interactions because scalar and tensor quark bilinears have non-vanishing anomalous dimensions in contrast to the vector case. Given the large uncertainties, and in order not to clutter up the expressions, we refrain from explicitly including the RGE corrections in the interpolation formula.

\bibliographystyle{utphysmod}
\bibliography{bibliography}

\providecommand{\href}[2]{#2}\begingroup\raggedright\begin{thebibliography}{100}

\bibitem{Arnaboldi:2002te}
C.~Arnaboldi {\em et~al.}, Phys. Lett. {\bfseries B557}, 167 (2003),
[\href{https://arxiv.org/abs/hep-ex/0211071}{{arXiv:hep-ex/0211071}}].

\bibitem{Umehara:2008ru}
S.~Umehara {\em et~al.}, Phys. Rev. {\bfseries C78}, 058501 (2008),
[\href{https://arxiv.org/abs/0810.4746}{{arXiv:0810.4746~[nucl-ex]}}].

\bibitem{Barabash:2010bd}
A.~S. Barabash and V.~B. Brudanin [NEMO Collaboration], Phys. Atom. Nucl.
  {\bfseries 74}, 312 (2011),
[\href{https://arxiv.org/abs/1002.2862}{{arXiv:1002.2862~[nucl-ex]}}].

\bibitem{Gando:2012zm}
A.~Gando {\em et~al.} [KamLAND-Zen Collaboration], Phys. Rev. Lett. {\bfseries
  110}, 062502 (2013),
[\href{https://arxiv.org/abs/1211.3863}{{arXiv:1211.3863~[hep-ex]}}].

\bibitem{Agostini:2013mzu}
M.~Agostini {\em et~al.} [GERDA Collaboration], Phys. Rev. Lett. {\bfseries
  111}, 122503 (2013),
[\href{https://arxiv.org/abs/1307.4720}{{arXiv:1307.4720~[nucl-ex]}}].

\bibitem{Albert:2014awa}
J.~B. Albert {\em et~al.} [EXO-200 Collaboration], Nature {\bfseries 510}, 229
  (2014),
[\href{https://arxiv.org/abs/1402.6956}{{arXiv:1402.6956~[nucl-ex]}}].

\bibitem{Andringa:2015tza}
S.~Andringa {\em et~al.} [SNO+ Collaboration], Adv. High Energy Phys.
  {\bfseries 2016}, 6194250 (2016),
[\href{https://arxiv.org/abs/1508.05759}{{arXiv:1508.05759~[physics.ins-det]}}].

\bibitem{Arnold:2015wpy}
R.~Arnold {\em et~al.} [NEMO-3 Collaboration], Phys. Rev. {\bfseries D92},
  072011 (2015),
[\href{https://arxiv.org/abs/1506.05825}{{arXiv:1506.05825~[hep-ex]}}].

\bibitem{Arnold:2016ezh}
R.~Arnold {\em et~al.} [NEMO-3 Collaboration], Phys. Rev. {\bfseries D93},
  112008 (2016),
[\href{https://arxiv.org/abs/1604.01710}{{arXiv:1604.01710~[hep-ex]}}].

\bibitem{KamLAND-Zen:2016pfg}
A.~Gando {\em et~al.} [KamLAND-Zen Collaboration], Phys. Rev. Lett. {\bfseries
  117}, 082503 (2016),
  [\href{https://arxiv.org/abs/1605.02889}{{arXiv:1605.02889~[hep-ex]}}],
[Addendum: Phys. Rev. Lett.117,no.10,109903(2016)].

\bibitem{Elliott:2016ble}
S.~R. Elliott {\em et~al.}, J. Phys. Conf. Ser. {\bfseries 888}, 012035 (2017),
[\href{https://arxiv.org/abs/1610.01210}{{arXiv:1610.01210~[nucl-ex]}}].

\bibitem{Arnold:2016qyg}
R.~Arnold {\em et~al.} [NEMO-3 Collaboration], Phys. Rev. {\bfseries D94},
  072003 (2016),
[\href{https://arxiv.org/abs/1606.08494}{{arXiv:1606.08494~[hep-ex]}}].

\bibitem{Arnold:2016bed}
R.~Arnold {\em et~al.} [NEMO-3 Collaboration], Phys. Rev. {\bfseries D95},
  012007 (2017),
[\href{https://arxiv.org/abs/1610.03226}{{arXiv:1610.03226~[hep-ex]}}].

\bibitem{Agostini:2017iyd}
M.~Agostini {\em et~al.},
  [\href{https://arxiv.org/abs/1703.00570}{{arXiv:1703.00570~[nucl-ex]}}],
[Nature544,47(2017)].

\bibitem{Aalseth:2017btx}
C.~E. Aalseth {\em et~al.} [Majorana Collaboration], Phys. Rev. Lett.
  {\bfseries 120}, 132502 (2018),
[\href{https://arxiv.org/abs/1710.11608}{{arXiv:1710.11608~[nucl-ex]}}].

\bibitem{Albert:2017owj}
J.~B. Albert {\em et~al.} [EXO Collaboration], Phys. Rev. Lett. {\bfseries
  120}, 072701 (2018),
[\href{https://arxiv.org/abs/1707.08707}{{arXiv:1707.08707~[hep-ex]}}].

\bibitem{Alduino:2017ehq}
C.~Alduino {\em et~al.} [CUORE Collaboration], Phys. Rev. Lett. {\bfseries
  120}, 132501 (2018),
[\href{https://arxiv.org/abs/1710.07988}{{arXiv:1710.07988~[nucl-ex]}}].

\bibitem{Agostini:2018tnm}
M.~Agostini {\em et~al.} [GERDA Collaboration], Phys. Rev. Lett. {\bfseries
  120}, 132503 (2018),
[\href{https://arxiv.org/abs/1803.11100}{{arXiv:1803.11100~[nucl-ex]}}].

\bibitem{Azzolini:2018dyb}
O.~Azzolini {\em et~al.},
[\href{https://arxiv.org/abs/1802.07791}{{arXiv:1802.07791~[nucl-ex]}}].

\bibitem{Arnold:2018tmo}
R.~Arnold {\em et~al.}, Eur. Phys. J. {\bfseries C78}, 821 (2018),
[\href{https://arxiv.org/abs/1806.05553}{{arXiv:1806.05553~[hep-ex]}}].

\bibitem{Adams:2019jhp}
D.~Q. Adams {\em et~al.} [CUORE Collaboration],
[\href{https://arxiv.org/abs/1912.10966}{{arXiv:1912.10966~[nucl-ex]}}].

\bibitem{Alvis:2019sil}
S.~I. Alvis {\em et~al.} [Majorana Collaboration], Phys. Rev. {\bfseries C100},
  025501 (2019),
[\href{https://arxiv.org/abs/1902.02299}{{arXiv:1902.02299~[nucl-ex]}}].

\bibitem{CANDLES_TAUP2019}
K.~Tetsuno {\em et~al.}, J. Phys. Conf. Ser. {\bfseries 1468}, 012132 (2020).

\bibitem{Agostini:2019hzm}
M.~Agostini {\em et~al.} [GERDA Collaboration], Science {\bfseries 365}, 1445
  (2019),
[\href{https://arxiv.org/abs/1909.02726}{{arXiv:1909.02726~[hep-ex]}}].

\bibitem{Azzolini:2019tta}
O.~Azzolini {\em et~al.} [CUPID Collaboration], Phys. Rev. Lett. {\bfseries
  123}, 032501 (2019),
[\href{https://arxiv.org/abs/1906.05001}{{arXiv:1906.05001~[nucl-ex]}}].

\bibitem{Alenkov:2019jis}
V.~Alenkov {\em et~al.}, Eur. Phys. J. {\bfseries C79}, 791 (2019),
[\href{https://arxiv.org/abs/1903.09483}{{arXiv:1903.09483~[hep-ex]}}].

\bibitem{Anton:2019wmi}
G.~Anton {\em et~al.} [EXO-200 Collaboration], Phys. Rev. Lett. {\bfseries
  123}, 161802 (2019),
[\href{https://arxiv.org/abs/1906.02723}{{arXiv:1906.02723~[hep-ex]}}].

\bibitem{Iida:2016vfi}
T.~Iida {\em et~al.},
Nucl. Part. Phys. Proc. {\bfseries 273-275}, 2633 (2016).

\bibitem{Abgrall:2017syy}
N.~Abgrall {\em et~al.} [LEGEND Collaboration], AIP Conf. Proc. {\bfseries
  1894}, 020027 (2017),
[\href{https://arxiv.org/abs/1709.01980}{{arXiv:1709.01980~[physics.ins-det]}}].

\bibitem{Patrick:2017eso}
C.~Patrick and F.~Xie, ``{Status of the SuperNEMO 0$\nu\beta\beta$
  experiment},'' in {\em {Proceedings, Prospects in Neutrino Physics
  (NuPhys2016): London, UK, December 12-14, 2016}}.
\newblock 2017.
\newblock
\href{https://arxiv.org/abs/1704.06670}{{arXiv:1704.06670~[physics.ins-det]}}.
\newblock

\bibitem{Salvio:2019agg}
A.~Salvio and F.~Sannino, eds.,
  \href{http://dx.doi.org/10.3389/978-2-88963-205-3}{{\em {From the Fermi Scale
  to Cosmology}}}.
\newblock Frontiers, 2019.
\newblock
\url{http://www.desy.de/~schwenn/9782889632053-1.PDF}.
\newblock

\bibitem{Adams:2018nek}
[CUORE Collaboration], D.~Q. Adams {\em et~al.}, ``{Update on the recent
  progress of the CUORE experiment},'' in {\em {28th International Conference
  on Neutrino Physics and Astrophysics (Neutrino 2018) Heidelberg, Germany,
  June 4-9, 2018}}.
\newblock 2018.
\newblock
  \href{https://arxiv.org/abs/1808.10342}{{arXiv:1808.10342~[nucl-ex]}}.
\newblock
\url{https://doi.org/10.5281/zenodo.1286904}.
\newblock

\bibitem{Paton:2019kgy}
[SNO+ Collaboration], J.~Paton, ``{Neutrinoless Double Beta Decay in the SNO+
  Experiment},'' in {\em {Prospects in Neutrino Physics (NuPhys2018) London,
  United Kingdom, December 19-21, 2018}}.
\newblock 2019.
\newblock
\href{https://arxiv.org/abs/1904.01418}{{arXiv:1904.01418~[hep-ex]}}.
\newblock

\bibitem{Albert:2017hjq}
J.~B. Albert {\em et~al.} [nEXO Collaboration], Phys. Rev. {\bfseries C97},
  065503 (2018),
[\href{https://arxiv.org/abs/1710.05075}{{arXiv:1710.05075~[nucl-ex]}}].

\bibitem{Gomez-Cadenas:2019sfa}
J.~J. Gomez-Cadenas, ``{Status and prospects of the NEXT experiment for
  neutrinoless double beta decay searches},''
\newblock 2019.
\newblock
\href{https://arxiv.org/abs/1906.01743}{{arXiv:1906.01743~[hep-ex]}}.
\newblock

\bibitem{Han:2017fol}
[PandaX-III Collaboration], K.~Han, ``{PandaX-III: Searching for Neutrinoless
  Double Beta Decay with High Pressure Gaseous Time Projection Chambers},'' in
  {\em {15th International Conference on Topics in Astroparticle and
  Underground Physics (TAUP 2017) Sudbury, Ontario, Canada, July 24-28, 2017}}.
\newblock 2017.
\newblock
\href{https://arxiv.org/abs/1710.08908}{{arXiv:1710.08908~[physics.ins-det]}}.
\newblock

\bibitem{CUPIDInterestGroup:2019inu}
W.~R. Armstrong {\em et~al.} [CUPID Collaboration],
[\href{https://arxiv.org/abs/1907.09376}{{arXiv:1907.09376~[physics.ins-det]}}].

\bibitem{Kobach:2016ami}
A.~Kobach, Phys. Lett. {\bfseries B758}, 455 (2016),
[\href{https://arxiv.org/abs/1604.05726}{{arXiv:1604.05726~[hep-ph]}}].

\bibitem{Weinberg:1979sa}
S.~Weinberg,
Phys. Rev. Lett. {\bfseries 43}, 1566 (1979).

\bibitem{Minkowski:1977sc}
P.~Minkowski,
Phys. Lett. {\bfseries 67B}, 421 (1977).

\bibitem{GellMann:1980vs}
M.~Gell-Mann, P.~Ramond, and R.~Slansky, Conf. Proc. {\bfseries C790927}, 315
  (1979),
[\href{https://arxiv.org/abs/1306.4669}{{arXiv:1306.4669~[hep-th]}}].

\bibitem{Mohapatra:1980yp}
R.~N. Mohapatra and G.~Senjanovic,
Phys. Rev. {\bfseries D23}, 165 (1981).

\bibitem{Zee:1980ai}
A.~Zee, Phys. Lett. {\bfseries 93B}, 389 (1980),
[Erratum: Phys. Lett.95B,461(1980)].

\bibitem{Zee:1985id}
A.~Zee,
Nucl. Phys. {\bfseries B264}, 99 (1986).

\bibitem{Babu:1988ki}
K.~S. Babu,
Phys. Lett. {\bfseries B203}, 132 (1988).

\bibitem{Babu:1988ig}
K.~S. Babu and E.~Ma,
Phys. Rev. Lett. {\bfseries 61}, 674 (1988).

\bibitem{Babu:1988wk}
K.~S. Babu, E.~Ma, and J.~T. Pantaleone,
Phys. Lett. {\bfseries B218}, 233 (1989).

\bibitem{Babu:2001ex}
K.~S. Babu and C.~N. Leung, Nucl. Phys. {\bfseries B619}, 667 (2001),
[\href{https://arxiv.org/abs/hep-ph/0106054}{{arXiv:hep-ph/0106054}}].

\bibitem{Meloni:2017cig}
D.~Meloni, Front.in Phys. {\bfseries 5}, 43 (2017),
[\href{https://arxiv.org/abs/1709.02662}{{arXiv:1709.02662~[hep-ph]}}].

\bibitem{Cirigliano:2017djv}
V.~Cirigliano, W.~Dekens, J.~de~Vries, M.~L. Graesser, and E.~Mereghetti, JHEP
  {\bfseries 12}, 082 (2017),
[\href{https://arxiv.org/abs/1708.09390}{{arXiv:1708.09390~[hep-ph]}}].

\bibitem{Cirigliano:2018yza}
V.~Cirigliano, W.~Dekens, J.~de~Vries, M.~L. Graesser, and E.~Mereghetti, JHEP
  {\bfseries 12}, 097 (2018),
[\href{https://arxiv.org/abs/1806.02780}{{arXiv:1806.02780~[hep-ph]}}].

\bibitem{Pas:1999fc}
H.~P\"{a}s, M.~Hirsch, H.~V. Klapdor-Kleingrothaus, and S.~G. Kovalenko, Phys.
  Lett. {\bfseries B453}, 194 (1999),
[,393(1999)].

\bibitem{Pas:2000vn}
H.~P\"{a}s, M.~Hirsch, H.~V. Klapdor-Kleingrothaus, and S.~G. Kovalenko, Phys.
  Lett. {\bfseries B498}, 35 (2001),
[\href{https://arxiv.org/abs/hep-ph/0008182}{{arXiv:hep-ph/0008182}}].

\bibitem{Graf:2018ozy}
L.~Graf, F.~F. Deppisch, F.~Iachello, and J.~Kotila, Phys. Rev. {\bfseries
  D98}, 095023 (2018),
[\href{https://arxiv.org/abs/1806.06058}{{arXiv:1806.06058~[hep-ph]}}].

\bibitem{Ghiglieri:2017gjz}
J.~Ghiglieri and M.~Laine, JHEP {\bfseries 05}, 132 (2017),
[\href{https://arxiv.org/abs/1703.06087}{{arXiv:1703.06087~[hep-ph]}}].

\bibitem{Hernandez:2016kel}
P.~Hern\'{a}ndez, M.~Kekic, J.~L\'{o}pez-Pav\'{o}n, J.~Racker, and J.~Salvado,
  JHEP {\bfseries 08}, 157 (2016),
[\href{https://arxiv.org/abs/1606.06719}{{arXiv:1606.06719~[hep-ph]}}].

\bibitem{Akhmedov:1998qx}
E.~K. Akhmedov, V.~A. Rubakov, and A.~{\relax Yu}. Smirnov, Phys. Rev. Lett.
  {\bfseries 81}, 1359 (1998),
[\href{https://arxiv.org/abs/hep-ph/9803255}{{arXiv:hep-ph/9803255}}].

\bibitem{Asaka:2005an}
T.~Asaka, S.~Blanchet, and M.~Shaposhnikov, Phys. Lett. {\bfseries B631}, 151
  (2005),
[\href{https://arxiv.org/abs/hep-ph/0503065}{{arXiv:hep-ph/0503065}}].

\bibitem{Asaka:2005pn}
T.~Asaka and M.~Shaposhnikov, Phys. Lett. {\bfseries B620}, 17 (2005),
[\href{https://arxiv.org/abs/hep-ph/0505013}{{arXiv:hep-ph/0505013}}].

\bibitem{Shaposhnikov:2006nn}
M.~Shaposhnikov, Nucl. Phys. {\bfseries B763}, 49 (2007),
[\href{https://arxiv.org/abs/hep-ph/0605047}{{arXiv:hep-ph/0605047}}].

\bibitem{Canetti:2012vf}
L.~Canetti, M.~Drewes, and M.~Shaposhnikov, Phys. Rev. Lett. {\bfseries 110},
  061801 (2013),
[\href{https://arxiv.org/abs/1204.3902}{{arXiv:1204.3902~[hep-ph]}}].

\bibitem{Boyarsky:2018tvu}
A.~Boyarsky, M.~Drewes, T.~Lasserre, S.~Mertens, and O.~Ruchayskiy, Prog. Part.
  Nucl. Phys. {\bfseries 104}, 1 (2019),
[\href{https://arxiv.org/abs/1807.07938}{{arXiv:1807.07938~[hep-ph]}}].

\bibitem{Adhikari:2016bei}
M.~Drewes {\em et~al.}, JCAP {\bfseries 1701}, 025 (2017),
[\href{https://arxiv.org/abs/1602.04816}{{arXiv:1602.04816~[hep-ph]}}].

\bibitem{Boser:2019rta}
S.~B{\"o}ser, C.~Buck, C.~Giunti, J.~Lesgourgues, L.~Ludhova, S.~Mertens,
  A.~Schukraft, and M.~Wurm,
[\href{https://arxiv.org/abs/1906.01739}{{arXiv:1906.01739~[hep-ex]}}].

\bibitem{Blennow:2010th}
M.~Blennow, E.~Fernandez-Martinez, J.~Lopez-Pavon, and J.~Menendez, JHEP
  {\bfseries 07}, 096 (2010),
[\href{https://arxiv.org/abs/1005.3240}{{arXiv:1005.3240~[hep-ph]}}].

\bibitem{Mitra:2011qr}
M.~Mitra, G.~Senjanovic, and F.~Vissani, Nucl. Phys. {\bfseries B856}, 26
  (2012),
[\href{https://arxiv.org/abs/1108.0004}{{arXiv:1108.0004~[hep-ph]}}].

\bibitem{Li:2011ss}
Y.~F. Li and S.-s. Liu, Phys. Lett. {\bfseries B706}, 406 (2012),
[\href{https://arxiv.org/abs/1110.5795}{{arXiv:1110.5795~[hep-ph]}}].

\bibitem{deGouvea:2011zz}
A.~de~Gouvea and W.-C. Huang, Phys. Rev. {\bfseries D85}, 053006 (2012),
[\href{https://arxiv.org/abs/1110.6122}{{arXiv:1110.6122~[hep-ph]}}].

\bibitem{Barea:2015zfa}
J.~Barea, J.~Kotila, and F.~Iachello, Phys. Rev. {\bfseries D92}, 093001
  (2015),
[\href{https://arxiv.org/abs/1509.01925}{{arXiv:1509.01925~[hep-ph]}}].

\bibitem{Giunti:2015kza}
C.~Giunti and E.~M. Zavanin, JHEP {\bfseries 07}, 171 (2015),
[\href{https://arxiv.org/abs/1505.00978}{{arXiv:1505.00978~[hep-ph]}}].

\bibitem{Asaka:2011pb}
T.~Asaka, S.~Eijima, and H.~Ishida, JHEP {\bfseries 04}, 011 (2011),
[\href{https://arxiv.org/abs/1101.1382}{{arXiv:1101.1382~[hep-ph]}}].

\bibitem{Asaka:2013jfa}
T.~Asaka and S.~Eijima, PTEP {\bfseries 2013}, 113B02 (2013),
[\href{https://arxiv.org/abs/1308.3550}{{arXiv:1308.3550~[hep-ph]}}].

\bibitem{Asaka:2016zib}
T.~Asaka, S.~Eijima, and H.~Ishida, Phys. Lett. {\bfseries B762}, 371 (2016),
[\href{https://arxiv.org/abs/1606.06686}{{arXiv:1606.06686~[hep-ph]}}].

\bibitem{delAguila:2008ir}
F.~del Aguila, S.~Bar-Shalom, A.~Soni, and J.~Wudka, Phys. Lett. {\bfseries
  B670}, 399 (2009),
[\href{https://arxiv.org/abs/0806.0876}{{arXiv:0806.0876~[hep-ph]}}].

\bibitem{Cirigliano:2012ab}
V.~Cirigliano, M.~Gonz\'alez-Alonso, and M.~L. Graesser, JHEP {\bfseries 02},
  046 (2013),
[\href{https://arxiv.org/abs/1210.4553}{{arXiv:1210.4553~[hep-ph]}}].

\bibitem{Liao:2016qyd}
Y.~Liao and X.-D. Ma, Phys. Rev. {\bfseries D96}, 015012 (2017),
[\href{https://arxiv.org/abs/1612.04527}{{arXiv:1612.04527~[hep-ph]}}].

\bibitem{Pati:1974yy}
J.~C. Pati and A.~Salam, Phys. Rev. {\bfseries D10}, 275 (1974),
[Erratum: Phys. Rev.D11,703(1975)].

\bibitem{Mohapatra:1974hk}
R.~N. Mohapatra and J.~C. Pati,
Phys. Rev. D {\bfseries 11}, 566 (1975).

\bibitem{Senjanovic:1975rk}
G.~Senjanovi\'c and R.~N. Mohapatra,
Phys. Rev. D {\bfseries 12}, 1502 (1975).

\bibitem{Perez:2013osa}
P.~Fileviez~Perez and M.~B. Wise, Phys. Rev. {\bfseries D88}, 057703 (2013),
[\href{https://arxiv.org/abs/1307.6213}{{arXiv:1307.6213~[hep-ph]}}].

\bibitem{Dorsner:2016wpm}
I.~Dor{\v s}ner, S.~Fajfer, A.~Greljo, J.~F. Kamenik, and N.~Ko{\v s}nik, Phys.
  Rept. {\bfseries 641}, 1 (2016),
[\href{https://arxiv.org/abs/1603.04993}{{arXiv:1603.04993~[hep-ph]}}].

\bibitem{Grinstein:2006cg}
B.~Grinstein, V.~Cirigliano, G.~Isidori, and M.~B. Wise, Nucl. Phys. {\bfseries
  B763}, 35 (2007),
[\href{https://arxiv.org/abs/hep-ph/0608123}{{arXiv:hep-ph/0608123}}].

\bibitem{Alcaide:2019pnf}
J.~Alcaide, S.~Banerjee, M.~Chala, and A.~Titov, JHEP {\bfseries 08}, 031
  (2019),
[\href{https://arxiv.org/abs/1905.11375}{{arXiv:1905.11375~[hep-ph]}}].

\bibitem{Butterworth:2019iff}
J.~M. Butterworth, M.~Chala, C.~Englert, M.~Spannowsky, and A.~Titov, Phys.
  Rev. {\bfseries D100}, 115019 (2019),
[\href{https://arxiv.org/abs/1909.04665}{{arXiv:1909.04665~[hep-ph]}}].

\bibitem{Caputo:2017pit}
A.~Caputo, P.~Hern\'{a}ndez, J.~L\'{o}pez-Pav\'{o}n, and J.~Salvado, JHEP
  {\bfseries 06}, 112 (2017),
[\href{https://arxiv.org/abs/1704.08721}{{arXiv:1704.08721~[hep-ph]}}].

\bibitem{Grzadkowski:2010es}
B.~Grzadkowski, M.~Iskrzynski, M.~Misiak, and J.~Rosiek, JHEP {\bfseries 10},
  085 (2010),
[\href{https://arxiv.org/abs/1008.4884}{{arXiv:1008.4884~[hep-ph]}}].

\bibitem{Lehman:2014jma}
L.~Lehman, Phys. Rev. {\bfseries D90}, 125023 (2014),
[\href{https://arxiv.org/abs/1410.4193}{{arXiv:1410.4193~[hep-ph]}}].

\bibitem{Liao:2016hru}
Y.~Liao and X.-D. Ma, JHEP {\bfseries 11}, 043 (2016),
[\href{https://arxiv.org/abs/1607.07309}{{arXiv:1607.07309~[hep-ph]}}].

\bibitem{Jenkins:2013wua}
E.~E. Jenkins, A.~V. Manohar, and M.~Trott, JHEP {\bfseries 01}, 035 (2014),
[\href{https://arxiv.org/abs/1310.4838}{{arXiv:1310.4838~[hep-ph]}}].

\bibitem{Jenkins:2013zja}
E.~E. Jenkins, A.~V. Manohar, and M.~Trott, JHEP {\bfseries 10}, 087 (2013),
[\href{https://arxiv.org/abs/1308.2627}{{arXiv:1308.2627~[hep-ph]}}].

\bibitem{Alonso:2013hga}
R.~Alonso, E.~E. Jenkins, A.~V. Manohar, and M.~Trott, JHEP {\bfseries 04}, 159
  (2014),
[\href{https://arxiv.org/abs/1312.2014}{{arXiv:1312.2014~[hep-ph]}}].

\bibitem{Graesser:2016bpz}
M.~L. Graesser, JHEP {\bfseries 08}, 099 (2017),
[\href{https://arxiv.org/abs/1606.04549}{{arXiv:1606.04549~[hep-ph]}}].

\bibitem{Prezeau:2003xn}
G.~Pr\'ezeau, M.~Ramsey-Musolf, and P.~Vogel, Phys. Rev. {\bfseries D68},
  034016 (2003),
[\href{https://arxiv.org/abs/hep-ph/0303205}{{arXiv:hep-ph/0303205}}].

\bibitem{Schechter:1980gr}
J.~Schechter and J.~W.~F. Valle,
Phys. Rev. {\bfseries D22}, 2227 (1980).

\bibitem{Chala:2020vqp}
M.~Chala and A.~Titov,
[\href{https://arxiv.org/abs/2001.07732}{{arXiv:2001.07732~[hep-ph]}}].

\bibitem{Cirigliano:2013xha}
V.~Cirigliano, S.~Gardner, and B.~Holstein, Prog. Part. Nucl. Phys. {\bfseries
  71}, 93 (2013),
[\href{https://arxiv.org/abs/1303.6953}{{arXiv:1303.6953~[hep-ph]}}].

\bibitem{Buras:2000if}
A.~J. Buras, M.~Misiak, and J.~Urban, Nucl. Phys. {\bfseries B586}, 397 (2000),
[\href{https://arxiv.org/abs/hep-ph/0005183}{{arXiv:hep-ph/0005183}}].

\bibitem{Buras:2001ra}
A.~J. Buras, S.~J\"{a}ger, and J.~Urban, Nucl. Phys. {\bfseries B605}, 600
  (2001),
[\href{https://arxiv.org/abs/hep-ph/0102316}{{arXiv:hep-ph/0102316}}].

\bibitem{Weinberg:1978kz}
S.~Weinberg,
Physica {\bfseries A96}, 327 (1979).

\bibitem{Gasser:1983yg}
J.~Gasser and H.~Leutwyler,
Annals Phys. {\bfseries 158}, 142 (1984).

\bibitem{Jenkins:1990jv}
E.~E. Jenkins and A.~V. Manohar,
Phys. Lett. {\bfseries B255}, 558 (1991).

\bibitem{Bernard:1995dp}
V.~Bernard, N.~Kaiser, and U.-G. Mei{\ss}ner, Int. J. Mod. Phys. {\bfseries
  E4}, 193 (1995),
[\href{https://arxiv.org/abs/hep-ph/9501384}{{arXiv:hep-ph/9501384}}].

\bibitem{Manohar:1983md}
A.~Manohar and H.~Georgi,
Nucl. Phys. {\bfseries B234}, 189 (1984).

\bibitem{Cirigliano:2017tvr}
V.~Cirigliano, W.~Dekens, E.~Mereghetti, and A.~Walker-Loud, Phys. Rev.
  {\bfseries C97}, 065501 (2018),
  [\href{https://arxiv.org/abs/1710.01729}{{arXiv:1710.01729~[hep-ph]}}],
[Erratum: Phys. Rev.C100,no.1,019903(2019)].

\bibitem{Hammer:2019poc}
H.~W. Hammer, S.~K\"{o}nig, and U.~van Kolck,
[\href{https://arxiv.org/abs/1906.12122}{{arXiv:1906.12122~[nucl-th]}}].

\bibitem{Valderrama:2014vra}
M.~Pavon~Valderrama and D.~R. Phillips, Phys. Rev. Lett. {\bfseries 114},
  082502 (2015),
[\href{https://arxiv.org/abs/1407.0437}{{arXiv:1407.0437~[nucl-th]}}].

\bibitem{Cirigliano:2018hja}
V.~Cirigliano, W.~Dekens, J.~De~Vries, M.~L. Graesser, E.~Mereghetti,
  S.~Pastore, and U.~Van~Kolck, Phys. Rev. Lett. {\bfseries 120}, 202001
  (2018),
[\href{https://arxiv.org/abs/1802.10097}{{arXiv:1802.10097~[hep-ph]}}].

\bibitem{Cirigliano:2019vdj}
V.~Cirigliano, W.~Dekens, J.~De~Vries, M.~L. Graesser, E.~Mereghetti,
  S.~Pastore, M.~Piarulli, U.~Van~Kolck, and R.~B. Wiringa, Phys. Rev.
  {\bfseries C100}, 055504 (2019),
[\href{https://arxiv.org/abs/1907.11254}{{arXiv:1907.11254~[nucl-th]}}].

\bibitem{Gasser:1984gg}
J.~Gasser and H.~Leutwyler,
Nucl. Phys. {\bfseries B250}, 465 (1985).

\bibitem{Gonzalez-Alonso:2013ura}
M.~Gonz\'alez-Alonso and J.~Martin~Camalich, Phys. Rev. Lett. {\bfseries 112},
  042501 (2014),
[\href{https://arxiv.org/abs/1309.4434}{{arXiv:1309.4434~[hep-ph]}}].

\bibitem{Menendez:2008jp}
J.~Menendez, A.~Poves, E.~Caurier, and F.~Nowacki, Nucl. Phys. {\bfseries
  A818}, 139 (2009),
[\href{https://arxiv.org/abs/0801.3760}{{arXiv:0801.3760~[nucl-th]}}].

\bibitem{Nicholson:2018mwc}
A.~Nicholson {\em et~al.},
[\href{https://arxiv.org/abs/1805.02634}{{arXiv:1805.02634~[nucl-th]}}].

\bibitem{Bhattacharya:2016zcn}
T.~Bhattacharya, V.~Cirigliano, S.~Cohen, R.~Gupta, H.-W. Lin, and B.~Yoon,
  Phys. Rev. {\bfseries D94}, 054508 (2016),
[\href{https://arxiv.org/abs/1606.07049}{{arXiv:1606.07049~[hep-lat]}}].

\bibitem{Gupta:2018qil}
R.~Gupta, Y.-C. Jang, B.~Yoon, H.-W. Lin, V.~Cirigliano, and T.~Bhattacharya,
  Phys. Rev. {\bfseries D98}, 034503 (2018),
[\href{https://arxiv.org/abs/1806.09006}{{arXiv:1806.09006~[hep-lat]}}].

\bibitem{Aoki:2019cca}
S.~Aoki {\em et~al.} [Flavour Lattice Averaging Group Collaboration],
[\href{https://arxiv.org/abs/1902.08191}{{arXiv:1902.08191~[hep-lat]}}].

\bibitem{Monge-Camacho:2019nby}
H.~Monge-Camacho {\em et~al.}, PoS {\bfseries LATTICE2018}, 263 (2019),
[\href{https://arxiv.org/abs/1904.12055}{{arXiv:1904.12055~[hep-lat]}}].

\bibitem{Pastore:2017ofx}
S.~Pastore, J.~Carlson, V.~Cirigliano, W.~Dekens, E.~Mereghetti, and R.~B.
  Wiringa, Phys. Rev. {\bfseries C97}, 014606 (2018),
[\href{https://arxiv.org/abs/1710.05026}{{arXiv:1710.05026~[nucl-th]}}].

\bibitem{Hyvarinen:2015bda}
J.~Hyv{\"a}rinen and J.~Suhonen,
Phys. Rev. {\bfseries C91}, 024613 (2015).

\bibitem{Menendez:2017fdf}
J.~Men\'endez,
J. Phys. {\bfseries G45}, 014003 (2018).

\bibitem{Barea:2015kwa}
J.~Barea, J.~Kotila, and F.~Iachello, Phys. Rev. {\bfseries C91}, 034304
  (2015),
[\href{https://arxiv.org/abs/1506.08530}{{arXiv:1506.08530~[nucl-th]}}].

\bibitem{Barea}
J.~Barea, private communication .

\bibitem{Olive:2016xmw}
C.~Patrignani {\em et~al.} [Particle Data Group],
Chin. Phys. {\bfseries C40}, 100001 (2016).

\bibitem{Mertig:1990an}
R.~Mertig, M.~Bohm, and A.~Denner,
Comput. Phys. Commun. {\bfseries 64}, 345 (1991).

\bibitem{Shtabovenko:2016sxi}
V.~Shtabovenko, R.~Mertig, and F.~Orellana, Comput. Phys. Commun. {\bfseries
  207}, 432 (2016),
[\href{https://arxiv.org/abs/1601.01167}{{arXiv:1601.01167~[hep-ph]}}].

\bibitem{Horoi:2017gmj}
M.~Horoi and A.~Neacsu,
[\href{https://arxiv.org/abs/1706.05391}{{arXiv:1706.05391~[hep-ph]}}].

\bibitem{Stoica:2013lka}
S.~Stoica and M.~Mirea, Phys. Rev. {\bfseries C88}, 037303 (2013),
[\href{https://arxiv.org/abs/1307.0290}{{arXiv:1307.0290~[nucl-th]}}].

\bibitem{Doi:1985dx}
M.~Doi, T.~Kotani, and E.~Takasugi,
Prog. Theor. Phys. Suppl. {\bfseries 83}, 1 (1985).

\bibitem{Kotila:2012zza}
J.~Kotila and F.~Iachello, Phys. Rev. {\bfseries C85}, 034316 (2012),
[\href{https://arxiv.org/abs/1209.5722}{{arXiv:1209.5722~[nucl-th]}}].

\bibitem{Stefanik:2015twa}
D.~Stefanik, R.~Dvornicky, F.~Simkovic, and P.~Vogel, Phys. Rev. {\bfseries
  C92}, 055502 (2015),
[\href{https://arxiv.org/abs/1506.07145}{{arXiv:1506.07145~[hep-ph]}}].

\bibitem{Knecht:1998sp}
M.~Knecht, S.~Peris, and E.~de~Rafael, Phys. Lett. {\bfseries B443}, 255
  (1998),
[\href{https://arxiv.org/abs/hep-ph/9809594}{{arXiv:hep-ph/9809594}}].

\bibitem{Braaten:1991qm}
E.~Braaten, S.~Narison, and A.~Pich,
Nucl. Phys. {\bfseries B373}, 581 (1992).

\bibitem{Tanabashi:2018oca}
M.~Tanabashi {\em et~al.} [Particle Data Group],
Phys. Rev. {\bfseries D98}, 030001 (2018) and update (2019).

\bibitem{Cirigliano:2019jig}
V.~Cirigliano, Z.~Davoudi, T.~Bhattacharya, T.~Izubuchi, P.~E. Shanahan,
  S.~Syritsyn, and M.~L. Wagman [USQCD Collaboration], Eur. Phys. J. {\bfseries
  A55}, 197 (2019),
[\href{https://arxiv.org/abs/1904.09704}{{arXiv:1904.09704~[hep-lat]}}].

\bibitem{Drischler:2019xuo}
C.~Drischler, W.~Haxton, K.~McElvain, E.~Mereghetti, A.~Nicholson, P.~Vranas,
  and A.~Walker-Loud, ``{Towards grounding nuclear physics in QCD},''
\newblock 2019.
\newblock
\href{https://arxiv.org/abs/1910.07961}{{arXiv:1910.07961~[nucl-th]}}.
\newblock

\bibitem{Feng:2018pdq}
X.~Feng, L.-C. Jin, X.-Y. Tuo, and S.-C. Xia, Phys. Rev. Lett. {\bfseries 122},
  022001 (2019),
[\href{https://arxiv.org/abs/1809.10511}{{arXiv:1809.10511~[hep-lat]}}].

\bibitem{Tuo:2019bue}
X.-Y. Tuo, X.~Feng, and L.-C. Jin, Phys. Rev. {\bfseries D100}, 094511 (2019),
[\href{https://arxiv.org/abs/1909.13525}{{arXiv:1909.13525~[hep-lat]}}].

\bibitem{Detmold:2018zan}
W.~Detmold and D.~Murphy, PoS {\bfseries LATTICE2018}, 262 (2019),
[\href{https://arxiv.org/abs/1811.05554}{{arXiv:1811.05554~[hep-lat]}}].

\bibitem{Barry:2011wb}
J.~Barry, W.~Rodejohann, and H.~Zhang, JHEP {\bfseries 07}, 091 (2011),
[\href{https://arxiv.org/abs/1105.3911}{{arXiv:1105.3911~[hep-ph]}}].

\bibitem{Abazajian:2012ys}
K.~N. Abazajian {\em et~al.},
[\href{https://arxiv.org/abs/1204.5379}{{arXiv:1204.5379~[hep-ph]}}].

\bibitem{Helo:2018qej}
J.~C. Helo, M.~Hirsch, and Z.~S. Wang, JHEP {\bfseries 07}, 056 (2018),
[\href{https://arxiv.org/abs/1803.02212}{{arXiv:1803.02212~[hep-ph]}}].

\bibitem{Chrzaszcz:2019inj}
M.~Chrzaszcz, M.~Drewes, T.~E. Gonzalo, J.~Harz, S.~Krishnamurthy, and
  C.~Weniger,
[\href{https://arxiv.org/abs/1908.02302}{{arXiv:1908.02302~[hep-ph]}}].

\bibitem{Bolton:2019pcu}
P.~D. Bolton, F.~F. Deppisch, and P.~S.~B. Dev,
[\href{https://arxiv.org/abs/1912.03058}{{arXiv:1912.03058~[hep-ph]}}].

\bibitem{Bryman:2019bjg}
D.~A. Bryman and R.~Shrock, Phys. Rev. {\bfseries D100}, 073011 (2019),
[\href{https://arxiv.org/abs/1909.11198}{{arXiv:1909.11198~[hep-ph]}}].

\bibitem{Giunti:2019aiy}
C.~Giunti and T.~Lasserre, Ann. Rev. Nucl. Part. Sci. {\bfseries 69}, 163
  (2019),
[\href{https://arxiv.org/abs/1901.08330}{{arXiv:1901.08330~[hep-ph]}}].

\bibitem{Donini:2012tt}
A.~Donini, P.~Hernandez, J.~Lopez-Pavon, M.~Maltoni, and T.~Schwetz, JHEP
  {\bfseries 07}, 161 (2012),
[\href{https://arxiv.org/abs/1205.5230}{{arXiv:1205.5230~[hep-ph]}}].

\bibitem{Mahanta:1999xd}
U.~Mahanta, Phys. Rev. D {\bfseries 62}, 073009 (2000),
  [\href{https://arxiv.org/abs/hep-ph/9909518}{{arXiv:hep-ph/9909518}}].

\bibitem{AristizabalSierra:2007nf}
D.~Aristizabal~Sierra, M.~Hirsch, and S.~Kovalenko, Phys. Rev. D {\bfseries
  77}, 055011 (2008),
  [\href{https://arxiv.org/abs/0710.5699}{{arXiv:0710.5699~[hep-ph]}}].

\bibitem{Babu:2019mfe}
K.~Babu, P.~B. Dev, S.~Jana, and A.~Thapa, JHEP {\bfseries 03}, 006 (2020),
  [\href{https://arxiv.org/abs/1907.09498}{{arXiv:1907.09498~[hep-ph]}}].

\bibitem{Aguilar:2001ty}
A.~Aguilar-Arevalo {\em et~al.} [LSND Collaboration], Phys. Rev. {\bfseries
  D64}, 112007 (2001),
[\href{https://arxiv.org/abs/hep-ex/0104049}{{arXiv:hep-ex/0104049}}].

\bibitem{AguilarArevalo:2008rc}
A.~A. Aguilar-Arevalo {\em et~al.} [MiniBooNE Collaboration], Phys. Rev. Lett.
  {\bfseries 102}, 101802 (2009),
[\href{https://arxiv.org/abs/0812.2243}{{arXiv:0812.2243~[hep-ex]}}].

\bibitem{Aguilar-Arevalo:2013pmq}
A.~A. Aguilar-Arevalo {\em et~al.} [MiniBooNE Collaboration], Phys. Rev. Lett.
  {\bfseries 110}, 161801 (2013),
[\href{https://arxiv.org/abs/1303.2588}{{arXiv:1303.2588~[hep-ex]}}].

\bibitem{Aguilar-Arevalo:2018gpe}
A.~A. Aguilar-Arevalo {\em et~al.} [MiniBooNE Collaboration], Phys. Rev. Lett.
  {\bfseries 121}, 221801 (2018),
[\href{https://arxiv.org/abs/1805.12028}{{arXiv:1805.12028~[hep-ex]}}].

\bibitem{Dentler:2018sju}
M.~Dentler, {\'A}.~Hern{\'a}ndez-Cabezudo, J.~Kopp, P.~A.~N. Machado,
  M.~Maltoni, I.~Martinez-Soler, and T.~Schwetz, JHEP {\bfseries 08}, 010
  (2018),
[\href{https://arxiv.org/abs/1803.10661}{{arXiv:1803.10661~[hep-ph]}}].

\bibitem{Abazajian:2001nj}
K.~Abazajian, G.~M. Fuller, and M.~Patel, Phys. Rev. {\bfseries D64}, 023501
  (2001),
[\href{https://arxiv.org/abs/astro-ph/0101524}{{arXiv:astro-ph/0101524}}].

\bibitem{Kusenko:2009up}
A.~Kusenko, Phys. Rept. {\bfseries 481}, 1 (2009),
[\href{https://arxiv.org/abs/0906.2968}{{arXiv:0906.2968~[hep-ph]}}].

\bibitem{Sirunyan:2018btu}
A.~M. Sirunyan {\em et~al.} [CMS Collaboration], Phys. Rev. {\bfseries D99},
  052002 (2019),
[\href{https://arxiv.org/abs/1811.01197}{{arXiv:1811.01197~[hep-ex]}}].

\bibitem{Aaboud:2019jcc}
M.~Aaboud {\em et~al.} [ATLAS Collaboration], Eur. Phys. J. {\bfseries C79},
  733 (2019),
[\href{https://arxiv.org/abs/1902.00377}{{arXiv:1902.00377~[hep-ex]}}].

\bibitem{Aaboud:2018spl}
M.~Aaboud {\em et~al.} [ATLAS Collaboration], JHEP {\bfseries 01}, 016 (2019),
[\href{https://arxiv.org/abs/1809.11105}{{arXiv:1809.11105~[hep-ex]}}].

\end{thebibliography}\endgroup

\end{document}